\pdfoutput=1

\documentclass[12pt,hyperref,compress]{JHEP3}

\let\ifpdf\relax
\usepackage{ifpdf}
\usepackage{multirow}
\usepackage{color}
\let\normalcolor\relax
\usepackage{mathdots}
\usepackage{mathtools}
\usepackage{graphicx}
\usepackage{cite}

\renewcommand{\bar}{\overline}
\renewcommand{\hat}{\widehat}
\newcommand{\newcap}{{\small\mathrm{\raisebox{0.95pt}{{$\,\bigcap\,$}}}}}
\newcommand{\eq}[1]{\begin{equation}#1\end{equation}}
\newcommand{\eqs}[1]{\begin{equation}\begin{split}#1\end{split}\end{equation}}
\newcommand{\mi}{{\rm\rule[2.4pt]{6pt}{0.65pt}}}
\newcommand{\pl}{\hspace{0.5pt}\text{{\small+}}\hspace{-0.5pt}}
\newcommand{\ab}[1]{\langle #1 \rangle}
\renewcommand{\sb}[1]{[ #1 ]}
\definecolor{perm}{rgb}{0.1,0.45,0.85}
\definecolor{deemph}{rgb}{0.7,0.7,0.7}

\title{\hspace{-0.0cm}{\LARGE Scattering Amplitudes} {\Large and the}\\ {\LARGE Positive Grassmannian}}
\author{\vspace{-.5cm}N. Arkani-Hamed$^{a}$, J. Bourjaily$^{b}$, F. Cachazo$^{c}$, A. Goncharov$^{d}$, A. Postnikov$^{e}$, and J. Trnka$^{a,f}$\\
{\footnotesize
{\it $^{\!\!\phantom{f}a}$ School of Natural Sciences, Institute for Advanced Study, Princeton, NJ}\\
{\it $^{\!\!\phantom{f}b}$ Department of Physics, Harvard University, Cambridge, MA}\\
{\it $^{\!\!\phantom{f}c}$ Perimeter Institute for Theoretical Physics, Waterloo, Ontario, CA}\\
{\it $^{\!\!\phantom{f}d}$ Department of Mathematics, Yale University, New Haven CT}\\
{\it $^{\!\!\phantom{f}e}$  Department of Mathematics, Massachusetts Institute of Technology, Cambridge, MA}\\
{\it $^{\!\!\phantom{f}f}$ Department of Physics, Princeton University, Princeton,
NJ}}\vspace{-.5cm}} 
\abstract{We establish a direct connection between scattering amplitudes in planar four-dimensional theories and a remarkable mathematical structure known as the {\it positive} Grassmannian. The central physical idea is to focus on on-shell diagrams as objects of fundamental importance to scattering amplitudes. We show that the all-loop integrand in $\mathcal{N}\!=\!4$ super Yang-Mills (SYM) is naturally represented in this way. On-shell diagrams in this theory are intimately tied to a variety of mathematical objects, ranging from a new graphical representation of permutations to a beautiful stratification of the Grassmannian $G(k,n)$ which generalizes the notion of a simplex in projective space. All physically important operations involving on-shell diagrams map to canonical operations on permutations---in particular, BCFW deformations correspond to simple adjacent transpositions. Each cell of the positive Grassmannian is naturally endowed with ``positive'' coordinates $\alpha_i$ and an invariant measure of the form $\prod_id\!\log\alpha_i$ which determines the on-shell function associated with the diagram. This understanding allows us to classify and compute all on-shell diagrams, and give a geometric understanding for all the non-trivial relations among them. The Yangian invariance of scattering amplitudes is transparently represented by diffeomorphisms of $G(k,n)$ which preserve the positive structure. Scattering amplitudes in $(1\pl1)$-dimensional integrable systems and the ABJM theory in $(2\pl1)$ dimensions can both be understood as special cases of these ideas.  On-shell diagrams in theories with less (or no) supersymmetry are associated with exactly the same structures in the Grassmannian, but with a measure deformed by a factor encoding ultraviolet singularities. The Grassmannian representation of on-shell processes also gives a new understanding of the all-loop integrand for scattering amplitudes---presenting all integrands in a novel ``$d\!\log$'' form which is a direct reflection of the underlying positive structure.}
\preprint{PUPT-2435}

\begin{document}

\newpage
\section{Introduction}\label{introduction_section}

The traditional formulation of quantum field theory---encoded in its very name---is built on the two pillars of {\it locality}  and {\it unitarity} \cite{Weinberg:1995mt}. The standard apparatus of Lagrangians and path integrals allows us to make these two fundamental principles manifest. This approach, however, requires the introduction of a large amount of unphysical redundancy in our description of physical processes. Even for the simplest case of scalar field theories, there is the freedom to perform field-redefinitions. Starting with massless particles of spin-one or higher, we are forced to introduce even larger, gauge redundancies, \cite{Weinberg:1995mt}.\index{Quantum Field Theory}

Over the past few decades, there has been a  growing realization that these redundancies hide amazing physical and mathematical structures lurking within the heart of quantum field theory. This has been seen dramatically at strong coupling in gauge/gauge (see, e.g., \cite{Seiberg:1994rs,Seiberg:1994pq,Kapustin:2006pk}) and gauge/gravity dualities, \cite{Maldacena:1997re}. The past decade has uncovered further remarkable new structures in field theory even at weak coupling, seen in the properties of scattering amplitudes in gauge theories and gravity (for reviews, see \cite{Mangano:1990by,Dixon:1996wi,Cachazo:2005ga,Bern:2007dw,Feng:2011np,GreyBook}). The study of scattering amplitudes is fundamental to our understanding of field theory, and  fueled its early development in the hands of Feynman, Dyson and Schwinger among others. It is therefore surprising to see that even here, by committing so strongly to particular, gauge-redundant descriptions of the physics, the usual formalism is completely blind to astonishingly simple and beautiful properties of the gauge-invariant physical observables of the theory.

Many of the recent developments have been driven by an intensive exploration of $\mathcal{N} \!=\!4$ supersymmetric Yang-Mills (SYM) in the planar limit, \cite{Beisert:2010jr,GreyBook}. The all-loop integrand for scattering amplitudes in this theory can be determined by a generalization of the BCFW recursion relations, \cite{ArkaniHamed:2010kv}, in a way that is closely tied to remarkable new structures in algebraic geometry, associated with contour integrals over the Grassmannian $G(k,n)$, \cite{ArkaniHamed:2009dn, Mason:2009qx, ArkaniHamed:2009vw,Bourjaily:2010kw}. This makes both the {\it conformal} and long-hidden {\it dual conformal} invariance of the theory (which together close into the infinite-dimensional Yangian symmetry) completely manifest, \cite{Drummond:2010km}. It is remarkable that a single function of external kinematical variables can be interpreted as a scattering amplitude in one space-time, and as a Wilson-loop in another (for a review, see \cite{GreyBook}). Each of these descriptions makes a commitment to locality in its own space-time, making it impossible to see the dual picture. By contrast, the Grassmannian picture makes no mention of locality or unitarity, and does not commit to any gauge-redundant description of the physics, allowing it to manifest {\it all} the symmetries of the theory.

There has also been extraordinary progress in determining  the amplitude itself beyond the integrand, using the technology of symbols of transcendental functions to powerfully constrain and control the polylogarithms occurring in the final results, \cite{Goncharov:2009kx,Goncharov:2010jf}. While a global picture is still missing, a huge amount of data has been generated. The symbol for all $2$-loop MHV amplitudes has been determined, \cite{CaronHuot:2011ky} (see also \cite{Alday:2010jz}), and a handful of $2$-loop NMHV and $3$-loop MHV symbols have been found, \cite{Dixon:2011pw,Heslop:2011hv,Dixon:2011nj}. Remarkable strategies have also been presented to bootstrap amplitudes to very high loop-orders, \cite{Bourjaily:2011hi,Eden:2012tu,CaronHuot:2011kk,Sever:2011da,Sever:2012qp}. Many of these ideas have a strong resonance with the explosion of progress in the last decade using integrability to find exact results in planar $\mathcal{N} \!=\!4$ SYM, starting with the spectacular solution of the spectral problem for anomalous dimensions, \cite{Gromov:2009tv, Beisert:2010jr}.

All of these developments have made it completely clear that there are powerful new mathematical structures underlying the extraordinary properties of scattering amplitudes in gauge theories. If history is any guide, formulating and understanding the physics in a way that makes the symmetries manifest should play a central role in the story. The Grassmannian picture does this, but up to this point there has been little understanding for why this formulation exists, exactly how it works,  or where it comes from physically. Our primary goal in this note is to resolve this unsatisfactory state of affairs.

We will derive the connection between scattering amplitudes and the Grassmannian, starting physically from first principles. This will lead us into direct contact with several beautiful and active areas of current research in mathematics \cite{Lusztig::1998,L2, FZ, FG1, FG2,FG3, P, KLS,GK,T}. The past few decades have seen vigorous interactions between physics and mathematics in a wide variety of areas, but what is going on here involves {\it new} areas of mathematics that have only very recently played any role in physics, involving  simple but deep ideas ranging from combinatorics to algebraic geometry. It is both startling and exciting that such elementary mathematical notions are found at the heart of the physics of scattering amplitudes. 

This new way of thinking about scattering amplitudes involves many novel physical and mathematical ideas. Our presentation will be systematic, and we have endeavored to make it self contained and completely accessible to physicists. While we will discuss a number of mathematical results---some of them new---we will usually be content with the physicist's level of rigor. While the essential ideas here are all very simple, they are tightly interlocking, and range over a wide variety of areas---most of which are unfamiliar to most physicists.  Thus, before jumping into the detailed exposition, as a guide to the reader we end this introductory section by giving a roadmap of the logical structure and content of the paper. 

In \mbox{section \ref{on_shell_diagrams_section}}, we introduce the central physical idea motivating our work, which is to focus on {\it on-shell diagrams}, obtained by gluing together fundamental $3$-particle amplitudes and integrating over the on-shell phase space of internal particles. These objects are of central importance to the understanding scattering amplitudes. We will see that scattering amplitudes in planar $\mathcal{N}\!=\!4$ SYM---to all loop orders---can be represented {\it directly} in terms of on-shell processes. In this picture, ``virtual particles'' make no appearance at all. We should emphasize that we are not merely using on-shell information to determine scattering amplitudes, but rather seeing that the amplitudes can be directly computed in terms of fully on-shell processes. The off-shell, virtual particles familiar from Feynman diagrams are replaced by internal, {\it on-shell} particles (with generally complex momenta).

In our study of on-shell diagrams, we will see that different diagrams related by certain elementary moves can be physically equivalent, leading to the natural question of how to invariantly characterize their physical content. Remarkably, the invariant content of on-shell diagrams turns out to be characterized by {\it combinatorial} data. We discuss this in detail in \mbox{section \ref{combinatorics_of_scattering_amplitudes_section}} where we show how a long-known and beautiful connection between permutations and scattering amplitudes in integrable $(1\pl1)$-dimensional theories generalizes to more realistic theories in $(3\pl1)$ dimensions.

In \mbox{section \ref{intro_to_grassmannian_section}} we turn to actually calculating on-shell diagrams and find that the most natural way of carrying out the computations is to associate each diagram with a certain differential form on an {\it auxiliary} Grassmannian. In \mbox{sections \ref{configuration_of_vectors_section} and \ref{boundary_configurations_section}} we show how the invariant, combinatorial content of an on-shell diagram is reflected in the Grassmannian directly. This is described in terms of a surprisingly simple stratification of the configurations of $k$-dimensional vectors endowed with a cyclic ordering, classified by the linear dependencies among consecutive chains of vectors. For the real Grassmannian, this stratification can be equivalently described in an amazingly simple and beautiful way as nested `boundaries' of the {\it positive part} of the Grassmannian, \cite{Lusztig::1998}, which is motivated by the theory of totally positive matrices, \cite{GKr,Sch,L2}. Each on-shell diagram can then be associated with a particular configuration or ``stratum'' among the boundaries of the positive Grassmannian.

In \mbox{section \ref{invariant_top_form_section}} we make contact with the Grassmannian contour integral of reference \cite{ArkaniHamed:2009dn}, which is now seen as a compact way of representing the natural, invariant top-form on the
positive Grassmannian. This form of the measure allows us to easily identify the conformal and dual conformal symmetries of the theory which are related by a simple mapping of permutations described in \mbox{section \ref{superconformal_and_dual_superconformal_invariance}}. In \mbox{section \ref{positive_diffeos_and_the_yangian_section}}, we show that the invariance of scattering amplitudes under the action of the level-one generators of the Yangian has a transparent  interpretation: these generators correspond to the leading, non-trivial diffeomorphisms that preserve all the cells of the positive Grassmannian.

In \mbox{section \ref{kinematical_support_section}} we begin a systematic classification of Yangian invariants and their relations by first describing a combinatorial test to determine whether an on-shell diagram has non-vanishing kinematical support (and if so, how many points of support exist). In \mbox{section \ref{geometric_origin_of_identities_section}} a geometric basis is given for all the myriad, highly non-trivial identities satisfied among Yangian-invariants. This completes the classification of {\it all} Yangian Invariants together with {\it all} their relations. In \mbox{section \ref{classification_section}}, we give a tour of this classification as it emerges through N$^4$MHV.

\newpage
In \mbox{section \ref{yang_baxter_and_abjm_section}} we show that the story for scattering amplitudes in integrable $(1\pl1)$-dimensional theories---in particular, the Yang-Baxter relation---can be understood as a special case of our general results regarding on-shell diagrams. We further show that scattering amplitudes for the ABJM theory in $(2\pl1)$ dimensions, \cite{Aharony:2008ug}, can also be computed in terms of a natural specialization of on-shell diagrams: those associated with the null orthogonal Grassmannian. And we initiate the study of on-shell diagrams in theories with less (or no) supersymmetry in \mbox{section \ref{less_supersymmetries_section}}.

The positive Grassmannian is naturally endowed with a rich mathematical structure known as a {\it cluster algebra}---the original theory of which was developed in \cite{FZ} and has since been generalized to the theory of {\it cluster varieties} in \cite{FG2,FG3}. Incredibly, this structure has made striking appearances in widely disparate parts of physics in the last decade---from conformal blocks for higher Toda theories \cite{FG1,FG4}, to wall-crossing phenomena \cite{KS,Gaiotto:2011tf}, to quiver gauge theories with $\mathcal{N}\!=\!1$ super-conformal symmetry \cite{Xie:2012dw,Franco:2012mm,Xie:2012mr,Xie:2012jd,Heckman:2012jh,Franco:2012wv}, to soliton solutions to the KP equation \cite{Kodama:2011ht,Kodama:2011iq,Kodama:2012zx}. We briefly review this story in \mbox{section \ref{cluster_coordinates_section}}, as well as summarize its various physical manifestations in hopes of stimulating a deeper understanding for these extremely surprising connections between physics and mathematics.

In \mbox{section \ref{on_shell_scattering_amplitudes_section}} we move beyond the discussion of individual on-shell diagrams and describe the particular combinations which represent scattering amplitudes. We present a self-contained direct proof---using on-shell diagrams alone---that the BCFW construction of the all-loop integrand generates an object with precisely those singularities dictated by quantum field theory. We then show that the Grassmannian representation of loop-integrands are always given in a remarkable, ``$d\!\log$'' form, which we illustrate using examples of simple, one- and two-loop amplitudes. We discuss the implications of this representation for the transcendental functions that arise after the loop integrands are integrated.

We conclude our story in \mbox{section \ref{outlook_section}} with a discussion of a number of the outstanding, open directions for further research.

\newpage
\section{On-Shell Diagrams}\label{on_shell_diagrams_section}

Theoretical explorations in field theory have been greatly advanced by focusing on interesting classes of observables---from local correlation functions and scattering amplitudes, to Wilson and 't Hooft loops, surface operators and line defects, to partition functions on various manifolds (see e.g.\ \cite{Pestun:2007rz,Gaiotto:2008ak}). The central physical idea of our work is to study {\it on-shell scattering processes} as a new set of objects of fundamental interest.

\subsection{Encoding External Kinematical Data}

We are interested in the scattering amplitude for $n$ massless particles with momenta $p_a$ and helicities $h_a$, for $a=1,\ldots, n$. Since the momenta are null, the $(2\!\times\!2)$-matrix,
\vspace{-.25cm}\eq{p_a^{\alpha\dot{\alpha}}\equiv p_a^\mu \sigma_\mu^{\alpha\dot{\alpha}} = \left(\begin{array}{cc} p_a^0 + p_a^3 & p_a^1 - i p_a^2 \\ p_a^1 + i p_a^2 & p_a^0 - p_a^3 \end{array} \right)\,,\vspace{-.1cm}}
has vanishing determinant; and so $p^{\alpha\dot{\alpha}}$ has (at most) rank 1. We can therefore write\\[-7pt]
\vspace{-.2cm}\eq{p_a^{\alpha\dot{\alpha}} = \lambda_a^\alpha \widetilde{\lambda}_a^{\dot{\alpha}},\vspace{-.15cm}}
where $\lambda, \widetilde \lambda$ are referred to as {\it spinor-helicity} variables \cite{vanderWaerden:1929,Berends:1981rb,DeCausmaecker:1981bg,Gunion:1985vca,Kleiss:1985yh}. If the momentum is real, we have $\widetilde \lambda_a = \pm\lambda_a^*$; but in general, we will allow the momenta to be complex and consider $\lambda, \widetilde \lambda$ as independent, complex variables.

The rescaling $\lambda_a\!\mapsto\! t_a \lambda_a, \, \widetilde \lambda_a \!\mapsto\! t_a^{-1} \widetilde \lambda_a$ leaves the momentum $p_a$ invariant and represents the action of the {\it little group} (for more details see e.g.\ \cite{Weinberg:1995mt, ArkaniHamed:2008gz}). All the information about the helicities $h_a$ of particles involved in a scattering amplitude $A_n$ is encoded by its weights under such rescaling:\\[-10pt]
\vspace{-.0cm}\eq{\label{weights}A_n(t_a \lambda_a, t_a^{-1} \widetilde \lambda_a; h_a) = t_a^{-2 h_a} A_n (\lambda_a,\widetilde \lambda_a; h_a).\vspace{-.0cm}}

Theories with maximal supersymmetry have the wonderful feature that particles of all helicities can be unified into a single super-multiplet, \cite{Nair:1988bq,Witten:2003nn,Bianchi:2008pu,Drummond:2008bq,ArkaniHamed:2008gz}. For $\mathcal{N}\!=\!4$ SYM, we can group all the helicity states into a single {\it Grassmann coherent state} labeled by Grassmann (anti-commuting) parameters $\widetilde{\eta}^I$ for $I= 1,\ldots, 4$:
\vspace{-.2cm}\eq{\left|\widetilde \eta \right> \equiv \left|+1 \right> + \widetilde \eta^I  \left|+\text{$\textstyle\frac{1}{2}$}\right>_I + \frac{1}{2!} \widetilde \eta^I \widetilde \eta^J \left|0 \right>_{IJ} + \frac{1}{3!} \epsilon_{IJKL} \widetilde \eta^I \widetilde \eta^J \widetilde \eta^K \left|-\text{$\textstyle\frac{1}{2}$} \right>^L + \frac{1}{4!} \epsilon_{IJKL} \widetilde \eta^I \widetilde \eta^J \widetilde \eta^K \widetilde \eta^L \left| -1 \right>. \nonumber\vspace{-.2cm}}
The complete amplitude, denoted $\mathcal{A}^{}_n(\lambda_a, \widetilde \lambda_a, \widetilde \eta_a)$, is then a polynomial in the $\widetilde{\eta}$'s. It is convenient to expand this according to,
\vspace{-.0cm}\eq{\mathcal{A}_n(\lambda_a, \widetilde \lambda_a, \widetilde \eta_a) = \sum_k \mathcal{A}^{(k)}_{n}(\lambda_a, \widetilde \lambda_a, \widetilde \eta_a)\,,\vspace{-.1cm}}
where $\mathcal{A}^{(k)}_{n}$ is a polynomial of degree $4k$ in the $\widetilde \eta$'s. Under the little group, $\widetilde{\eta}$ transforms like $\widetilde{\lambda}$, so $\widetilde{\eta}_a\!\mapsto\!t_a^{-1}\widetilde\eta_a$; with this, the ``super-amplitude'' $\mathcal{A}_n^{(k)}$ transforms uniformly according to:
\vspace{-.0cm}\eq{\mathcal{A}_n^{(k)}(t_a\lambda_a,t_a^{-1}\widetilde{\lambda}_a,t_a^{-1}\widetilde{\eta}_a) =t_a^{-2}\mathcal{A}_n^{(k)}(\lambda_a,\widetilde{\lambda}_a,\widetilde{\eta}_a).\vspace{-.0cm}}
The $\mathcal{A}^{(k)}_n$ super-amplitude contains among its components those amplitudes which involve $k$ `negative helicity' ($h_a\!=\!\mi1$) and $(n\,\mi\,k)$ `positive-helicity' ($h_a\!=\!\pl1$) gluons---particles for which $h_a\!=\!\pm1$. $\mathcal{A}^{(k)}_n$ is often referred to as an ``N$^{(k-2)}$MHV amplitude'', where `MHV' stands for `maximal helicity violating' and `N' denotes `next-to'---$\mathcal{A}_n^{(k=2)}$ are considered `MHV' because $\mathcal{A}_n^{(k<2)}$ have vanishing kinematical support. 
\subsection{On-Shell Building Blocks: the Three-Particle Amplitudes}

The fundamental building blocks for all on-shell scattering processes are the three-particle amplitudes, which are completely determined (up to an overall coupling constant) by Poincar\'{e} invariance and little group rescaling. This is a consequence of the unique simplicity of three-particle kinematics. It is very easy to show that momentum conservation can only be satisfied if either: (A) all the $\lambda$'s are proportional to each other, or (B) all the $\widetilde \lambda$'s are proportional:
\vspace{-.4cm}\eq{\lambda_1 \widetilde \lambda_1 + \lambda_2 \widetilde \lambda_2 + \lambda_3 \widetilde \lambda_3 = 0\quad\raisebox{-2.2pt}{\text{{\LARGE$\Leftrightarrow$}}}\quad{\rm }\left\{\begin{array}{@{}l@{$\quad$}c}(A):& \lambda_1 \propto \lambda_2 \propto \lambda_3\\ (B):&\widetilde \lambda_1 \propto \widetilde \lambda_2 \propto \widetilde \lambda_3\end{array}\right\}\;.\vspace{-.3cm}}
Because of this, in the kinematic configuration where all the $\lambda$'s are proportional, the amplitude can {\it only} depend non-trivially on the $\widetilde \lambda$'s, and vice-versa. The dependence on $\lambda\,(\widetilde \lambda)$ is fully determined by the weights according to equation (\ref{weights}), together with the requirement that the amplitude is non-singular in the limit where the momenta are taken real (see equation (\ref{three_point_functions_for_general_helicities})).

We will denote the three-particle amplitude associated with the configuration where all the $\lambda$'s ($\widetilde \lambda$'s) are parallel with a white (black) three-point vertex. In a non-supersymmetric theory, i.e.\ with only gluons, these are associated with helicity configurations involving one (two) negative-helicity gluons:
\vspace{-0.3cm}\eq{\raisebox{-42pt}{\includegraphics[scale=1]{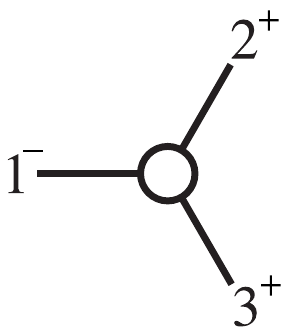}}\qquad\mathrm{and}\qquad\raisebox{-42pt}{\includegraphics[scale=1]{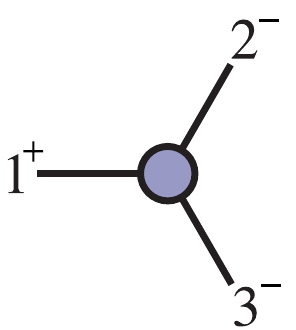}}\vspace{-0.2cm}\label{3pt_helicity_amplitudes}}
The corresponding helicity amplitudes are given by, \vspace{-.15cm}\eq{\begin{array}{ll}A^{(1)}_3(-,+,+) &=\displaystyle\,\frac{\phantom{{}^3}\sb{2\,3}^3}{\sb{1\,2}\sb{3\,1}} \,\,\delta^{2\times2}\big(\lambda_1 \widetilde \lambda_1 + \lambda_2 \widetilde \lambda_2 + \lambda_3 \widetilde \lambda_3\big);  \\ A^{(2)}_3(+,-,-) &=\displaystyle \frac{\phantom{{}^3}\ab{2\,3}^3}{\ab{1\,2}\ab{3\,1}} \delta^{2\times2}\big(\lambda_1 \widetilde \lambda_1 + \lambda_2 \widetilde \lambda_2 + \lambda_3 \widetilde \lambda_3\big).\end{array}\vspace{-.15cm}\label{non_supersymmetric_three_point_amplitudes}}
Here, we have made use of the Lorentz-invariants constructed out of the spinors,
\vspace{-.2cm}\eq{\ab{a\,b}\equiv\det\{\lambda_a,\lambda_b\}\quad\mathrm{and}\quad\sb{a\,b}\equiv\det\{\widetilde{\lambda}_a,\widetilde{\lambda}_b\}.\vspace{-.2cm}}

These amplitudes are of course what we get from the two-derivative Yang-Mills Lagrangian. Amplitudes involving all-plus or all-minus helicities are also fixed by Poincar\'{e} invariance in the same way, but arise only in theories with higher-dimension operators like $F^3$ or $R^3$. In general, Poincar\'{e} invariance fixes the kinematical dependence of the three-particle amplitude involving massless particles with arbitrary helicities to be, \cite{Benincasa:2007xk}:
\vspace{-.2cm}\eq{A_3(h_1,h_2,h_3)\propto\left\{\begin{array}{@{}l@{}l@{}ll}\hspace{1.35pt}\sb{12}^{\,h_1+h_2-h_3}& \hspace{1.35pt}\sb{23}^{\,h_2+h_3-h_1}&\hspace{1.35pt}\sb{31}^{\,h_3+h_1-h_2}&\qquad\sum h_a>0;\\ \ab{12}^{h_3-h_1-h_2}&\ab{23}^{h_1-h_2-h_3}&\ab{31}^{h_2-h_3-h_1}&\qquad\sum h_a<0.\end{array}\right.\vspace{-.2cm}\label{three_point_functions_for_general_helicities}}

As mentioned above, in maximally supersymmetric theories all helicity states are unified in a  single super-multiplet, and so there is no need to distinguish among the particular helicities of particles involved; and so, we may consider the simpler, cyclically-invariant amplitudes:
\vspace{-0.4cm}\eq{\raisebox{-42pt}{\includegraphics[scale=1]{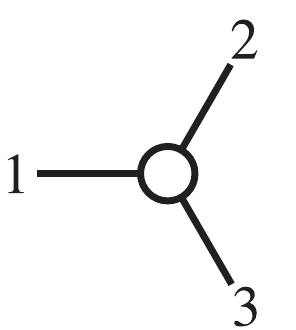}}\qquad\hspace{6pt}\mathrm{and}\qquad\raisebox{-42pt}{\includegraphics[scale=1]{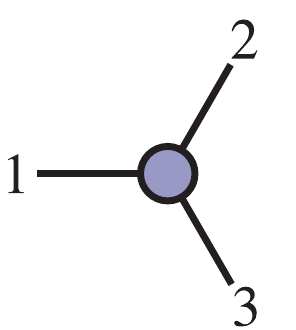}}\vspace{-0.3cm}\label{three_particle_vertices}}
The first includes among its components the $(-,+,+)$ amplitude of \mbox{(\ref{3pt_helicity_amplitudes})}, while the latter includes the $(+,-,-)$ amplitude. These super-amplitudes are given by,
\vspace{-.2cm}\eq{\begin{array}{ll}\mathcal{A}^{(1)}_3&=\displaystyle\,\frac{\delta^{1\times4}\big(\sb{2\,3}\widetilde{\eta}_1+\sb{3\,1}\widetilde{\eta}_2+\sb{1\,2}\widetilde{\eta}_3\big)}{\sb{1\,2}\sb{2\,3}\sb{3\,1}}\delta^{2\times2}\big(\lambda_1 \widetilde \lambda_1 + \lambda_2 \widetilde \lambda_2 + \lambda_3 \widetilde \lambda_3\big);  \\ \mathcal{A}^{(2)}_3&\displaystyle= \frac{\delta^{2\times4}\big(\lambda_1\widetilde{\eta}_1+\lambda_2\widetilde{\eta}_2+\lambda_3\widetilde{\eta}_3\big)}{\ab{1\,2}\ab{2\,3}\ab{3\,1}} \delta^{2\times2}\big(\lambda_1 \widetilde \lambda_1 + \lambda_2 \widetilde \lambda_2 + \lambda_3 \widetilde \lambda_3\big). \end{array}\vspace{-.1cm}}

(Although not essential for our present considerations, it may be of some interest that these objects can be made fully {\it permutation} invariant by including also a prefactor $f^{c_1,c_2,c_3}$ depending on the `{\it colors}' $c_a$ of the particles involved (where  `color' is simply a label denoting the possible distinguishable states in the theory). General considerations of quantum mechanics and locality (see e.g.\ \cite{Benincasa:2007xk}) require that any such prefactor must be fully antisymmetric and satisfy a Jacobi identity---implying that color labels combine to form the adjoint representation of a Lie algebra. The most physically interesting case is when this is the algebra of $U(N)$; in this case, $N$ can be viewed as a {\it parameter} of the theory, and scattering amplitudes can be expanded in powers of $1/N$ to all orders of perturbation theory, \cite{tHooft:1973jz}. In this paper, we will mostly concern ourselves with the leading-terms in $1/N$---the {\it planar} sector of the theory.)

\newpage
\subsection{Gluing Three-Particle Amplitudes Into On-Shell Diagrams}\label{gluing_intro_section}

It is remarkable that  three-particle amplitudes are totally fixed by Poincar\'{e} symmetry; they carry  all the essential information about the particle content and obvious symmetries of the physical theory. It is natural to ``glue'' these elementary building blocks together to generate more complicated objects we will call {\it on-shell diagrams}. Such objects will be our primary interest in this paper; examples of these include:
\vspace{-0.2cm}\eq{\raisebox{-80pt}{\includegraphics[scale=1]{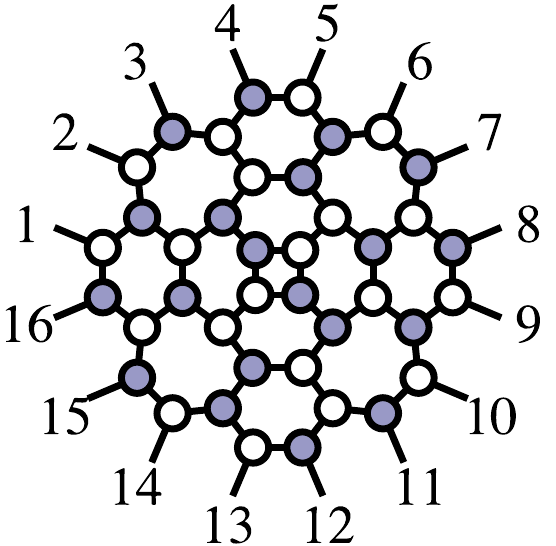}}\quad\mathrm{and}\quad\raisebox{-80pt}{\includegraphics[scale=1]{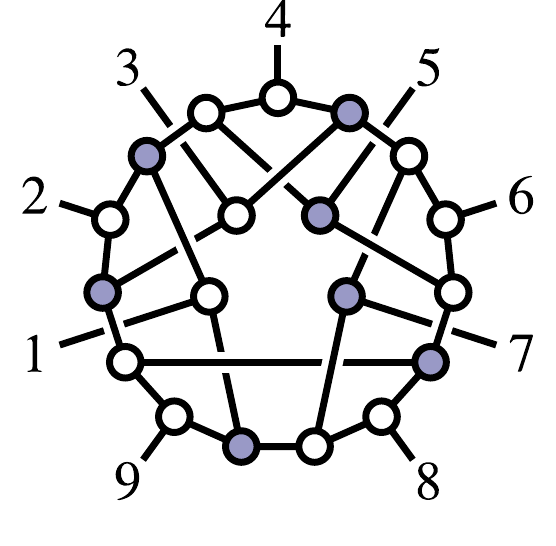}}\hspace{-1.2cm}\vspace{-0.2cm}\label{example_on_shell_graphs}}
We draw both planar and non-planar examples here to stress that on-shell diagrams have nothing  to do with planarity. In this paper, however, we will focus on the case of planar $\mathcal{N}\!=\!4$; we leave a systematic exploration of non-planar on-shell diagrams to future work.

Note that on-shell diagrams such as those of (\ref{example_on_shell_graphs})  are {\it not} Feynman diagrams! There are no ``virtual'' or ``off-shell'' internal particles involved: {\it all the lines in these pictures are on-shell} (meaning that their momenta are null). Each internal line represents a sum over all possible particles which can be exchanged in the theory, with (often complex) momenta constrained by momentum conservation at each vertex---integrating over the on-shell phase space of each. If $I$ denotes an internal particle with momentum $p_I=\lambda_I\widetilde{\lambda}_I$ and helicity $h_I$, then $p_I$ flows into one vertex with helicity $h_I$, and $(\mi\,p_I)$ flows into the other with helicity $(\mi\,h_I)$. In pure (non-supersymmetric) Yang-Mills we would have, \cite{Nair:1988bq},
\vspace{-.1cm}\eq{\sum_{h_I = \pm} \int\!\! \frac{d^2 \lambda_I d^2 \widetilde \lambda_I}{\mathrm{vol}(GL(1))}\,,\vspace{-.2cm}}
for each internal line; in a theory with maximal supersymmetry we would have,
\vspace{-.3cm}\eq{\int\!\!\!d^4\widetilde{\eta}\,\int\!\!\frac{d^2 \lambda_I d^2 \widetilde \lambda_I}{\mathrm{vol}(GL(1))}\,.\vspace{-.2cm}}
Here, the on-shell phase-space integral is clearly over $\lambda, \widetilde \lambda$, modulo the $GL(1)$-redundancy of the little group---rescaling $\lambda_I\!\mapsto\!t_I \lambda_I$ and $\widetilde \lambda_I\!\mapsto\!t_I^{-1} \widetilde \lambda_I$.

In general, we have some number of integration variables corresponding to the (on-shell) internal momenta, and $\delta$-functions enforcing momentum-conservation at each vertex. We may have just enough $\delta$-functions to {\it fully localize} all the internal momenta; in this case the on-shell diagram becomes an ordinary {\it function} of the external data, which has historically been called a ``leading singularity'' in the literature \cite{ELOP,GreyBook}. If there are more $\delta$-functions than necessary to fix the internal momenta, the left-over constraints will impose conditions on the external momenta; such an object is said to be a {\it singularity} or to have ``singular support''. If there are fewer $\delta$-functions than necessary to fix the internal momenta, there will be some degrees of freedom left over; the on-shell diagram then leaves us with some {\it differential form} on these extra degrees of freedom which we are free to integrate over any contour we please. But there is no fundamental distinction between these cases; and so we will generally think of an on-shell diagram as providing us with an ``on-shell form''---a differential form defined on the space of external {\it and} internal on-shell momenta. If we define the (super) phase space factor of the on-shell particle denoted $a$ by,
\vspace{-.2cm}\eq{\Omega_a= \frac{d^2 \lambda_a d^2 \widetilde \lambda_a}{\mathrm{vol}(GL(1))} d^4 \widetilde \eta_a\,,\vspace{-.2cm}}
then we can think of the $3$-particle amplitude involving particles $a,b,c$ also as a {\it form}:\\[-10pt]
\vspace{-.2cm}\eq{\mathcal{A}_3\,\Omega_a\,\Omega_b\,\Omega_c\,.\vspace{-.2cm}}

Putting all the $3$-particle amplitudes in an on-shell diagram together gives rise to a (typically high-dimensional) differential form on the space of external and internal momenta. The on-shell form associated with a diagram is then obtained by taking residues of this high-dimensional form on the support of all the $\delta$-function constraints (thought of {\it holomorphically}---as representing poles which enforce their arguments to vanish); this produces a lower-dimensional form defined on the support of any remaining $\delta$-functions.

Individual Feynman diagrams are {\it not} gauge invariant  and thus don't have any physical meaning. By contrast, each on-shell diagram {\it is} physically meaningful and corresponds to some particular on-shell scattering process. Note that although on-shell diagrams almost always involve `loops' of internal particles, these internal particles often have momenta {\it fixed} by the constraints (or are otherwise free). On-shell forms are simply the products of on-shell $3$-particle amplitudes; as such, they are always well-defined, finite objects---free from either infrared or ultraviolet divergences. This makes them ideal for exposing symmetries of a theory which are often obscured by such divergences.

\newpage
\subsection{The ``BCFW-Bridge''}\label{BCFW_bridge_section}

One particularly simple way of building-up more complicated on-shell diagrams from simpler ones will play an important role in our story. Starting from any on-shell diagram, we can pick two external lines, and attach a ``BCFW-bridge'' to make a new diagram as follows:
\vspace{-0.7cm}\eq{\mbox{\hspace{-1.1cm}\raisebox{-65pt}{\includegraphics[scale=1]{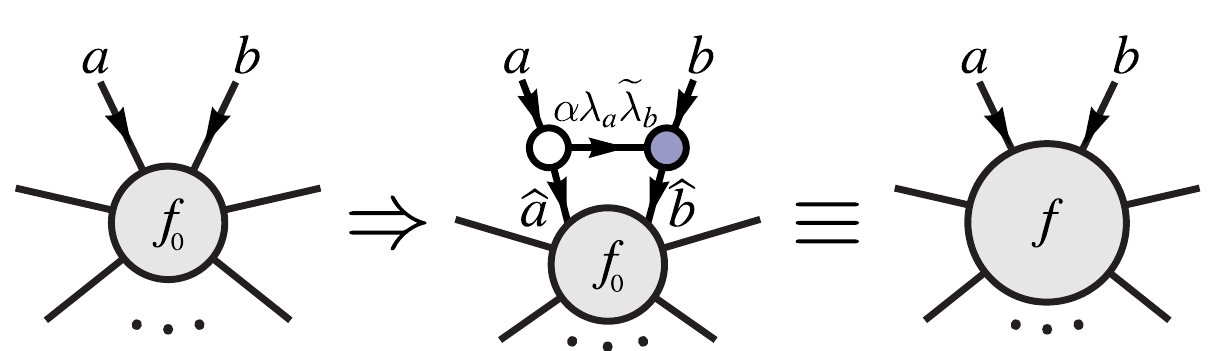}}\hspace{-1.1cm}}\nonumber\label{adding_a_bridge_with_lambdas}\vspace{-.4cm}}
Note that the momentum $\lambda_I \widetilde \lambda_I$ flowing through the ``{\it bridge}'', as indicated by the arrow, is very special: the white vertex on the left forces $\lambda_I \!\propto\! \lambda_a$, and the black vertex on the right forces $\widetilde \lambda_I \!\propto\! \widetilde \lambda_{b}$; thus, $\lambda_I \widetilde \lambda_I = \alpha \lambda_a \widetilde \lambda_{b}$ for some $\alpha$. The momenta entering the rest of the graph through legs $(a\,b)$ are deformed according to:
\vspace{-.3cm}\eq{\left\{\rule{0pt}{15pt}\right.\hspace{-3pt}\begin{array}{@{}ll@{}}\lambda_a\!\mapsto\!\lambda_{\hat{a}}&=\lambda_a\\\widetilde\lambda_a\!\mapsto\!\widetilde{\lambda}_{\hat{a}}&=\widetilde{\lambda}_a\,\mi\,\alpha\widetilde{\lambda}_b\end{array}\left.\hspace{-3pt}\rule{0pt}{15pt}\right\}\qquad\mathrm{and}\qquad\left\{\rule{0pt}{15pt}\right.\hspace{-3pt}\begin{array}{@{}ll@{}}\lambda_b\!\mapsto\!\lambda_{\hat{b}}&=\lambda_b\pl\,\alpha\lambda_a\\\widetilde\lambda_b\!\mapsto\!\widetilde{\lambda}_{\hat{b}}&=\widetilde{\lambda}_b\end{array}\left.\hspace{-3pt}\rule{0pt}{15pt}\right\}.\vspace{-.2cm}}
For theories with supersymmetry, there is also a deformation of $\widetilde \eta_a$ according to $\widetilde \eta_a \mapsto \widetilde \eta_a - \alpha\,\widetilde \eta_{b}$. (It is useful to remember that $\widetilde{\eta}$ {\it always} transforms as $\widetilde{\lambda}$ does.)

Thus, attaching a BCFW-bridge adds one new variable, $\alpha$,  to an on-shell form $f_0$, and gives rise to  a new on-shell form $f$ given by,
\vspace{-1cm}\eq{\hspace{-3cm}\begin{array}{ll}\\[5pt]\hspace{-0cm}f(\ldots;\lambda_a,\widetilde{\lambda}_a,\widetilde{\eta}_a;\lambda_{b},\widetilde{\lambda}_{b},\widetilde{\eta}_{b};\ldots)&=\displaystyle\frac{d\alpha}{\alpha}f_0(\ldots;\lambda_{\hat{a}},\widetilde{\lambda}_{\hat{a}},\widetilde{\eta}_{\hat{a}};\lambda_{\hat{b}},\widetilde{\lambda}_{\hat{b}},\widetilde{\eta}_{\hat{b}};\ldots);\\~\\[-12pt]
&=\displaystyle\frac{d\alpha}{\alpha}f_0(\ldots;\lambda_a,\widetilde{\lambda}_a\,\mi\,\alpha\,\widetilde{\lambda}_{b},\widetilde{\eta}_a\,\mi\,\alpha\,\widetilde{\eta}_{b};\lambda_{b}\pl\,\alpha\,\lambda_a,\widetilde{\lambda}_{b},\widetilde{\eta}_{b};\ldots).\end{array}\hspace{-3cm}}
Notice that {\it very} complex on-shell diagrams (both planar and non-planar alike) can be generated by successively attaching BCFW-bridges to a small set of `simple' diagrams. As we will soon understand, it turns out that {\it all} (physically-relevant) on-shell diagrams can be constructed in this way.

\subsection{On-Shell Recursion for All-Loop Amplitudes}\label{onshell_diagrams_for_amplitudes_section}

While on-shell diagrams are interesting in their own right,  for planar \mbox{$\mathcal{N}\!=\!4$} SYM, we will see that they are of much more than purely formal interest. Scattering amplitudes to all loop orders can be directly represented and computed as on-shell scattering processes. This is quite remarkable, considering the ubiquity of ``off-shell'' data in the more familiar Feynman expansion.

Of course by now we have become accustomed to the idea that amplitudes  can be `determined' using on-shell data---as evidenced, for instance, by the BCFW recursion relations at tree-, \cite{Britto:2004ap,Britto:2005fq}, and loop-levels, \cite{ArkaniHamed:2010kv} (see also \cite{Bern:2005hs,Dunbar:2010wu,Boels:2010nw,Alston:2012xd}). But our statement goes beyond this: the claim is not just that an off-shell object such as ``the loop {\it integral}'' can be {\it determined} using only on-shell information, but rather that loop {\it integrands} can be {\it directly represented} by fully on-shell objects.

Before discussing loops, let us look at some examples of ``tree-level'' amplitudes. Recall from \cite{Britto:2004nj} that the four-particle tree-amplitude $\mathcal{A}^{(2)}_4$ can be represented by a single on-shell diagram---its ``BCFW representation'':
\vspace{-0.1cm}\eq{\raisebox{-40.5pt}{\includegraphics[scale=1]{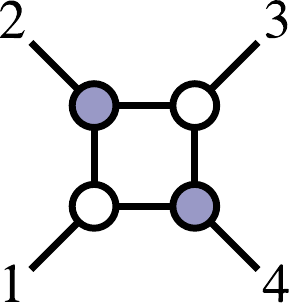}}\vspace{-0.1cm}\label{four_point_box}}
This is very far from what would be obtained using Feynman diagrams which would have represented (\ref{four_point_box}) as the sum of three terms,
\vspace{-0.0cm}\eq{~\hspace{0.7cm}\raisebox{-31.5pt}{\includegraphics[scale=1]{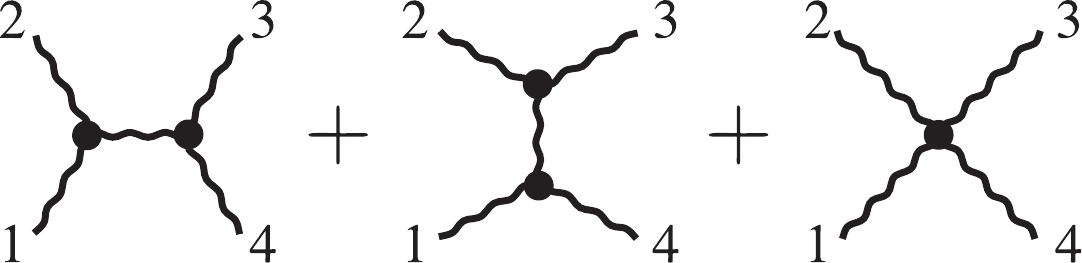}}\vspace{-0.0cm}}
the first two of which involve {\it off-shell} gluon exchange. (The terms ``tree-amplitude'' and ``loop-amplitude'' are artifacts of such Feynman-diagrammatic expansions.) Another  striking difference is that, despite the fact that we're discussing a {\it tree}-amplitude, the on-shell diagram (\ref{four_point_box}) looks like a loop!  To emphasize this distinction, consider a (possibly more familiar) ``tree-like'' on-shell diagram such as:
\vspace{-0.cm}\eq{\raisebox{-44.5pt}{\includegraphics[scale=1]{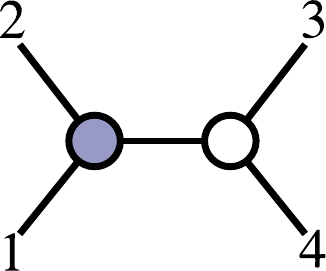}}\vspace{-0.cm}\label{four_point_factorization}}
Since the internal line must be on-shell, the diagram imposes a $\delta$-function constraint, $\delta((p_1\,\pl\,p_2)^2)$, on the external momenta; and so, (\ref{four_point_factorization}) corresponds to a singularity---a {\it factorization channel}. The extra leg in (\ref{four_point_box}) that makes the ``loop'' allows for a non-vanishing result for {\it generic} (on-shell, momentum-conserving) external momenta. It is interesting to note that we can interpret (\ref{four_point_box}) as having been obtained by attaching a ``BCFW-bridge'' to {\it any} of the factorization channels of the four-particle amplitude---such as that of (\ref{four_point_factorization}). This makes it possible for the single diagram (\ref{four_point_box}) to {\it simultaneously} exhibit {\it all} the physical factorization channels.

This simple example illustrates the fundamental physical idea behind the BCFW description of an amplitude---not just at tree-level, but at all loop orders: any amplitude can be {\it fully} reconstructed from the knowledge of its singularities; and the singularities of an amplitude are determined by entirely by on-shell data. At tree-level, the singularities are simply the familiar factorization channels,
\vspace{-0.0cm}\eq{\raisebox{-28.5pt}{\includegraphics[scale=1]{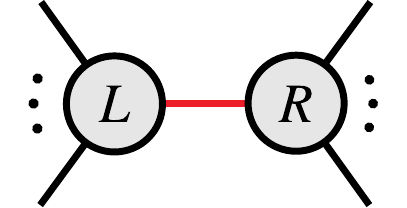}}\vspace{-0.0cm}\label{factorization_channel}}
where the left- and right-hand sides are both fully on-shell scattering amplitudes. At loop-level, all the singularities of the integrand can be understood as factorizations like that of (\ref{factorization_channel}), or those for which an {\it internal} particle is put on-shell; at least for $\mathcal{N}\!=\!4$ SYM in the planar limit, these singularities are given by the ``forward-limit'' \cite{CaronHuot:2010zt} of an on-shell amplitude with one fewer loop and two extra particles, where any two adjacent particles have equal and opposite momenta, denoted:
\vspace{-0.0cm}\eq{\raisebox{-37.5pt}{\includegraphics[scale=1]{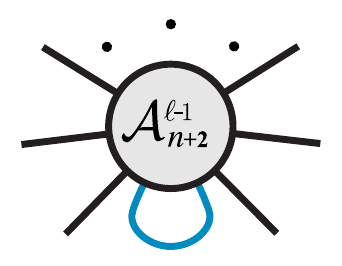}}\vspace{-0.0cm}\label{forward_limit}}

Combining these two terms, the singularities of the {\it full amplitude} are, \cite{ArkaniHamed:2010kv}:
\vspace{-0.0cm}\eq{\hspace{-4.5cm}\raisebox{-39.5pt}{\includegraphics[scale=.95]{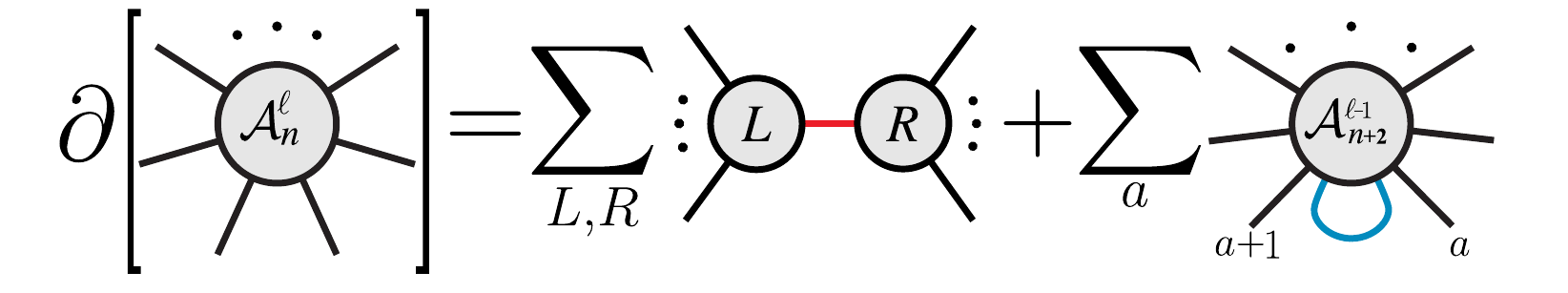}}\hspace{-3cm}\vspace{-0.0cm}\label{general_amplitude_boundary}}
Here we have suggestively used the symbol ``$\partial$'' to signify ``singularity of''. Of course, the symbol $\partial$ is often used to denote ``boundary'' or ``derivative''; we will soon see that all of these senses are appropriate.

Equation (\ref{general_amplitude_boundary}) can be understood as defining a ``differential equation'' for scattering amplitudes; and it turns out to be possible to `integrate' it directly. This is precisely what is accomplished by the BCFW recursion relations. For planar $\mathcal{N}\!=\!4$ SYM, the all-loop BCFW recursion relations, when represented in terms of on-shell diagrams are simply:
\vspace{-0.0cm}\eq{\hspace{-4.25cm}\raisebox{-47.5pt}{\includegraphics[scale=.9]{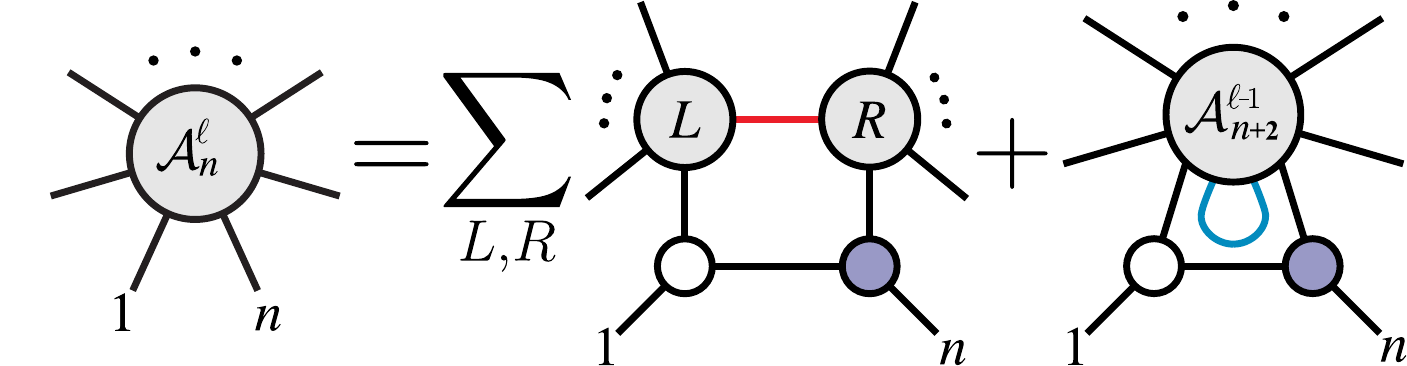}}\hspace{-3cm}\label{all_loop_recursion}\vspace{-0.05cm}}
The structure of this solution will be discussed in much greater detail in \mbox{section \ref{on_shell_scattering_amplitudes_section}}. For instance, notice that this presentation only makes {\it some} of the factorization channels and forward-limits manifest, and seems to break the cyclic symmetry of the amplitude by singling-out legs $(1\,n)$. In other words, working intrinsically with on-shell diagrams, it is not obvious that the sum (\ref{all_loop_recursion}) includes {\it all} the required singularities of an amplitude. Of course Feynman diagrams {\it do} make it manifest that such an object exists; but it would be nice to understand this more directly, without recourse to the usual formalism of field theory. We will show how this works in \mbox{section \ref{proof_of_the_recursion_relations}}, demonstrating that (\ref{all_loop_recursion}) has all the necessary singularities purely from within the framework of on-shell diagrams.

The seed of loop integrands in the recursion relation are the ``forward-limit'' terms as the three-point amplitudes are fixed by Poincar\'{e} invariance to all loop-orders. Each loop is accompanied by four integration variables: three of these are given by the phase space of the forward-limit momentum $\lambda_{AB}\widetilde{\lambda}_{AB}$ (from merging legs `$A$' and `$B$'), and the BCFW deformation parameter $\alpha$ is the fourth. Of course, all the objects appearing in these expressions are completely on-shell, and so do not seem to contain anything that looks like the conventional``$\int\!d^4\ell$'' with which we are accustomed (where $\ell$ is the momentum of a generally off-shell, {\it virtual} particle). However, it is easy to convert the parameters of the on-shell forward-limit to the more familiar one via the identification:
\vspace{-.3cm}\eq{\ell \equiv\lambda_{AB} \widetilde \lambda_{AB} + \alpha\,\lambda_1\widetilde \lambda_n\quad\mathrm{with}\quad d^4\ell =  \frac{d^2 \lambda_{AB} d^2 \widetilde \lambda_{AB}}{\mathrm{vol}(GL(1))}d \alpha  \, \ab{1\,\lambda_{AB}} \sb{n\,\widetilde \lambda_{AB}}\,.\label{loop_momenta_from_on_shell_data}\vspace{-.3cm}}
At $L$ loops, the all-loop recursion relation produces a $4L$-form on internal phase-space, and we can identify the $4L$ integration variables with loop momenta at each order via (\ref{loop_momenta_from_on_shell_data}). Integrating these on-shell forms over a contour which restricts each loop-momentum to be {\it real} (i.e.\ in $\mathbb{R}^{3,1}$) generates the final, physical amplitude.

Thus, as advertised, on-shell diagrams are of much more than mere academic interest: they {\it fully} determine the amplitude in planar $\mathcal{N}\!=\!4$ SYM to {\it all loop-orders}.

\subsection{Physical Equivalences Among On-Shell Diagrams}

We have seen that on-shell diagrams are objects of fundamental importance to the physics of scattering amplitudes. It is therefore natural to try and compute the forms associated with on-shell diagrams more explicitly, and better understand their structure. At first sight, the class of on-shell diagrams may look as complicated as Feynman diagrams. For instance, even for a fixed number of external particles, there are obviously an infinite number of such diagrams (by continuously adding BCFW bridges, for example). As we will see however, at least for $\mathcal{N}\!=\!4$ SYM in the planar limit, this complexity is entirely illusory. The reason is that apparently very different graphs actually give rise to exactly the same differential form---differing only by a change of variables.

The first instance of this phenomenon is extremely simple and trivial. Consider an analog of the ``factorization channel'' diagram (\ref{four_point_factorization}), but connecting two black vertices. Because these vertices require that all the $\widetilde{\lambda}$'s be parallel, it makes no physical difference how they are connected. And so, on-shell diagrams related by,
\vspace{-0.2cm}\eq{\mbox{\hspace{-2cm}\raisebox{-46.5pt}{\includegraphics[scale=1]{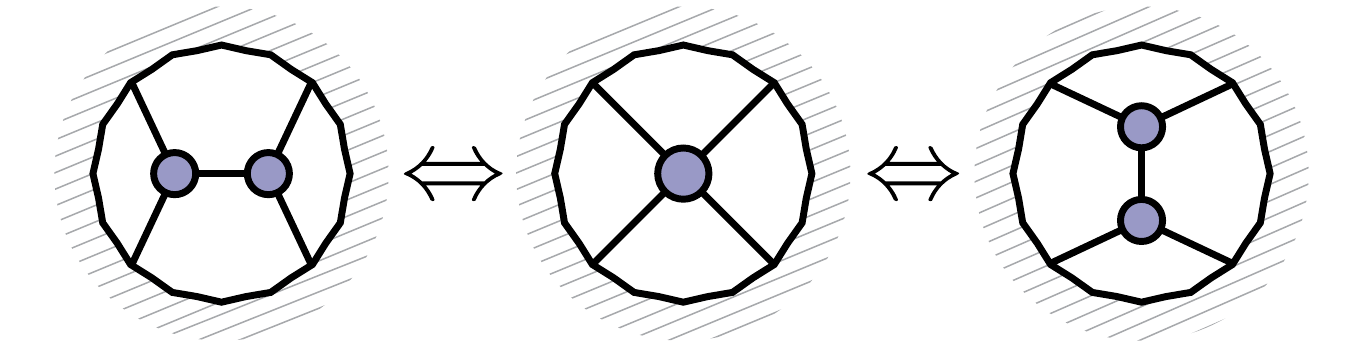}}\hspace{-2cm}}\label{plabic_graph_merger_rule}\vspace{-.2cm}}
represent the same on-shell form. Thus, we can collapse and re-expand any chain of connected black vertices in anyway we like; the same is obviously true for white vertices. Because of this, for some purposes it may be useful to define composite black and white vertices with any number of legs. By grouping black and white vertices together in this way, on-shell diagrams can always be made {\it bipartite}---with (internal) edges only connecting white with black vertices. We will, however, preferentially draw trivalent diagrams because of the fundamental role played by the three-particle amplitudes.

There is also a more interesting equivalence between on-shell diagrams that will play an important role in our story. We can see this already in the BCFW representation of the four-particle amplitude given above, (\ref{four_point_box}). The {\it picture} is obviously not cyclically invariant---as a rotation would exchange its black and white vertices. But the four-particle {\it amplitude} of course {\it is} cyclically invariant; and so there is another generator of equivalences among on-shell diagrams, the ``square move'', \cite{Hodges:2005aj}:
\vspace{-0.2cm}\eq{\mbox{\hspace{-1cm}\raisebox{-47.5pt}{\includegraphics[scale=1]{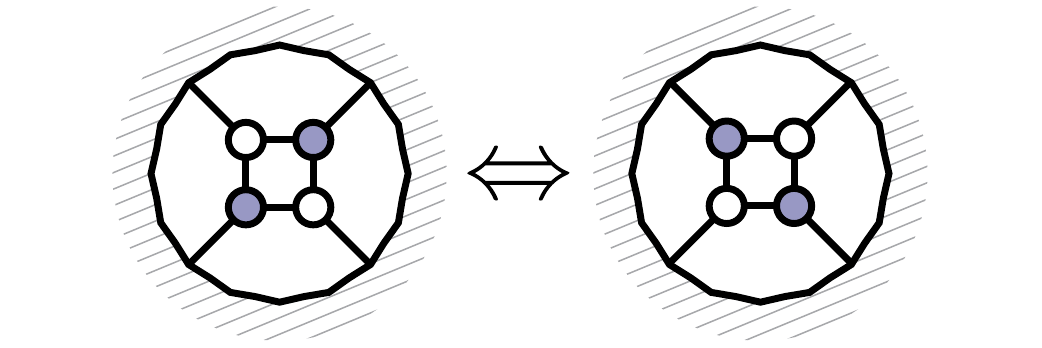}}\hspace{-1.0cm}}\label{square_move}\vspace{-.2cm}}

The merger and square moves can be used to show the physical equivalence of many seemingly different on-shell diagrams. For instance, the following two diagrams generate physically equivalent on-shell forms:
\vspace{-0.2cm}\eq{\hspace{-0.0cm}\raisebox{-44pt}{\includegraphics[scale=1]{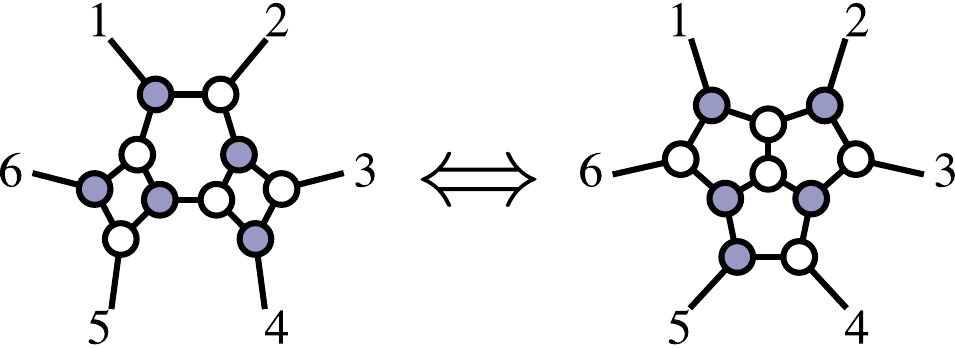}}\label{equivalent_graphs_in_g36}\vspace{-0.3cm}}
We can see this by explicitly constructing the chain of moves which brings one graph into the other:
\vspace{-0.0cm}\eq{\hspace{-0.0cm}\includegraphics[scale=.875]{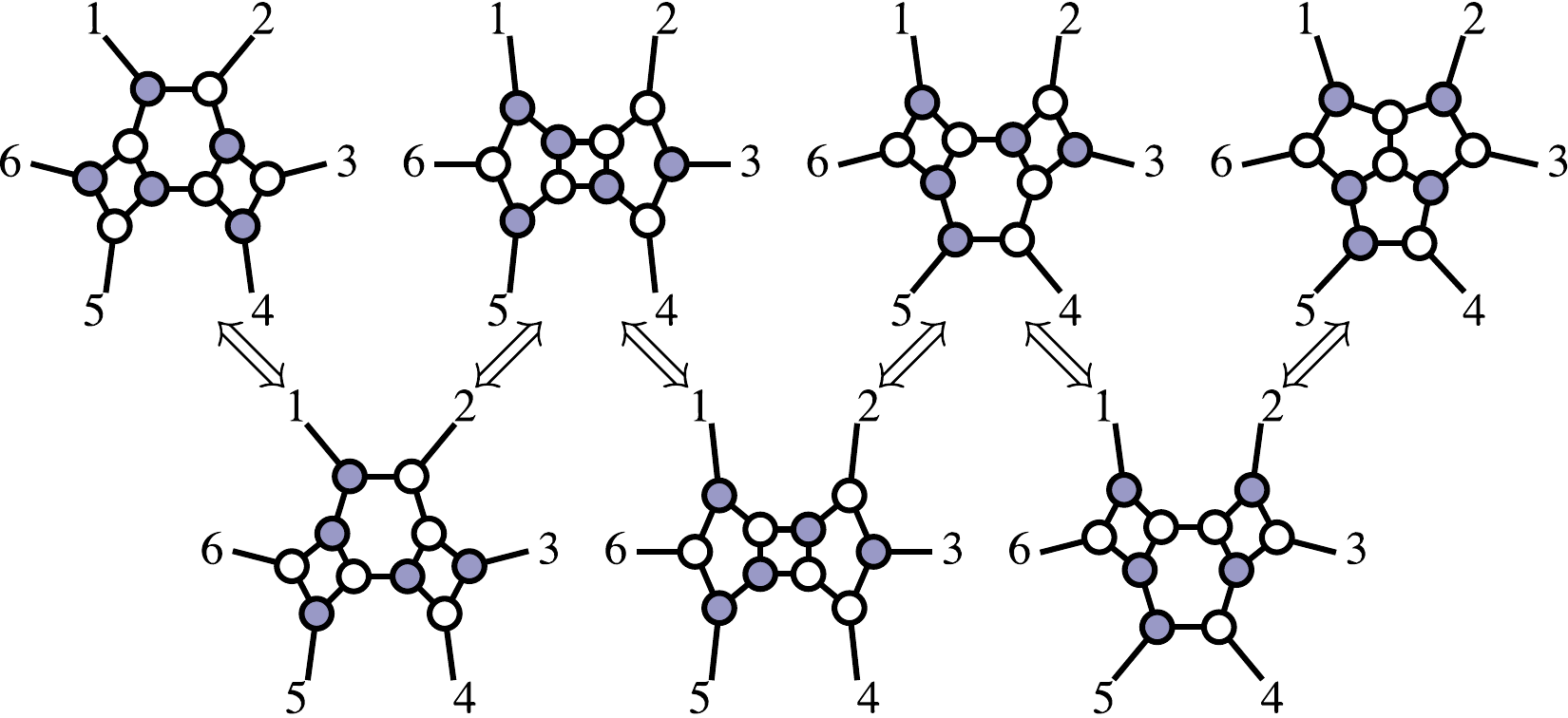}\nonumber\vspace{-0.2cm}}
Here, each step down involves one or more square-moves, and each step up involves one or more mergers.

To give another example, the on-shell diagram representing the one-loop four-particle amplitude---as obtained directly from BCFW recursion---is given by:
\vspace{-0.0cm}\eq{\raisebox{-55pt}{\includegraphics[scale=.85]{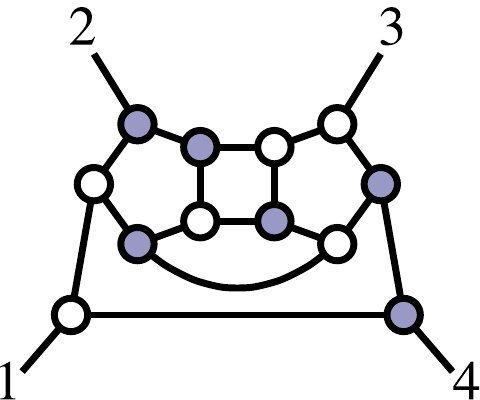}}\label{fl_of_six_point}\vspace{-.0cm}}
Using a series of mergers and square moves, it can be brought to the beautifully symmetric, bipartite form:
\vspace{-0.0cm}\eq{\raisebox{-40pt}{\includegraphics[scale=.9]{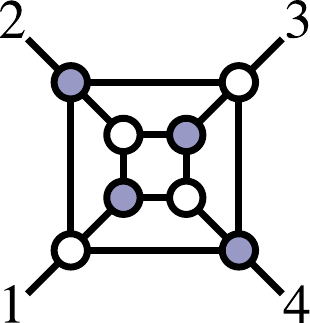}}\label{four_point_one_loop}\vspace{-.0cm}}
These forms are completely equivalent, but suggest very different physical interpretations. The first, (\ref{fl_of_six_point}), clearly exposes its origin as a forward-limit---arising through the gluing of two of the external particles of the six-particle tree-amplitude. The second form, (\ref{four_point_one_loop}), does not look like this at all; instead, it appears ast four BCFW-bridges attached to an internal square---which is of course the four-particle tree-amplitude. Thus, in this picture, we can think of the one-loop amplitude as an integral over a four-parameter deformation of the tree-amplitude!

This is more than mere amusement. It immediately tells us that with an appropriate choice of variables representing the BCFW-shifts, the one-loop amplitude can be represented in a remarkably simple form:
\vspace{-.0cm}\eq{\mathcal{A}^{\ell=1}_4\propto \mathcal{A}^{\ell=0}_4\times \int\!\!\frac{d\alpha_1}{\alpha_1} \frac{d\alpha_2}{\alpha_2} \frac{d\alpha_3}{\alpha_3} \frac{d\alpha_4}{\alpha_4}\,.\vspace{-.0cm}\label{nice_form_of_four_point_one_loop_integrand}}
Of course, this does not look anything like the more familiar expression, \cite{Green:1982sw},
\vspace{-.0cm}\eq{\mathcal{A}_4^{\ell=1}\propto\mathcal{A}_4^{\ell=0} \times\raisebox{-30pt}{\includegraphics[scale=1]{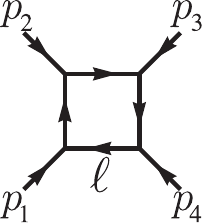}}=\mathcal{A}_4^{\ell=0} \times\int\!\!\frac{d^4\ell\;\; (p_1 + p_2)^2 (p_1 + p_3)^2}{\ell^2 (\ell+p_1)^2(\ell+p_1 + p_2)^2 (\ell - p_4)^2}\,.\label{olde_form_of_one_loop_four_point}\vspace{-.0cm}}
In this form, it is not at all obvious that there is any change of variables that reduces the integrand to the ``$d\!\log$''-form of (\ref{nice_form_of_four_point_one_loop_integrand}). However, following the rule for identifying off-shell loop momenta in terms of on-shell data, (\ref{loop_momenta_from_on_shell_data}), we may easily identify the map which takes us from the $\ell$ of (\ref{olde_form_of_one_loop_four_point}) to the $\alpha_i$ of (\ref{nice_form_of_four_point_one_loop_integrand}):
\begin{align}\nonumber\\[-20pt]\hspace{-0cm}&\hspace{-1cm}\frac{d^4\ell\;\; (p_1 + p_2)^2 (p_1 + p_3)^2}{\ell^2 (\ell+p_1)^2(\ell+p_1 + p_2)^2 (\ell - p_4)^2}\\\hspace{-0cm}=& d\!\log\!\left(\!\frac{\ell^2}{(\ell -\ell^*)^2}\!\right)d\!\log\!\left(\!\frac{(\ell + p_1)^2}{(\ell- \ell^*)^2}\!\right)d\!\log\!\left(\!\frac{(\ell + p_1 + p_2)^2}{(\ell -\ell^*)^2}\!\right)d\!\log\!\left(\!\frac{(\ell-p_4)^2}{(\ell-\ell^*)^2}\!\right)\,,\nonumber\\[-20pt]\nonumber\end{align}
where $\ell^*$ is either of the two points null separated from all four external momenta. This expression will be derived in detail in \mbox{section \ref{canonical_coordinates_for_loop_integrands}}.

As we will see, the existence of this ``$d\!\log$'' representation for loop integrands is a completely general feature of all amplitudes at all loop-orders. But the possibility of such a form even existing was never anticipated from the more traditional formulations of field theory. Indeed, even for the simple example of the four-particle one-loop amplitude, the existence of a change of variables converting $d^4\ell$ to four $d\!\log$'s went unnoticed for decades. We will see that these ``$d\!\log$''-forms follow directly from the on-shell diagram description of scattering amplitudes generated by the BCFW recursion relations, (\ref{all_loop_recursion}). Beyond their elegance, these $d\!\log$-forms suggest a completely new way of carrying out loop integrations, and more directly expose an underlying, ``motivic'' structure of the final results which will be a theme pursued in a later, more extensive work.

The equivalence of on-shell diagrams related by mergers and square-moves clearly represents a major simplification in the structure on-shell diagrams; but these alone cannot reduce the seemingly infinite complexities of graphs with arbitrary numbers of `loops' (faces) as neither of these operations affect the number of faces of a graph. However, using mergers and square-moves, it may be possible to represent an on-shell diagram in a way that exposes a ``bubble'' on an internal line. As one might expect, there is a sense in which such diagrams can be {\it reduced} by eliminating bubbles:
\vspace{-0.25cm}\eq{\hspace{-1.4cm}\raisebox{-47.5pt}{\includegraphics[scale=1]{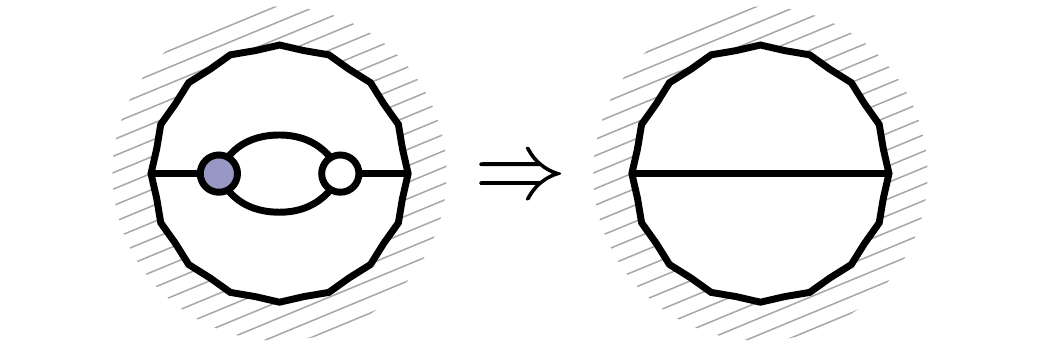}}\label{bubble_deletion}\hspace{-1.4cm}\vspace{-.25cm}}
Of course this can't literally be true: there is one more integration variable in the diagram with the bubble than the one without.  What ``reduction'' actually means is that there is a concrete {\it and simple} change of variables for which this extra degree of freedom, say $\alpha$, factors-out of the on-shell form cleanly as $d\!\log\alpha$---which, upon taking the residue on a contour around $\alpha=0$, yields the reduced diagram and the associated on-shell form.

Before completing our discussion, it is worth mentioning that there are other---somewhat trivial---operations on diagrams which leave the corresponding on-shell form invariant; these include, adding or deleting a bivalent vertex (of either color) along a line, or exchanging the colors involved in a bubble such as that in (\ref{bubble_deletion}).

It turns out that using mergers, square-moves {\it and} bubble-deletion, all planar on-shell diagrams involving $n$ external particles can be reduced to a {\it finite} number of diagrams. This shows that the essential content of on-shell diagrams are encapsulated by the finite list of {\it reduced} objects. And as we will see, the extra, ``irrelevant'' variables associated with bubble-deletion also have a purpose in life: they represent the loop integration variables.

{\it Reduced} diagrams are still not unique of course: they can still be transmuted into each other using mergers and square-moves. Given that the same on-shell form can be represented by many different on-shell diagrams, it is natural to ask for some  {\it invariant} way to characterize them. For instance, if we are given two complicated on-shell diagrams such as those of (\ref{equivalent_graphs_in_g36}), how can we decide whether they can be morphed into each other using the merge and square-moves? The answer  to this question ends up being simple and striking: the invariant data associated with reduced on-shell diagram is encoded by a {\it permutation} of the particle labels! We will describe this connection in detail in the next section.

It is amazing that a connection between scattering amplitudes in $(3\pl1)$ dimensions and combinatorics exists at all, let alone that it will play a central role in the story. This is the tip of an iceberg of remarkable connections between on-shell diagrams and rich mathematical structures only recently explored in the literature. We will spend much of the rest of this paper outlining these connections in greater detail. But we will start by recalling that this is not the first time scattering theory has been related to permutations in an important way: a classic example of such a connection is for integrable theories in $(1\pl1)$ dimensions. In addition to providing us with some historical context, revisiting this story will give us an interesting perspective on recent developments.

\newpage
\section{Permutations and Scattering Amplitudes}\label{combinatorics_of_scattering_amplitudes_section}
\subsection{Combinatorial Descriptions of Scattering Processes}\label{combinatorial_descriptions_of_scattering_processes_subsection}

To a physicist, scattering is perhaps the most fundamental physical process; but scattering amplitudes are rather sophisticated functions of the helicities and momenta of the external particles. If we strip-away all of this data, all that would be left would be the arbitrary labels identifying the particles involved, which we will denote simply by $(1,\ldots,n)$. The simplest kind of ``interaction'' that could be associated with just this data would be a {\it permutation}; because of the central role played by permutations in combinatorics, we might fancifully say that a permutation is the combinatorial analog of the physicists' $S$-matrix.

At first sight, it certainly seems as if a ``combinatorial $S$-matrix'' would be far too simple an object to capture anything remotely resembling the richness of physical scattering amplitudes. However, we will see that this is not the case: in a specific sense, our study of on-shell diagrams will be fully determined by a novel way of thinking about permutations.

Indeed something very much like this happens for integrable theories in $(1\pl1)$ dimensions, \cite{Yang:1967bm,Baxter:1972hz}. Consider for instance the permutation given by
\vspace{-.2cm}\eq{\Bigg(\begin{array}{@{}cccccc@{}}1&2&3&4&5&6\\[-5pt]\downarrow&\downarrow&\downarrow&\downarrow&\downarrow&\downarrow\\[-2.5pt]{\color{black}5}&{\color{black}3}&{\color{black}2}&{\color{black}6}&{\color{black}1}&{\color{black}4}\end{array}\Bigg)\,.\vspace{-.2cm}}
Its relationship to physics can be seen by representing it graphically as:
\vspace{-0.3cm}\eq{\hspace{-0cm}\raisebox{-35pt}{\includegraphics[scale=1]{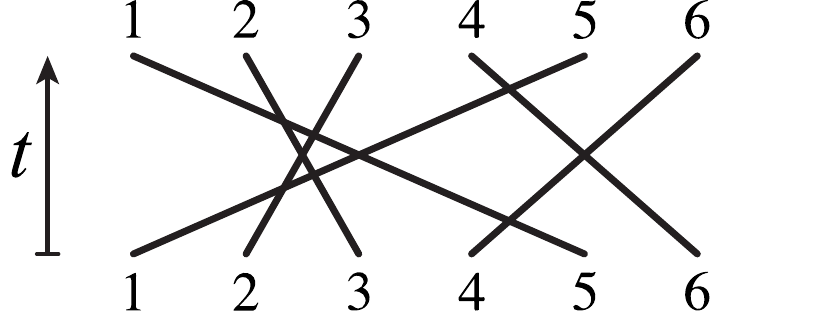}}\vspace{-0.2cm}}
This can be thought of as a space-time picture for a scattering process in $(1\pl1)$ dimensions, where time flows upwards. First, particles 4 and 5 scatter, then 1 and 2, then 2 and 3, and so on. The time-ordering of these scatterings corresponds to one way of representing the permutation as a product of adjacent transpositions. Of course, this decomposition is not unique: there are many ways of drawing the same picture with different time-orderings for the various $2\!\to\!2$ processes. In a general theory with only $4$-point interactions, the amplitude for different orderings would be different, and therefore the amplitude for the scattering process would not be completely determined by the permutation alone. For the amplitude to depend {\it only on the permutation} and {\it nothing else}, the $2\!\to\!2$ amplitudes must satisfy the famous Yang-Baxter relation, \cite{Yang:1967bm,Baxter:1972hz}:
\vspace{-0.5cm}\eq{\hspace{-0cm}\raisebox{-40pt}{\includegraphics[scale=1]{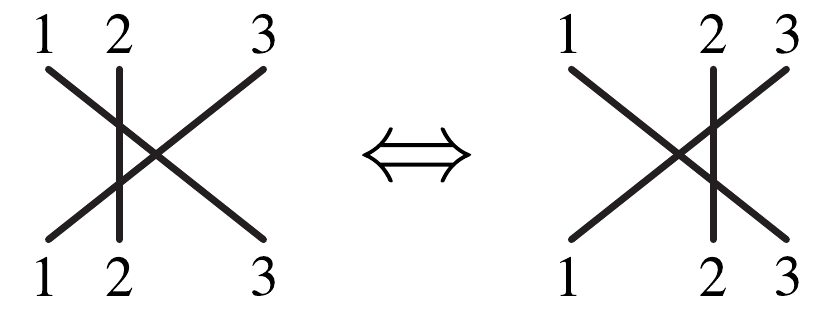}}\vspace{-0.2cm}\label{yang_baxter_move}}
\indent It is natural to ask whether such a picture can be generalized to more realistic theories in higher dimensions. This seems impossible at first sight, since the pictures drawn above only make physical sense in $(1\pl1)$ dimensions (not only because they are drawn on a plane). The fact that particles can only move in one spatial dimension is what makes it possible to describe all interactions as a sequence of local \mbox{$2\!\to\!2$} scattering processes. Also important is the absence of any particle creation or destruction, allowing us to label the final-states by the same labels as the initial-states. Neither of these features hold for the higher-dimensional theories in which we are primarily interested: for planar $\mathcal{N}\!=\!4$ SYM, particle creation and destruction plays a fundamental role; and the most primitive processes are not \mbox{$2\!\to\!2$} amplitudes, but rather the $3$-particle amplitudes discussed above, (\ref{three_particle_vertices}).

An important starting-point for describing higher-dimensional scattering processes is to forgo the traditional meaning of the ``$S$-matrix''---an operator which maps initial states to final states. Rather, we find it much more convenient to treat all the external particles on equal footing, using crossing symmetry to formulate the $S$-matrix as a process for which {\it all} the external particles are taken to be {\it incoming}.

One lesson we can take from $(1\pl1)$ dimensions is that any connection between scattering and permutations {\it must} involve {\it on-shell} processes. In $(3\pl1)$ dimensions, this leads us to trivalent, on-shell diagrams with black and white vertices discussed in the previous section. And so we are led to try and associate a permutation with these diagrams. As it turns out, just such a connection exists between two-colored, planar graphs and permutations, and has recently been studied in the mathematical literature, \cite{P} (see also \cite{T}).

Let's jump-in and describe how it works. The way to read-off a permutation from an on-shell graph is as follows. For each external leg $a$ (with clockwise ordering), follow the graph inward from $a$, turning left at each white vertex, and turning right at each black vertex; this ``left-right path'' will terminate at some external leg, denoted $\sigma(a)$. For example, the three-particle building blocks of $\mathcal{N}\!=\!4$, (\ref{three_particle_vertices}), are associated with permutations in the following way:
\vspace{-0.2cm}\eq{\hspace{-0cm}\begin{array}{c}\\[-20pt]\raisebox{-40pt}{\includegraphics[scale=1]{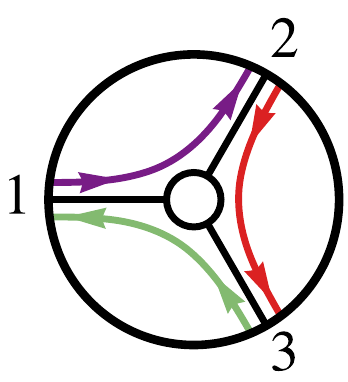}}\\[0pt]\end{array}\raisebox{-2pt}{{\LARGE$\Leftrightarrow$}}\Bigg(\begin{array}{@{}ccc@{}}1&2&3\\[-5pt]\downarrow&\downarrow&\downarrow\\[-2.5pt]{\color{black}2}&{\color{black}3}&{\color{black}1}\end{array}\Bigg)\quad\mathrm{and}\quad\begin{array}{c}\\[-20pt]\raisebox{-30pt}{\includegraphics[scale=1]{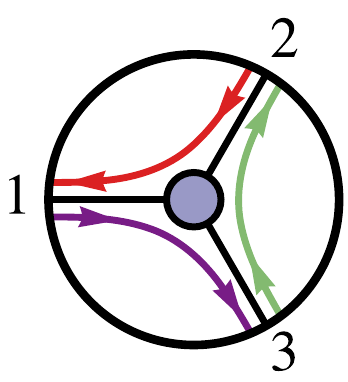}}\\[0pt]\end{array}\raisebox{-2pt}{{\LARGE$\Leftrightarrow$}}\Bigg(\begin{array}{@{}ccc@{}}1&2&3\\[-5pt]\downarrow&\downarrow&\downarrow\\[-2.5pt]{\color{black}3}&{\color{black}1}&{\color{black}2}\end{array}\Bigg)\vspace{-0.05cm}}

Of course, this works equally-well for more complex on-shell graphs; for example, the graph which gives the four-particle tree-amplitude, (\ref{four_point_box}), is associated with the following permutation:
\vspace{-0.2cm}\eq{\begin{array}{c}\\[-10pt]\hspace{-0cm}\raisebox{-50pt}{\includegraphics[scale=1]{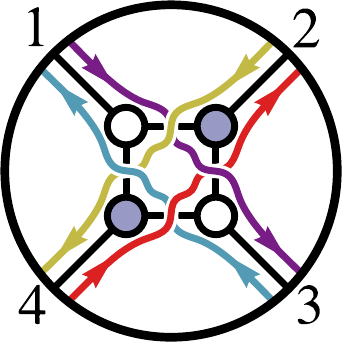}}\\{\color[rgb]{1,1,1}\{3,4,1,2\}}\end{array}\raisebox{-2pt}{{\LARGE$\Leftrightarrow$}}\Bigg(\begin{array}{@{}cccc@{}}1&2&3&4\\[-5pt]\downarrow&\downarrow&\downarrow&\downarrow\\[-2.5pt]{\color{black}3}&{\color{black}4}&{\color{black}1}&2\end{array}\Bigg)\vspace{-0.5cm}\label{four_point_box_with_LR_paths}}

It is very easy to see that such ``left-right paths'' allow us to define a permutation for any planar graph constructed with black and white vertices (not only those which are trivalent). Starting from any external leg of such a graph, this path will always lead back out to the boundary; and because any path can be trivially reversed (by exchanging the roles of black and white), it is clear that every external leg is the terminus of some such path. And so, the left-right paths do indeed define a {\it permutation} of the external legs.

Actually, left-right paths associate each graph with a slight generalization of an ordinary permutation known as a {\it decorated} permutation---a generalization which allows for two types of fixed-points. By convention, we always consider a left-right path to permute each label `to its right'---in other words, we think of the paths as being associated with a map $\sigma\!:\!\{1,\ldots,n\}\!\mapsto\!\{1,\ldots,2n\}$ such that  \mbox{$a\leq\sigma(a)\leq a\pl\,n$} and taking $\sigma(a)\!\!\mod\!n$ would be an ordinary permutation. The two types of fixed points correspond to the cases of $\sigma(a)=a$ or $\sigma(a)= a \pl\,n$. For the sake of simplicity, for the rest of this paper we will refer to these {\it decorated} permutations simply as `permutations' and denote them by ``${\color{perm}\{\sigma(1),\ldots,\sigma(n)\}}$''.

This allows us to differentiate between $2^n$ possible `decorations' of the trivial permutation. Such `decorations' arise for graphs such as,
\vspace{-.2cm}\eq{\raisebox{-47pt}{\includegraphics[scale=.85]{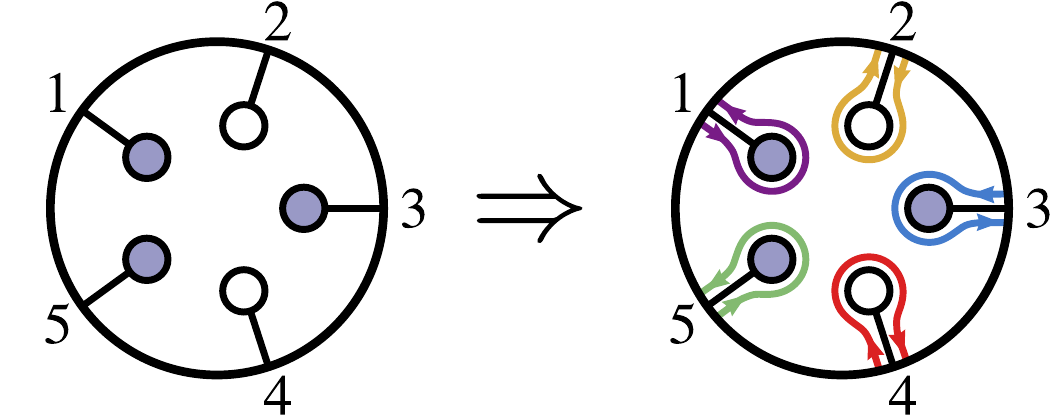}}\vspace{-.2cm}\label{g25_lolli_examples}}
which would be labeled by a `permutation' ${\color{perm}\{1,7,3,9,5\}}$. Although such empty graphs are themselves of little direct relevance to physics, they will play an important role in the general toolbox---as we will see in the following subsection. 

Associated with any permutation is a number, $k$, which is the number of $a\in\{1,\ldots,n\}$ which are mapped `beyond $n$' by $\sigma$---that is, for which $\sigma(a)>n$. This number is also given by the mean value of $\sigma(a)-a$: $k\equiv\frac{1}{n}\sum_{a}(\sigma(a)-a).$ To see this, notice that while the mean of any {\it ordinary} permutation always vanishes, our requirement that $a\leq\sigma(a)\leq a+n$ means that $\sigma$ must be shifted by $n$ relative to an ordinary permutation for some $k$ elements. For example, both the $4$-point graph, (\ref{four_point_box_with_LR_paths}), and the $5$-particle graph, (\ref{g25_lolli_examples}), have $k=2$.

The reason why the permutations associated with on-shell graphs are so important is that in many cases they {\it invariantly} encode the physical information about the graph and the on-shell form associated with it. Recall that graphs related by mergers, (\ref{merger_rule}), or square-moves, (\ref{square_move}), represent the same physical form. These operations also leave permutations invariant:
\vspace{-0.2cm}\eq{\begin{array}{c}\mbox{\hspace{-2cm}\raisebox{-46.5pt}{\includegraphics[scale=1]{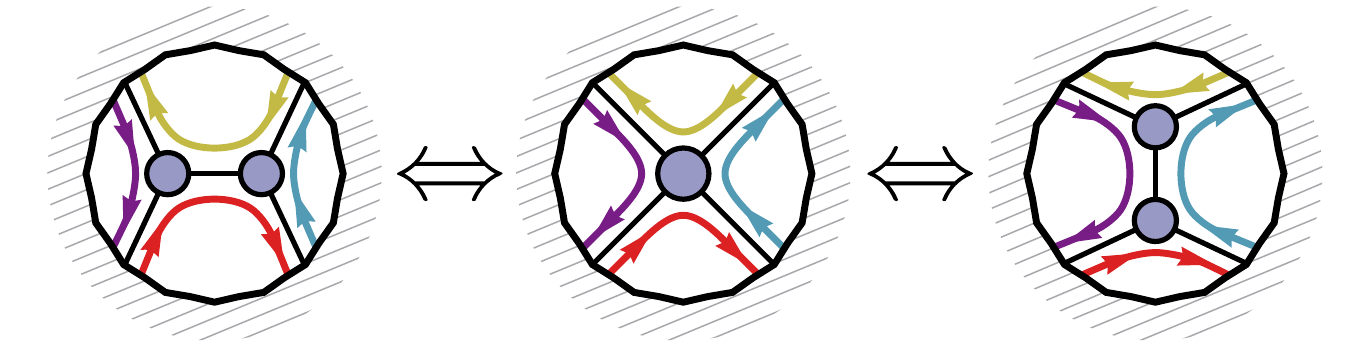}}\hspace{-2cm}}\\[-15pt]\mbox{\hspace{-1cm}\raisebox{-61.5pt}{\includegraphics[scale=1]{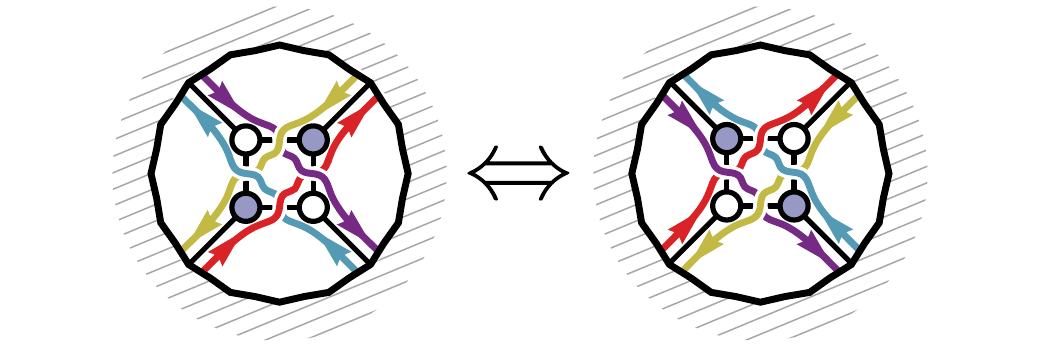}}\hspace{-1.0cm}}\end{array}\label{merger_and_square_moves_with_LR_paths}\vspace{-.2cm}}
Bubble-deletion, however, {\it does} change the permutation associated with an on-shell diagram; it also changes the number of faces. But by deleting bubbles, any graph can be `reduced'---and any two {\it reduced} graphs labeled by the same permutation always represent the same physical form. More explicitly, all physical information in reduced graphs is captured by the corresponding permutation. To see a simple example of this, recall the pair of inequivalent graphs given in (\ref{equivalent_graphs_in_g36}) which were related by a rather long sequence of mergers and square-moves; it is much easier to test the equivalence of the permutations which label them: \\[-10pt]
\vspace{0.1cm}\eq{\hspace{-0.0cm}\begin{array}{c}\\[-20pt]\includegraphics[scale=1]{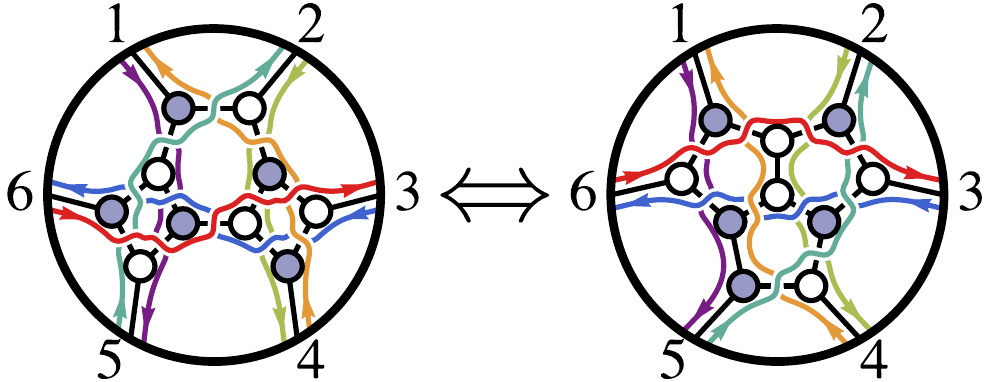}\\[-25pt]{\color{perm}\{5,4,6,7,8,9\}}\\[3pt]\end{array}\vspace{0.0cm}}

We should note in passing that there is something very special about $\mathcal{N}\!=\!4$ SYM and integrability which allows us to {\it fully} characterize on-shell diagrams in this way. Just as the Yang-Baxter relation (\ref{yang_baxter_move}) was the prerequisite for $(1\pl1)$-dimensional theories to be `combinatorial' in nature, it is the square-move (\ref{square_move}) which does this for $\mathcal{N}\!=\!4$: recall that in a non-supersymmetric theory, all $3$-particle vertices would need to be dressed by the helicities of the particles involved---such as in (\ref{3pt_helicity_amplitudes}); this dressing represents {\it extra} data which must be supplied in order to specify the physical process, and this data is {\it not} left invariant under square-moves. That being said, however, the purely combinatorial story of $\mathcal{N}\!=\!4$ will play a central role even for non-supersymmetric theories. This will be described more completely in \mbox{section \ref{less_supersymmetries_section}}.

\subsection{The BCFW-Bridge Construction of Representative Graphs}\label{BCFW_bridge_decomposition_subsection}
We have seen that every on-shell graph is associated with a permutation; quite beautifully, the converse is also true: {\it all} permutations can be represented by an on-shell graph. A constructive procedure for building a representative graph for any permutation was described in \cite{P} (and in somewhat different terms by D. Thurston in \cite{T}). Here, we will describe a different method---motivated by simple physical and combinatorial considerations and by analogy with physics in $(1\pl1)$ dimensions---where graphs are constructed out of simple, adjacent transpositions. Of course, in $(3\pl1)$ dimensions, there is no space-time evolution analogue of successive $2\!\to\!2$ scattering; and so we must find some way to `build-up' on-shell objects directly from the ``vacuum'' (a trivial permutation).

The key is understanding what an adjacent transposition means in terms of on-shell graphs. The answer is extremely simple: an adjacent transposition is nothing but the addition of the BCFW-bridge:
\vspace{-0.3cm}\eq{\mbox{\hspace{-1.5cm}\raisebox{-47pt}{\includegraphics[scale=.9]{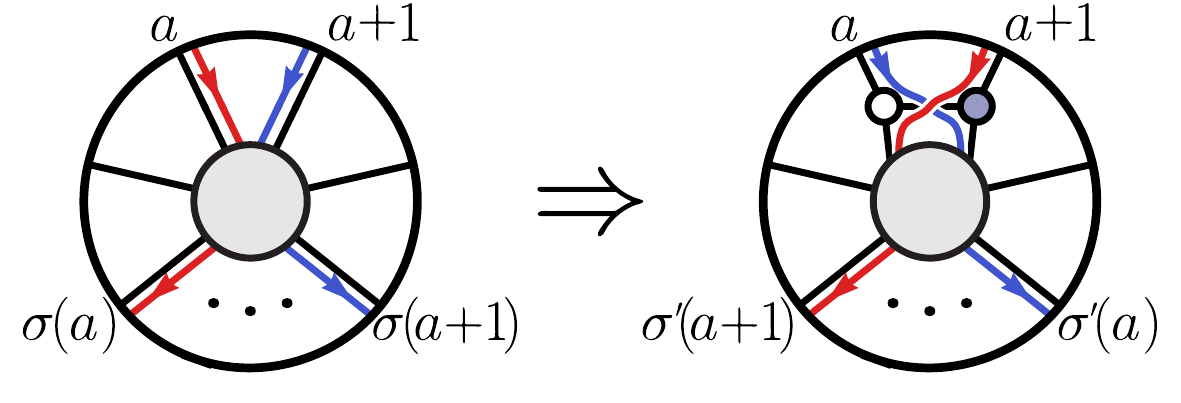}}\hspace{-1.5cm}}\label{adding_a_BCFW_bridge_to_a_graph}\vspace{-.4cm}}
Notice that any number of `hanging legs'---those which map to themselves under $\sigma$---can be inserted between $a$ and ``$a\pl1$'' without consequence; and so, we will consider any transposition $(a\,c)$ to be ``adjacent'' so long as for all $b$ between $a$ and $c$, \mbox{$\sigma(b)=b\!\!\mod\! n$}. (Although the bridge drawn in (\ref{adding_a_BCFW_bridge_to_a_graph}) will be sufficient for most applications, the oppositely-colored bridge---where black and white vertices are exchanged---could also be used; the principle difference being that such a bridge would transpose the {\it pre}-images of $a$ and $a\pl1$ under $\sigma$ instead of the images).

Because adjacent transpositions simply correspond to adding BCFW-bridges, any decomposition of a permutation $\sigma$ into a sequence of such transpositions acting on a trivial permutation can be read as {\it instructions} for building-up a {\it representative} on-shell graph for $\sigma$ by successively adding BCFW-bridges to an empty graph like that of (\ref{g25_lolli_examples}).

Of course, adding a BCFW bridge may potentially give us a reducible on-shell diagram. However, it turns out that when adding a bridge to a reduced graph, so long as $\sigma(a\pl1)<\sigma(a)$---that is, the are paths arranged as drawn in (\ref{adding_a_BCFW_bridge_to_a_graph})---then the resulting graph is guaranteed to be reduced. We will not prove this statement now, but its proof will become trivial after the discussions in \mbox{section \ref{configuration_of_vectors_section}}. 

And so, when breaking-down a permutation into adjacent transpositions, we want to find pairs $(a\, c)$ with $a<c$ (separated only by external legs $b$ self-identified under $\sigma$) such that $\sigma(a)<\sigma(c)$; then when we decompose $\sigma$ as $(a\, c)\circ\sigma'$ with $\{\sigma(a),\sigma(c)\}=\{\sigma'(c),\sigma'(a)\}$, adding a BCFW-bridge to a {\it reduced} on-shell diagram labeled by $\sigma'$ will result in a {\it reduced} on-shell diagram labeled by $\sigma$. Of course, there are {\it many} ways of decomposing a permutation $\sigma$ into such a chain of adjacent transpositions, and {\it any} such decomposition will result in a representative, reduced graph whose left-right permutation is $\sigma$. But for the sake of concreteness, let us describe one very specific, {\it canonical} procedure to decompose any permutation---one which will turn out to have rather special properties discussed in \mbox{section \ref{boundaries_in_canonical_coordinates}}.\\[-10pt]

\noindent{\bf BCFW-Bridge Decomposition:} Starting with any permutation $\sigma$, if $\sigma$ is not a decoration of the identity, then decompose $\sigma$ as $(a\,c)\circ\sigma'$ where $1\leq a<c\leq n$ is the {\it lexicographically-first} pair separated only by legs $b$ which are self-identified under $\sigma$ and for which $\sigma(a)<\sigma(c)$; repeat until $\sigma$ is the identity.\\[-10pt]

To illustrate this procedure, let's see how it generates a representative, reduced on-shell diagram which is labeled by the permutation ${\color{perm}\{4,6,5,7,8,9\}}$:\\[-15pt]

\noindent\scalebox{.875}{\mbox{\hspace{1.07cm}\begin{minipage}[h]{\textwidth}\vspace{.2cm}\eq{\hspace{-4cm}\hspace{2.5cm}\mbox{\raisebox{-80pt}{\includegraphics[scale=1]{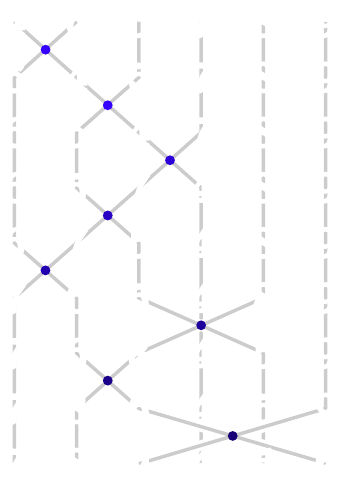}}\hspace{-129.52pt}}\begin{array}{|c@{$\,$}|@{$\,$}cccccc@{$\,$}|c|}\cline{1-7}\multicolumn{1}{|c@{$\,$}|@{$\,$}}{}&1&2&3&4&5&6
\\[-4pt]\multicolumn{1}{|c@{$\,$}|@{$\,$}}{\tau}&\,\,\downarrow\,\,&\,\,\downarrow\,\,&\,\,\downarrow\,\,&\,\,\downarrow\,\,&\,\,\downarrow\,\,&\,\,\downarrow\,\,
\\\cline{1-7}\multirow{2}{*}{({\color[rgb]{0.2,0,1.00}1\,2})}&{\color[rgb]{0.2,0,1.00}4}&{\color[rgb]{0.2,0,1.00}6}&5&7&8&9
\\\multirow{2}{*}{({\color[rgb]{0.184615,0,0.923}2\,3})}&6&{\color[rgb]{0.184615,0,0.923}4}&{\color[rgb]{0.184615,0,0.923}5}&7&8&9
\\\multirow{2}{*}{({\color[rgb]{0.169231,0,0.846}3\,4})}&6&5&{\color[rgb]{0.169231,0,0.846}4}&{\color[rgb]{0.169231,0,0.846}7}&8&9
\\\multirow{2}{*}{({\color[rgb]{0.153846,0,0.769}2\,3})}&6&{\color[rgb]{0.153846,0,0.769}5}&{\color[rgb]{0.153846,0,0.769}7}&{\color{deemph}4}&8&9
\\\multirow{2}{*}{({\color[rgb]{0.138462,0,0.692}1\,2})}&{\color[rgb]{0.138462,0,0.692}6}&{\color[rgb]{0.138462,0,0.692}7}&5&{\color{deemph}4}&8&9
\\\multirow{2}{*}{({\color[rgb]{0.123077,0,0.615}3\,5})}&{\color{deemph}7}&6&{\color[rgb]{0.123077,0,0.615}5}&{\color{deemph}4}&{\color[rgb]{0.123077,0,0.615}8}&9
\\\multirow{2}{*}{({\color[rgb]{0.107692,0,0.538}2\,3})}&{\color{deemph}7}&{\color[rgb]{0.107692,0,0.538}6}&{\color[rgb]{0.107692,0,0.538}8}&{\color{deemph}4}&{\color{deemph}5}&9
\\\multirow{2}{*}{({\color[rgb]{0.0923077,0,0.462}3\,6})}&{\color{deemph}7}&{\color{deemph}8}&{\color[rgb]{0.0923077,0,0.462}6}&{\color{deemph}4}&{\color{deemph}5}&{\color[rgb]{0.0923077,0,0.462}9}
\\&{\color{deemph}7}&{\color{deemph}8}&{\color{deemph}9}&{\color{deemph}4}&{\color{deemph}5}&{\color{deemph}6}
\\\cline{1-7}\end{array}\raisebox{-20pt}{{\Huge$\Leftrightarrow$}}\raisebox{-83.35pt}{\includegraphics[scale=1]{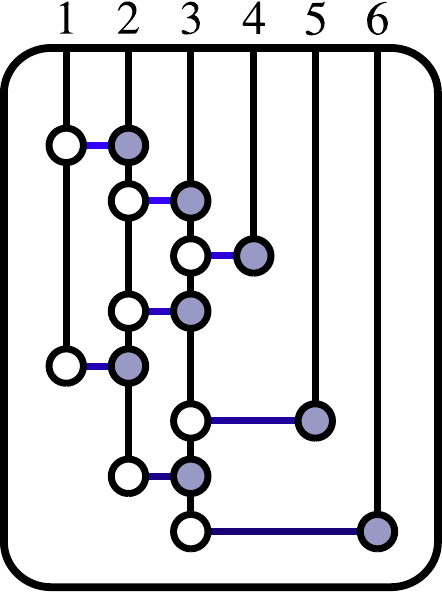}}\;\raisebox{-20pt}{{\Huge$\approx$}}\hspace{-0.45cm}\begin{array}{c}\\[-15pt]\raisebox{-75pt}{\includegraphics[scale=1]{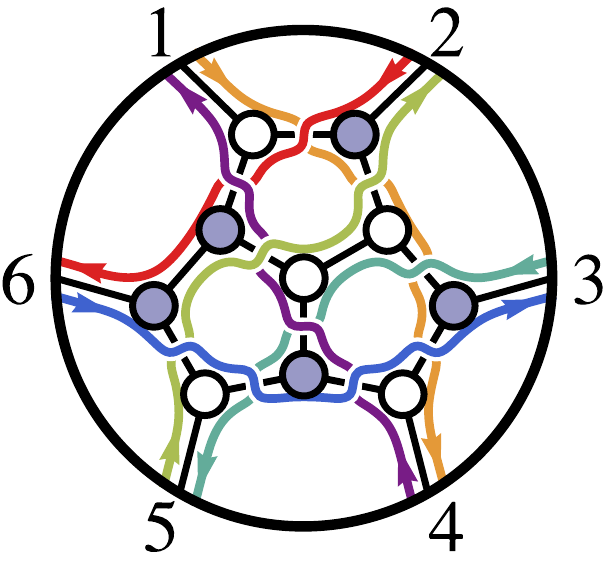}}\\[-10pt]{\color{perm}\phantom{\{4,6,5,7,8,9\}}}\end{array}\nonumber\vspace{-1.05cm}\hspace{-4cm}}\eq{\hspace{-0.795cm}\begin{array}{@{}c@{}c@{}c@{}c@{}c@{}c@{}c@{}c@{}}\\[-10pt]&&&&&&\mbox{\hspace{15pt}({\color[rgb]{0.2,0,1.00}12})}\raisebox{-2pt}{\hspace{-10pt}\includegraphics[scale=1.5]{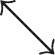}}\,{\color[rgb]{1,1,1}\alpha_8}\;~\\[-0.45cm]
\begin{array}{c}\\[-5pt]\raisebox{-47pt}{\includegraphics[scale=1]{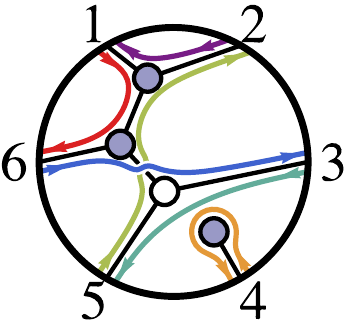}}\\[-3pt]{\color{perm}\{}{\color{red}\mathbf{6}}{\color{perm},\,}{\color{red}\mathbf{7}}{\color{perm},5,4,8,9\}}\end{array}&\raisebox{0pt}{$\xleftrightarrow[\text{{\normalsize${\color[rgb]{1,1,1}\alpha_5}$}}]{\text{{\normalsize$\!({\color[rgb]{0.153846,0,0.769}23})\!$}}}$}&\begin{array}{c}\\[-5pt]\raisebox{-47pt}{\includegraphics[scale=1]{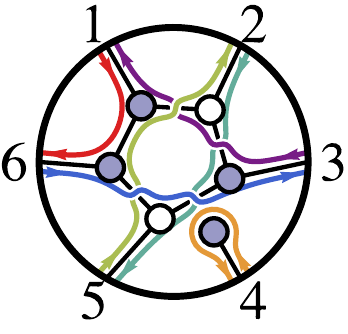}}\\[-3pt]{\color{perm}\{6,\,}{\color{red}\mathbf{5}}{\color{perm},\,}{\color{red}\mathbf{7}}{\color{perm},4,8,9\}}\end{array}&\raisebox{0pt}{$\xleftrightarrow[\text{{\normalsize${\color[rgb]{1,1,1}\alpha_6}$}}]{\text{{\normalsize$\!({\color[rgb]{0.169231,0,0.846}34})\!$}}}$}&\begin{array}{c}\\[-5pt]\raisebox{-47pt}{\includegraphics[scale=1]{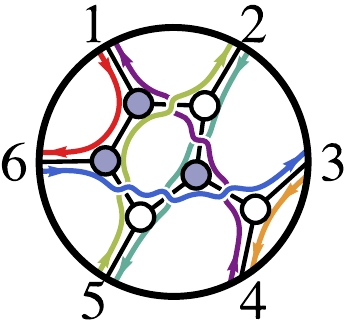}}\\[-3pt]{\color{perm}\{6,5,\,}{\color{red}\mathbf{4}}{\color{perm},\,}{\color{red}\mathbf{7}}{\color{perm},8,9\}}\end{array}&\raisebox{0pt}{$\xleftrightarrow[\text{{\normalsize${\color[rgb]{1,1,1}\alpha_7}$}}]{\text{{\normalsize$\!({\color[rgb]{0.184615,0,0.923}23})\!$}}}$}&\begin{array}{c}\\[-5pt]\raisebox{-47pt}{\includegraphics[scale=1]{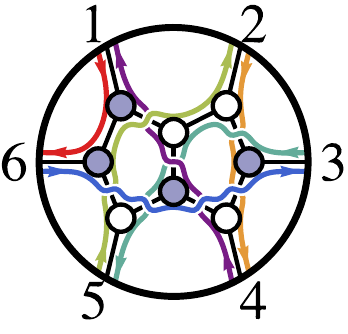}}\\[-3pt]{\color{perm}\{6,\,}{\color{red}\mathbf{4}}{\color{perm},\,}{\color{red}\mathbf{5}}{\color{perm},7,8,9\}}\end{array}\\[0.25cm]
({\color[rgb]{0.138462,0,0.692}12})\text{{\Large$\updownarrow$}}\,{\color[rgb]{1,1,1}\alpha_4}\;~
\\[-0.45cm]
\begin{array}{c}\\[-5pt]\raisebox{-52pt}{\includegraphics[scale=1]{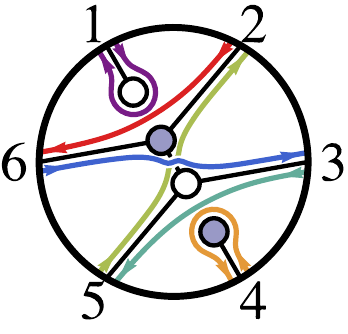}}\\[-3pt]{\color{perm}\{7,6,\,}{\color{red}\mathbf{5}}{\color{perm},4,\,}{\color{red}\mathbf{8}}{\color{perm},9\}}\end{array}&\raisebox{0pt}{$\xleftrightarrow[\text{{\normalsize${\color[rgb]{1,1,1}\alpha_3}$}}]{\text{{\normalsize$\!({\color[rgb]{0.123077,0,0.615}35})\!$}}}$}&\begin{array}{c}\\[-5pt]\raisebox{-47pt}{\includegraphics[scale=1]{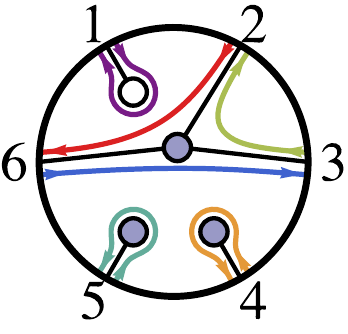}}\\[-3pt]{\color{perm}\{7,\,}{\color{red}\mathbf{6}}{\color{perm},\,}{\color{red}\mathbf{8}}{\color{perm},4,5,9\}}\end{array}&\raisebox{0pt}{$\xleftrightarrow[\text{{\normalsize${\color[rgb]{1,1,1}\alpha_2}$}}]{\text{{\normalsize$\!({\color[rgb]{0.107692,0,0.538}23})\!$}}}$}&\begin{array}{c}\\[-5pt]\raisebox{-47pt}{\includegraphics[scale=1]{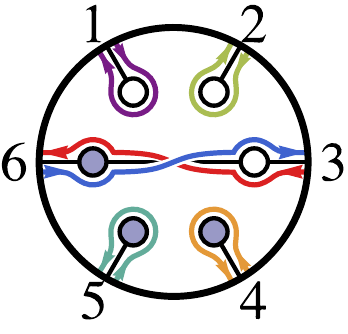}}\\[-3pt]{\color{perm}\{7,8,\,}{\color{red}\mathbf{6}}{\color{perm},4,5,}{\color{red}\mathbf{9}}{\color{perm}\}}\end{array}&\raisebox{0pt}{$\xleftrightarrow[\text{{\normalsize${\color[rgb]{1,1,1}\alpha_1}$}}]{\text{{\normalsize$\!({\color[rgb]{0.0923077,0,0.462}36})\!$}}}$}&\begin{array}{c}\\[-5pt]\raisebox{-47pt}{\includegraphics[scale=1]{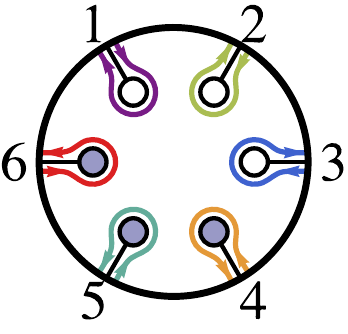}}\\[-3pt]{\color{perm}\{7,8,9,4,5,6\}}\end{array}\\[-40pt]\end{array}\nonumber}\end{minipage}}}

\noindent In the sequence of figures drawn above, we often made use of the fact that any bivalent or (non-boundary) monovalent vertex can be deleted without changing the permutation. So, for example, adding the  BCFW bridge `$({\color[rgb]{0.107692,0,0.538}23})$' to the second graph (from the bottom-right) results in the succeeding graph drawn via the sequence of (essentially trivial) moves:
\vspace{-.2cm}\eq{\begin{array}{c}\raisebox{-47pt}{\includegraphics[scale=1]{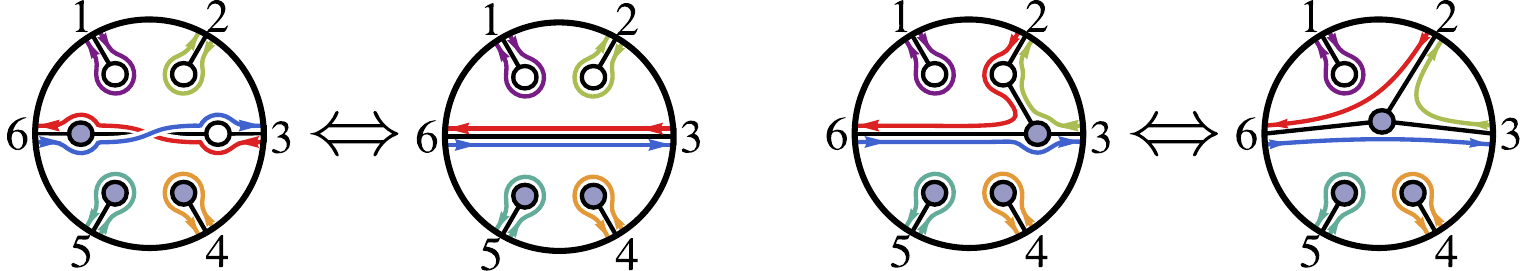}}\\[-57.pt]\text{{\footnotesize${\color{perm}\{7,8,6,4,5,9\}}$}}\hspace{2.6cm}\raisebox{31pt}{{\Large$\xleftrightarrow{\text{{\normalsize$\!({\color[rgb]{0.107692,0,0.538}23})\!$}}}$}}\hspace{2.6cm}\text{{\footnotesize${\color{perm}\{7,\,}{\color{red}\mathbf{6}}{\color{perm},\,}{\color{red}\mathbf{8}}{\color{perm},4,5,9\}}$}}\end{array}\vspace{-.2cm}\nonumber}

This procedure provides us with a combinatorial test of a graph's reducibility: because the BCFW-bridge construction always produces a {\it reduced} representative graph for any permutation, and each step in the construction adds one face to the graph as it is built, a graph is reduced if and only if the number of its faces minus one is equal to the number of steps in the BCFW-bridge decomposition of the permutation which labels it. If not, then the graph is reducible, and has some number of faces which can be deleted by bubble reduction:
\vspace{-0.4cm}\eq{\hspace{-1.4cm}\raisebox{-47.5pt}{\includegraphics[scale=1]{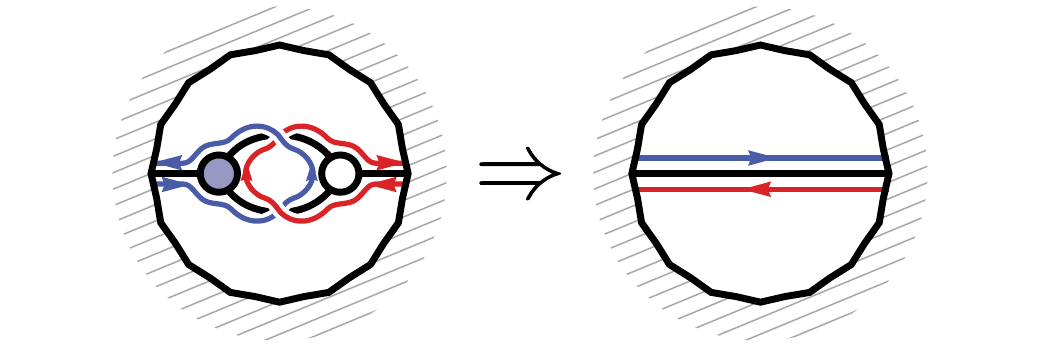}}\label{bubble_deletion_with_LR_paths}\hspace{-1.4cm}\vspace{-.2cm}}

A more intrinsic way to identify a reducible graph is if any pair of left-right paths $a\!\to\!\sigma(a)$ and $b\!\to\!\sigma(b)$ cross each other along more than one edge in the graph in the manner known as a ``bad double crossing'', or if there is any purely-internal path.
\vspace{-.3cm}\eq{\raisebox{-47.5pt}{\includegraphics[scale=1]{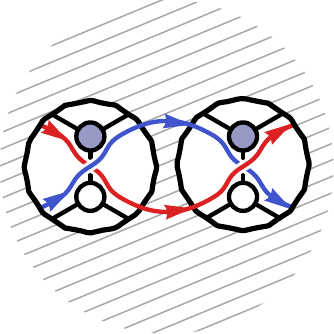}}\quad\mathrm{or}\quad\raisebox{-47.5pt}{\includegraphics[scale=1]{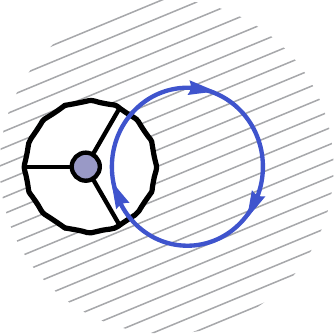}}\vspace{-.3cm}\label{bad_double_crossing}}
A bad double-crossing is distinguished from those double-crossings of the form:
\vspace{-.3cm}\eq{\raisebox{-47.5pt}{\includegraphics[scale=1]{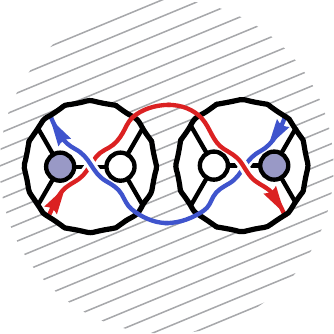}}\vspace{-.3cm}\label{not_bad_double_crossing}}
Double-crossings such as that above do not indicate that a graph is reducible.

We thus have a complete dictionary between (reduced) on-shell graphs and permutations. As we will discuss in \mbox{section \ref{yang_baxter_and_abjm_section}}, this new picture actually contains the $(1\pl1)$-dimensional story as a special case. Another closely related special case is relevant for describing on-shell diagrams (and all-loop amplitudes)  of the ABJM theory in $(2\pl1)$ dimensions!

But let us now move beyond the purely combinatorial aspects of the story, and turn towards actually {\it computing} on-shell diagrams. This will lead us to uncover beautiful structures in algebraic geometry also described by decorated permutations, ultimately connecting on-shell graphs to the ``positive'' Grassmannian of our title.

\newpage
\section{From On-Shell Diagrams to the Grassmannian}\label{intro_to_grassmannian_section}

In this section we will show that the computation of on-shell diagrams is most efficiently and transparently carried out by associating each diagram with an auxiliary structure: a matrix $C$ representing an element of the Grassmannian $G(k,n)$. But let us begin by reviewing some elementary properties about Grassmannian manifolds in general, and describe the first appearance of these spaces in the story of scattering amplitudes, as they arise in the description of external kinematical data.

\subsection{The Grassmannian of $k$-Planes in $n$ Dimensions, $G(k,n)$}\label{intro_to_the_grassmannian_subsection}
The Grassmannian $G(k,n)$ is the space of $k$-dimensional planes passing through the origin in an $n$-dimensional space (see e.g.\ \cite{GriffithsHarris}). We can specify a $k$-plane in $n$ dimensions by giving $k$ vectors $C_{\alpha}\!\in\!\mathbb{C}^n$, whose span defines the plane. We can assemble these vectors into a $(k\! \times\!n)$-matrix $C$, whose components are $c_{\alpha a}$ for \mbox{$\alpha\!=\!1,\ldots, k$} and $a=1,\ldots, n$.

Under $GL(k)$-transformations, $C\!\mapsto\!\Lambda\!\cdot\!C$---with $\Lambda\!\in\! GL(k)$---the row vectors will change, but the plane spanned by them is obviously unchanged. Thus, the Grassmannian $G(k,n)$ can be thought of as the space of $(k\! \times\!n)$-matrices modulo this $GL(k)$ ``gauge'' redundancy. From this, we see that the dimension of $G(k,n)$ is $k\!\times\!n\,\mi\,k^2\!=\!k(n\,\mi\,k)$. In practice, we can ``gauge-fix'' the $GL(k)$ redundancy by choosing any $k$ of the columns of the matrix to form the $(k\!\times\!k)$ identity matrix. For instance, we can represent a generic point in $G(2,5)$ in the following gauge-fixed form:
\vspace{-.3cm}\eq{C=\left(\begin{array}{@{}ccccc@{}}\;1\;&\;0\;& c_{1\,3} & c_{1\,4} & c_{1\,5} \\ 0 & 1 & c_{2\,3} & c_{2\,4} & c_{2\,5} \end{array} \right).\vspace{-.2cm}\label{gauge_fixed_rep_of_g25}}
This coordinate chart does not cover the {\it entire} Grassmannian---though of course the collection of all $\binom{n}{k}$ such charts would obviously suffice.

The $GL(k)$-invariant information associated with $C$ is easily specified. First, notice that the only $SL(k)$-invariants of $C\!\in\! G(k,n)$ are the minors constructed out of the {\it columns} of $C$,
\vspace{-.2cm}\eq{(a_1\,\cdots\,a_k)\equiv\mathrm{det}\{c_{a_1},\ldots,c_{a_k}\}\,.\vspace{-.2cm}}
$GL(k)$-invariants are then simply ratios of these:
\vspace{-.2cm}\eq{\frac{(a_1 \cdots a_k)}{(b_1 \cdots b_k)}.\vspace{-.2cm}}
While the (ratios of) minors are $GL(k)$-invariant, the number of these, $n \choose k$, is much greater than the dimensionality of the Grassmannian, $\dim(G(k,n))\!=\!k(n\,\mi\,k)$, and so the minors represent a highly-redundant set of data to describe $C$. The identities among minors arise from the simple fact that any $k$-vector can be expanded in a basis of any $k$ linearly-independent $k$-vectors---a statement that is equivalent to the identity known as {\it Cramer's rule}:
\vspace{-.2cm}\eq{c_{a_1}(a_2\cdots a_{k+1})-c_{a_2}(a_1\,a_3\cdots a_{k+1})+\cdots+(-1)^{k}c_{a_{k+1}}(a_1\cdots a_k)=0,\vspace{-.2cm}\label{Cramers_rule}}
for any $c_a\!\in\!\mathbb{C}^k$. Contracting each of the vectors in (\ref{Cramers_rule}) with another set of vectors $c_{b_1},\ldots,c_{b_{k-1}}$ generates the identities known as the Pl\"{u}cker relations,
\vspace{-.2cm}\eq{(b_1\,\cdots\,b_{k-1}\,a_1)(a_2\cdots a_{k+1})+\cdots+(-1)^{k-1}(b_1\,\cdots\,b_{k-1}\,a_{k+1})(a_1\cdots a_k)=0.\vspace{-.2cm}\label{Plucker_relations}}

Associated with any $k$-plane $C$ is a natural $(n\,\mi\,k)$-plane denoted $C^\perp$, the ``orthogonal complement'' of $C$, which is defined by,
\vspace{-.2cm}\eq{C^{\perp}\!\!\cdot C=0.\vspace{-.2cm}}
Therefore, there is a natural isomorphism between $G(k,n)$ and $G(n\,\mi\,k,n)$, which is reflected in the invariance of $\dim(G(k,n))=k(n\,\mi\,k)$ under the exchange $k\leftrightarrow(n\,\mi\,k)$. The minors of $C^{\perp}$ are fully determined by the minors of $C$ in the obvious way: for any complementary sets $\{a_1,\ldots,a_k\}$ and $\{b_1,\ldots,b_{n-k}\}$ (whose union is $\{1,\ldots,n\}$), we have
\vspace{-.2cm}\eq{\left.(a_1 \cdots a_k)\right|_{C}=\left.\pm(b_1\cdots b_{n-k})\right|_{C^{\perp}}.\vspace{-.1cm}}
To be completely explicit, suppose we represent $C$ in a gauge where columns $c_A$ with $A\equiv\{a_1,\ldots,a_k\}$ are taken as the identity; then the $n\,\mi\,k$ columns of $C$ in the complementary set $B\equiv A^c$, $c_{b}$ for $b\!\in\! B$---whose components we write as $c_{a\,b}$---encode the $k(n\,\mi\,k)$ degrees of freedom of $C$; then the matrix $C^{\perp}$ has components,
\vspace{-.2cm}\eq{c^{\perp}_{a\,b}=-c_{b\,a}.\vspace{-.2cm}}
For example, the plane $C^\perp\!\in\! G(3,5)$ orthogonal to $C\!\in\! G(2,5)$ given in (\ref{gauge_fixed_rep_of_g25}) is:
\vspace{-.2cm}\eq{
C^\perp= \left(\begin{array}{@{}ccccc@{}}\mi\,c_{1\,3}\,& \mi\,c_{2\,3}\,&\;1\;&\;0\;&\;0\;\\\mi\,c_{1\,4}\,& \mi\,c_{2\,4}\,& 0 & 1 & 0 \\ \mi\,c_{1\,5}\,& \mi\,c_{2\,5}\,&
0 & 0 & 1 \end{array} \right)
\vspace{-.2cm}}

Finally, we will eventually be talking about a certain top-dimensional differential form on the Grassmannian, so it is useful to discuss what general forms on the Grassmannian look like in the coordinates $c_{a\,b}$. Consider first the familiar example of a form on the projective space $G(1,2)$. We can think of this as a $(1\!\times\! 2)$ matrix $C = (c_1 \, c_2)$, modulo the $GL(1)$-action of $C\!\to\!t C$. Any top-form can be written as
\vspace{-.2cm}\eq{\Omega =\frac{ d^2 C}{{\rm vol }(GL(1))} \frac{1}{f(C)},\vspace{-.2cm}}
where $f(C)$ must have homogeneity $(\pl\,2)$ under rescaling $C$; that is, $f(t C) = t^{2} f(C)$. In practice, modding-out by the $GL(1)$-action is trivial: one can simply gauge-fix the $GL(1)$ so that, say, $C\!\mapsto\!C^*=(1 \, c_2)$; and then $\Omega = dc_2 /f (C^*)$. We can also say this more invariantly, by writing,
\vspace{-.2cm}\eq{\Omega = \langle C d C \rangle \frac{1}{f(C)}.\vspace{-.2cm}}
The generalization of this simple case to an arbitrary Grassmannian is straightforward. We can write,
\vspace{-.2cm}\eq{\Omega = \frac{d^{k \times n} C}{{\rm vol}(GL(k))} \frac{1}{f(C)},\vspace{-.0cm}}
where  $GL(k)$-invariance implies, in particular, that $f(C)$ must be a function of the minors of $C$ with homogeneity under rescaling
\vspace{-.2cm}\eq{f(t C) = t^{k\times n} f(C).\vspace{-.2cm}}
In the coordinate chart where we gauge-fix $k$ of the columns to the identity as above, then $\Omega = d^{k\times (n-k)}c_{a,b}/f(C)$. Said more invariantly, we have
\vspace{-.2cm}\eq{\Omega = \langle C_1 \cdots C_k (dC_1)^{(n-k)} \rangle \cdots \langle C_1 \cdots C_k (dC_k)^{(n-k)} \rangle \frac{1}{f(C)},\vspace{-.2cm}}
where $C_\alpha$ is a row-vector of $C$ and, e.g.,
\vspace{-.2cm}\eq{\langle C_1 \cdots C_k (dC_1)^{(n-k)} \rangle \equiv \epsilon^{a_1a_2\ldots a_n} c_{1,a_1}\cdots c_{k,a_k}dc_{1,a_{k+1}}\wedge\cdots \wedge dc_{1,a_{n}}.\vspace{-.2cm}}

\subsection{Grassmannian Description of Kinematical Data: the 2-Planes $\lambda$ and $\tilde{\lambda}$}
In a moment, we will establish a very direct connection between on-shell diagrams and the Grassmannian; but let us first pause to point out an even more basic way in which the Grassmannian makes an appearance in scattering amplitudes: in the very way we encode external kinematical data. We normally think of this data as simply being specified by $n$ $2$-component spinors $\lambda_a^{\alpha}$ and $\widetilde \lambda_a^{\dot{\alpha}}$; but of course we may also think of this data as given by a pair of $(2\!\times\!n)$-matrices---which we denote collectively by $\lambda$ and $\widetilde\lambda$. For example, the $\lambda$'s are naturally associated with the $(2\!\times\!n)$-matrix, \vspace{-.2cm}\eq{\lambda\equiv\left(\begin{array}{@{}cccc@{}}\lambda_1^1&\lambda_2^1&\cdots&\lambda^1_n\\\lambda_1^2&\lambda_2^2&\cdots&\lambda^2_n\end{array}\right)\raisebox{-.75pt}{\text{{\Large$\Leftrightarrow$}}}\left(\begin{array}{@{}cccc@{}}\lambda_1&\lambda_2&\cdots&\lambda_n\end{array}\right).\vspace{-.2cm}} Instead of focusing on the {\it columns} of the matrix $\lambda$, let us think about it as two row-vectors. Each of these is a vector in an $n$-dimensional space. Under Lorentz transformations, these two vectors change, but since Lorentz transformations act on the $\lambda$'s by $SL(2)$-transformations on their $\alpha$ indices, the two new vectors will simply be a linear combination of the original ones. Therefore, while the vectors themselves change, the {\it plane} that is spanned by them is invariant under Lorentz transformations. Quite beautifully then, the Lorentz-invariant information encoded by the $\lambda$'s is really just this $2$-{\it plane} in $n$ dimensions---an element of $G(2,n)$ as realized in \cite{ArkaniHamed:2009dn}. The same is obviously true for the $\widetilde \lambda$'s. Of course, the Lorentz group is only the $SL(2)$ part of $GL(2)$ and on-shell forms do transform under ``global'' little group transformations which correspond to the $GL(1)$ subgroup of $GL(2)$.

In terms of spinor helicity variables, momentum conservation is simply,
\vspace{-.2cm}\eq{\sum_a\lambda_a^{\alpha}\widetilde\lambda_a^{\dot{\alpha}}=0,\vspace{-.2cm}}
which has the geometric interpretation that the plane $\lambda$ is {\it orthogonal to} the plane $\widetilde\lambda$, \cite{ArkaniHamed:2009dn}:
\vspace{-.4cm}\eq{\raisebox{-36pt}{\includegraphics[scale=1]{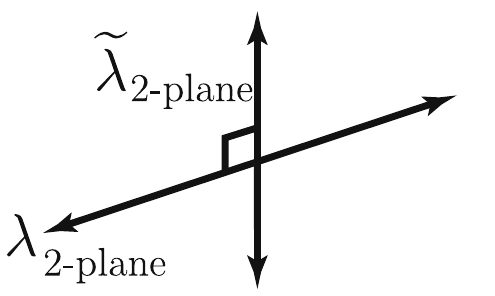}}\label{momentum_conservation_as_planes}\vspace{-.2cm}}
This geometric understanding of momentum-conservation also nicely explains the unique nature of its application to the case of three-particles: two $2$-planes in $3$ dimensions {\it cannot} be orthogonal in general. The only solution, therefore, is for one of the planes to actually be a $1$-plane in disguise. For example, suppose that we have three generic $\widetilde\lambda$'s. Momentum conservation requires that $\lambda\subset\widetilde\lambda^\perp$, but $\widetilde\lambda^{\perp}$ is a $1$-plane! A $GL(1)$-representative of $\widetilde\lambda^{\perp}$ is given by
\vspace{-.2cm}\eq{\widetilde\lambda^\perp\equiv\left(\begin{array}{@{}ccc@{}}\sb{2\,3}&\sb{3\,1}&\sb{1\,2}\end{array}\right),\vspace{-.2cm}\label{dual_to_lambda_tilde}}
for which $\widetilde\lambda^\perp\!\!\cdot\!\widetilde\lambda=0$ follows as a trivial instance of Cramer's rule, (\ref{Cramers_rule}):
\vspace{-.2cm}\eq{\widetilde\lambda^\perp\!\!\cdot\!\widetilde\lambda=\sb{2\,3}\widetilde\lambda_1+\sb{3\,1}\widetilde\lambda_2+\sb{1\,2}\widetilde\lambda_3=0.\vspace{-.2cm}}
Because this is the {\it unique} plane orthogonal to $\widetilde\lambda$, momentum conservation requires that the $\lambda$-plane be spanned by it. In particular, this means that all the $\lambda$'s must be proportional: in a Lorentz frame where $\lambda_1=${\footnotesize$\left(\begin{array}{@{}c@{}}\sb{2\,3}\\[-0pt]0\end{array}\right)$}, we have
\vspace{-.2cm}\eq{\lambda\equiv\left(\begin{array}{@{}ccc@{}}\lambda_1&\lambda_2&\lambda_3\end{array}\right)=\left(\begin{array}{@{}ccc@{}}\sb{2\,3}&\sb{3\,1}&\sb{1\,2}\\0&0&0\end{array}\right).\vspace{-.2cm}}

\subsection{Grassmannian Representation of On-Shell Diagrams}
Let us begin to more explicitly calculate the differential form associated with a given on-shell diagram. We use the momentum-conserving $\delta$-functions at the vertices to localize as many of the internal momenta as we can. This looks highly non-trivial because momentum conservation is a quadratic constraint on the $\lambda, \widetilde \lambda$ in general. But a moment's reflection suggests that the situation may be easier to understand. We know that for $3$-particle amplitudes, momentum conservation implies a very simple geometric situation---where either the $\lambda$'s or the $\widetilde \lambda$'s are forced to be parallel to each other. However, our representation of the three-particle amplitude, simple and elegant though it is, does not make this simple fact manifest. This motivates us to try to express the 3-particle amplitude in a slightly different form---one which makes the geometry of the $\lambda$'s and $\widetilde\lambda$'s in each case as transparent as possible.

Let's start with the $\mathcal{A}_3^{(1)}$ vertex:
\vspace{-.4cm}\eq{\raisebox{-47.5pt}{\includegraphics[scale=1]{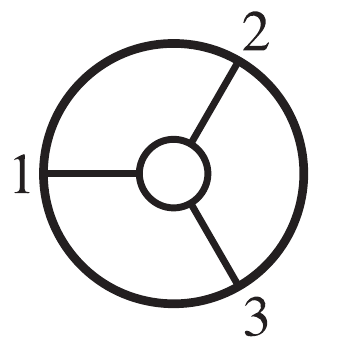}}\hspace{-10pt}\raisebox{-2.5pt}{\text{{\LARGE$\;\Leftrightarrow\;$}}}\mathcal{A}^{(1)}_3=\,\frac{\delta^{1\times4}(\sb{2\,3}\widetilde{\eta}_1+\sb{3\,1}\widetilde{\eta}_2+\sb{1\,2}\widetilde{\eta}_3)}{\sb{1\,2}\sb{2\,3}\sb{3\,1}}\delta^{2\times2}\big(\lambda\!\cdot\!\widetilde{\lambda}\big).\vspace{-.3cm}}
Notice that the coefficients of the $\widetilde{\eta}$'s are the same as the factors that appear in the denominator of $\mathcal{A}_3^{(1)}$, and coincide with the 1-plane $\widetilde \lambda^{\perp}$ orthogonal to $\widetilde \lambda$.
We can make this geometry manifest by introducing an {\it auxiliary} $1$-plane $W\!\in\! G(1,3)$, and demand that it be orthogonal to $\widetilde \lambda$ and that it contains the plane $\lambda$. This latter constraint is equivalent to the somewhat less concise condition that the orthogonal complement of $W^\perp$ is orthogonal to $\lambda$.
\noindent Thus, we can represent,
\vspace{-0.2cm}\eq{\mathcal{A}^{(1)}_{3}=\!\int\!\!\frac{d^{1\times3}W}{\mathrm{vol}(GL(1))}\frac{\delta^{1\times4}\big(W\!\cdot\!\widetilde{\eta}\big)}{(1)(2)(3)} \delta^{1\times2}\big(W\!\cdot\!\widetilde{\lambda}\big) \delta^{2\times2}\big(\lambda\!\cdot\!W^\perp\!\big),\vspace{-.2cm}\label{g13_version_of_vertex}}
where $W\!\in\! G(1,3)$ is given by the $(1\!\times\!3)$-matrix
\vspace{-.2cm}\eq{
W \equiv \left(\begin{array}{@{}ccc@{}}w_1&w_2&w_3\end{array}\right),
\vspace{-.2cm}}
$(a)\equiv\det\{w_a\}$ is a $(1\!\times\!1)$-`minor' of the matrix $W$ and $\widetilde{\eta}\equiv(\widetilde\eta_1\,\,\widetilde\eta_2\,\,\widetilde\eta_3)$. The $\delta$-function \mbox{$\delta^{1\times2}\big(W\!\cdot\!\widetilde{\lambda}\big)$} fixes \mbox{$W\mapsto W^*=\widetilde\lambda^{\perp}$} (written above, in (\ref{dual_to_lambda_tilde})).
On the support of the point $W^*\!\in\! G(1,3)$, the remaining $\delta$-functions in (\ref{g13_version_of_vertex}),
\vspace{-.2cm}\eq{\delta^{1\times4}\big(W^*\!\!\cdot\!\widetilde{\eta}\big)\delta^{2\times2}\big(\lambda\!\cdot\!(W^*)^\perp\!\big),\vspace{-.2cm}}
simply become ordinary super-momentum conservation.

A comment is in order here. To make the invariance of the integrand under $GL(1)$ manifest one has to find a $GL(1)$ invariant way of writing $\delta^{2\times2}\big(\lambda\!\cdot\!W^\perp\!\big)$. As usual, this is achieved by introducing auxiliary variables as explained in detail (and more generality) in \mbox{section \ref{twistor_space_and_super_conformal_invariance_subsection}}.

We can of course make the same generalization for the $\mathcal{A}_3^{(2)}$ vertex:
\vspace{-.4cm}\eq{\hspace{-56.25pt}\raisebox{-47.5pt}{\includegraphics[scale=1]{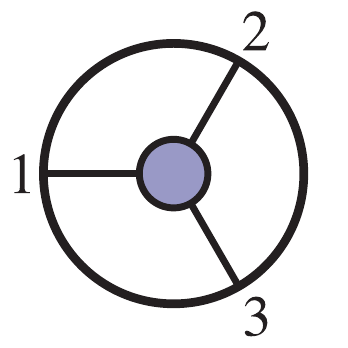}}\hspace{-10pt}\raisebox{-2.5pt}{\text{{\LARGE$\;\Leftrightarrow\;$}}}\mathcal{A}^{(2)}_3=\,\frac{\delta^{2\times4}(\lambda_1\widetilde{\eta}_1+\lambda_2\widetilde{\eta}_2+\lambda_3\widetilde{\eta}_3)}{\ab{1\,2}\ab{2\,3}\ab{3\,1}}\delta^{2\times2}\big(\lambda\!\cdot\!\widetilde\lambda\big).\vspace{-.3cm}}
We can think of this as an integral over an auxiliary $2$-plane \mbox{$B\!\in\! G(2,3)$} according to:
\vspace{-0.2cm}\eq{\mathcal{A}^{(2)}_{3}=\!\int\!\!\frac{d^{2\times3}B}{\mathrm{vol}(GL(2))}\frac{\delta^{2\times4}\big(B\!\cdot\!\widetilde{\eta}\big)}{(12)(23)(31)} \delta^{2\times2}\big(B\!\cdot\!\widetilde{\lambda}\big) \delta^{2\times1}\big(\lambda\!\cdot\!B^\perp\!\big).\vspace{-.2cm}\label{g23_version_of_vertex}}
In this case, we can use the constraint $\delta^{2\times1}\big(\lambda\!\cdot\!B^\perp\!\big)$ to localize the integral over $B$, (somewhat trivially) fixing $B\mapsto B^*=\lambda$, and the minors in the measure trivially become $(1\,2)(2\,3)(3\,1)\!\mapsto\!\ab{1\,2}\ab{2\,3}\ab{3\,1}$. As before, the remaining $\delta$-functions in (\ref{g23_version_of_vertex}),
\vspace{-.2cm}\eq{\delta^{2\times4}\big(B^*\!\!\cdot\!\widetilde{\eta}\big)\delta^{2\times2}\big(B^*\!\!\cdot\!\widetilde\lambda\big),\vspace{-.2cm}\label{lambda_tilde_system_of_constraints}}
encode super-momentum conservation.

The crucial feature of these Grassmannian representations of the three-particle amplitudes is that the constraints on the kinematical data $\lambda$ and $\widetilde \lambda$ are now decoupled, and occur {\it linearly} in the $\delta$-function constraints. This makes it essentially trivial to perform the phase space integral over the internal lines, making any on-shell graph simply a collection of auxiliary $1$-planes $W\!\in\! G(1,3)$ and $2$-planes $B\!\in\! G(2,3)$ associated with the white and black vertices---each carrying with it all the constraints to impose momentum-conservation.

To summarize, for each white vertex involving the (possibly internal) legs $(a,b,c)$ we introduce a $1$-plane \mbox{$W\!\in\! G(1,3)$},
\vspace{-.3cm}\eq{\raisebox{-36.5pt}{\includegraphics[scale=1]{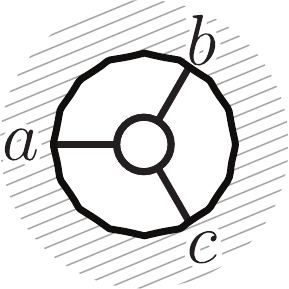}}\quad\raisebox{-1.5pt}{\text{{\LARGE$\;\Leftrightarrow\;$}}}\quad W\equiv\left(\begin{array}{@{}ccc@{}}w_a&w_b&w_c\end{array}\right),\vspace{-.3cm}}
carrying with it an integration measure,
\vspace{-.2cm}\eq{d\Omega_w\equiv\frac{d^{1\times3}W}{\mathrm{vol}{(GL(1))}}\frac{1}{(a)(b)(c)},\vspace{-.2cm}}
and corresponding constraints; similarly, for each black vertex involving legs $(a,b,c)$ we have a plane \mbox{$B\!\in\! G(2,3)$},
\vspace{-.3cm}\eq{\hspace{-13.25pt}\raisebox{-36.5pt}{\includegraphics[scale=1]{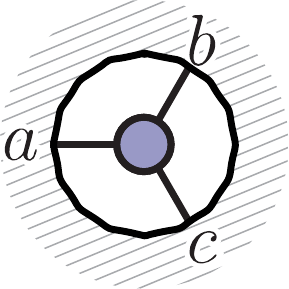}}\quad\raisebox{-1.5pt}{\text{{\LARGE$\;\Leftrightarrow\;$}}}\quad B\equiv\left(\begin{array}{@{}ccc@{}}b_a&b_b&b_c\end{array}\right),\vspace{-.3cm}}
together with its associated integration measure,
\vspace{-.2cm}\eq{d\Omega_b\equiv\frac{d^{2\times3}B}{\mathrm{vol}{(GL(2))}}\frac{1}{(a\,b)(b\,c)(c\,a)}\vspace{-.2cm},}
and corresponding constraints. Each white vertex imposes {\it one} relation among $\widetilde\lambda$'s:
\vspace{-.2cm}\eq{W\!\cdot\!\widetilde\lambda=w_a\widetilde\lambda_a+w_b\widetilde\lambda_b+w_c\widetilde\lambda_c=0;\vspace{-.2cm}}
and each black vertex imposes {\it two} relations (as the columns $b_a$ of $B$ are two-vectors):\\[-2pt]
\vspace{-.4cm}\eq{B\!\cdot\!\widetilde\lambda=b_a\widetilde\lambda_a+b_b\widetilde\lambda_b+b_c\widetilde\lambda_c=0.\vspace{-.2cm}}

Thus, for a graph with $n_b$ black vertices, $n_w$ white vertices, and $n_I$ internal edges, we have a total of $2n_b + n_w$ constraints; from these, one constraint is needed to fix (and eliminate) each internal $\widetilde \lambda_I$---leaving us with a total of:
 \vspace{-.2cm}\eq{k\equiv  2n_b +n_w- n_I\vspace{-.2cm}}
linear constraints relating the external $\widetilde\lambda$'s for any given graph. We may write this collection of constraints as $C\!\cdot\!\widetilde\lambda=0$ for some $(k\!\times\!n)$-matrix $C$, where
\vspace{-.4cm}\eq{n=3n_V-2n_I,\vspace{-.4cm}}with $n_V=n_b+n_w$.
Because these are {\it linear} constraints among the $\widetilde\lambda$'s, the matrix $C$ is of course only well-defined up to an arbitrary re-shuffling of its $k$ equations (a $GL(k)$-transformation of $C$); and so, $C$ actually represents a point in $G(k,n)$! Of course, integrating-out the internal $\widetilde\eta$'s follows identically to the $\widetilde\lambda$, giving us the same final constraints among the external $\widetilde \eta$'s as for the $\widetilde\lambda$'s.

Thus, eliminating the internal $\widetilde \lambda_I$ and $\widetilde\eta_I$ combines all the ``little Grassmannians'' \mbox{$W\!\in\! G(1,3)$} and \mbox{$B\!\in\! G(2,3)$} associated with the vertices, and gives us finally a point in the Grassmannian $G(k,n)$ represented by some matrix $C$ which encodes the relations satisfied among the $\widetilde\lambda$'s and $\widetilde\eta$'s via the $\delta$-functions,
\vspace{-.2cm}\eq{\delta^{k\times4}\big(C\!\cdot\!\widetilde \eta\big) \delta^{k\times2}\big(C\!\cdot\!\widetilde \lambda\big).\vspace{-.2cm}}
Following the same logic, but exchanging each plane $B$ and $W$ for their orthogonal complements, gives us the complementary set of relations involving the $\lambda$'s. Not surprisingly, these are simply given by the $\delta$-functions,
\vspace{-.2cm}\eq{\delta^{2\times(n-k)}\big(\lambda\!\cdot\!C^\perp\!\big).\vspace{-.2cm}}
Geometrically, the ordinary $\delta$-functions constrain the matrix $C$ to be {\it orthogonal to} $\widetilde\lambda$ and to {\it contain} $\lambda$:
\vspace{-.4cm}\eq{\raisebox{-40pt}{\includegraphics[scale=.8]{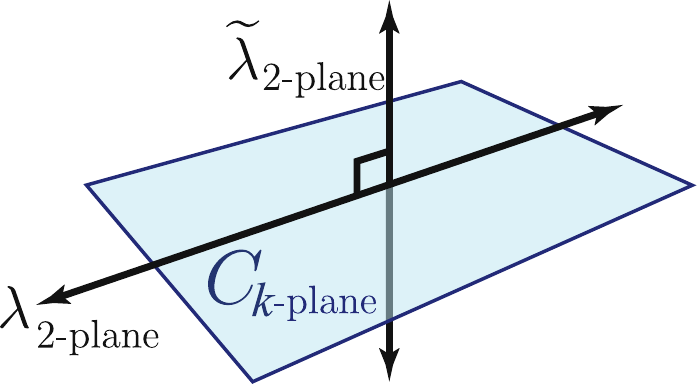}}\label{momentum_space_constraints_first_appearance}\vspace{-.0cm}}

Putting everything together, each on-shell diagram is associated with a differential-form obtained by integration over,
\vspace{-.1cm}\eq{\prod_{\substack{\mathrm{internal}\\\mathrm{edges~}e}}\Big(\frac{1}{{\rm vol}(GL(1)_e)}\Big) \prod_w d \Omega_w \prod_b d \Omega_b\;\;\delta^{k\times4}\big(C\!\cdot\!\widetilde \eta\big)\delta^{k\times2}\big(C\!\cdot\!\widetilde \lambda\big)\delta^{2\times(n-k)}\big(\lambda\!\cdot \!C^\perp\!\big). \vspace{-.4cm}\label{first_form_of_general_on_shell_form}}
Notice that while freely using the $\delta$-functions to fix each internal $\lambda_I$ and $\widetilde\lambda_I$, we have not modded-out by the $GL(1)$-redundancies acting on these momenta (which explains the appearance of the $1/\mathrm{vol}(GL(1))$ factors in (\ref{first_form_of_general_on_shell_form})). It is natural to refer to the net number of auxiliary variables---after modding-out by all these $GL(1)$-redundancies---as the {\it dimension} of the space of configurations $C\!\in\!G(k,n)$. As each vertex carries two auxiliary degrees of freedom, and each $GL(1)$ from the internal lines  can be used to remove one of them, the `dimension' associated with an on-shell graph is simply:
\vspace{-.2cm}\eq{\dim(C)=2n_V-n_I.\vspace{-.2cm}}
We should mention that this can be counted in a more direct way from the graph as follows. Because each on-shell graph is trivalent, we have \mbox{$3n_V=2n_I\pl\,n$} so that \mbox{$\dim(C)=2n_V\,\mi\,n_I=n_I\,\mi\,n_V\pl\,n$}; and restricting our attention to planar graphs, Euler's formula tells us that $(n_F\,\mi\,n)\,\mi\,n_I\pl\,n_V=1$ (where $n_F$ is the number of faces of the graph {\it including} the $n$ faces of the boundary). Putting these two facts together shows that:
\vspace{-.2cm}\eq{\dim(C)=n_F-1.\vspace{-.2cm}}
We will soon see that this is not an accident: there is a natural way in which the degrees of freedom associated with a graph are encoded by its {\it faces}.

So far, we've described in general terms how to compute the differential-form associated with a given on-shell graph. In the next subsection, we will describe how this can be done systematically using only two very simple, elementary operations; and in \mbox{section \ref{boundary_measurements_section}}, we'll show how these two operations can be efficiently automated to construct an explicit representative of the plane $C$ expressed in terms of variables associated with either a graph's {\it edges} or {\it faces}.

\newpage
\subsection{Amalgamation of On-Shell Diagrams}\label{amalgamation_subsection}

General on-shell diagrams can be built-up in steps from more elementary ones using two simple operations: {\it direct-products} and {\it projections}. Collectively, we refer to this step-wise construction of more complicated diagrams from simpler ones as {\it amalgamation} (see \cite{FG3} for a mathematical construction and \cite{Kaplan:2009mh,Mason:2009qx} for some early steps in the physical setup). In this subsection, we describe both operations in turn, and show how they completely determine the $k$-plane $C$ associated with any on-shell graph. Since all the $GL(k)$-invariant information about $C$ is given by the ratios of its minors, $(a_1\cdots a_k)/(b_1\cdots b_k)$, it suffices for us to simply describe how these two primitive
operations act on the minors of $C$.

The first operation is rather trivial: starting with any two diagrams, we can take their {\it direct-product}:
\vspace{-.2cm}\eq{\hspace{1.05cm}\raisebox{-50pt}{\includegraphics[scale=1]{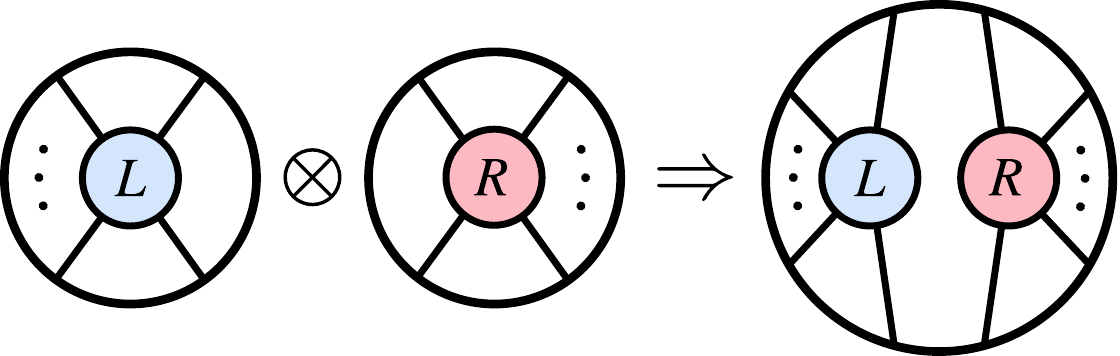}}\vspace{-.2cm}}
If the left-graph is associated with the plane $C_L\!\in\!G(k_L,n_L)$, and the right-graph is associated with the plane $C_R\!\in\!G(k_R,n_R)$, the direct-product produces a plane \mbox{$C_L\otimes C_R\mapsto C\!\in\!G(k_L\pl\,k_R,n_L\pl\,n_R)$} according to:
\vspace{-.2cm}\eq{\hspace{0.85cm}\left(\raisebox{-16.5pt}{\includegraphics[scale=1]{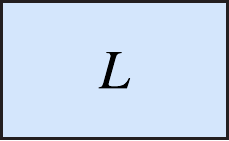}}\right)\!\bigotimes\!\left(\raisebox{-16.5pt}{\includegraphics[scale=1]{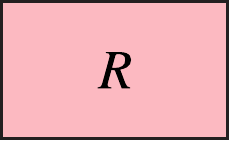}}\right)\text{{\Huge$\Rightarrow$}}\left(\begin{array}{@{}c@{$\hspace{-1pt}$}c@{}}\raisebox{-16.5pt}{\includegraphics[scale=1]{left_matrix}}&\text{{\LARGE$0$}}\\[-1pt]\text{{\LARGE$0$}}&\raisebox{-16.5pt}{\includegraphics[scale=1]{right_matrix}}\end{array}\right)\hspace{-1.05cm}\vspace{-.2cm}}
The non-vanishing minors of $C$ are easily expressed in terms of those of $C_L$ and $C_R$:\\[-10pt]
\vspace{-.1cm}\eq{\left.(a_1\cdots a_{k_L}\,b_1\cdots b_{k_R})\right|_{C}=\left.(a_1\cdots a_{k_L})\right|_{C_L}\!\!\times\left.(b_1\cdots b_{k_R})\right|_{C_R}.\vspace{-.1cm}}

The second operation, {\it projection}, is more interesting. It corresponds to the identification of two (external) legs---say $A$ and $B$---of a graph:
\vspace{-.2cm}\eq{\raisebox{-50pt}{\includegraphics[scale=1]{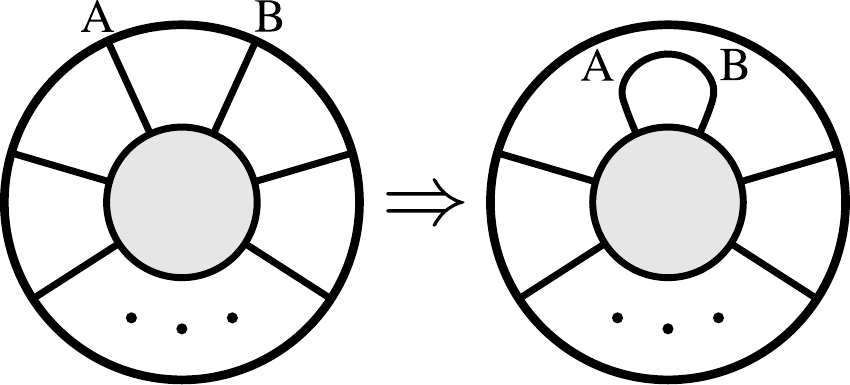}}\vspace{-.2cm}}
We call this operation ``projection'' because it takes a plane $C\!\in\!G(k\pl1,n\pl2)$, and produces a plane $\hat{C}\!\in\!G(k,n)$, which is the {\it projection of} $C$ onto the quotient of the column space of $C$ modulo $(c_A\,\mi\,c_B)$. This follows directly from how the plane $C$ associated with an on-shell graph is interpreted geometrically as constraints imposed on the external momenta.

For convenience, let us suppose that the $n\,\pl\,2$ particles of the configuration before projection are ordered $(A,B,1,\ldots,n)$. Then the minors of the projection's image $\hat{C}\!\in\!G(k,n)$ will be given in terms of the minors of \mbox{$C\!\in\!G(k\pl1,n\pl2)$} according to:
\vspace{-.2cm}\eq{\left.(a_1\cdots a_k)\right|_{\hat{C}}=\left.(A\,a_1\cdots a_k)\right|_{C}+\!\left.(B\,a_1\cdots a_k)\right|_{C}.\vspace{-.2cm}}

Let us consider a simple case where these two operations are used to construct an on-shell graph. For example, consider the sequence,
\vspace{-.2cm}\eq{\hspace{-1.5cm}\raisebox{-50pt}{\includegraphics[scale=1]{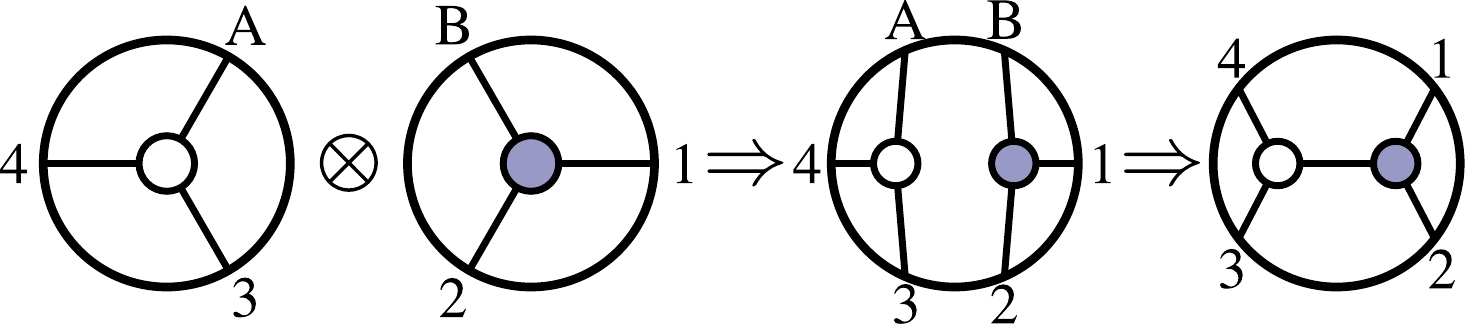}}\hspace{-1cm}\nonumber\vspace{-.2cm}}
which builds-up the $4$-particle factorization graph by first taking the direct-product of $W\!\in\!G(1,3)$ and $B\!\in\!G(2,3)$ to produce a graph associated with a plane \mbox{$\hat{C}\!\in\!G(3,6)$}, then merge legs $A$ and $B$ to produce the final graph associated with a plane \mbox{$C\!\in\!G(2,4)$}. As we have described, minors of the final plane $C\!\in\!G(2,4)$ are fully specified by those of its constituents; e.g.,
\vspace{-.2cm}\eqs{&\left.(13)\right|_{C}=\left.(A13)\right|_{\hat{C}}+\!\left.(B13)\right|_{\hat{C}}=0+\!\left.(B1)\right|_{B}\!\!\times\! \left.(3)\right|_{W};\\\hspace{-.75cm}\mathrm{and}\quad&\left.(24)\right|_{C}=\left.(A24)\right|_{\hat{C}}+\!\left.(B24)\right|_{\hat{C}}=0+\!\left.(B2)\right|_{B}\!\!\times\! \left.(4)\right|_{W}.\vspace{-.2cm}}

Let us look at one more interesting example: the amalgamation of diagrams generating the $4$-particle tree-amplitude:
\vspace{-.2cm}\eq{\raisebox{-50pt}{\includegraphics[scale=1]{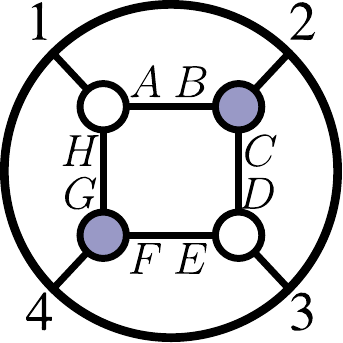}}\vspace{-.2cm}}
Following the amalgamation rules described above, we find, for example that\vspace{.0cm}
\vspace{-.2cm}\eq{\frac{(24)}{(13)} = \left(\frac{(F4) (H)}{(GF)(1)} \right) \left(\frac{(B2)(D)}{(CB)(3)}\right) + \left(\frac{(C2)(A)}{(BC)(1)} \right) \left(\frac{(G4)(E)}{(GF) (3)} \right).\vspace{-.2cm}}

Notice that the amalgamation picture makes it clear that $C$ will only depend on special combinations of the minors of the matrices associated with its constituent vertices. This ultimately stems from the fact that the only $GL(k)$-invariant data associated with the vertices themselves are the ratios of minors. These appear, for example, as the face variables of the three-particle diagrams:
\vspace{-0.4cm}\eq{\raisebox{-1.75cm}{\includegraphics[scale=1]{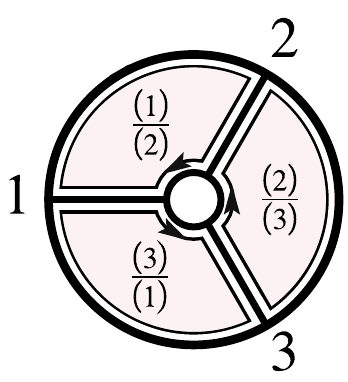}}\qquad\mathrm{and}\qquad\raisebox{-1.75cm}{\includegraphics[scale=1]{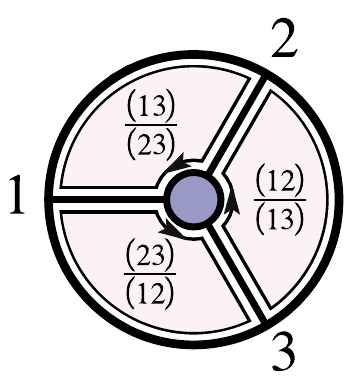}}\vspace{-0.35cm}\label{3pt_face_variables}}
Here, we have used arrows to show how the ratios transform under the little group.

Now, a very simple but important observation is that the final point in $G(k,n)$ obtained from amalgamation must obviously be completely invariant under the little group rescaling of any internal line. This means that only combinations of minors that are invariant under these scaling are ultimately relevant to our description of $C$. Graphically, it is clear that these are given by products of such ratios, as following along the boundary of a face we form a closed path. A face variable, then, can be built as the product of these variables along its boundary. To illustrate the point, consider the following graph---associated with a generic plane in $G(2,5)$:
\vspace{-0.2cm}\eq{\raisebox{-72pt}{\includegraphics[scale=1]{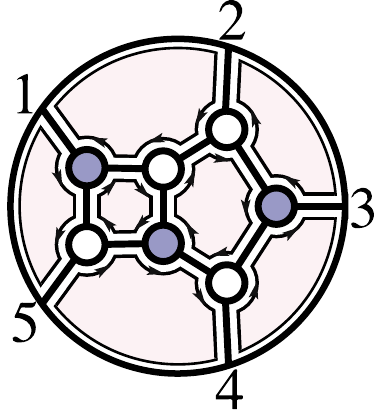}}\label{g25_faces_and_plaquette_boundaries}\vspace{-.2cm}}

Thus, while the variables describing the matrix $C$ can be constructed from the variables of the planes $B$ and $W$ attached to each vertex, we may alternatively view $C$ as being described by variables $f_i$ associated with its {\it faces}. Note that the product of the face variables for each $G(1,3)$ $G(2,3)$ vertex is manifestly equal to 1 (see (\ref{3pt_face_variables})); and so, it is easy to see that there are only {\it two} independent degrees of freedom per vertex---matching our calculation that $\dim(C)=n_F\,\mi\,1$. This clearly persists to larger diagrams, ensuring that $\prod_if_i=1$, which always accounts for the ``minus 1'' in the formula for the dimension of $C$. And so, the degrees of freedom are all but one of the face variables, say $f_*$. Rescaling $f_i\!\mapsto\!\hat{f}_i\equiv f_i/f_*$, the integration measure (\ref{first_form_of_general_on_shell_form}) for the auxiliary parameters  in $C$ becomes simply,
\vspace{-.2cm}\eq{\prod_{\substack{\mathrm{internal}\\\mathrm{edges~}e}}\Big(\frac{1}{{\rm vol}(GL(1)_e)}\Big) \prod_w d \Omega_w \prod_b d \Omega_b=\prod_{\substack{\text{rescaled}\\\text{faces}~\hat{f}_i}}\frac{d\hat{f}_i}{\hat{f}_i}.\vspace{-.2cm}}

\newpage
\subsection{``Boundary Measurements'' and Canonical Coordinates}\label{boundary_measurements_section}
Let us now turn to the problem of explicitly determining a matrix representative $C$ associated with a given on-shell graph. We will first do this in a very efficient---but somewhat overly redundant---way by attaching variables $\alpha_e$ to all the {\it edges} of a graph; and then, we will see how this procedure can be translated (with less redundancy) in terms of variables attached to a graph's {\it faces}.

One strategy for explicitly constructing the $k$-plane $C$ encoding the system of constraints (\ref{momentum_space_constraints_first_appearance}) associated with an on-shell graph is to put the degrees of freedom associated with each vertex in a way which allows us to eliminate all internal momenta as efficiently as possible. Of course, each vertex carries with it only two degrees of freedom. But it turns out to be useful to introduce an additional $GL(1)$-redundancy at each vertex, so that every {\it leg} attached to a given vertex carries its own degree of freedom (making it easier to pair-up the degrees of freedom attached to internal lines between vertices). To further simplify the elimination of internal momenta from the ultimate system of equations relating the $\widetilde\lambda$, it will be helpful also to provide an orientation to each edge, so that each white (black) has one (two) edges directed inward. With these decorations, each white vertex corresponds to:
\vspace{-0.4cm}\eq{\hspace{-0.05cm}\raisebox{-1.7cm}{\includegraphics[scale=1]{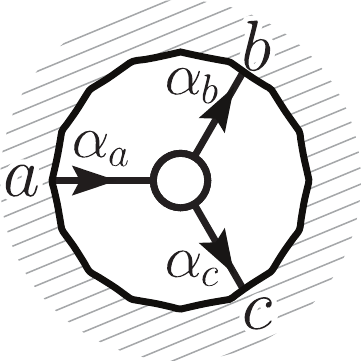}}\hspace{15pt}\text{{\Large$\Leftrightarrow$}}\hspace{10pt} W\equiv\big(\begin{array}{@{}cc@{$\;$}c@{}}a\;&b&c\;\\[-3pt]\hline \alpha_a^{-1}&\mi\alpha_b\,\,&\!\mi\alpha_c\!\,\,\\[-3pt]&\end{array}\big)\hspace{5.5pt}\text{{\Large$\Rightarrow$}}\hspace{11pt}\widetilde\lambda_a\!=\!\alpha_a(\alpha_b\widetilde\lambda_b+\alpha_c\widetilde\lambda_c);\vspace{-0.0cm}\hspace{-0cm}\label{g13_directed_vertex}\vspace{-.3cm}}
and each black-vertex corresponds to:
\vspace{-0.4cm}\eq{\hspace{-0.05cm}\raisebox{-1.7cm}{\includegraphics[scale=1]{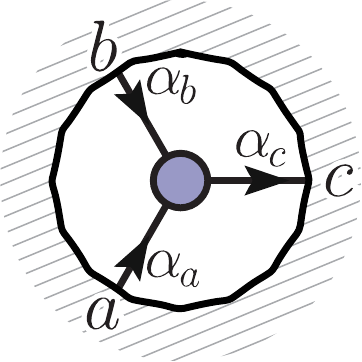}}\hspace{15pt}\text{{\Large$\Leftrightarrow$}}\hspace{10pt} B\equiv\left(\rule{0pt}{15pt}\right.\!\!\begin{array}{@{}cc@{$\;$}c@{}}a\;&b&c\;\\[-3pt]\hline\alpha_a^{-1}&0&\mi\alpha_c\,\,\\0&\!\,\,\alpha_b^{-1}&\mi\alpha_c\,\,\\[-3pt]&\end{array}\!\!\!\left.\rule{0pt}{15pt}\right)\hspace{0pt}\text{{\Large$\Rightarrow$}}\hspace{0pt}\left\{\!\begin{array}{c}\widetilde\lambda_a\!=\!\alpha_a\alpha_c\widetilde\lambda_c\\\widetilde\lambda_b\!=\!\alpha_b\alpha_c\widetilde\lambda_c\end{array}\right\}. \vspace{-0.3cm}\hspace{-0cm}\label{g23_directed_vertex}}

Decorating a graph in this way is called giving it a {\it perfect orientation}; and it is a general fact that all two-colored, trivalent graphs {\it relevant to physics} can be given a perfect orientation.

(The only graphs which cannot be given a perfect orientation are those which contain a sub-graph with $k\leq0$ or $k\geq \nu$ (where $\nu$ denotes the number of legs of the {\it sub}-graph). This obstruction is closely tied to an inability to eliminate some internal line's $\lambda_I$ or $\widetilde\lambda_I$ from the complete system of equations. But this subtlety plays no role in our story, as the differential-form associated with such a graph {\it always} vanishes due to the $\widetilde\eta_I$ integration. And so, these `pathological' diagrams never contribute to physically-relevant processes.)

Once we have given a perfect orientation, the system of equations $C\!\cdot\!\widetilde{\lambda}$ becomes trivial to construct: each vertex can be viewed as giving an equation which expands the $\widetilde{\lambda}$'s of the vertex's {\it sources} in terms of those of its {\it sinks}. Combining all such equations then gives us an expansion of the external sources' $\widetilde\lambda$'s in terms of those of the external sinks. Notice that when identifying two legs, $(I_{in},I_{out})$ during amalgamation the degree of freedom lost in the process is accounted for via the replacement of the pair $(\alpha_{I_{in}},\alpha_{I_{out}})$ with the single variable $\alpha_I\equiv \alpha_{I_{in}}\alpha_{I_{out}}$.

If we denote the external sources of a graph by $\{a_1,\ldots,a_k\}\equiv A$, then the final linear relations imposed on the $\widetilde{\lambda}$'s can easily be seen to be given by,
\vspace{-.2cm}\eq{\widetilde{\lambda}_A+c_{A\,a}\widetilde{\lambda}_{a}=0,\vspace{-.2cm}}
with
\vspace{-.1cm}\eq{c_{A\,a}=-\sum_{\Gamma\in\{A\rightsquigarrow a\}}\prod_{e\in\Gamma}\alpha_e\;,\vspace{-.2cm}\label{general_link_in_terms_of_edges}}
and where $\Gamma\in\{A\!\rightsquigarrow\!a\}$ is any (directed) path from $A$ to $a$ in the graph. (If there is a closed, directed loop, then the geometric series should be summed---we will see an example of this in (\ref{g24_edge_variables_under_square_move}).) The entries of the matrix  $c_{A\,a}$ are called the ``boundary measurements'' of the on-shell graph. The on-shell form on $C(\alpha)\!\in\!G(k,n)$ can then be written in terms of the variables $c_{A\,a}$ according to:
\vspace{-.1cm}\eq{\left(\prod_{\mathrm{vertices~}v}\!\!\frac{1}{\mathrm{vol}(GL(1)_v)}\right)\left(\prod_{\mathrm{edges~}e}\frac{d\alpha_e}{\alpha_e}\right)\delta^{k\times4}(C\!\cdot\!\widetilde{\eta})\delta^{k\times2}(C\!\cdot\!\widetilde{\lambda})\delta^{2\times(n-k)}(\lambda\!\cdot\!C^{\perp}\!)\,.\vspace{-.2cm}\label{general_on_shell_form_in_edge_variables}}

Let us consider a simple example to see how this works. Consider the following perfectly oriented graph:
\vspace{-0.15cm}\eq{\raisebox{-45pt}{\includegraphics[scale=1]{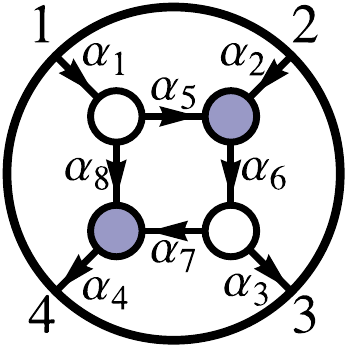}}\label{g24_boundary_measurement_example}\vspace{-.15cm}}
Using the equations for each directed $3$-particle vertex, we can easily expand the $\widetilde\lambda$ of each source---legs $1$ and $2$---in terms of those of the sinks---legs $3$ and $4$; e.g.,
\vspace{-.2cm}\eq{\widetilde{\lambda}_2=\alpha_2\alpha_6(\alpha_3\widetilde{\lambda}_3+\alpha_7(\alpha_4\widetilde{\lambda}_4)).\vspace{-.2cm}}
Such expansions obviously result in (\ref{general_link_in_terms_of_edges}): the coefficient $c_{A\,a}$ of $\widetilde{\lambda}_a$ in the expansion of $\widetilde{\lambda}_A$ is simply (minus) the product of all edge-variables $\alpha_e$ along any path \mbox{$\Gamma\in\{A\!\rightsquigarrow\! a\}$}. Doing this for all the $c_{A\,a}$ of our example above, we find,
\vspace{-.5cm}\eq{\hspace{-.2cm}\begin{array}{cccc}\\[5pt]\raisebox{-45pt}{\includegraphics[scale=1]{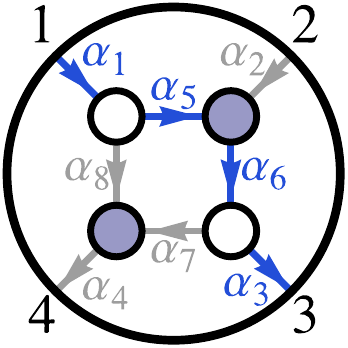}}&\raisebox{-45pt}{\includegraphics[scale=1]{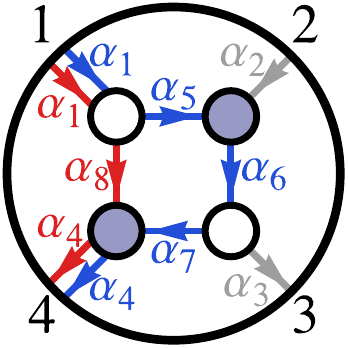}}&\raisebox{-45pt}{\includegraphics[scale=1]{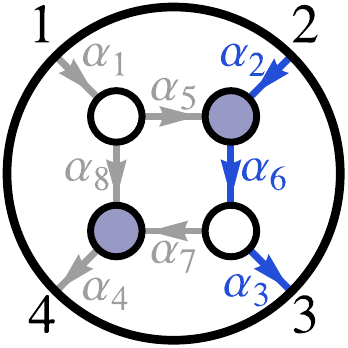}}&\raisebox{-45pt}{\includegraphics[scale=1]{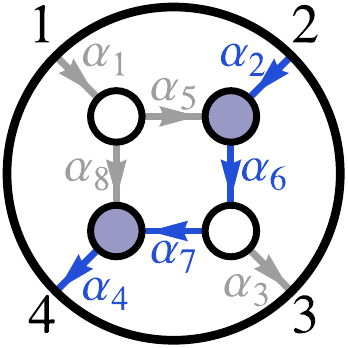}}\\[-10pt]\mi\,c_{1\,3}={\color[rgb]{.13725,.30588,.847}\alpha_1\,\alpha_5\,\alpha_6\,\alpha_3}&\begin{array}{l}\\\mi\,c_{1\,4}={\color[rgb]{.13725,.30588,.847}\alpha_1\,\alpha_5\,\alpha_6\,\alpha_7\,\alpha_4}\\[-5pt]\phantom{\mi\,c_{1\,4}=}+{\color[rgb]{.8588,.1294,.1333}\alpha_1\,\alpha_8\,\alpha_4}\end{array}&\mi\,c_{2\,3}={\color[rgb]{.13725,.30588,.847}\alpha_2\,\alpha_6\,\alpha_3}&\mi\,c_{2\,4}={\color[rgb]{.13725,.30588,.847}\alpha_2\,\alpha_6\,\alpha_7\,\alpha_4}\\[-30pt]\end{array}\nonumber\vspace{-.25cm}}
Thus, the final relations involving the $\widetilde\lambda$'s is encoded by the matrix $C\equiv\left(\begin{array}{@{}cccc@{}}1&0&c_{1\,3}&c_{1\,4}\\0&1&c_{2\,3}&c_{2\,4}\end{array}\right)$.

Notice that only certain combinations of edge-weights appear in the equations. This happens for a very simple---and by now familiar---reason. Think of the $GL(1)$-redundancy of each vertex as a gauge-group, with the variable of a directed edge charged as a ``bi-fundamental'' of the $GL(1)\!\times\!GL(1)$ of the vertices it connects. Since the configuration $C$ must be invariant under these ``gauge groups'',  only gauge-invariant combinations of the edge variables can appear. And just as we saw in the previous subsection, these combinations are those familiar from lattice gauge theory and can be viewed as encoding the flux though each closed loop in the graph---that is, each of its faces. Fixing the orientation of each face to be clockwise, the flux through it is given by the product of $\alpha_e$ ($\alpha_e^{-1}$) for each aligned (anti-aligned) edge along its boundary. For future convenience, we define the face {\it variables} $f_i$ to be {\it minus} this product.

Applying this to the example above, we find:
\vspace{-0.0cm}\eq{\hspace{-.75cm}\raisebox{-45pt}{\includegraphics[scale=1]{g24_with_bcfw_orientations_and_edge_variables}}\;\text{{\LARGE$\Leftrightarrow$}}\;\raisebox{-45pt}{\includegraphics[scale=1]{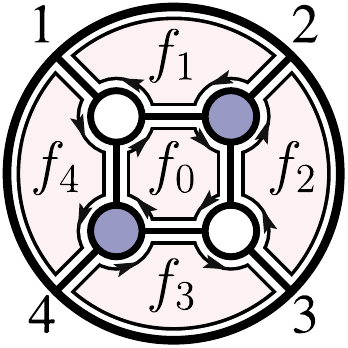}}\;\;\mathrm{with}\quad\begin{array}{lll}\cline{2-2}&\multicolumn{1}{|c|}{\begin{array}{c} \!\!\!\!\phantom{=}f_1\!=\!\\[-0pt]\mi\,\alpha_1^{-1}\alpha_5^{-1}\alpha_2\phantom{\mi}\end{array}}\\\hline\multicolumn{1}{|c|}{\begin{array}{c} \!\!\!\!\phantom{=}f_4\!=\!\\[-0pt]\mi\,\alpha_4\,\alpha_8\,\alpha_1\phantom{\mi}\end{array}}&
\multicolumn{1}{c|}{\begin{array}{c} \!\!\!\!\phantom{=}f_0\!=\!\\[-0pt]\mi\,\alpha_5\,\alpha_6\,\alpha_7\,\alpha_8^{-1}\phantom{\mi}\end{array}}&\multicolumn{1}{c|}{\begin{array}{c} \!\!\!\!\phantom{=}f_2\!=\!\\[-0pt]\mi\,\alpha_2^{-1}\alpha_6^{-1}\alpha_3^{-1}\phantom{\mi}\end{array}}\\\hline&\multicolumn{1}{|c|}{\begin{array}{c} \!\!\!\!\phantom{=}f_3\!=\!\\[-0pt]\mi\,\alpha_3\,\alpha_7^{-1}\alpha_4^{-1}\phantom{\mi}\end{array}}\\\cline{2-2}\end{array}\nonumber\vspace{-.0cm}}
The boundary-measurements $c_{A\,a}$ can then be expressed in terms of the faces by
\vspace{-.2cm}\eq{c_{A\,a}=-\sum_{\Gamma\in\{A\rightsquigarrow a\}}\prod_{f\in\hat{\Gamma}}(-f)\;,\vspace{-.4cm}\label{general_link_in_terms_of_faces}}
where $\hat{\Gamma}$ is the `clockwise' closure of $\Gamma$. (If there are any closed, directed loops, the geometric series of faces enclosed should be summed.)
The faces of course over-count the degrees of freedom by one, and this is reflected by the fact that $\prod_i (-f_i)=1$.
\eq{\hspace{-.0cm}\begin{array}{cccc}\raisebox{-45pt}{\includegraphics[scale=1]{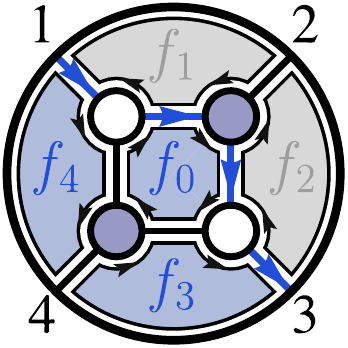}}&\raisebox{-45pt}{\includegraphics[scale=1]{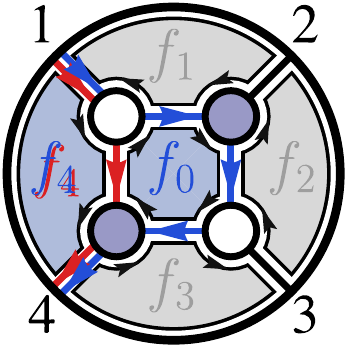}}&\raisebox{-45pt}{\includegraphics[scale=1]{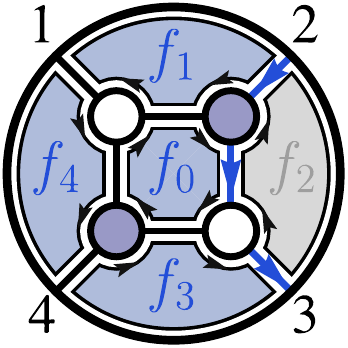}}&\raisebox{-45pt}{\includegraphics[scale=1]{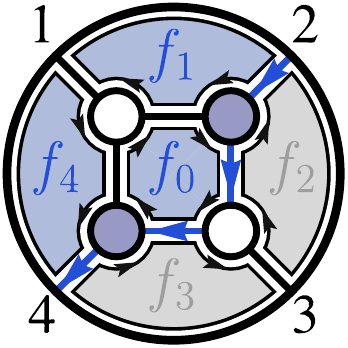}}\\[-10pt]\mi\,c_{1\,3}=\mi\,{\color[rgb]{.13725,.30588,.847}f_0\,f_3\,f_4}&\begin{array}{l}\\\mi\,c_{1\,4}={\color[rgb]{.13725,.30588,.847}f_0\,f_4}\\[-5pt]\phantom{\mi\,c_{1\,4}=}-{\color[rgb]{.8588,.1294,.1333}f_4}\end{array}&\mi\,c_{2\,3}={\color[rgb]{.13725,.30588,.847}f_0\,f_1\,f_3\,f_4}&\mi\,c_{2\,4}={\color[rgb]{.13725,.30588,.847}f_0\,f_1\,f_4}\end{array}\nonumber}

\newpage
\subsection{Coordinate Transformations Induced by Moves and Reduction}\label{coordinate_transformations_induced_by_moves}
Let us now examine how the identification of diagrams via merge-operations, square-moves, and bubble-deletion is reflected in the coordinates---the edge- or face-variables ---used to parameterize cells $C\!\in\!G(k,n)$. As usual, the simplest of these is the merge/un-merge operation which trivially leaves any set of coordinates unchanged. For example, in terms of the face variables, it is easy to see that
\vspace{-0.2cm}\eq{\mbox{\hspace{-2cm}\raisebox{-46.5pt}{\includegraphics[scale=1]{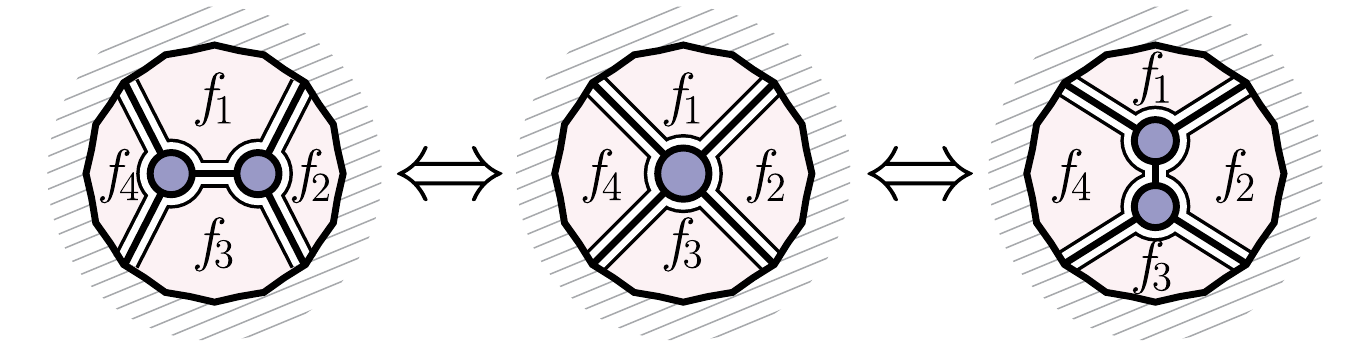}}\hspace{-2cm}}\label{merger_rule}\vspace{-.2cm}}
The square-move is more interesting. It is obvious that squares with opposite coloring both give us a generic configuration in $G(2,4)$, but (as we will soon see), the square-move acts rather non-trivially on coordinates used to parameterize a cell,
\vspace{-0.2cm}\eq{\mbox{\hspace{-0.0cm}\raisebox{-46.5pt}{\includegraphics[scale=1]{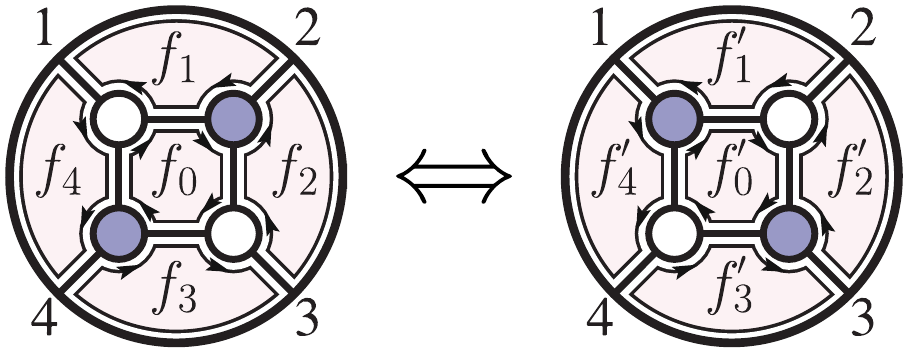}}\hspace{-00cm}}\label{g24_face_variables_under_square_move}\vspace{-.2cm}}

Let us start by determining the precise way the face-variables $f_i$ and $f_i'$ of square-move related diagrams are related to one another. To do this, we will provide perfect orientations (decorated with edge variables) for both graphs, allowing us to compare the resulting boundary-measurement matrices in each case.  Because these two boundary measurement matrices must represent the same point in $G(2,4)$, we will be able to explicitly determine how all the various coordinate charts are related---including the relationship between the variables $f_i$ and $f_i'$. Our work will be considerably simplified if we remove the $GL(1)$-redundancies from each vertex, leaving us with a non-redundant set of edge-variables. Of course, {\it any} choice of perfect orientations for the graphs, and any fixing of the $GL(1)$-redundancies would suffice for our purposes; but for the sake of concreteness, let us consider the following:
\vspace{-1.25cm}\eq{\mbox{\hspace{-0.15cm}$\begin{array}{c@{$\qquad$}c}\\[15pt]\multicolumn{2}{l}{\raisebox{-45pt}{\includegraphics[scale=1]{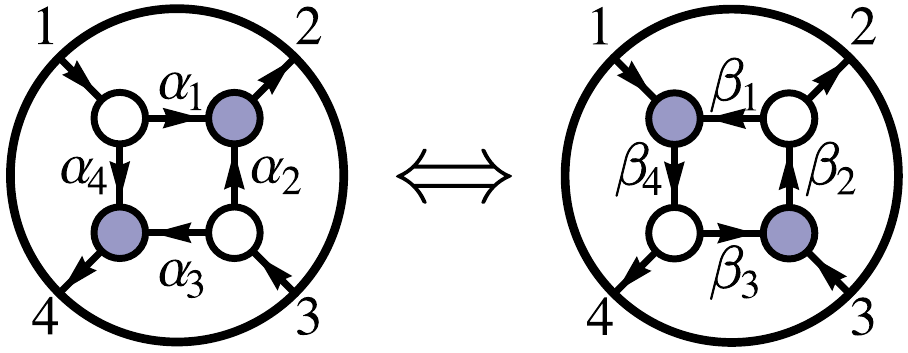}}}\\[1pt]~\hspace{1pt}\left(\begin{array}{@{}c@{$\;\;$}c@{$\;\;$}c@{$\;\;$}c@{}}1&\mi\alpha_1\,\,&0&\mi\alpha_4\\0&\mi\alpha_2\,\,&1&\mi\alpha_3\end{array}\right)&~\hspace{8.5pt}\left(\begin{array}{@{}c@{$\;\;$}c@{$\;\;$}c@{$\;\;$}c@{}}1&\mi\beta_2\beta_3\beta_4\Delta\,\,&0&\mi\beta_4\Delta\\0&\mi\beta_2\Delta\,\,&1&\mi\beta_1\beta_2\beta_4\Delta\end{array}\right)\\[-20pt]\end{array}$\hspace{-1cm}}\label{g24_edge_variables_under_square_move}\vspace{-.2cm}}
Here, we have written the matrices $C(\alpha)$ and $C(\beta)$ obtained as boundary-measurements as discussed in \mbox{section \ref{boundary_measurements_section}}. The factor $\Delta$ in $C(\beta)$ is given by, \vspace{-.2cm}\eq{\Delta\equiv \frac{1}{1-\beta_1\beta_2\beta_3\beta_4},\vspace{-.05cm}}
and arises from summing the infinite geometric series of paths which circle-around the internal loop of the perfectly-oriented graph. The edge-variables in (\ref{g24_edge_variables_under_square_move}) used as coordinates in $G(2,4)$ are closely-related to the face-variables in (\ref{g24_face_variables_under_square_move}).

It is not hard to express the face variables in terms of the edge variables for the two orientations in (\ref{g24_face_variables_under_square_move}). It is easy to see that,
\vspace{-.3cm}\eq{\begin{array}{ll@{$\!\!,\quad$}ll@{$\!\!,\quad$}ll@{$,\quad$}ll@{$\!\!,\quad$}ll}f_0=&\mi\,\alpha_1\,\alpha_2^{-1}\alpha_3\,\alpha_4^{-1}&f_1=&\mi\,\alpha_1^{-1}&f_2=&\mi\,\alpha_2&f_3=&\mi\,\alpha_3^{-1}&f_4=&\mi\,\alpha_4^{-1};\\f_0'=&\mi\,(\beta_1\beta_2\beta_3\beta_4)^{-1}&f_1'=&\mi\,\beta_1&f_2'=&\mi\,\beta_2&f_3'=&\mi\,\beta_3&f_4'=&\mi\,\beta_4\;\;.\end{array}\vspace{-.2cm}}
Because the boundary-measurements must represent the same point in the Grassmannian regardless of whether we use $\alpha$ or $\beta$ coordinates, we see that:
\vspace{-.2cm}\eq{~\quad\left\{\begin{array}{ll}\alpha_1=&\beta_2\beta_3\beta_4\Delta\\\alpha_2=&\beta_2\Delta\\\alpha_3=&\beta_1\beta_2\beta_4\Delta\\\alpha_4=&\beta_4\Delta\end{array}\right\}\raisebox{-3pt}{\text{\Huge$\;\;\Rightarrow\;\;$}}\left\{\begin{array}{lllll}\mi\,\beta_1=&f_1'=&\mi\,\alpha_2^{-1}\alpha_3\,\alpha_4^{-1}\Delta&=f_1f_0\Delta\\\mi\,\beta_2=&f_2'=&\mi\,\alpha_2\Delta^{-1}&=f_2\Delta^{-1}\\\mi\,\beta_3=&f_3'=&\mi\,\alpha_1\,\alpha_2^{-1}\alpha_4^{-1}\Delta&=f_3f_0\Delta\\\mi\,\beta_4=&f_4'=&\mi\,\alpha_4\Delta^{-1}&=f_4\Delta^{-1}\\\phantom{\beta_1}\therefore&f_0'=&\mi\,\alpha_1^{-1}\alpha_2\,\alpha_3^{-1}\alpha_4&=f_0^{-1}\end{array}\right\}.\vspace{-.2cm}}
Observing that $\Delta=(1+f_0'^{-1})^{-1}=(1+f_0)^{-1}$, we therefore conclude that a square-move alters face-variables according to:
\vspace{-.2cm}\eq{\hspace{-1cm}\raisebox{-61.5pt}{\includegraphics[scale=1]{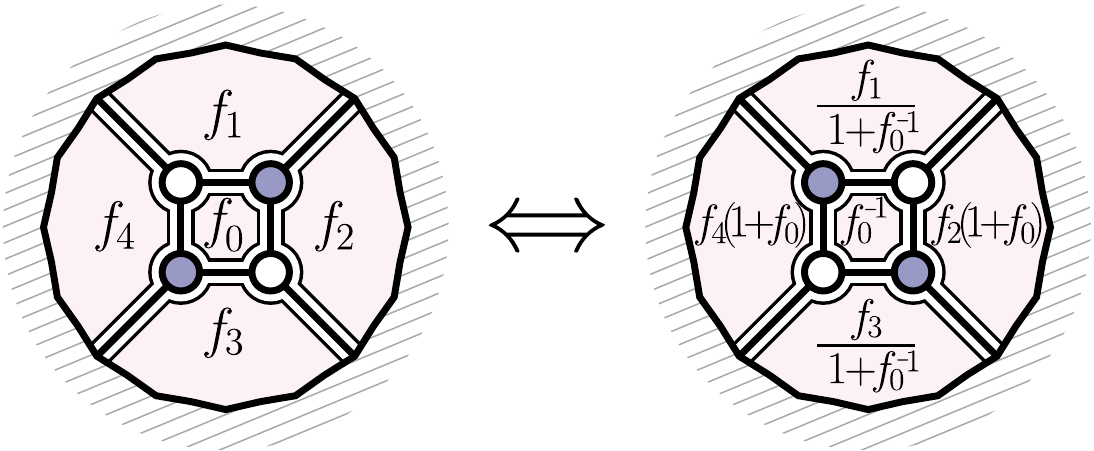}}\hspace{-1.0cm}\vspace{-.1cm}}
This transformation of the face variables is an example of a more general operation related to {\it cluster transformations} as described in section \ref{cluster_coordinates_and_mutations_subsection}. Note that, crucially, our form is invariant under this transformation:
\vspace{-.3cm}\eq{\prod_f \frac{df}{f} = -\prod_{f^\prime} \frac{df^\prime}{f^\prime}\vspace{-.1cm}}
The invariance of the measure (modulo an overall sign) guarantees that the on-shell forms associated with diagrams related by square moves are the same---differing only by a change of coordinates used.

Let us now turn to bubble-deletion. It is easy to see that the following oriented subdiagrams always lead to exactly the same boundary-measurements:
\vspace{-0.2cm}\eq{\hspace{-0.4cm}\raisebox{-61.5pt}{\includegraphics[scale=1]{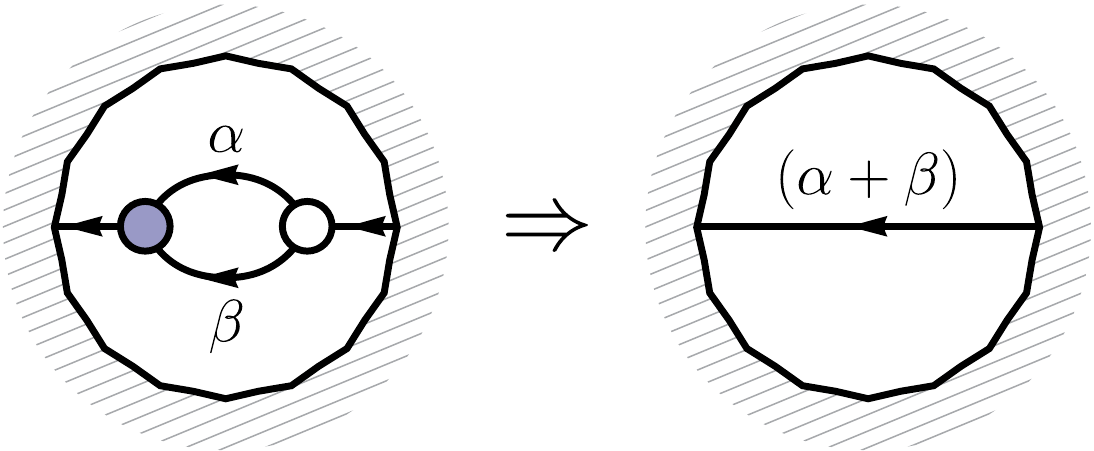}}\label{bubble_deletion_with_edge_variables}\hspace{-.2cm}\vspace{-.2cm}}
Following the same logic used to analyze the square-move, we find that the face-variables of these two diagrams are related by:
\vspace{-0.2cm}\eq{\hspace{-0.4cm}\raisebox{-61.5pt}{\includegraphics[scale=1]{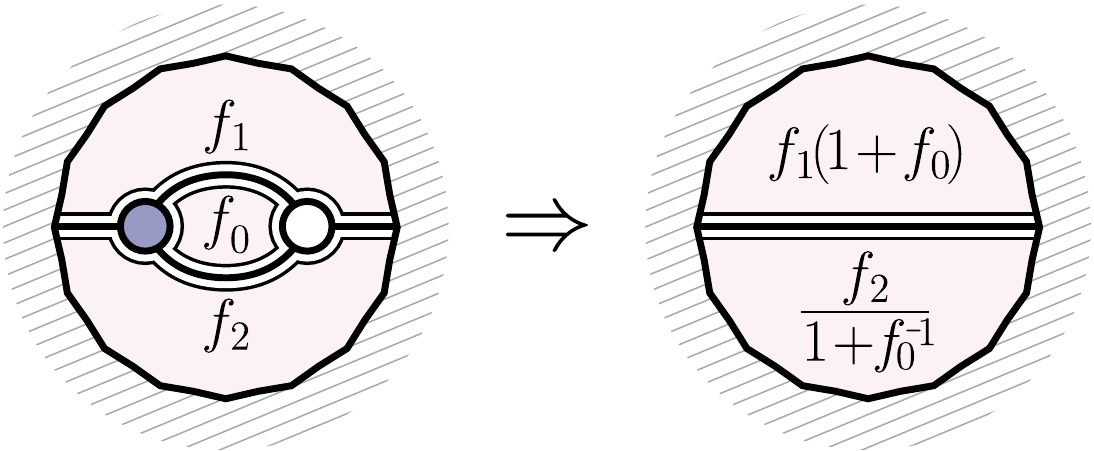}}\label{bubble_deletion_with_face_variables}\hspace{-.2cm}\vspace{-.2cm}}
Note again the crucial fact that the measure is invariant under this transformation:
\vspace{-.2cm}\eq{\frac{df_0}{f_0}\wedge \frac{df_1}{f_1}\wedge \frac{d f_2}{f_2} =-\frac{df_0'}{f_0'}\wedge\frac{df_1^\prime}{f_1^\prime}\wedge\frac{df_2^\prime}{f_2^\prime}\,,\vspace{-.2cm}}
where $f_0'=f_0^{-1}$. The change of variables from $f\!\to\!f^\prime$ eliminates all dependence on $f_0$ associated with the bubble from the final point in the Grassmannian. Of course, the variable $f_0$ {\it remains in the measure}, but it cleanly factors out as an overall prefactor of $d\!\log(f_0)$. As we will see later on, MHV amplitude integrands---to all loop-orders---are always the tree-amplitude, dressed with many additional $d\!\log$-factors arising from bubble-deletion. These ``irrelevant'' factors in the measure encode the internal degrees of freedom of the loop-momenta.

If instead of the {\it integrand} for scattering amplitudes, we were interested in the {\it residues} of the on-shell differential form---to compute, e.g.\ ``leading singularities''---then these ``irrelevant'' $d\!\log$-factors {\it really are irrelevant}: {\it any} residue involving them will give either one or zero.

Due to reduction, then, the number of interesting {\it residues} of general (non-reduced) on-shell diagrams turns is in fact finite despite the seemingly-infinite number of possible diagrams. Notice that in our way of thinking about `leading singularities' and on-shell diagrams, we've made no distinction whatsoever between what have historically been called ``ordinary'' versus ``composite'' objects, \cite{Buchbinder:2005wp,Cachazo:2008dx}. Historically, {\it reducible} on-shell diagrams were those with ``irrelevant'' additional degrees of freedom which could be systematically trivialized-away.

One example of such an on-shell form is the `double-box' involving four-particles; this on-shell diagram has been known to include {\it one} unfixed degree-of-freedom which factorizes-out of diagram trivially upon bubble-deletion:
\vspace{-.0cm}\eq{\hspace{-0.5cm}\raisebox{-32.5pt}{\includegraphics[scale=.75]{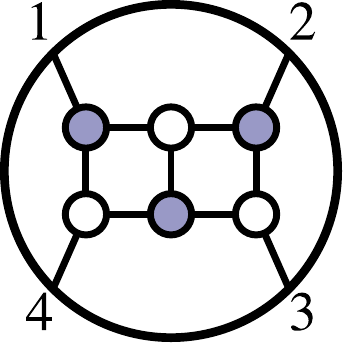}}\text{{\LARGE$\Rightarrow$}}\raisebox{-32.5pt}{\includegraphics[scale=.75]{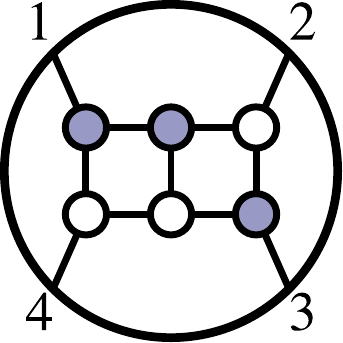}}\text{{\LARGE$\Rightarrow$}}\raisebox{-32.5pt}{\includegraphics[scale=.75]{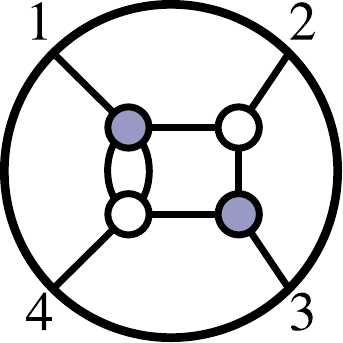}}\text{{\LARGE$\Rightarrow$}}
\raisebox{-32.5pt}{\includegraphics[scale=.75]{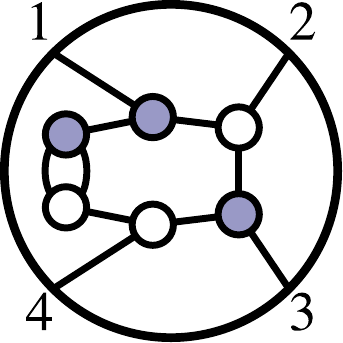}}\text{{\LARGE$\Rightarrow$}}\raisebox{-32.5pt}{\includegraphics[scale=.75]{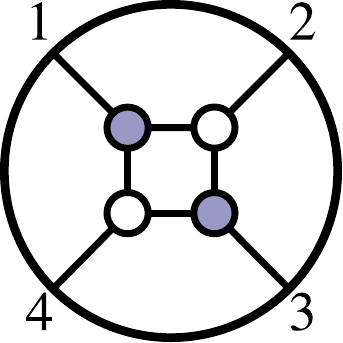}}\hspace{-.5cm}\nonumber\vspace{-.0cm}}
As discussed in generality above, the variable ``lost'' during bubble-deletion is in reality just a bare $d\!\log(\alpha)$ in the measure.


\subsection{Relation to Composite Leading Singularities}\label{Composite leading singularities}
When all the auxiliary degrees of freedom of an on-shell form can be localized by kinematical constraints,  we can think of it as having been obtained by starting with the $(n_F\mi\,n)$-loop integrand for the scattering amplitude, and successively putting (off-shell) Feynman propagators on-shell (`cutting them') until the on-shell diagram is obtained. Such on-shell diagrams are referred to as ``leading singularities''. Thought of in this way,
they are secondary---derived---quantities obtained from the `primary' object, the loop integrand. An important physical point of our present work (discussed more thoroughly in \mbox{section \ref{on_shell_scattering_amplitudes_section}})
is that it is much more fruitful to take the opposite viewpoint: that `loop-integrands' are in fact `derived' from on-shell diagrams. However, since the concept of a ``leading singularity'' will likely be more familiar to most readers, in this subsection we will briefly review how leading singularities have been used to inform us about scattering amplitudes, and discuss in particular the subtle issue of {\it composite} leading singularities---which is closely
related to reducibility. (This discussion is meant only to make contact with this point in previous literature, and isn't especially germane to the rest of our paper.)

The reduction procedure is related to what was called the ``computation of composite leading singularities'' in the physics literature, \cite{Buchbinder:2005wp,Cachazo:2008dx,Cachazo:2008vp,Cachazo:2008hp} (see \cite{Kosower:2011ty,CaronHuot:2012ab,Mastrolia:2012wf} for recent developments). In order to make the connection between the modern and the old procedures transparent let us explain what a composite leading singularity means for the four-point example already examined above. Starting with the diagram with two faces one realizes that any of the two squares actually represents a full four-particle amplitude. Choose the left one for example and draw the equivalent figure,
\vspace{-.2cm}\eq{\raisebox{-46.5pt}{\includegraphics[scale=1]{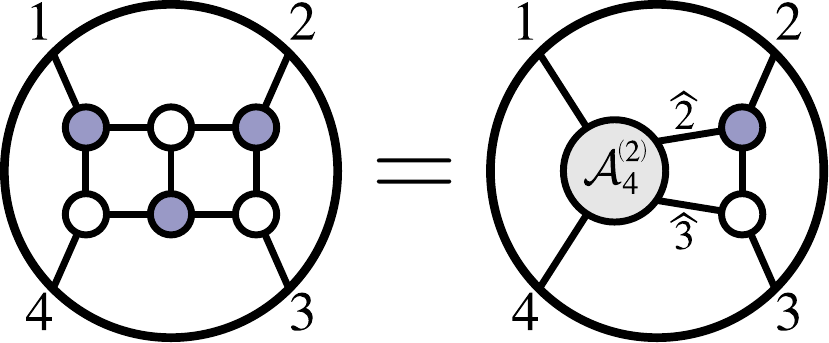}}\vspace{-.2cm}}
At this point the attentive reader can recognize this as a BCFW bridge on a physical scattering amplitude and it is given by the differential form
\vspace{-.2cm}\eq{\raisebox{-46.5pt}{\includegraphics[scale=1]{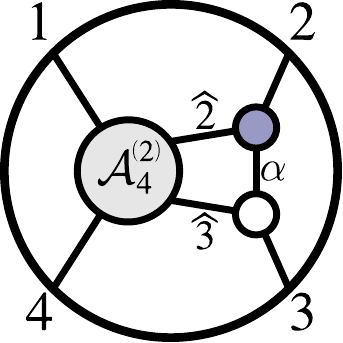}}=\frac{d\alpha}{\alpha}{\cal A}_4^{(2)}\!(\alpha),\vspace{-.2cm}\label{g24_composite}}
where the $\alpha$-dependence of $\mathcal{A}_4^{(2)}$ results from that of the shifted momenta $\hat{2}$ and $\hat{3}$. This on-shell form has only two poles in $\alpha$: a trivial pole at $\alpha=0$, and another where the $\mathcal{A}_4^{(2)}$ factorizes. Of course, as there are {\it only two} poles in the $\alpha$-plane, their residues sum to zero, and hence differ only by a sign; as the $\alpha=0$ residue is manifestly the {\it undeformed} tree-amplitude $\mathcal{A}_4^{(2)}(\alpha=0)$, so is the other (up to a sign).

The composite leading singularity technique was based on the observation that the pole at $(p_{1}+p_{\hat{2}})^2=0$ is guaranteed to be there simply as a pole of the physical  $\mathcal{A}^{(2)}_4(\alpha)$ tree  amplitude. Therefore the pole at $(p_{1}+p_{\hat{2}})^2=0$ , in combination with the other three on-shell conditions on the loop momenta already in the figure, can be used to determine a residue. This gives rise to,
\vspace{-.2cm}\eq{\raisebox{-46.5pt}{\includegraphics[scale=1]{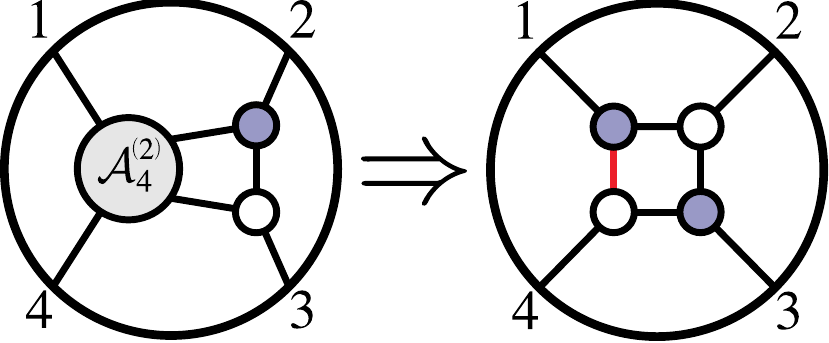}}\vspace{-.2cm}}
which is nothing but the on-shell diagram for a four-point amplitude $\mathcal{A}^{(2)}_4$.

We note in passing that this gives yet another ideal use of bubbles. Suppose that one is given an on-shell diagram corresponding to a leading singularity, i.e., an on-shell diagram which evaluates to an algebraic function of external momenta (conditions for this to happen are discussed in \mbox{section \ref{geometric_origin_of_identities_section}}). Next, apply a BCFW bridge to the diagram and ask what its possible poles and corresponding residues are as a function of the BCFW variable $\alpha$. Let again return to discussing to the same four-particle example. We can ask how could we have known that there was a pole in the `$s_{12}(\alpha)\!\to\!0$ channel' and not it any other channel,  by only manipulating the graph. The answer is already in figure at the end of the previous subsection: find a bubble and the channel of the bubble becomes the pole required by unitarity!

Composite leading singularities were first developed in order to compute two-loop amplitudes following a technique that was very successful at one loop \cite{Britto:2004nc}. While Feynman diagrams are even hard to write down explicitly for loop amplitudes, it is known that loop integrals can be reduced to a linear combination of basic standard integrals \cite{vanNeerven:1983vr}. The idea is then to start with the most general linear combination of such basic integrals and find ways of computing the coefficients. This is known as the ``unitarity-based method'', \cite{Bern:1994zx,Bern:1994cg,Bern:1997nh,Bern:2005iz,Bern:2007ct} (for recent applications of these techniques, see e.g.\ \cite{Ossola:2007ax,Mastrolia:2012wf}). In more modern language, the key idea is to use contour integrals to compute the coefficients. At one loop, $\mathcal{N}\!=\!4$ super Yang-Mills only requires integrals with four propagators. Thus, the four dimensional contour for computing a given coefficient is then obviously defined by the four propagators of the given integral.

At two loops and four particles the basis of integrals must include one such as,
\vspace{-.2cm}\eq{\raisebox{-30.5pt}{\includegraphics[scale=1.35]{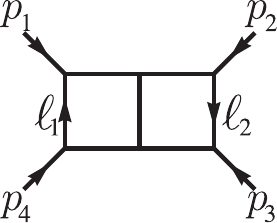}}.\vspace{-.2cm}\label{double_box_integral}}
Now there are eight integration variables but only seven propagators. Naively it seems that this integral does not have any non-vanishing residues. The key observation is that the propagators are non-linear functions of the integration variables and therefore computing the $\ell_1$ integral using the $T^4$ contour defined by the left box gives rise to $1/\big(s_{12}(\ell_2)s_{41}\big)$, which is  $\ell_2$-dependent. This can then be used together with the three-propagators already present on the right to define a second $T^4$ contour and hence a non-vanishing residue. The $\ell_2$-dependent pole, $1/\big(s_{12}(\ell_2)\big)$, generated in this form is precisely what is needed for the new computation to be that of a single scalar box on-shell diagram.

In this way of thinking about things, the existence of composite residues is unexpected, and are made possible from ``hidden'' poles  that are produced by Jacobian factors which appear as residues are taken. In our new picture,
{\it all} the singularities are manifestly exposed in our ``$d\!\log$'' measure for edge or face variables. There is no distinction between ``composite'' and ``ordinary'' singularities, and they are all treated together in a systematic and unified way.

\newpage
\section{Configurations of Vectors and the Positive Grassmannian}\label{configuration_of_vectors_section}

We have seen that every on-shell graph is associated with a $(k\!\times\!n)$-matrix $C$, where a reduced graph with $n_F$ faces gives us an $(n_F\,\mi\,1)$-dimensional sub-manifold of the Grassmannian $G(k,n)$.  We have also seen that the invariant content of an on-shell diagram is given by the permutation which labels it. We will now link these two observations by showing that the sub-manifold in the Grassmannian associated with an on-shell graph is also characterized---for geometric reasons---by the {\it same} permutation which labels the graph.

Our discussion will be most transparent if we think of the Grassmannian in a complementary way to our presentation so far: instead of viewing the $k \times n$ matrix $C$ {\it horizontally}, as a $k$-plane spanned by its rows, we want to now view $C$ {\it vertically}---as a collection of $n$, $k$-dimensional columns.
The $GL(k)$-invariant data to describe any configuration are ratios of minors:
\vspace{-0.25cm}\eq{\frac{(a_1\cdots a_k)}{(b_1\cdots b_k)},\vspace{-0.15cm}}
 Intuitively, a {\it generic} plane $C$ would be one for which none of its minors vanish. Such a configuration would have $k(n\,\mi\,k)$ degrees of freedom. The vanishing of any minor of $C$ implies some linear-dependence among its columns.
Allowing for all possible linear-dependencies among the columns of $C$ leads to the ``matroid stratification'' \cite{GGMS} of configurations, which is known to be arbitrarily complicated.  Indeed, it was proven
in \cite{mnev} that {\it all} algebraic varieties are part of this matroid stratification, so understanding this amounts to {\it completely} taming the entire category of algebraic varieties!
However, if we impose one small restriction on the set of admissible linear-dependencies, we will find that a rich, simple, and very beautiful structure emerges.

\subsection{The Geometry and Combinatorics of the Positroid Stratification}\label{geometry_of_the_positroid_stratification_subsection}

Notice that any configuration $C$ associated with an on-shell, planar graph is endowed with a cyclic-ordering for the columns $\{c_1,\ldots,c_n\}$. It is therefore natural to consider a stratification of $G(k,n)$ that involves only linear-dependencies among (cyclically) {\it consecutive} chains of columns. This is known as the {\it positroid stratification}, \cite{P,KLS} (see also \cite{L2,Rietsch}), and will turn out to be precisely what is relevant to the physics of on-shell diagrams. In order to understand the connection most clearly, we will first discuss the stratification in some detail on its own, and show how these configurations are characterized by permutations. We will then see how the geometrically-defined permutation which characterizes $C$ is precisely the one which would label the graph.

Before describing the stratification generally,  it may help to consider some simple examples. Since the kinematical data describing the external particles enjoys a rescaling symmetry, we often find it useful to transfer this symmetry to the columns of $C$, identifying $c_a\sim t_ac_a$, so that (non-vanishing) columns $c_a$ can be thought of as elements in $\mathbb{P}^{(k-1)}$ (vanishing columns simply being absent from the space). This makes it a little easier to visualize configurations---at least for small $k$. Consider a generic configuration $C\!\in\!G(3,6)$, whose $6$ columns---viewed as points in $\mathbb{P}^2$---are arranged according to:
\vspace{-0.25cm}\eq{\raisebox{-37pt}{\includegraphics[scale=.8]{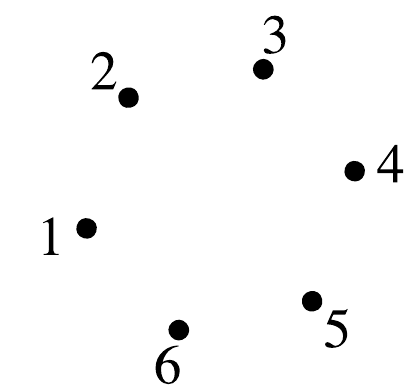}}\vspace{-0.0cm}\label{g36_d_eq_9_configuration}}
As no three of the columns are linearly-dependent, this indeed represents a generic configuration in $G(3,6)$, and has \mbox{$3(6\,\mi\,3)=9$} degrees of freedom.

The simplest {\it consecutive} constraint we could impose on (\ref{g36_d_eq_9_configuration}) would be to force any 3 consecutive columns to become linearly-dependent---projectively, collinear. For example, we could require that the minor $(123)$ vanish:
\vspace{-0.2cm}\eq{\raisebox{-37pt}{\includegraphics[scale=.8]{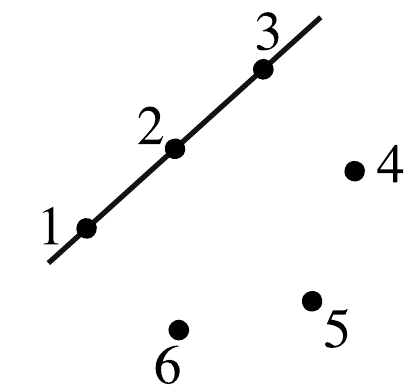}}\vspace{-0.1cm}\label{g36_d_eq_8_configuration}}
From this configuration, seven possible further restrictions are possible, including:
\vspace{-0.44cm}\eq{\hspace{-0.9cm}\raisebox{-37pt}{\includegraphics[scale=.8]{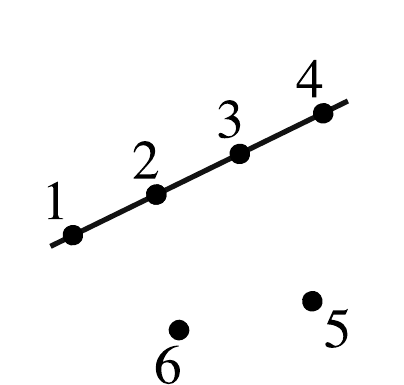}}\hspace{-0.0cm}\raisebox{-37pt}{\includegraphics[scale=.8]{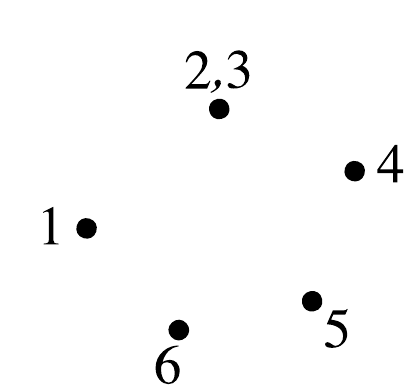}}\;\;\;\raisebox{-50pt}{\includegraphics[scale=.8]{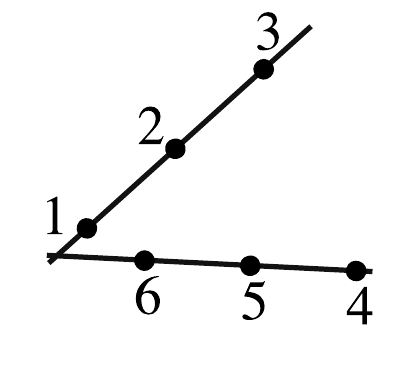}}\;\raisebox{-45pt}{\includegraphics[scale=.8]{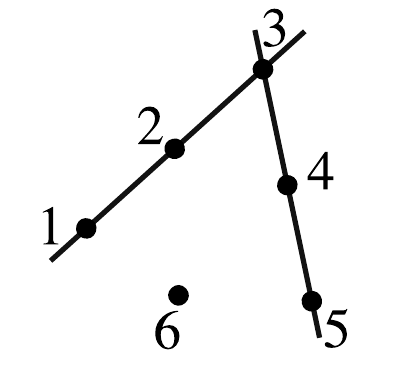}}\vspace{-0.6cm}\label{g36_d_eq_7_configurations}\nonumber}

For $k\leq3$, it is easy to describe such configurations geometrically---being easily visualizable. But such geometric descriptions rapidly become cumbersome as $k$ increases: even for $k\!=\!4$---which is still possible to visualize in three-dimensional space---configurations obtainable using only consecutive constraints can become impressively complex. Consider for example the following configuration in $G(4,8)$:
\vspace{.0cm}\eq{\hspace{1.cm}\begin{array}{c}\raisebox{-50pt}{\includegraphics[scale=1]{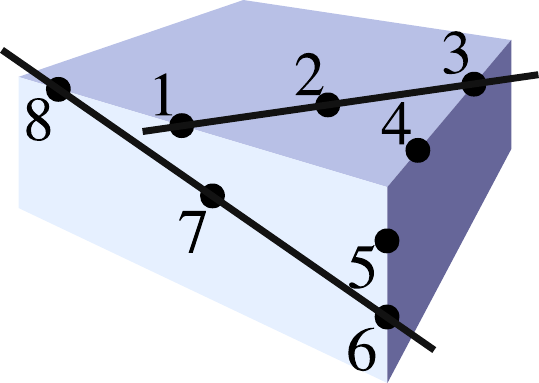}}\end{array}\,\qquad\begin{array}{c|c}\multicolumn{2}{c}{}\\[-32pt]\text{consec. chains of columns}&\text{span}\\\hline{\color{deemph}(1)\,}(2){\color{deemph}\,(3)\,(4)\,(5)\,(6)\,}(7){\color{deemph}\,(8)}&\mathbb{P}^0\\(123)\,{\color{deemph}(34)\,(45)\,(56)}\,(678)\,{\color{deemph}(81)}&\mathbb{P}^1\\(56781)\,(81234)\,(3456)&\mathbb{P}^2\\[-20pt]\end{array}\label{g48_configuration_example}\vspace{.2cm}}
A more systematic way to describe any configuration in this stratification would be to list the {\it ranks} of spaces spanned by all consecutive chains of columns. Labeling columns mod $n$, let us define,
\vspace{-.2cm}\eq{r[a;b]\equiv\mathrm{rank}\{c_a,c_{a+1},\ldots,c_{b}\};\label{chain_rank}\vspace{-.2cm}}
then knowing $r[a;b]$ for all $n^2$ pairs of columns $a\leq b$ would suffice to reconstruct any particular configuration. This data is obviously highly redundant: for example, \mbox{$r[a;a\pl n\mi1]=k$} for all $a$. We can discover how this data can be encoded more efficiently if by first organizing it in a clever way (we thank Pierre Deligne for suggesting this to us): \\[-6pt]
\newsavebox{\nmtwo}\savebox{\nmtwo}{$ n\!+\!1\!$}\newlength{\rnkwidth}\settowidth{\rnkwidth}{\usebox{\nmtwo}}\newcommand{\rnk}[2]{r[\!\begin{array}{c@{}c@{}c}$\text{\parbox[b]{\rnkwidth}{$~$}}$&;&\text{\parbox[b]{\rnkwidth}{$\hfill$}}\\[-16.5pt]#1&&#2\end{array}\!]}\newcommand{\rnkhat}[2]{\hat{r}[\!\begin{array}{c@{}c@{}c}$\text{\parbox[b]{\rnkwidth}{$~$}}$&;&\text{\parbox[b]{\rnkwidth}{$\hfill$}}\\[-16.5pt]#1&&#2\end{array}\!]}
\noindent\vspace{-0.1cm}\scalebox{0.98}{\begin{minipage}[h]{\textwidth}\vspace{0.2cm}\eq{\hspace{-3.5cm}\phantom{2n\pl1\,}
\begin{array}{|cc@{$\quad$}c@{$\quad$}cc@{$\quad$}c@{$\quad$}|cl}\cline{1-6}
~&~&~&~&\rnk{n}{2n\mi1}&\iddots&&2n\mi1\\
~&~&&\rnk{n\mi1}{2n\mi2}&\vdots&\iddots&&2n\mi2\\
~&~&\iddots&\vdots&\vdots&\iddots&&\vdots\\
&\rnk{2}{n\pl1}&\cdots&\rnk{n\mi1}{n\pl1}&\rnk{n}{n\pl1}&\iddots&&n\pl1\\
\rnk{1}{n}&\vdots&\cdots&\rnk{n\mi1}{n}&\rnk{n}{n}&&&n\\
\vdots&\vdots&\cdots&\rnk{n\mi1}{n\mi1}&~&&&n\mi1\\
\rnk{1}{3}&\rnk{2}{3}&\iddots&~&~&&&\vdots\\
\rnk{1}{2}&\rnk{2}{2}&~&~&~&&&2\\
\rnk{1}{1}&~&~&~&~&&&1\\\cline{1-6}
\multicolumn{1}{c}{1}&2&\cdots&n\mi1&\multicolumn{1}{c}{n}&\multicolumn{1}{@{$\quad$}c@{$\quad$}}{\cdots}
\end{array}\hspace{-3cm}
\label{deligne_table}\vspace{0.3cm}}\end{minipage}}

\noindent The advantages of arranging the ranks in this way will become clear momentarily. Notice that for each pair of adjacent columns $(a\,a\pl1)$ there is some $b$ sufficiently large  such that $r[a;b]=r[a\pl1;b]$, as $r[a;b]$ is bounded above by $k$ and strictly increases with $b$ (moving vertically in (\ref{deligne_table})). Moreover, it is easy to see that if $r[a;b]=r[a\pl1;b]$ for some $b$, then $r[a;c]=r[a\pl1;c]$ for all $c\geq b$, as we would have $c_a\in\mathrm{span}\{c_{a+1},\ldots,c_b\}$, and so $\mathrm{span}\{c_a,\ldots,c_b\}\subset\mathrm{span}\{c_a,\ldots,c_c\}$ for all $c\geq b$. The same argument shows that, moving from right to left along each pair of consecutive rows in (\ref{deligne_table}), for any $c$ there exists a $b$ such that $r[b;c]=r[b;c\pl1]$, and that for all $a<b$, $r[a;c]=r[a;c\pl1]$.

Because $r[a;b]\geq r[a\pl1;b]$ in general, for each $a$ there must be a {\it nearest} column, which we will denote (suggestively) as `$\sigma(a)$'$\geq a$ such that $r[a;\sigma(a)]=r[a\pl1;\sigma(a)]$. Notice that this implies that $r[a;\sigma(a)]=r[a;\sigma(a)\mi1]>r[a\pl1;\sigma(a)\mi1]$, as otherwise $\sigma(a)$ would not be the nearest. Similarly, we see that $a$ must be the {\it maximal} column $a\leq \sigma(a)$ such that $r[a;\sigma(a)]=r[a;\sigma(a)\mi1]$. Thus, there is a {\it unique} point vertically along each pair of consecutive columns and a {\it unique} point horizontally along each pair of consecutive rows where the table locally looks like:
\vspace{-0.1cm}\eq{~\qquad\begin{array}{|c|c|}\hline~&\\[-10pt]\rnk{a}{\;\,\sigma(a)\,\;}&\rnk{a\pl1}{\;\,\sigma(a)\,\;}\rule{0pt}{0pt}\\[5pt]\hline~&\\[-10pt]\rnk{a}{\sigma(a)\mi1}&\rnk{a\pl1}{\sigma(a)\mi1}\\[5pt]\hline\end{array}\text{{\LARGE$\quad\Leftrightarrow\quad$}}\begin{array}{|c|c|}\hline~&\\[-10pt]r&r\rule{0pt}{0pt}\\[5pt]\hline\phantom{\rnk{a}{\sigma(a)}}&\phantom{\rnk{a}{\sigma(a)}}\\[-10pt]r&r-1\\[5pt]\hline\end{array}\hspace{-2.4cm}\raisebox{-32.5pt}{\includegraphics[scale=1]{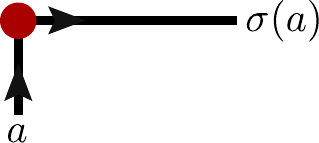}}\;.\vspace{-0.2cm}}
These ``hooks'' show that $\sigma$ is in fact a {\it permutation} among the labels $\{1,\ldots,n\}$ of the column-vectors. Actually, because this definition of $\sigma$ differentiates between $\sigma(a)=a$ (which occurs whenever $r[a;a]=0$) and $\sigma(a)=a\,\pl\,n$, $\sigma$ is automatically a {\it decorated} permutation as defined in \mbox{section \ref{combinatorial_descriptions_of_scattering_processes_subsection}}.

We can see how the permutation encoded by these hooks can be read-off from the table of ranks, (\ref{deligne_table}), by considering the example configuration given above, (\ref{g48_configuration_example}):
\vspace{-0.6cm}\eq{\hspace{-1.15cm}\begin{array}{c}\\[38pt]\raisebox{-55pt}{\includegraphics[scale=1]{g48_configuration_small}}\\[10pt]\\[10pt]\phantom{{\color{perm}\{3,7,6,10,9,8,13,12\}}}\end{array}\quad\raisebox{-100.5pt}{\includegraphics[scale=0.65]{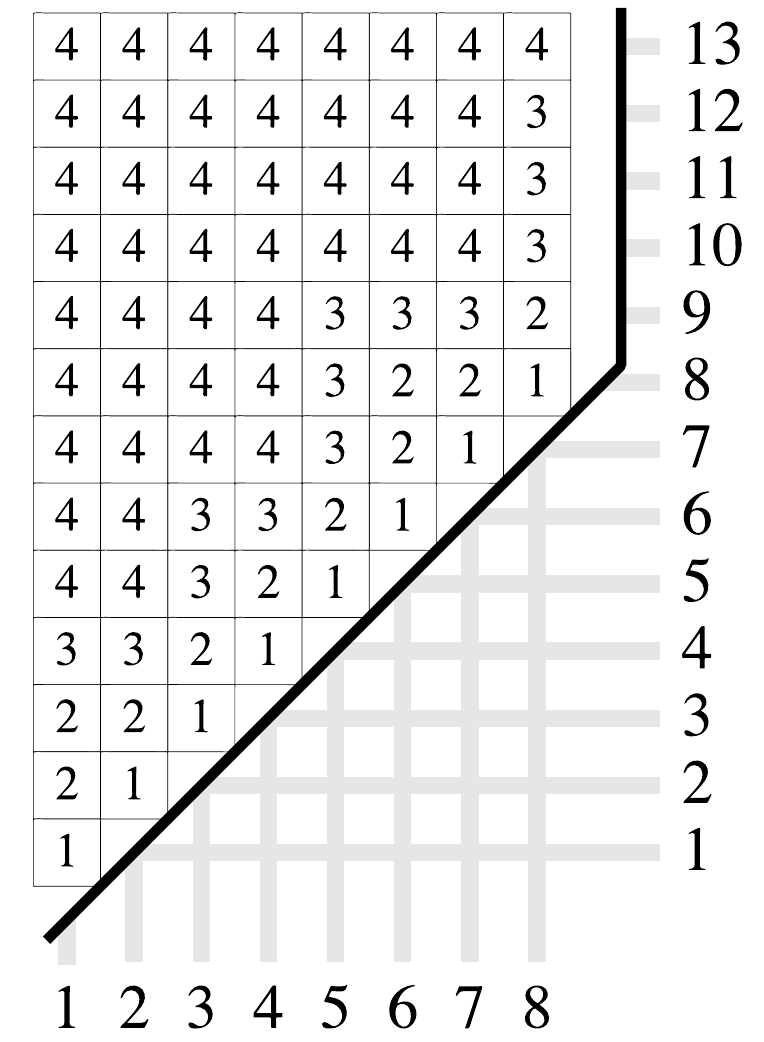}}\text{\Huge$\Rightarrow$}\raisebox{-100.5pt}{\includegraphics[scale=0.65]{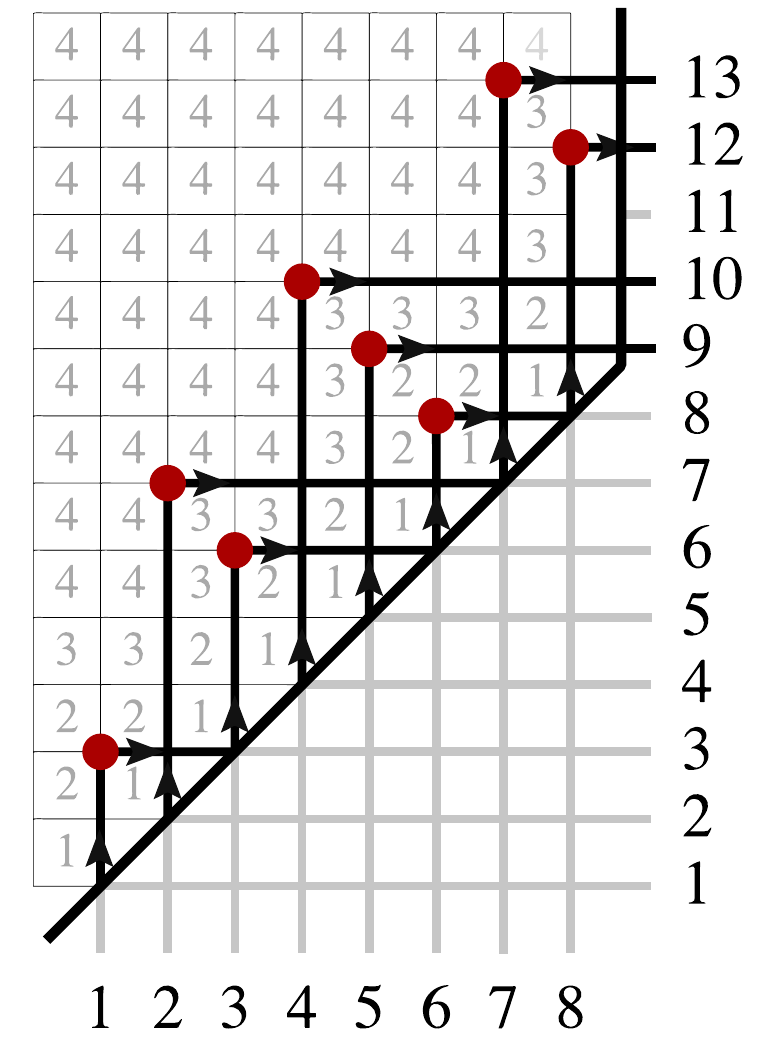}}\nonumber\vspace{-0.0cm}}
(This picture of the permutation $\sigma$ is similar to the ``juggling patterns'' illustrated in \cite{KLS}.) And so this configuration is associated with the permutation,
\vspace{-0.2cm}\eq{\sigma\equiv\left(\begin{array}{@{$\!$}cccccccc@{$\!$}}1&2&3&4&5&6&7&8\\[-0.2cm]\,\,\downarrow\,\,&\,\,\downarrow\,\,&\,\,\downarrow\,\,&\,\,\downarrow\,\,&\,\,\downarrow\,\,&\,\,\downarrow\,\,&\,\,\downarrow\,\,&\,\,\downarrow\,\,\\[-0.1cm]3&7&6&10&9&8&13&12\end{array}\right).\vspace{-0.2cm}}

The definition of $\sigma$ can be restated in an equivalent, but more transparently geometric form:\\[-7pt]

\noindent{\bf Definition:} For each $a\in\{1,\ldots,n\}$, the permutation $\sigma(a)\!\geq\!a$ labels the {\it first} column $c_{\sigma(a)}$ such that \mbox{$c_a\in\mathrm{span}\left\{c_{a+1},\ldots,c_{\sigma(a)}\right\}$}.

\noindent (Notice that if $c_a=\vec{0}$, then $\sigma(a)=a$, as $\vec{0}$ is spanned by the {\it empty} chain `$\{c_{a+1},\ldots,c_{a}\}$'.)\\[-7pt]

This definition is  useful in practice. For example, it makes it easy to understand how the dimensionality of a configuration is encoded by its permutation. Notice that because $c_a\in\mathrm{span}\{c_{a+1},\ldots,c_{\sigma(a)}\}$, we may expand $c_a$ into the $r[a;\sigma(a)]$-dimensional space spanned by $\{c_{a+1},\ldots,c_{\sigma(a)}\}$; therefore, specifying $c_{a}$ requires $r[a;\sigma(a)]$ degrees of freedom. And so, remembering to subtract the $k^2$ degrees of freedom absorbed by the overall $GL(k)$-redundancy, we find that:
\vspace{-.2cm}\eq{\dim(C_\sigma)=\Bigg(\sum_{a=1}^nr[a;\sigma(a)]\Bigg)-k^2\,.\label{dimension_by_ranks}\vspace{-.2cm}}
Notice that $r[a;\sigma(a)]$ is nothing but the number of other hooks which intersect the vertical (or horizontal) part of any particular hook $a\mapsto\sigma(a)$. Thus, for our example in $G(4,8)$ given above, the ranks $r[a,\sigma(a)]$ can be read-off as the number of intersections (marked in green) along each vertical (or horizontal) line:
\vspace{-.4cm}\eq{\raisebox{-100.5pt}{\includegraphics[scale=0.65]{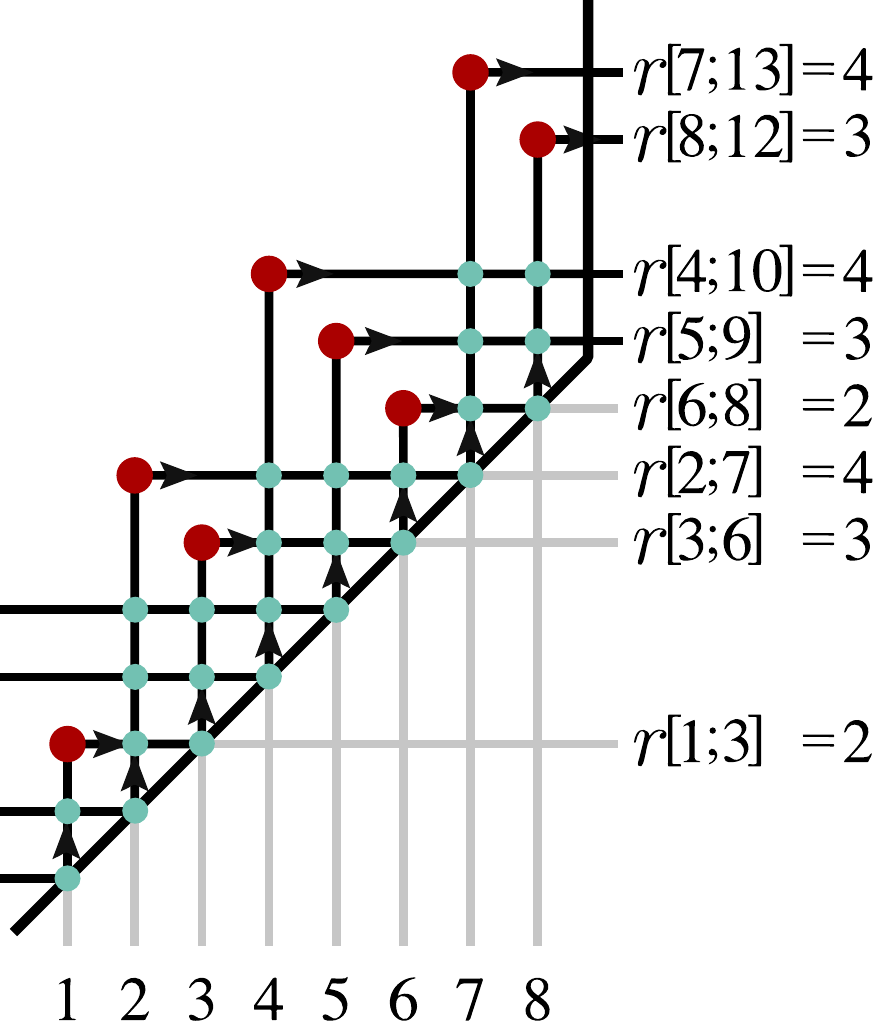}}\vspace{-.0cm}}
which shows that this configuration has $25-4^2=9$ degrees of freedom.

It is not hard to see how the permutation encodes {\it all} the ranks $r[a;b]$, thereby demonstrating that $\sigma$ {\it fully} characterizes any configuration in the positroid stratification. If we let $q[a;b]$ denote the number of \mbox{$c\in\{b\,\mi\, n,\ldots,a\}$} such that $\sigma(c)\in\{b,\ldots,a\pl\, n\}$, then $r[a;b]=k\,\mi\,q[a;b]$. Graphically, $q[a;b]$ is the number of hooks whose corners are above and to the left of $r[a;b]$ in the table (\ref{deligne_table}).

The permutation is the most compact, most invariant way of describing the consecutive linear dependencies of a configuration of vectors. A more redundant, but sometimes useful alternative characterization of a configuration is known as the {\it Grassmannian necklace}, \cite{P}: a list of $n$, $k$-tuples $A^{(a)}\equiv(A^{(a)}_1,\ldots,A^{(a)}_k)$ denoting the lexicographically-minimal non-vanishing minors starting from each of the $n$ columns. Geometrically, $A^{(a)}$ encodes the labels of the {\it first} $k$ column-vectors beyond (or possibly including) $c_a$, for which $\mathrm{rank}\{c_{A^{(a)}_{1}},\ldots,c_{A^{(a)}_{k}}\}=k$. In terms of the hooks described above, $A^{(a)}$ simply lists the $k$ horizontal lines which pass above the $a^{\mathrm{th}}$ column (which often do not cross the hook going from $a\mapsto\sigma(a)$). In the $G(4,8)$ example above, (\ref{g48_configuration_example}), the Grassmannian necklace can be read-off as follows:\\[-7.5pt]
\vspace{-.2cm}\eq{\begin{array}{c@{$\!\!\!$}c}\raisebox{-70.5pt}{\includegraphics[scale=.9]{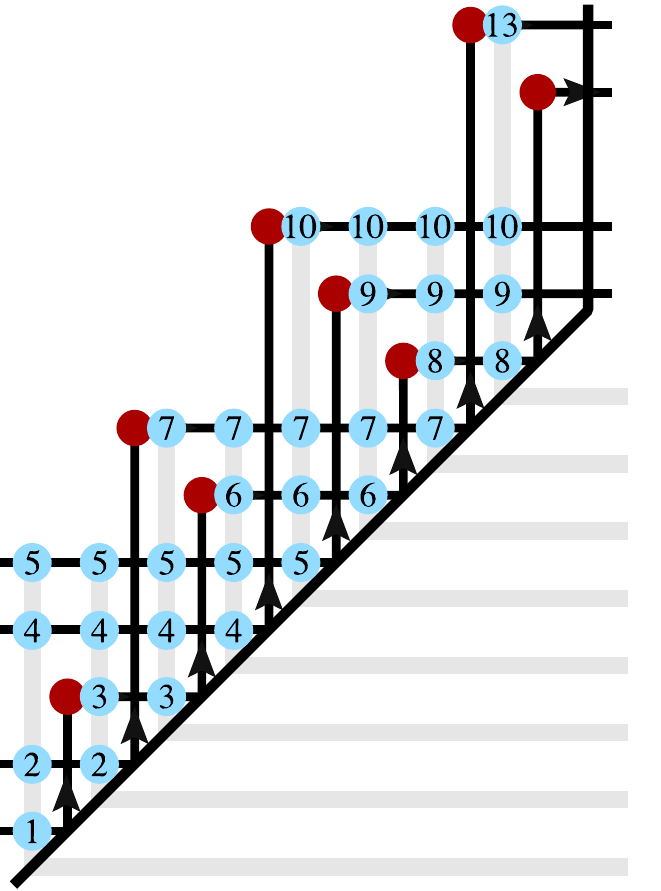}}&\begin{array}{l}~\\[-05.5pt]A^{(8)}=(8\,9\,10\,13)\\[1.8pt]A^{(7)}=(7\,8\,9\,10)\\[1.8pt]A^{(6)}=(6\,7\,9\,10)\\[1.8pt]A^{(5)}=(5\,6\,7\,10)\\[1.8pt]A^{(4)}=(4\,5\,6\,7)\\[1.8pt]A^{(3)}=(3\,4\,5\,7)\\[1.8pt]A^{(2)}=(2\,3\,4\,5)\\[1.8pt]A^{(1)}=(1\,2\,4\,5)\end{array}\\[-80pt]\end{array}\vspace{-.2cm}}

\newpage
\subsection{\mbox{Canonical Coordinates and the Equivalence of Permutation Labels}}\label{bcfw_coordinates_section}
In \mbox{section \ref{intro_to_grassmannian_section}}, we saw that every on-shell graph is associated with both a permutation (via left-right paths) and also a $k$-plane in $n$ dimensions $C\!\in\!G(k,n)$ encoding the linear-relations involving the external data. And we have just seen that any such plane $C$, viewed as a configuration of column-vectors, can {\it also} be labeled by a permutation. We will now demonstrate that these permutation labels match---that the configuration $C\!\in\!G(k,n)$ associated with an on-shell graph labeled by the left-right-path permutation $\sigma$, is labeled {\it geometrically} by the {\it same} permutation $\sigma$.

The proof of the equivalence of these permutation labels is both simple and constructive. Recall from section \ref{BCFW_bridge_decomposition_subsection} that a representative, reduced on-shell graph can be constructed for any permutation $\sigma$ by decomposing it into a sequence of `adjacent' transpositions acting on a trivial permutation, where each successive transposition in the decomposition adds a BCFW-bridge to the graph according to:
\vspace{-0.2cm}\eq{\mbox{\hspace{-0.195cm}\raisebox{-55pt}{\includegraphics[scale=1]{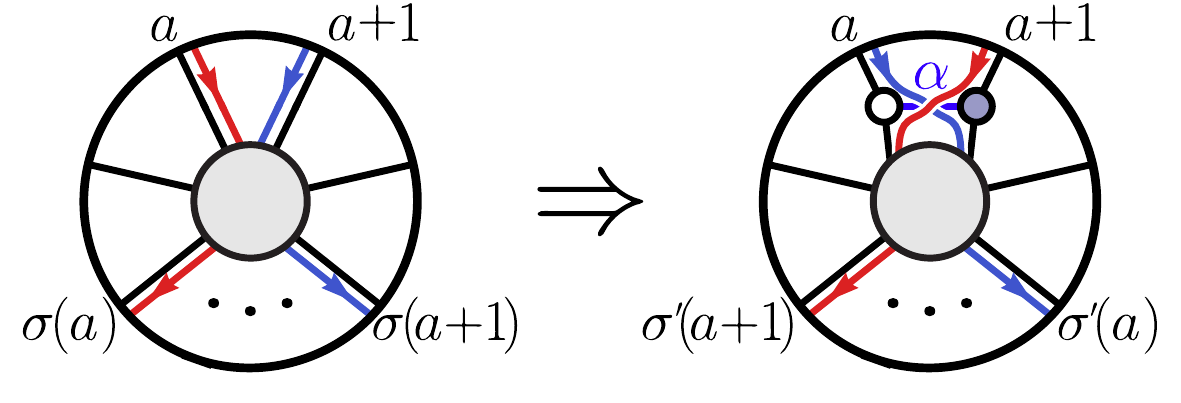}}\hspace{-0.9cm}}\label{adding_a_bridge_to_a_graph_with_variable}\vspace{-0.2cm}}
(As before, recall that two columns are to be considered `adjacent' if separated only by columns which are self-identified under $\sigma$.) Now, just as we can build-up a representative on-shell graph in this way for any permutation, we can also build-up a representative matrix $C_{\sigma}\!\in\!G(k,n)$, which we will find to be labeled {\it geometrically} by the same permutation. As a bonus, this construction will provide us with explicit coordinates for any cell of the positroid, and these coordinates will have many nice properties.

What action on the columns of $C$ corresponds to adding a BCFW bridge, (\ref{adding_a_bridge_to_a_graph_with_variable})? In terms of the matrices associated with on-shell diagrams, adding a bridge shifts,
\vspace{-0.2cm}\eq{c_{a+1}\mapsto c_{\widehat{a+1}}\equiv c_{a+1}+\alpha\,c_{a};\label{bcfw_shift_on_columns}\vspace{-0.15cm}}
recall also this shift changes the measure on the Grassmannian by adding a factor of $d\!\log(\alpha)$.

Notice that if we take a residue about $\alpha=0$, we restore the original configuration; thus, $\alpha\!\mapsto\!0$ can viewed as deleting the new edge from the graph in (\ref{adding_a_bridge_to_a_graph_with_variable}). Of course, in terms of the left-right path permutations, the BCFW bridge transposes the images of $a$ and $a\pl1$ under $\sigma$. What we need to show, therefore, is that the shift (\ref{bcfw_shift_on_columns}) has this same effect on the geometric permutation defined by the columns of $C$:
\vspace{-0.3cm}\eq{\mbox{\hspace{-0.195cm}\raisebox{-42.5pt}{\includegraphics[scale=1]{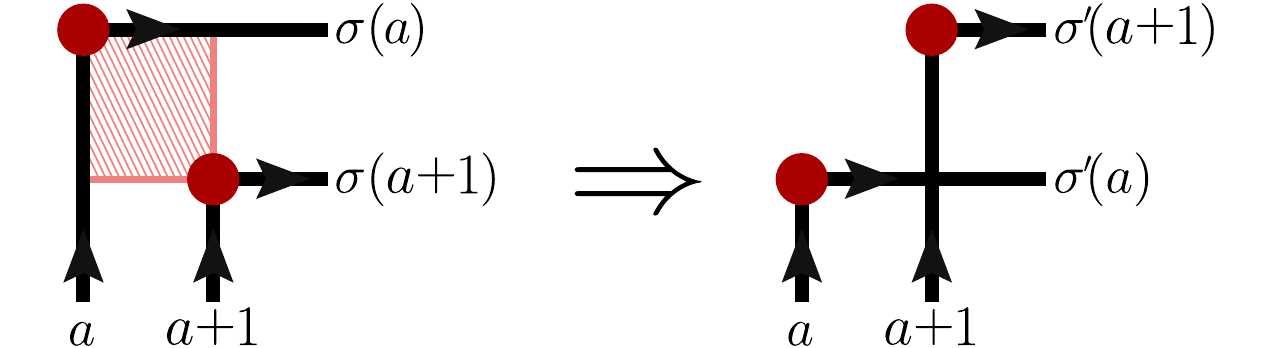}}\hspace{-0.9cm}}\label{deligne_adjacent_transposition}\vspace{-.2cm}}

Let us now show that this is indeed the change induced by (\ref{bcfw_shift_on_columns}). Clearly, the transformation (\ref{bcfw_shift_on_columns}) can at most affect the ranks of chains which {\it include} $c_{a+1}$ and {\it not} $c_a$. After the shift, $c_{\widehat{a+1}}$ is no longer spanned by $\{c_{a+2}\,\ldots,c_{\sigma(a+1)}\}$, because $c_{a}$ is not; but $c_{\widehat{a+1}}$ {\it is} spanned by $\{c_{a+2},\ldots,c_{\sigma(a)}\}$; and so, $\sigma(a\pl1)\mapsto\sigma'(a\pl1)=\sigma(a)$. Similarly, after the shift $c_a$ is trivially in the span of $\{c_{\widehat{a+1}},\ldots,c_{\sigma(a+1)}\}$ as $\mathrm{span}\{c_{\widehat{a+1}},\ldots,c_{\sigma(a+1)}\}=\mathrm{span}\{c_a,c_{a+1},\ldots,c_{\sigma(a+1)}\}$; and so, $\sigma(a)\mapsto\sigma'(a)=\sigma(a\pl1)$. And we are done.

Therefore, just as successive BCFW-bridges, (\ref{adding_a_bridge_to_a_graph_with_variable}), can be used to construct a representative, reduced on-shell graph for any permutation, they also provide us with a representative matrix for the configuration---and the BCFW-shift parameters, denoted $\alpha_i$, provide us with coordinates.

We can see how this works explicitly by revisiting the example given in \mbox{section \ref{BCFW_bridge_decomposition_subsection}} where we used successive BCFW-bridges to construct a representative on-shell graph for the permutation ${\color{perm}\{4,6,5,7,8,9\}}$.
Repeating the same construction as before, but now decorating each BCFW-bridge with its corresponding shift-parameter $\alpha_i$ gives rise to the following:\\[-15pt]

\mbox{\hspace{-0.67cm}\begin{minipage}[h]{\textwidth}\eq{\hspace{1.5cm}\mbox{\raisebox{-80pt}{\includegraphics[scale=1]{bridge_chain_paths}}\hspace{-129.52pt}}\begin{array}{|c@{$\,$}|@{$\,$}cccccc@{$\,$}|c|}\cline{1-8}\multicolumn{1}{|c@{$\,$}|@{$\,$}}{}&1&2&3&4&5&6&
\\[-4pt]\multicolumn{1}{|c@{$\,$}|@{$\,$}}{\tau}&\,\,\downarrow\,\,&\,\,\downarrow\,\,&\,\,\downarrow\,\,&\,\,\downarrow\,\,&\,\,\downarrow\,\,&\,\,\downarrow\,\,&\text{BCFW shift}
\\\cline{1-8}\multirow{2}{*}{({\color[rgb]{0.2,0,1.00}1\,2})}&{\color[rgb]{0.2,0,1.00}4}&{\color[rgb]{0.2,0,1.00}6}&5&7&8&9&\multirow{2}{*}{$c_{2\phantom{}}\mapsto c_{2}+{\color[rgb]{0.2,0,1.00}\alpha_{8}}c_{1}$}
\\\multirow{2}{*}{({\color[rgb]{0.184615,0,0.923}2\,3})}&6&{\color[rgb]{0.184615,0,0.923}4}&{\color[rgb]{0.184615,0,0.923}5}&7&8&9&\multirow{2}{*}{$c_{3\phantom{}}\mapsto c_{3}+{\color[rgb]{0.184615,0,0.923}\alpha_{7}}c_{2}$}
\\\multirow{2}{*}{({\color[rgb]{0.169231,0,0.846}3\,4})}&6&5&{\color[rgb]{0.169231,0,0.846}4}&{\color[rgb]{0.169231,0,0.846}7}&8&9&\multirow{2}{*}{$c_{4\phantom{}}\mapsto c_{4}+{\color[rgb]{0.169231,0,0.846}\alpha_{6}}c_{3}$}
\\\multirow{2}{*}{({\color[rgb]{0.153846,0,0.769}2\,3})}&6&{\color[rgb]{0.153846,0,0.769}5}&{\color[rgb]{0.153846,0,0.769}7}&{\color{deemph}4}&8&9&\multirow{2}{*}{$c_{3\phantom{}}\mapsto c_{3}+{\color[rgb]{0.153846,0,0.769}\alpha_{5}}c_{2}$}
\\\multirow{2}{*}{({\color[rgb]{0.138462,0,0.692}1\,2})}&{\color[rgb]{0.138462,0,0.692}6}&{\color[rgb]{0.138462,0,0.692}7}&5&{\color{deemph}4}&8&9&\multirow{2}{*}{$c_{2\phantom{}}\mapsto c_{2}+{\color[rgb]{0.138462,0,0.692}\alpha_{4}}c_{1}$}
\\\multirow{2}{*}{({\color[rgb]{0.123077,0,0.615}3\,5})}&{\color{deemph}7}&6&{\color[rgb]{0.123077,0,0.615}5}&{\color{deemph}4}&{\color[rgb]{0.123077,0,0.615}8}&9&\multirow{2}{*}{$c_{5\phantom{}}\mapsto c_{5}+{\color[rgb]{0.123077,0,0.615}\alpha_{3}}c_{3}$}
\\\multirow{2}{*}{({\color[rgb]{0.107692,0,0.538}2\,3})}&{\color{deemph}7}&{\color[rgb]{0.107692,0,0.538}6}&{\color[rgb]{0.107692,0,0.538}8}&{\color{deemph}4}&{\color{deemph}5}&9&\multirow{2}{*}{$c_{3\phantom{}}\mapsto c_{3}+{\color[rgb]{0.107692,0,0.538}\alpha_{2}}c_{2}$}
\\\multirow{2}{*}{({\color[rgb]{0.0923077,0,0.462}3\,6})}&{\color{deemph}7}&{\color{deemph}8}&{\color[rgb]{0.0923077,0,0.462}6}&{\color{deemph}4}&{\color{deemph}5}&{\color[rgb]{0.0923077,0,0.462}9}&\multirow{2}{*}{$c_{6\phantom{}}\mapsto c_{6}+{\color[rgb]{0.0923077,0,0.462}\alpha_{1}}c_{3}$}
\\&{\color{deemph}7}&{\color{deemph}8}&{\color{deemph}9}&{\color{deemph}4}&{\color{deemph}5}&{\color{deemph}6}&
\\\cline{1-8}\end{array}\qquad\hspace{0.05cm}\raisebox{-80pt}{\includegraphics[scale=1]{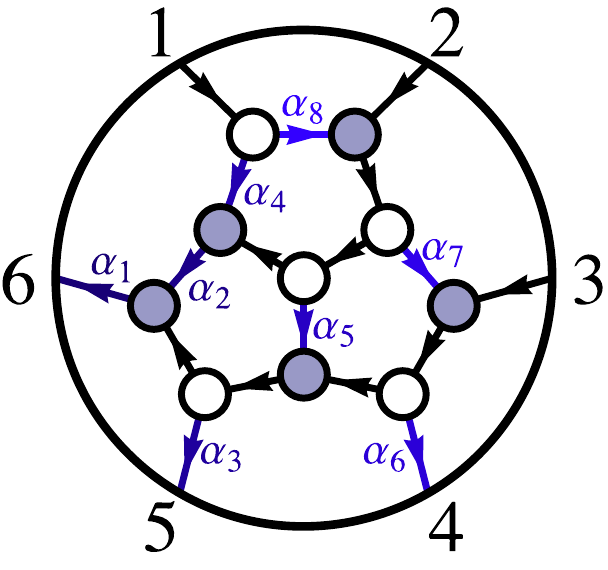}}\nonumber\vspace{0.25cm}}
\end{minipage}}\\
Starting with the zero-dimensional configuration labeled by ${\color{perm}\{7,8,9,4,5,6\}}$ and performing each successive BCFW-shift generates the following representation of $C$:\vspace{-0.2cm}
{\normalsize\eq{C(\vec{\alpha})\equiv\left(\begin{array}{@{}cccccc@{}}1&({\color[rgb]{0.138462,0,0.692}\alpha_{4}}\pl{\color[rgb]{0.2,0,1.00}\alpha_{8}})&{\color[rgb]{0.138462,0,0.692}\alpha_{4}}\,({\color[rgb]{0.153846,0,0.769}\alpha_{5}}\pl{\color[rgb]{0.184615,0,0.923}\alpha_{7}})&{\color[rgb]{0.138462,0,0.692}\alpha_{4}} {\color[rgb]{0.153846,0,0.769}\alpha_{5}} {\color[rgb]{0.169231,0,0.846}\alpha_{6}}&0&0\\
0&1&({\color[rgb]{0.107692,0,0.538}\alpha_{2}}\pl{\color[rgb]{0.153846,0,0.769}\alpha_{5}}\pl{\color[rgb]{0.184615,0,0.923}\alpha_{7}})&({\color[rgb]{0.107692,0,0.538}\alpha_{2}}\pl{\color[rgb]{0.153846,0,0.769}\alpha_{5}})\,{\color[rgb]{0.169231,0,0.846}\alpha_{6}}&{\color[rgb]{0.107692,0,0.538}\alpha_{2}} {\color[rgb]{0.123077,0,0.615}\alpha_{3}}&0\\
0&0&1&{\color[rgb]{0.169231,0,0.846}\alpha_{6}}&{\color[rgb]{0.123077,0,0.615}\alpha_{3}}&{\color[rgb]{0.0923077,0,0.462}\alpha_{1}}\end{array}\right).\label{bcfw_coordinates_for_g36_example}}

\noindent For the sake of illustration and completeness, below we give the complete sequence of coordinatized cells generated along the chain of BCFW-shifts which build-up $C(\alpha)$:
{\small\eq{\hspace{-1.175cm}\begin{array}{c}\\[-15pt]
\begin{array}{c}\\\left(\begin{array}{@{}cccccc@{}}1&0&0&0&0&0\\
0&1&0&0&0&0\\
0&0&1&0&0&0\end{array}\right)\\{\color{perm}\{7,8,\,}{\color{red}\mathbf{9}}{\color{perm},4,5,\,}{\color{red}\mathbf{6}}{\color{perm}\}}\end{array}
\,\,\xrightarrow[\text{{\normalsize${\color[rgb]{0.0923077,0,0.462}\alpha_{1}}$}}]{\text{{\normalsize$({\color[rgb]{0.0923077,0,0.462}36})$}}}\,\,
\begin{array}{c}\\\left(\begin{array}{@{}cccccc@{}}1&0&0&0&0&0\\
0&1&0&0&0&0\\
0&0&1&0&0&{\color[rgb]{0.0923077,0,0.462}\alpha_{1}}\end{array}\right)\\{\color{perm}\{7,\,}{\color{red}\mathbf{8}}{\color{perm},\,}{\color{red}\mathbf{6}}{\color{perm},4,5,9\}}\end{array}
\,\,\xrightarrow[\text{{\normalsize${\color[rgb]{0.107692,0,0.538}\alpha_{2}}$}}]{\text{{\normalsize$({\color[rgb]{0.107692,0,0.538}23})$}}}\,\,
\begin{array}{c}\\\left(\begin{array}{@{}cccccc@{}}1&0&0&0&0&0\\
0&1&{\color[rgb]{0.107692,0,0.538}\alpha_{2}}&0&0&0\\
0&0&1&0&0&{\color[rgb]{0.0923077,0,0.462}\alpha_{1}}\end{array}\right)\\{\color{perm}\{7,6,\,}{\color{red}\mathbf{8}}{\color{perm},4,\,}{\color{red}\mathbf{5}}{\color{perm},9\}}\end{array}
\,\,\xrightarrow[\text{{\normalsize${\color[rgb]{0.123077,0,0.615}\alpha_{3}}$}}]{\text{{\normalsize$({\color[rgb]{0.123077,0,0.615}35})$}}}\,\,
\begin{array}{c}\\\left(\begin{array}{@{}cccccc@{}}1&0&0&0&0&0\\
0&1&{\color[rgb]{0.107692,0,0.538}\alpha_{2}}&0&{\color[rgb]{0.107692,0,0.538}\alpha_{2}} {\color[rgb]{0.123077,0,0.615}\alpha_{3}}&0\\
0&0&1&0&{\color[rgb]{0.123077,0,0.615}\alpha_{3}}&{\color[rgb]{0.0923077,0,0.462}\alpha_{1}}\end{array}\right)
\\{\color{perm}\{}{\color{red}\mathbf{7}}{\color{perm},\,}{\color{red}\mathbf{6}}{\color{perm},5,4,8,9\}}\end{array}\\[-10pt]\\
~\hspace{357.5pt}\text{{\normalsize$({\color[rgb]{0.138462,0,0.692}12})$}}\hspace{-12pt}\text{\raisebox{-11pt}{\rotatebox{90}{$\xleftarrow[\text{{\normalsize$\phantom{{\color[rgb]{1,1,1}\alpha_{5}}}$}}]{\text{{\normalsize$\phantom{{\color[rgb]{1,1,1}(23)}}$}}}$}}}\hspace{-12pt}\,\text{{\normalsize${\color[rgb]{0.138462,0,0.692}\alpha_4}$}}\\[-10pt]
\begin{array}{c}\\\left(\begin{array}{@{}cccccc@{}}1&{\color[rgb]{0.138462,0,0.692}\alpha_{4}}&{\color[rgb]{0.138462,0,0.692}\alpha_{4}}\,{\color[rgb]{0.153846,0,0.769}\alpha_{5}}&{\color[rgb]{0.138462,0,0.692}\alpha_{4}} {\color[rgb]{0.153846,0,0.769}\alpha_{5}} {\color[rgb]{0.169231,0,0.846}\alpha_{6}}&0&0\\
0&1&({\color[rgb]{0.107692,0,0.538}\alpha_{2}}\pl{\color[rgb]{0.153846,0,0.769}\alpha_{5}})&({\color[rgb]{0.107692,0,0.538}\alpha_{2}}\pl{\color[rgb]{0.153846,0,0.769}\alpha_{5}})\,{\color[rgb]{0.169231,0,0.846}\alpha_{6}}&{\color[rgb]{0.107692,0,0.538}\alpha_{2}} {\color[rgb]{0.123077,0,0.615}\alpha_{3}}&0\\
0&0&1&{\color[rgb]{0.169231,0,0.846}\alpha_{6}}&{\color[rgb]{0.123077,0,0.615}\alpha_{3}}&{\color[rgb]{0.0923077,0,0.462}\alpha_{1}}\end{array}\right)\\
{\color{perm}\{6,\,}{\color{red}\mathbf{5}}{\color{perm},\,}{\color{red}\mathbf{4}}{\color{perm},7,8,9\}}\end{array}
\!\!\!\xleftarrow[\text{{\normalsize${\color[rgb]{0.169231,0,0.846}\alpha_{6}}$}}]{\text{{\normalsize$({\color[rgb]{0.169231,0,0.846}34})$}}}\!\!\!
\begin{array}{c}\\\left(\begin{array}{@{}cccccc@{}}1&{\color[rgb]{0.138462,0,0.692}\alpha_{4}}&{\color[rgb]{0.138462,0,0.692}\alpha_{4}}\,{\color[rgb]{0.153846,0,0.769}\alpha_{5}}&0&0&0\\
0&1&({\color[rgb]{0.107692,0,0.538}\alpha_{2}}\pl{\color[rgb]{0.153846,0,0.769}\alpha_{5}})&0&{\color[rgb]{0.107692,0,0.538}\alpha_{2}} {\color[rgb]{0.123077,0,0.615}\alpha_{3}}&0\\
0&0&1&0&{\color[rgb]{0.123077,0,0.615}\alpha_{3}}&{\color[rgb]{0.0923077,0,0.462}\alpha_{1}}\end{array}\right)
\\{\color{perm}\{6,5,\,}{\color{red}\mathbf{7}}{\color{perm},\,}{\color{red}\mathbf{4}}{\color{perm},8,9\}}\end{array}
\!\!\!\xleftarrow[\text{{\normalsize${\color[rgb]{0.153846,0,0.769}\alpha_{5}}$}}]{\text{{\normalsize$({\color[rgb]{0.153846,0,0.769}23})$}}}\!\!\!
\begin{array}{c}\\\left(\begin{array}{@{}cccccc@{}}1&{\color[rgb]{0.138462,0,0.692}\alpha_{4}}&0&0&0&0\\
0&1&{\color[rgb]{0.107692,0,0.538}\alpha_{2}}&0&{\color[rgb]{0.107692,0,0.538}\alpha_{2}} {\color[rgb]{0.123077,0,0.615}\alpha_{3}}&0\\
0&0&1&0&{\color[rgb]{0.123077,0,0.615}\alpha_{3}}&{\color[rgb]{0.0923077,0,0.462}\alpha_{1}}\end{array}\right)
\\{\color{perm}\{6,\,}{\color{red}\mathbf{7}}{\color{perm},\,}{\color{red}\mathbf{5}}{\color{perm},4,8,9\}}\end{array}\\[-10pt]\\
\text{{\normalsize$({\color[rgb]{0.184615,0,0.923}23})$}}\hspace{-12pt}\text{\raisebox{-11pt}{\rotatebox{90}{$\xleftarrow[\text{{\normalsize$\phantom{{\color[rgb]{1,1,1}\alpha_{5}}}$}}]{\text{{\normalsize$\phantom{{\color[rgb]{1,1,1}(23)}}$}}}$}}}\hspace{-12pt}\,\text{{\normalsize${\color[rgb]{0.184615,0,0.923}\alpha_7}$}}~\hspace{310pt}\\[-10pt]
\begin{array}{c}\\\left(\begin{array}{@{}cccccc@{}}1&{\color[rgb]{0.138462,0,0.692}\alpha_{4}}&{\color[rgb]{0.138462,0,0.692}\alpha_{4}}\,({\color[rgb]{0.153846,0,0.769}\alpha_{5}}\pl{\color[rgb]{0.184615,0,0.923}\alpha_{7}})&{\color[rgb]{0.138462,0,0.692}\alpha_{4}} {\color[rgb]{0.153846,0,0.769}\alpha_{5}} {\color[rgb]{0.169231,0,0.846}\alpha_{6}}&0&0\\
0&1&({\color[rgb]{0.107692,0,0.538}\alpha_{2}}\pl{\color[rgb]{0.153846,0,0.769}\alpha_{5}}\pl{\color[rgb]{0.184615,0,0.923}\alpha_{7}})&({\color[rgb]{0.107692,0,0.538}\alpha_{2}}\pl{\color[rgb]{0.153846,0,0.769}\alpha_{5}})\,{\color[rgb]{0.169231,0,0.846}\alpha_{6}}&{\color[rgb]{0.107692,0,0.538}\alpha_{2}} {\color[rgb]{0.123077,0,0.615}\alpha_{3}}&0\\
0&0&1&{\color[rgb]{0.169231,0,0.846}\alpha_{6}}&{\color[rgb]{0.123077,0,0.615}\alpha_{3}}&{\color[rgb]{0.0923077,0,0.462}\alpha_{1}}\end{array}\right)
\\{\color{perm}\{}{\color{red}\mathbf{6}}{\color{perm},\,}{\color{red}\mathbf{4}}{\color{perm},5,7,8,9\}}\end{array}
\,\,\xrightarrow[\text{{\normalsize${\color[rgb]{0.2,0,1.00}\alpha_{8}}$}}]{\text{{\normalsize$({\color[rgb]{0.2,0,1.00}12})$}}}\,\,
\begin{array}{c}\\\left(\begin{array}{@{}cccccc@{}}1&({\color[rgb]{0.138462,0,0.692}\alpha_{4}}\pl{\color[rgb]{0.2,0,1.00}\alpha_{8}})&{\color[rgb]{0.138462,0,0.692}\alpha_{4}}\,({\color[rgb]{0.153846,0,0.769}\alpha_{5}}\pl{\color[rgb]{0.184615,0,0.923}\alpha_{7}})&{\color[rgb]{0.138462,0,0.692}\alpha_{4}} {\color[rgb]{0.153846,0,0.769}\alpha_{5}} {\color[rgb]{0.169231,0,0.846}\alpha_{6}}&0&0\\
0&1&({\color[rgb]{0.107692,0,0.538}\alpha_{2}}\pl{\color[rgb]{0.153846,0,0.769}\alpha_{5}}\pl{\color[rgb]{0.184615,0,0.923}\alpha_{7}})&({\color[rgb]{0.107692,0,0.538}\alpha_{2}}\pl{\color[rgb]{0.153846,0,0.769}\alpha_{5}})\,{\color[rgb]{0.169231,0,0.846}\alpha_{6}}&{\color[rgb]{0.107692,0,0.538}\alpha_{2}} {\color[rgb]{0.123077,0,0.615}\alpha_{3}}&0\\
0&0&1&{\color[rgb]{0.169231,0,0.846}\alpha_{6}}&{\color[rgb]{0.123077,0,0.615}\alpha_{3}}&{\color[rgb]{0.0923077,0,0.462}\alpha_{1}}\end{array}\right)
\\{\color{perm}\{4,6,5,7,8,9\}}\end{array}
\end{array}\nonumber}}

Coordinates generated in this way enjoy many nice properties. For example, the physically-relevant measure on the Grassmannian (integration over which generates the on-shell differential forms of interest) is {\it maximally} simple in these coordinates: because each BCFW-shift simply adds a factor of $d\!\log(\alpha)$ to the measure, the final measure is simply,
\eq{\frac{d\alpha_1}{\alpha_1}\wedge\cdots\wedge\frac{d\alpha_d}{\alpha_d}=d\!\log(\alpha_1)\wedge\cdots\wedge d\!\log(\alpha_d)\,.\label{dLog_coordinates}}
Another important property---to be described more fully in \mbox{section \ref{boundaries_in_canonical_coordinates}}---is that these coordinates make it possible to access each of the lower-dimensional boundaries of $C$ as the zero-loci of some of the $\alpha_i$ (using an atlas of at most $n$ coordinate charts).

\subsection{Positroid Cells and the Positive Part of the Grassmannian}\label{positroid_subsection}
So far in of our discussion of configurations of vectors we have only discussed basic, linear dependencies. Let us now consider the case where these vectors are real. This will expose a natural and beautiful object, known as the {\it positive Grassmannian}, denoted $G_+(k,n)$. As in the previous subsection, let us first jump ahead and describe this object intrinsically, and then return to on-shell diagrams and show how the amalgamation picture described in \mbox{section \ref{amalgamation_subsection}} makes it obvious that on-shell diagrams---whether reduced or not---are always associated with points in $G_+(k,n)$, and demonstrate how this works explicitly for the reduced graphs obtained via the BCFW-bridge decomposition described in the previous section.

Perhaps the best way to motivate the positive Grassmannian is by starting with the simplest case, $G_{\mathbb{R}}(1,n)\simeq\mathbb{RP}^{n-1}\!\!$. Here, the column `vectors' $c_a$ of a $1$-plane $C\equiv(c_1,\ldots,c_n)$ are simply homogeneous coordinates on $\mathbb{RP}^{n-1}\!\!$, and the `positive part' of $\mathbb{RP}^{n-1}$ is simply the part of projective space where all the homogeneous coordinates are positive, which is nothing but a simplex.  Consider for example $\mathbb{RP}^2$ corresponding to the $1$-plane $C=(c_1,c_2,c_3)$:
\vspace{-.0cm}\eq{\raisebox{-55pt}{\includegraphics[scale=1]{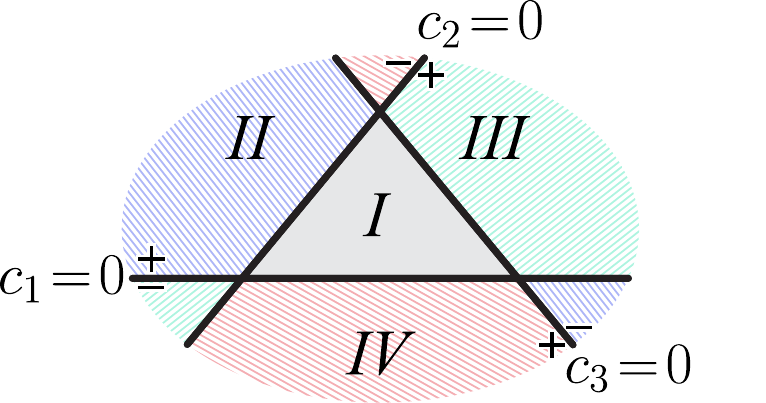}}\vspace{-.0cm}\label{positive_part_of_cp2}}
The `positive part' of $\mathbb{RP}^2$ is defined by the region where all the homogeneous coordinates $c_a$ are positive---corresponding to the (open) region labeled ``$I$'' above.  Of course, because we often allow ourselves to rescale each $c_a\sim t_ac_a$, {\it any} relative signs among the homogeneous coordinates will describe an open-region of $\mathbb{RP}^2$ essentially equivalent to region $I$, dividing $\mathbb{RP}^2$ into four ``positive parts'' as indicated in (\ref{positive_part_of_cp2}). Continuing this logic to higher $n$, it is clear that the ``positive part'' of $\mathbb{RP}^{n-1}$ should be defined as the (open) simplex for which all homogeneous coordinates are positive.

For higher $k$, the``positive part'' of $G(k,n)$ is a natural generalization of the notion of a simplex in $G(1,n)$. Thinking of the homogeneous coordinates $c_a$ as $(1\!\times\!1)$-`minors' of $C\!\in\!G(1,n)$, it is natural to define the positive part of $G(k,n)$ to be the region for which all {\it ordered} minors $(a_1\cdots a_k)$, with $a_1<\cdots<a_k$, are positive. (Notice that without a fixed ordering of the columns, it would be meaningless to discuss the positivity of minors as they are antisymmetric with respect to ordering.)

Although this definition of the positive part of $G(k,n)$ requires an ordering of the columns, no reference was made to any {\it cyclic} structure. But cyclicity emerges automatically. Na\"{i}vely, it would seem that there could be a distinct positive part for each of the $n!$ orderings of the columns, but some of these are actually the same. Suppose that $C\!\in\!G_+(k,n)$ for columns ordered according to $\{c_1,\ldots,c_n\}$. Then the change
\vspace{-.0cm}\eq{c_1\!\to\! c_2, \, c_2\!\to\!c_3,\, \cdots \, , c_n\!\to\! (-1)^{k+1} c_1,\vspace{-.0cm}}
gives a positive configuration in the rotated ordering. This is referred to as a ``twisted'' cyclic symmetry.

Notice that the definition of $G_+(k,n)$ has so far made no reference to {\it consecutivity} of the constraints involved in its boundary configurations (where some minors are allowed to vanish). The reason why consecutivity plays a role is that not all minors are independent---recall from \mbox{section \ref{intro_to_the_grassmannian_subsection}} that they satisfy Pl\"{u}cker relations following from Cramer's rule, (\ref{Cramers_rule}). The relevance of this will become clear in a simple example. Consider the case of $G(2,4)$, where we have
\vspace{-.2cm}\eq{(1\,3)(2\,4)=(1\,2)(3\,4)+(1\,4)(2\,3).\vspace{-.2cm}}
Notice the presence of the {\it plus} sign on the right-hand side. It implies that if we start with a configuration in $G_+(2,4)$, the minor $(1\,3)$ can only vanish if at least two other ordered minors also vanish.

We can see how consecutivity matters more generally for $G(2,n)$ by thinking of the column-vectors projectively as points in $\mathbb{RP}^1$. If we rescale the columns to be of the form $c_a\sim \left(\text{{\footnotesize$\begin{array}{c}\beta_a\\[-4pt]1\end{array}$}}\right)$, then $(a\,b)=(\beta_a\,\mi\,\beta_b)$, and so a positive configuration is simply one for which $\beta_a>\beta_b$ for all $a<b$. That is, the positive part of $G(2,n)$ is nothing but configurations of {\it ordered} points on a circle:
\vspace{-.4cm}\eq{\raisebox{-37pt}{\includegraphics[scale=1]{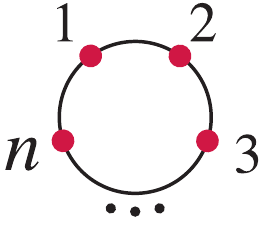}}\vspace{-.6cm}}
As such, it is clear that co-dimension one boundaries should correspond to the vanishing of only {\it consecutive} minors---the collision of adjacent points in $\mathbb{RP}^1$. In $G(2,4)$, for example, the following sequence of boundaries connect a generic configuration to one without any degrees of freedom:
\vspace{-.5cm}\eq{\hspace{-.45cm}\begin{array}{@{}c@{}}\\[-22.5pt]\raisebox{-32.5pt}{\includegraphics[scale=1]{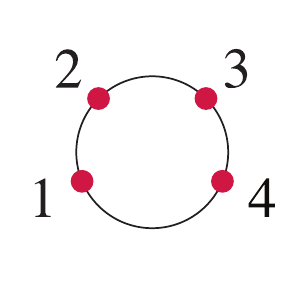}}\\[-15.5pt]{\color{perm}\{3,4,5,6\}}\\[-7pt]\end{array}\hspace{-10pt}\text{{\Large$\Rightarrow$}}\hspace{-10pt}\begin{array}{@{}c@{}}\\[-22.5pt]\raisebox{-32.5pt}{\includegraphics[scale=1]{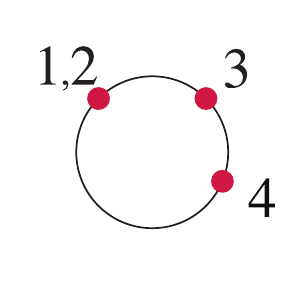}}\\[-15.5pt]{\color{perm}\{}{\color{red}\mathbf{2}}{\color{perm},4,5,}{\color{red}\mathbf{7}}{\color{perm}\}}\\[-7pt]\end{array}\hspace{-10pt}\text{{\Large$\Rightarrow$}}\hspace{-10pt}\begin{array}{@{}c@{}}\\[-22.5pt]\raisebox{-32.5pt}{\includegraphics[scale=1]{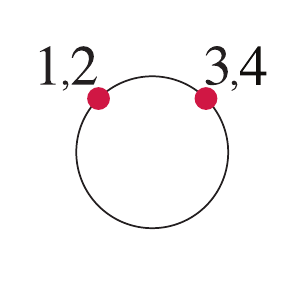}}\\[-15.5pt]{\color{perm}\{2,\,}{\color{red}\mathbf{5}}{\color{perm},\,}{\color{red}\mathbf{4}}{\color{perm},7\}}\\[-7pt]\end{array}\hspace{-10pt}\text{{\Large$\Rightarrow$}}\hspace{-10pt}\begin{array}{@{}c@{}}\\[-22.5pt]\raisebox{-32.5pt}{\includegraphics[scale=1]{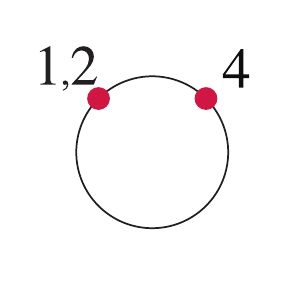}}\\[-15.5pt]{\color{perm}\{2,5,\,}{\color{red}\mathbf{3}}{\color{perm},\,}{\color{red}\mathbf{8}}{\color{perm}\}}\\[-7pt]\end{array}\hspace{-10pt}\text{{\Large$\Rightarrow$}}\hspace{-10pt}\begin{array}{@{}c@{}}\\[-22.5pt]\raisebox{-32.5pt}{\includegraphics[scale=1]{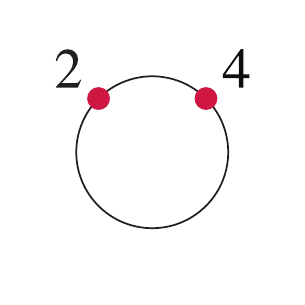}}\\[-15.5pt]{\color{perm}\{}{\color{red}\mathbf{1}}{\color{perm},\,}{\color{red}\mathbf{6}}{\color{perm},3,8\}}\\[-7pt]\end{array}\hspace{-1cm}
\vspace{-.0cm}}

In order to see that this phenomenon is not peculiar to $G(2,n)$, and to get a better picture for what is going on, let us look again at $G(3,n)$. We may use the rescaling symmetry to write each column as $c_a\sim \left(\text{{\footnotesize$\begin{array}{c}\hat{c}_a\\[-3.5pt]1\end{array}$}}\right)$, where each $\hat{c}_a$ is in $\mathbb{R}^2$. It is then easy to check that the requirement of positivity for all ordered minors  translates into the geometric statement that the points $\hat{c}_a$ form the vertices of a {\it convex} polygon in the plane.

Because of convexity, it is easy to see that going to boundaries can only involve linear relations between {\it consecutive} chains of columns. For instance, below we draw a projective representation of a generic configuration $G(3,6)$, and some of the boundaries obtainable while preserving convexity:
\vspace{-.2cm}\eq{\hspace{-.00cm}\begin{array}{c}\\[-10pt]\raisebox{-55pt}{\includegraphics[scale=1]{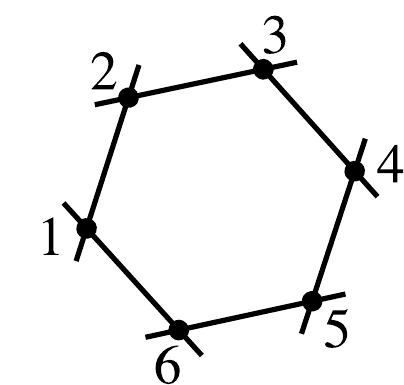}}\\{\color{perm}\{4,5,6,7,8,9\}}\end{array}\hspace{-15pt}\text{{\LARGE$\quad\Rightarrow$}}\hspace{-5pt}\begin{array}{c}\\[-10pt]\raisebox{-55pt}{\includegraphics[scale=1]{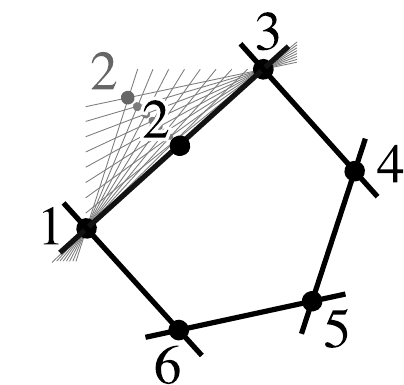}}\\{\color{perm}\{}{\color{red}\mathbf{3}}{\color{perm},5,6,7,8,\,}{\color{red}\mathbf{10}}{\color{perm}\}}\end{array}\hspace{-15pt}\text{{\LARGE$\quad\Rightarrow$}}\hspace{-5pt}\begin{array}{c}\\[-10pt]\raisebox{-55pt}{\includegraphics[scale=1]{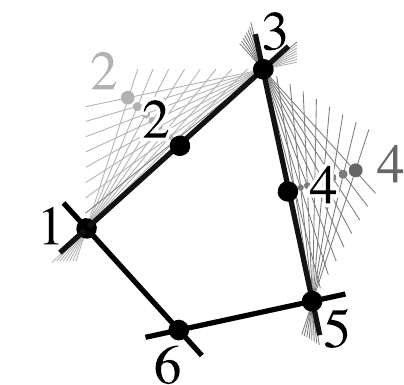}}\\{\color{perm}\{3,\,}{\color{red}\mathbf{6}}{\color{perm},\,}{\color{red}\mathbf{5}}{\color{perm},7,8,10\}}\end{array}\hspace{-1.5cm}\vspace{-.2cm}}
From the generic configuration, it is possible to make any consecutive minor vanish such as $(1\,2\,3)$ shown above. Projectively, a minor will vanish whenever three points become collinear. However, note that for instance the non-consecutive minor $(1\,3\,5)$ cannot be made to vanish without either: 1.\ destroying convexity, or 2.\ forcing {\it additional} minors to vanish along the way. And so, we find the same stratification of successive boundaries as those obtained by consecutive constraints.

These examples suffice to motivate a remarkable connection, which we will shortly understand in a simple and general way. In the first part of this section, we discussed a stratification of the {\it complex} Grassmannian, in terms of specified linear dependencies between consecutive column vectors. We now see that this structure is beautifully characterized by the structure of the {\it real} Grassmannian: the cell decomposition of the positive Grassmannian is precisely specified by giving linear dependencies between consecutive vectors.

But first, let us step back and understand the simple and direct connection between on-shell diagrams and the positive Grassmannian. Recall that we can construct the configuration $C\!\in\!G(k,n)$ for any on-shell diagram by simply ``amalgamating'' the $1$- and $2$-planes associated with the white, and black vertices, respectively. We saw in \mbox{section \ref{amalgamation_subsection}} that only two operations were needed to construct the plane $C\!\in\!G(k,n)$ for any  on-shell graph: combining graphs via {\it direct-products}, and gluing legs together by {\it projecting-out} on-shell pairs of particles. Let us briefly recall how these two operations act on the minors of the planes involved, and verify the wonderful fact that {\it amalgamation preserves positivity}.

The proof is simple. First, observe that we can always use rescaling symmetry to make any configuration in $G(1,3)$ or $G(2,3)$ positive (see, e.g.\ (\ref{positive_part_of_cp2})). Therefore, an on-shell graph can always be constructed by attaching these positive cells to each vertex, and then proceeding with amalgamation as described in \mbox{section \ref{amalgamation_subsection}}. Recall that the simplest of the two operations, taking direct-products, acts trivially on minors: suppose that the columns of $C_L\!\in\!G(k_L,n_L)$  are ordered $\{c_1,\ldots,c_{n_L}\}$, and that those of $C_R\!\in\!G(k_R,n_R)$ are ordered $\{c_{n_L+1},\ldots,c_{n_L+n_R}\}$, then all  the non-vanishing minors $C_L\bigotimes C_R\mapsto C\!\in\!G(k_L\pl\,k_R,n_L\pl\,n_R)$ will be given by,
\vspace{-.2cm}\eq{\left.(a_1\cdots a_{k_L}\,b_1\cdots b_{k_R})\right|_{C}=\left.(a_1\cdots a_{k_L})\right|_{C_L}\!\!\times\left.(b_1\cdots b_{k_R})\right|_{C_R}\,;\vspace{-.2cm}}
and so, if $C_L$ and $C_R$ are both {\it positive}, then $C$ will be as well.

The second fundamental operation, {\it projection}, takes a configuration $C\!\in\!G(k\pl1,n\pl2)$ and produces a configuration $\hat{C}\!\in\!G(k,n)$, obtained by projecting $C$ into the orthogonal-complement of $(c_A-c_B)$, for two {\it adjacent} legs $(A\,B)$. In terms of minors, this operation acts according to:
\vspace{-.2cm}\eq{\left.(a_1\cdots a_k)\right|_{\hat{C}}=\left.(A\,a_1\cdots a_k)\right|_{C}+\!\left.(B\,a_1\cdots a_k)\right|_{C}.\vspace{-.2cm}}
If $(A\,B)$ are the {\it first two} labels for the columns of $C\!\in\!G_+(k\pl1,n\pl2)$, then both terms on the right hand side are trivially positive; if $(A\,B)$ are not the first two columns, then they can always be brought to this position at the trivial cost of rescaling some columns by $(-1)$ as described during our discussion of the twisted cyclic structure of $G_+(k,n)$ in \mbox{section \ref{positroid_subsection}}.

\subsection{Canonically Positive Coordinates for Positroids}\label{polygon_coordinates_section}
We have seen many ways to describe the configuration $C\!\in\!G(k,n)$ associated with an on-shell diagram, including procedures which explicitly generate a matrix representative of $C$ parameterized by variables attached to the {\it faces} or the {\it edges} of a graph (see section \ref{boundary_measurements_section}).  And in \mbox{section \ref{bcfw_coordinates_section}}, we saw that ``canonical'' coordinates
for any cell $C\!\in\!G(k,n)$ in the positroid stratification can be systematically generated (along with a representative, reduced graph) by applying  successive BCFW-shifts. In this subsection, we demonstrate that a slight-modification of these BCFW-bridge coordinates (see \mbox{equation (\ref{signed_bcfw_shift}})) have the remarkable property that when the coordinates $\alpha_i$ are themselves positive, then $C(\alpha_i)\!\in\!G_+(k,n)$! We will refer to any such coordinates which have this property as ``positive''.

Before we describe how the BCFW-bridge coordinates make positivity manifest in this way, let us first describe a more intuitive way to parameterize generic configurations in $G(k,n)$ with coordinates which share this property. It will turn out that this geometrically-motivated parameterization of $G(k,n)$ will be essentially identical to that which is generated by the BCFW-bridge construction, and so this slight detour will prove itself quite useful later (see \mbox{section \ref{invariant_top_form_section}}).

Observe that any homogeneous coordinates for $G(1,n)\simeq\mathbb{P}^{n-1}$ are trivially positive:\vspace{-.2cm}\eq{C^{(1,n)}\equiv\left(\begin{array}{ccccc}{\color[rgb]{0,0,0.9}\beta_{1,1}}&{\color[rgb]{0,0,0.8}\beta_{1,2}}&\cdots&{\color[rgb]{0,0,0.7}\beta_{1,n-1}}&{\color[rgb]{0,0,0.6}\beta_{1,n}}\end{array}\right)\,,\vspace{-.2cm}\label{g1n_polygon_coordinates}}
because $C^{(1,n)}(\beta)\!\in\!G_+(1,n)$ whenever all the variables $\beta_{1,a} > 0$.

The first non-trivial case is for $G(2,n)$. Recall from our discussion above that if we rescale all the column vectors of $C\!\in\!G(2,n)$ to be of the form $c_a\sim \left(\text{{\footnotesize$\begin{array}{c}\hat{c}_a\\[-4pt]1\end{array}$}}\right)$, then $(a\,b)=\hat{c}_a-\hat{c}_b$; and so any set of ordered numbers $\hat{c}_1>\cdots>\hat{c}_n$ will parameterize a point in $G_+(2,n)$. One natural way to create such an ordered list of positive numbers would be to have $\hat{c}_{a}=\hat{c}_{a+1}+\beta_{1,a+1}$ for arbitrary, positive $\beta_{1,a+1}$---where we have intentionally named these `arbitrary' positive parameters according to our parameterization of $G_+(1,n)$ in (\ref{g1n_polygon_coordinates}). Restoring the degrees of freedom which rescale each column vector, we obtain the following:
\vspace{-.2cm}\eq{\hspace{.32cm}C^{(2,n)}\equiv\left(\begin{array}{ccccc}{\color[rgb]{0.9,0,0}\beta_{2,1}}({\color[rgb]{0,0,0.8}\beta_{1,2}}+\cdots+{\color[rgb]{0,0,0.7}\beta_{1,n}})&{\color[rgb]{0.8,0,0}\beta_{2,2}}({\color[rgb]{0,0,0.7}\beta_{1,3}}+\cdots+{\color[rgb]{0,0,0.6}\beta_{1,n}})&\cdots&{\color[rgb]{0.7,0,0}\beta_{2,n-1}}({\color[rgb]{0,0,0.6}\beta_{1,n}})&0\\{\color[rgb]{0.9,0,0}\beta_{2,1}}&{\color[rgb]{0.8,0,0}\beta_{2,2}}&\cdots&{\color[rgb]{0.7,0,0}\beta_{2,n-1}}&{\color[rgb]{0.6,0,0}\beta_{2,n}}\end{array}\right)\!.\vspace{-.2cm}\hspace{-1cm}\label{g2n_polygon_coordinates}}
It is easy to verify that if $\beta_{\alpha,a}>0$, then $C^{(2,n)}(\beta)\!\in\!G_+(2,n)$.

This construction naturally continues recursively, generating positive coordinates for any (generic) configuration in $G(k,n)$ as follows:
\vspace{-.2cm}\eq{C^{(k,n)}\equiv\left(\begin{array}{ccccc}{\color[rgb]{0.9,0,0}\beta_{k,1}}{\color[rgb]{0,0,0.7}\hat{c}^{\,(k,n)}_1}&\cdots&{\color[rgb]{0.7,0,0}\beta_{k,n-1}}{\color[rgb]{0,0,0.7}\hat{c}^{\,(k,n)}_{n-1}}&0\\{\color[rgb]{0.9,0,0}\beta_{k,1}}&\cdots&{\color[rgb]{0.7,0,0}\beta_{k,n-1}}&{\color[rgb]{0.6,0,0}\beta_{k,n}}\end{array}\right)\quad\mathrm{with}\quad {\color[rgb]{0,0,0.75}\hat{c}^{\,(k,n)}_a}\equiv\sum_{j=(a+1)}^{n}c^{(k-1,n)}_j\!.\vspace{-.2cm}}

Surprisingly, after using $GL(k)$-redundancy to remove the excess degrees of freedom in the parameterization of $C^{(k,n)}(\beta)$, it turns out that these are (essentially) identical to the coordinates produced by the BCFW-bridge construction described in \mbox{section \ref{bcfw_coordinates_section}}. Indeed, the only distinction is a relabeling of bridge-variables $\alpha_1,\ldots,\alpha_d$ (where $d\!\equiv\!\dim(G(k,n))\!=\!k(n\,\mi\,k)$) according to: 
\vspace{-.75cm}\eq{\hspace{-1.25cm}\begin{array}{@{}|ccccc|@{}}\multicolumn{5}{c}{\phantom{n-k}}\\\hline\beta_{1,k+1}&\beta_{1,k+2}&\cdots&\beta_{1,n-1}&\beta_{1,n}\\[-3.5pt]
\beta_{2,k+1}&\beta_{2,k+2}&\cdots&\beta_{2,n-1}&\beta_{2,n}\\[-3pt]
\vdots&\vdots&\ddots&\vdots&\vdots\\[-5pt]
\beta_{k,k+1}&\beta_{k,k+2}&\cdots&\beta_{k,n-1}&\beta_{k,n}\\\hline\multicolumn{1}{c}{~}\\
\end{array}\raisebox{-2.5pt}{$\left.\rule{0pt}{34pt}\right.\!\text{{\LARGE$\Leftrightarrow$}}$}
\raisebox{-2.5pt}{$\left.\rule{0pt}{34pt}\right.$}\begin{array}{@{}|rccccccccl|@{}}\multicolumn{10}{c}{~}\\\hline{\color[rgb]{0.0,0.4,0.3}\alpha_d}&{\color[rgb]{0.8,0.1,0.2}\alpha_{d-2}}&\cdots&\cdots&\cdots&\cdots&{\color[rgb]{0.2,0.1,0.8}\alpha_{\ell}}&\cdots&\cdots&\alpha_{k(k-1)/2+1}\\[-5pt]
\raisebox{3pt}{${\color[rgb]{0.8,0.1,0.2}\alpha_{d-1}}$}&\iddots&\iddots&\iddots&\iddots&\raisebox{3pt}{${\color[rgb]{0.2,0.1,0.8}\alpha_{\ell+1}}$}&\iddots&\iddots&\iddots&\vdots\\[-8pt]
\vdots&\iddots&\iddots&\iddots&{\color[rgb]{0.2,0.1,0.8}\iddots}&\iddots&\iddots&\iddots&\iddots&\raisebox{2pt}{${\color[rgb]{0.8,0.1,0.2}\alpha_2}$}\\[-5pt]
\alpha_{d-k(k-1)/2}&\cdots&\cdots&{\color[rgb]{0.2,0.1,0.8}\alpha_{\ell+k-1}}&\cdots&\cdots&\cdots&\cdots&{\color[rgb]{0.8,0.1,0.2}\alpha_3}&{\color[rgb]{0.0,0.4,0.3}\alpha_1}\\\hline\multicolumn{1}{c}{~}\\
\end{array}\label{general_form_of_bcfw_variables_first}\hspace{-.75cm}\nonumber\vspace{-.7cm}}

Let us now show that positivity is a {\it manifest} property of the BCFW-bridge coordinates for {\it all} positroid cells. This will also complete the connection between on-shell graphs, the stratification of configurations of vectors given by prescribing linear dependencies between consecutive vectors, and the cell decomposition of the {\it positive} Grassmannian.

We begin by observing that the minors of $C$ transform nicely under BCFW-shifts:\vspace{-.2cm}\eq{(\cdots a\pl1\cdots)\mapsto(\cdots\widehat{a\pl1}\cdots)=(\cdots a\pl1\cdots)+\alpha\,(\cdots a\cdots).\label{bcfw_shift_on_minors}\vspace{-.2cm}} And so, if we start with a configuration $C$ in the positive Grassmannian, and if $a$ and $a\pl1$ are {\it strictly} adjacent---with no columns between them self-identified under $\sigma$---then the BCFW-shift preserves positivity, because whenever $(\cdots a\pl1\cdots)$ is ordered, so is $(\cdots a\cdots)$.

However, we must remember that the decomposition of a permutation into `adjacent' transpositions allows for $a$ and ``$a\pl1$'' to be separated by any number of columns which map to themselves ($\mathrm{mod}\,n$) under $\sigma$. Because $\sigma(b)=b$ (as opposed to $\sigma(b)=b\,\pl\,n$) implies that $c_b=0$, all minors involving $b$ vanish; and so, skipping-over these columns will not affect any non-vanishing minors. However, $\sigma(b)=b\,\pl\,n$ if and only if $c_b\notin\mathrm{span}\{c_{b+1},\ldots,c_{b+n-1}\}$, implying that $c_b$ is not spanned by the rest of the columns of $C$; as such, $\sigma(b)=b\,\pl\,n$ implies that $b$ {\it must} be involved in any non-vanishing $(k\!\times\!k)$-minor of $C$. And so, when this happens, the shift in (\ref{bcfw_shift_on_minors}) may not preserve ordering for both of the terms.

To illustrate this minor subtlety, consider the very simplest case in which it arises: the one-dimensional configuration $C\!\in\!G(2,3)$ labeled by the permutation $\sigma\equiv{\color{perm}\{3,5,4\}}$.
The decomposition of $\sigma$ into `adjacent' transpositions involves only one step: $\!({\color[rgb]{.2,0,1}1\,3})\!$ ---an `adjacent' transposition which skips-over column $c_2$ because $\sigma(2)=2\pl3$. Explicitly, the BCFW-coordinates of $C_\sigma$ would be generated as follows:\\[-10pt]
\noindent\mbox{\hspace{0.05cm}\begin{minipage}[h]{\textwidth}\eq{\hspace{2.cm}\mbox{\raisebox{-22.87pt}{\includegraphics[scale=1]{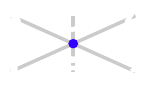}}\hspace{-75.12pt}}\begin{array}{|c@{$\,$}|@{$\,$}ccc@{$\,$}|c|}\cline{1-5}\multicolumn{1}{|c@{$\,$}|@{$\,$}}{}&1&2&3&
\\[-4pt]\multicolumn{1}{|c@{$\,$}|@{$\,$}}{\tau}&\,\,\downarrow\,\,&\,\,\downarrow\,\,&\,\,\downarrow\,\,&\text{BCFW shift}
\\\cline{1-5}\multirow{2}{*}{({\color[rgb]{0.2,0,1.00}1\,3})}&{\color[rgb]{0.2,0,1.00}3}&{\color{deemph}5}&{\color[rgb]{0.2,0,1.00}4}&\multirow{2}{*}{$c_{3}\mapsto c_{3}+{\color[rgb]{0.2,0,1.00}\alpha_{1}}c_{1}$}
\\&{\color{deemph}4}&{\color{deemph}5}&{\color{deemph}3}&
\\\cline{1-5}\end{array}\quad\phantom{\Longleftrightarrow}\quad\begin{array}{c}\\\text{{\normalsize$\left(\begin{array}{@{}ccc@{}}1&0&0\\0&1&0\end{array}\right)$}}\\{\color{perm}\{}{\color{red}\mathbf{4}}{\color{perm},5,\,}{\color{red}\mathbf{3}}{\color{perm}\}}\end{array}\,\,\xrightarrow[\text{{\normalsize${\color[rgb]{0.2,0,1.00}\alpha_{1}}$}}]{\text{{\normalsize$({\color[rgb]{0.2,0,1.00}13})$}}}\,\,\begin{array}{c}\\\text{{\normalsize$\left(\begin{array}{@{}ccc@{}}1&0&{\color[rgb]{0.2,0,1.00}\alpha_{1}}\\0&1&0\end{array}\right)$}}\\{\color{perm}\{3,5,4\}}\end{array}
}\end{minipage}}\\[-2pt]

\noindent Notice that the minor $(23)$, which vanishes before the shift, becomes $(23)\mapsto(2\hat{3})=(23)\pl\,{\color[rgb]{0.2,0,1.00}\alpha_{1}} (21)=\mi\,{\color[rgb]{0.2,0,1.00}\alpha_{1}} (12)$ after the shift. And so, if we wish to make the final configuration $C$ {\it positive}, we must take ${\color[rgb]{0.2,0,1.00}\alpha_{1}}$ to be negative; alternatively, we could redefine the rule for BCFW-shifts so that the transposition $({\color[rgb]{0.2,0,1.00}13})$ actually corresponds to a shift $c_3\!\mapsto\!c_3\,\mi\,{\color[rgb]{0.2,0,1.00}\alpha_{1}}c_1$. Of the two alternatives, we prefer the latter as then positivity of the BCFW-shift coordinates would directly imply that a configuration were positive.

It is easy to see how this simple example generalizes: in order to preserve the positivity of minors {\it and} the coordinates, we should redefine the BCFW-shift so that the transposition of $a$ and ``$a\pl1$'' changes the columns of $C$ according to
\vspace{-.2cm}\eq{c_{a+1}\mapsto c_{a+1}+(-1)^q\alpha\,c_a,\label{signed_bcfw_shift}\vspace{-.2cm}}
where $q$ is the number of columns $b$ between $a$ and ``$a\pl1$'' such that $\sigma(b)=b\,\pl\,n$. In this modified form, the BCFW-shift is {\it guaranteed} to preserve positivity. And so, restricting all the coordinates $\alpha_i$ to be positive will always result in a configuration $C(\vec{\alpha})$ in the {\it positive} Grassmannian $G_+(k,n)$.

To see how these {\it signed} BCFW-shifts make positivity manifest---and as one further example of the BCFW-bridge construction described in \mbox{section \ref{bcfw_coordinates_section}}---consider the following coordinates constructed for the configuration in $G(4,8)$ given in (\ref{g48_configuration_example}):\\[5pt]
\noindent\mbox{\hspace{0.40cm}\begin{minipage}[h]{\textwidth}\eq{\hspace{-0.75cm}\mbox{\raisebox{-86.66pt}{\includegraphics[scale=1]{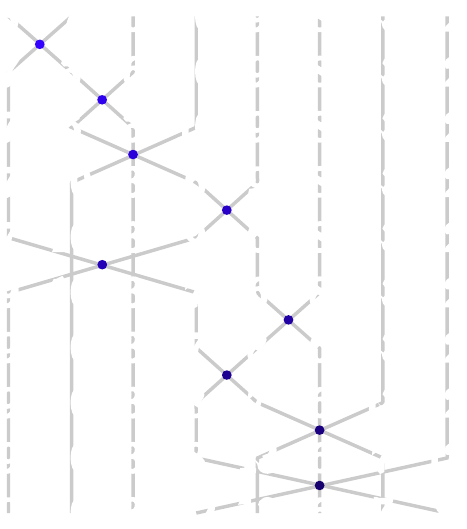}}\hspace{-182.175pt}}\begin{array}{|c|c|cccccccc@{$\,$}|c|}\cline{1-11}\multicolumn{1}{|c}{}&\multicolumn{1}{c|}{}&1&2&3&4&5&6&7&8&
\\[-4pt]\cline{1-2}\multicolumn{1}{|c}{\tau}&\multicolumn{1}{|c|}{q}&\,\,\downarrow\,\,&\,\,\downarrow\,\,&\,\,\downarrow\,\,&\,\,\downarrow\,\,&\,\,\downarrow\,\,&\,\,\downarrow\,\,&\,\,\downarrow\,\,&\,\,\downarrow\,\,&\text{BCFW shift}
\\\cline{1-11}\multirow{2}{*}{({\color[rgb]{0.2,0,1.00}1\,2})}&\multirow{2}{*}{\,0\,}&{\color[rgb]{0.2,0,1.00}3}&{\color[rgb]{0.2,0,1.00}7}&6&10&9&8&13&12&\multirow{2}{*}{$c_{2\phantom{}}\mapsto c_{2}+{\color[rgb]{0.2,0,1.00}\alpha_{9}}c_{1}$}
\\\multirow{2}{*}{({\color[rgb]{0.185714,0,0.929}2\,3})}&\multirow{2}{*}{0}&7&{\color[rgb]{0.185714,0,0.929}3}&{\color[rgb]{0.185714,0,0.929}6}&10&9&8&13&12&\multirow{2}{*}{$c_{3\phantom{}}\mapsto c_{3}+{\color[rgb]{0.185714,0,0.929}\alpha_{8}}c_{2}$}
\\\multirow{2}{*}{({\color[rgb]{0.171429,0,0.857}2\,4})}&\multirow{2}{*}{0}&7&{\color[rgb]{0.171429,0,0.857}6}&{\color{deemph}3}&{\color[rgb]{0.171429,0,0.857}10}&9&8&13&12&\multirow{2}{*}{$c_{4\phantom{}}\mapsto c_{4}+{\color[rgb]{0.171429,0,0.857}\alpha_{7}}c_{2}$}
\\\multirow{2}{*}{({\color[rgb]{0.157143,0,0.786}4\,5})}&\multirow{2}{*}{0}&7&{\color{deemph}10}&{\color{deemph}3}&{\color[rgb]{0.157143,0,0.786}6}&{\color[rgb]{0.157143,0,0.786}9}&8&13&12&\multirow{2}{*}{$c_{5\phantom{}}\mapsto c_{5}+{\color[rgb]{0.157143,0,0.786}\alpha_{6}}c_{4}$}
\\\multirow{2}{*}{({\color[rgb]{0.142857,0,0.714}1\,4})}&\multirow{2}{*}{1}&{\color[rgb]{0.142857,0,0.714}7}&{\color{deemph}10}&{\color{deemph}3}&{\color[rgb]{0.142857,0,0.714}9}&6&8&13&12&\multirow{2}{*}{$c_{4\phantom{}}\mapsto c_{4}{\color{red}\,-\,}{\color[rgb]{0.142857,0,0.714}\alpha_{5}}c_{1}$}
\\\multirow{2}{*}{({\color[rgb]{0.128571,0,0.643}5\,6})}&\multirow{2}{*}{0}&{\color{deemph}9}&{\color{deemph}10}&{\color{deemph}3}&7&{\color[rgb]{0.128571,0,0.643}6}&{\color[rgb]{0.128571,0,0.643}8}&13&12&\multirow{2}{*}{$c_{6\phantom{}}\mapsto c_{6}+{\color[rgb]{0.128571,0,0.643}\alpha_{4}}c_{5}$}

\\\multirow{2}{*}{({\color[rgb]{0.114286,0,0.571}4\,5})}&\multirow{2}{*}{0}&{\color{deemph}9}&{\color{deemph}10}&{\color{deemph}3}&{\color[rgb]{0.114286,0,0.571}7}&{\color[rgb]{0.114286,0,0.571}8}&{\color{deemph}6}&13&12&\multirow{2}{*}{$c_{5\phantom{}}\mapsto c_{5}+{\color[rgb]{0.114286,0,0.571}\alpha_{3}}c_{4}$}
\\\multirow{2}{*}{({\color[rgb]{0.1,0,0.500}5\,7})}&\multirow{2}{*}{0}&{\color{deemph}9}&{\color{deemph}10}&{\color{deemph}3}&8&{\color[rgb]{0.1,0,0.500}7}&{\color{deemph}6}&{\color[rgb]{0.1,0,0.500}13}&12&\multirow{2}{*}{$c_{7\phantom{}}\mapsto c_{7}+{\color[rgb]{0.1,0,0.500}\alpha_{2}}c_{5}$}
\\\multirow{2}{*}{({\color[rgb]{0.0857143,0,0.429}4\,8})}&\multirow{2}{*}{1}&{\color{deemph}9}&{\color{deemph}10}&{\color{deemph}3}&{\color[rgb]{0.0857143,0,0.429}8}&{\color{deemph}13}&{\color{deemph}6}&{\color{deemph}7}&{\color[rgb]{0.0857143,0,0.429}12}&\multirow{2}{*}{$c_{8\phantom{}}\mapsto c_{8}{\color{red}\,-\,}{\color[rgb]{0.0857143,0,0.429}\alpha_{1}}c_{4}$}
\\&&{\color{deemph}9}&{\color{deemph}10}&{\color{deemph}3}&{\color{deemph}12}&{\color{deemph}13}&{\color{deemph}6}&{\color{deemph}7}&{\color{deemph}8}&
\\\cline{1-11}\end{array}\hspace{0.3cm}\begin{array}{@{}c@{}}\raisebox{-3.25cm}{\includegraphics[scale=0.85]{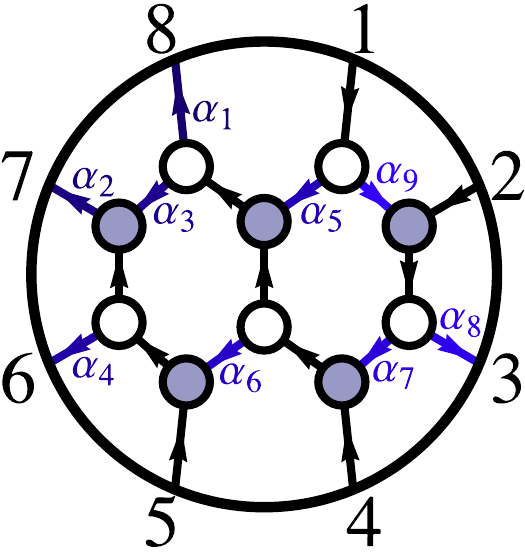}}\\[-3pt]\text{{\small$\displaystyle\left(\begin{array}{@{}cccccccc@{}}1&{\color[rgb]{0.2,0,1.00}\alpha_{9}}&0&\!\!{\color{red}\mi}\,{\color[rgb]{0.142857,0,0.714}\alpha_{5}}\,&\!\!{\color{red}\mi}\,{\color[rgb]{0.142857,0,0.714}\alpha_{5}} {\color[rgb]{0.157143,0,0.786}\alpha_{6}}\,&0&0&0\\[-2pt]
0&1&{\color[rgb]{0.185714,0,0.929}\alpha_{8}}&{\color[rgb]{0.171429,0,0.857}\alpha_{7}}&0&0&0&0\\[-2pt]
0&0&0&1&{\color[rgb]{0.114286,0,0.571}\alpha_{3}}\pl\,{\color[rgb]{0.157143,0,0.786}\alpha_{6}}&{\color[rgb]{0.114286,0,0.571}\alpha_{3}} {\color[rgb]{0.128571,0,0.643}\alpha_{4}}&0&\!\!{\color{red}\mi}\,{\color[rgb]{0.0857143,0,0.429}\alpha_{1}}\\[-2pt]
0&0&0&0&1&{\color[rgb]{0.128571,0,0.643}\alpha_{4}}&{\color[rgb]{0.1,0,0.500}\alpha_{2}}&0\end{array}\right)$}}\end{array}\hspace{-2cm}\nonumber}\end{minipage}}\\[-4pt]

\noindent It is easy to verify that all the non-vanishing minors of $C(\alpha)\!\in\!G(4,8)$ are positive when $\alpha_i\in\mathbb{R}_+$. For example, consider the minor,
\vspace{-.2cm}\eq{(2\,4\,5\,7)={\color[rgb]{0.1,0,0.500}\alpha_{2}}\,{\color[rgb]{0.114286,0,0.571}\alpha_{3}}\,{\color[rgb]{0.142857,0,0.714}\alpha_{5}}+{\color[rgb]{0.1,0,0.500}\alpha_{2}}\,{\color[rgb]{0.114286,0,0.571}\alpha_{3}}\,{\color[rgb]{0.171429,0,0.857}\alpha_{7}}\,{\color[rgb]{0.2,0,1.00}\alpha_{9}}+{\color[rgb]{0.1,0,0.500}\alpha_{2}}\,{\color[rgb]{0.157143,0,0.786}\alpha_{6}}\,{\color[rgb]{0.171429,0,0.857}\alpha_{7}}\,{\color[rgb]{0.2,0,1.00}\alpha_{9}},\vspace{-.2cm}}
the positivity of which requires, for example, the {\it signed} BCFW-shift $c_4\mapsto c_4\,{\color{red}\mi}\,{\color[rgb]{0.142857,0,0.714}\alpha_{5}} c_1$.

\section{Boundary Configurations, Graphs, and Permutations}\label{boundary_configurations_section}
\subsection{Physical Singularities and Positroid Boundaries}
Recall that an on-shell diagram labeled by the permutation $\sigma$ corresponds to a differential form $f_\sigma$ obtained via integration over the configuration $C_\sigma(\alpha)\!\in\!G(k,n)$ subject to the constraints that $C_\sigma$ be orthogonal to $\widetilde\lambda$ and contain $\lambda$:
\vspace{-.2cm}\eq{f_\sigma=\int\limits_{C_{\sigma}}\!\!\frac{d\alpha_1}{\alpha_1}\wedge\cdots\wedge\frac{d\alpha_d}{\alpha_d}\;\delta^{k\times4}\big(C_\sigma\!\cdot\!\widetilde\eta\big)\delta^{k\times2}\big(C_\sigma\!\cdot\!\widetilde\lambda\big)\delta^{2\times(n-k)}\big(\lambda\!\cdot\!C_\sigma^{\perp}\!\big),\vspace{-.4cm}\label{physical_functions_in_canonical_coords_one}}
where $\alpha_i$ are {\it canonical} (e.g.\ BCFW-bridge) coordinates for the configuration $C_\sigma$. Because the $\delta$-functions encode $(2n\,\mi\,4)$ constraints in general (together with the $4$ constraints of momentum-conservation), cells with $(2n\,\mi\,4)$ degrees of freedom can be fully-localized, while those of lower dimension leave-behind further $\delta$-functions which impose constraints on the external kinematical data.

On-shell differential forms which impose constraints on the external data (beyond momentum conservation) represent physical {\it singularities}: places in the space of kinematical data where higher-degree forms develop poles. As we saw in \mbox{section \ref{onshell_diagrams_for_amplitudes_section}}, such singularities are of primary physical interest: for example, knowing the singularity-structure of scattering amplitudes suffices to fix them completely to all loop-orders via the BCFW recursion relations, (\ref{all_loop_recursion}).

The physical singularities of on-shell differential forms, therefore, correspond to the {\it boundaries} of the corresponding configurations in the Grassmannian. Suppose we consider a {\it reduced} graph with $n_F$ faces; then, because such a graph is associated with an $(n_F\,\mi\,1)$-dimensional configuration $C$, it is easy to see that its boundaries are those graphs obtained by deleting edges (reducing the number of faces by one). However, sometimes a graph obtained in this way is no longer reduced, and actually corresponds to a configuration in the Grassmannian whose dimension has been lowered by more than one. This raises the question: which edges in a graph can be removed while keeping a graph reduced? Such edges will be called {\it removable}. It turns out that this question is easiest to answer not in terms of on-shell graphs directly, but in terms of the geometry of their corresponding configurations in the Grassmannian and the combinatorics of their permutations.

\subsection{Boundary Configuration Combinatorics in the Positroid Stratification}
The boundaries of a configuration $C$, denoted $\partial(C)$, in the positroid stratification are those configurations obtained by imposing any one additional constraint involving consecutive chains of columns. Before describing the combinatorial rule for finding boundary configurations, let us first build some intuition through simple examples. Recall from \mbox{section \ref{geometry_of_the_positroid_stratification_subsection}} the configuration in $G_+(3,6)$ whose boundaries included:\\[-02pt]
\vspace{-.2cm}{\small\eq{\hspace{-1.35cm}\begin{array}{c}\begin{array}{c}\\[-20pt]\raisebox{37.5pt}{{\LARGE$\partial\!$}}\raisebox{-00pt}{\includegraphics[scale=.75]{g36_configurations_8d}}\\[-5pt]{\color{perm}\{3,5,6,7,8,10\}}\\[-50pt]\end{array}\raisebox{-22pt}{{\LARGE$=\!$}}\raisebox{-20pt}{$\,\,\left\{\rule{0pt}{35pt}\right.\!\!\!\!\!\!\!$}\begin{array}{c}\\[-22pt]\raisebox{7pt}{\includegraphics[scale=.75]{g36_configurations_7d_2}}\\[-5pt]{\color{perm}\{3,\,}{\color{red}\mathbf{4}}{\color{perm},6,7,8,\,}{\color{red}\mathbf{11}}{\color{perm}\}}\\[-50pt]\end{array}\raisebox{-22pt}{{\LARGE$\!,\!\!$}}\begin{array}{c}\\[-22pt]\raisebox{7pt}{\includegraphics[scale=.75]{g36_configurations_7d_1}}\\[-5pt]{\color{perm}\{}{\color{red}\mathbf{5}}{\color{perm},\,}{\color{red}\mathbf{3}}{\color{perm},6,7,8,10}{\color{perm}\}}\\[-50pt]\end{array}\raisebox{-22pt}{{\LARGE$\!,\!\!$}}\begin{array}{c}\\[-12pt]\raisebox{-3pt}{\includegraphics[scale=.75]{g36_configurations_7d_4}}\\[-5pt]{\color{perm}\{3,\,}{\color{red}\mathbf{6}}{\color{perm},\,}{\color{red}\mathbf{5}}{\color{perm},7,8,10}{\color{perm}\}}\\[-50pt]\end{array}\raisebox{-22pt}{{\LARGE$\!,\!$}}\begin{array}{c}\\[-15pt]\raisebox{-00pt}{\includegraphics[scale=.75]{g36_configurations_7d_3}}\\[-5pt]{\color{perm}\{3,5,\,}{\color{red}\mathbf{7}}{\color{perm},\,}{\color{red}\mathbf{6}}{\color{perm},8,10}{\color{perm}\}}\\[-50pt]\end{array}\raisebox{-22pt}{{\LARGE$\!\!\!,$}{\large$\ldots$}}\raisebox{-20pt}{$\!\!\left.\rule{0pt}{35pt}\right\}$}\\[-20pt]\end{array}\vspace{-.2cm}\nonumber}}
where we have highlighted how the permutation changes for each boundary-element.

And so---if it weren't sufficiently obvious already---this example makes it clear that boundary elements of a configuration labeled by $\sigma$ are those labeled by $\sigma'$ which are related to $\sigma$ by a transposition of its images. However, not all transpositions lower the dimension of the configuration, and some transpositions lower the dimensionality by more than one. The way to identify the transpositions which lower the dimension by precisely one is easily understood from the way dimensionality is encoded by a configuration's permutation: if we view the permutation as given by the `hooks' described in \mbox{section \ref{geometry_of_the_positroid_stratification_subsection}}, then the dimension of a configuration is counted by the number of intersections of its hooks (minus $k^2$). Therefore, boundaries are those transpositions which eliminate any {\it one} such intersection:
\vspace{-.2cm}\eq{\hspace{0.0cm}\mbox{\raisebox{-40pt}{\includegraphics[scale=.9]{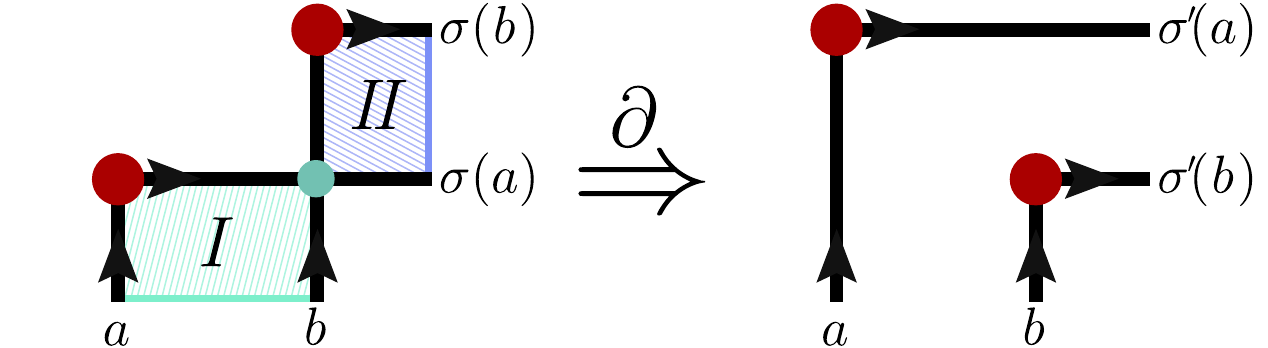}}}\hspace{-0.0cm}\vspace{-.1cm}}
Here, it is important that $a<b\leq\sigma(a)<\sigma(b)\leq(a\pl\,n)$, and that there are no hooks from $c\!\in\!I$ to $\sigma(c)\!\in\!I\!I$ as otherwise the dimensionality would be lowered by more than one:
\vspace{-.3cm}\eq{\hspace{0.0cm}\mbox{\raisebox{-45pt}{\includegraphics[scale=.9]{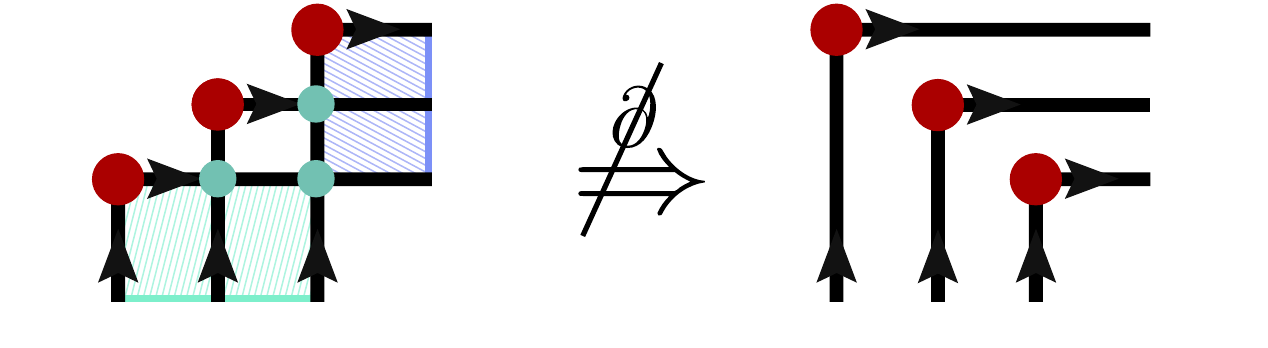}}}\hspace{-0.0cm}\vspace{-.4cm}}

Restated in terms of on-shell graphs decorated by left-right paths, this rule identifies {\it removable} edges as those along which two paths cross, $a\!\to\!\sigma(a)$ and $b\!\to\!\sigma(b)$ with $a<b\leq\sigma(b)<\sigma(a)\leq(a\pl\,n)$, provided that there is no path $c\!\to\!\sigma(c)$ with $c\!\in\!I$ and $\sigma(c)\!\in\!I\!I$:\\[-0pt]
\vspace{-.4cm}\eq{\hspace{0.0cm}\mbox{\raisebox{-67pt}{\includegraphics[scale=.9]{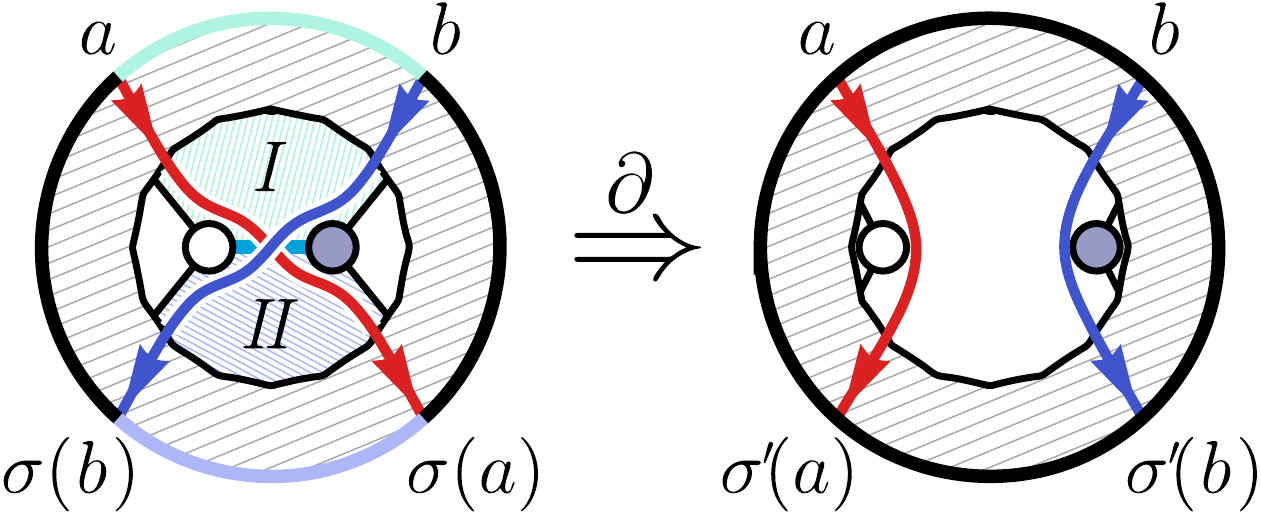}}}\hspace{-0.0cm}\vspace{-.0cm}}
These two definitions of the boundary elements of a configuration are of course equivalent; but without the combinatorial rule for counting dimensions, it would have been considerably more difficult to see that these---and only these---edges are removable. \\[-0pt]

\newpage
\subsection{(Combinatorial) Polytopes in the Grassmannian}\label{combinatorial_polytope_subsection}
The boundary operator $\partial$ given above {\it defines} the positroid stratification of $G(k,n)$; and this stratification is a very special one, with many nice features. For one thing, it allows us to view every positroid configuration in $G_+(k,n)$ is {\it something like} a `polytope' in $G(k,n)$. By this we mean that the inclusions induced by $\partial$ (viewed as a strong Bruhat covering relation) define an {\it Eulerian poset}---the key combinatorial property of the poset of faces of an ordinary polytope.

We will not prove that $\partial$ defines an Eulerian poset (this was proven in \cite{williams:eulerian}), but let us at least demonstrate that $\partial^2=0$ (mod 2)---which is of course a prerequisite for $\partial$ to actually have the meaning of a homological `boundary' operator. It turns out that every configuration in $\partial^2(C)$ is found as the boundary of precisely {\it two} configurations in $\partial(C)$ (a fact which follows trivially from the more complete statement that $\partial$ defines an Eulerian poset). This is not hard to prove, and it trivially implies that $\partial^2=0$ (mod 2). To see this, notice that each configuration in $\partial[C_{\sigma}]$ is labeled by $\sigma'$ related to $\sigma$ by a transposition. It is easy to see that the pair of transpositions must involve at least three distinct labels. If the pair involved four labels, say $(a\,b)$ and $(c\,d)$, then obviously the two transpositions can be taken in either order. When the pair involves three labels, say $\{a\,b\,c\}$, then there are only four possible scenarios to check:
\vspace{-.2cm}\eq{\begin{array}{l@{$\qquad$}l}(a\,b)\!\circ\!(a\,c)\simeq(b\,c)\!\circ\!(a\,b)&(a\,b)\!\circ\!(b\,c)\simeq(b\,c)\!\circ\!(a\,c)\\
(a\,b)\!\circ\!(b\,c)\simeq(a\,c)\!\circ\!(a\,b)&(a\,c)\!\circ\!(b\,c)\simeq(b\,c)\!\circ\!(a\,b)\end{array};\vspace{-.2cm}}
the first of these, for example, can be understood graphically in terms of hooks as,
\eq{\hspace{-3.0cm}\raisebox{-50pt}{\includegraphics[scale=.8]{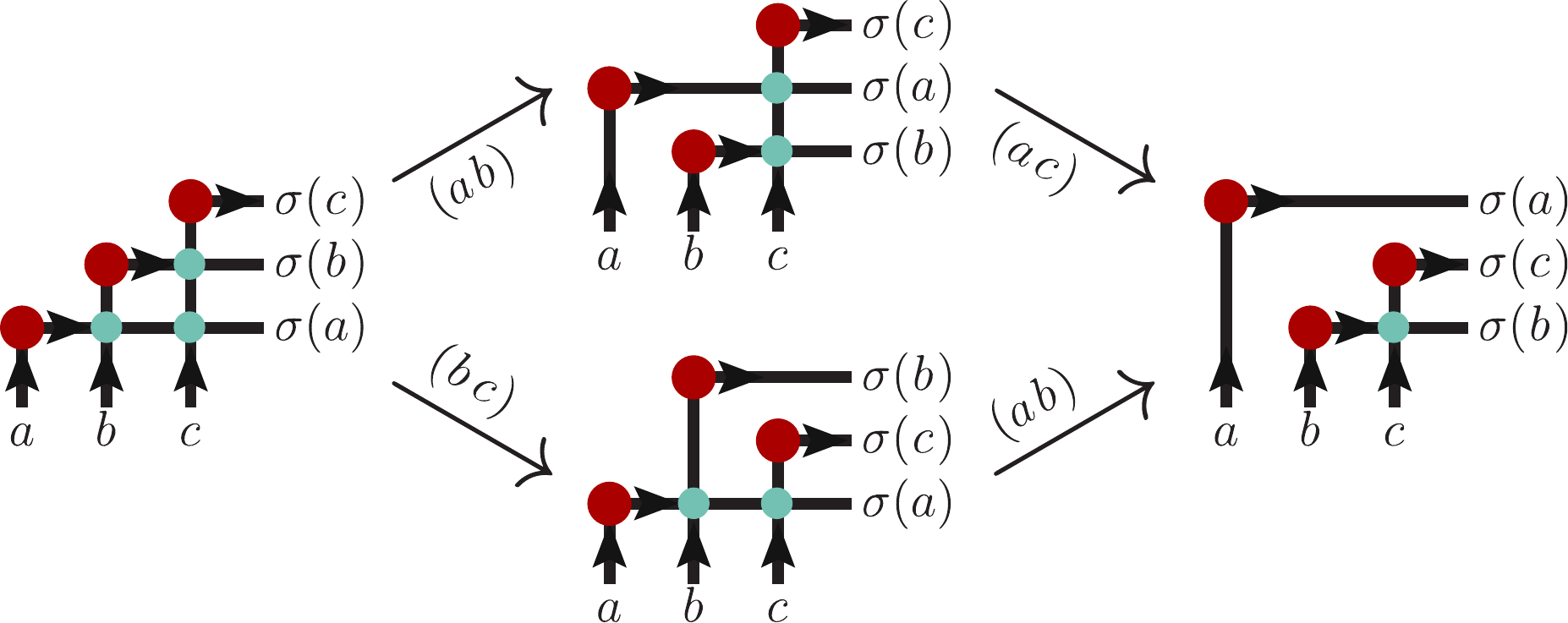}}\hspace{-3cm}\nonumber}

A more immediate, but somewhat indirect proof of this fact follows from the association of each permutation $\sigma$ with a reduced, on-shell graph. Recall that the graphs in the boundary of an on-shell graph labeled by $\sigma$ are those for which {\it one} edge has been removed. Because each pair of left-right paths $a\!\to\!\sigma(a)$ and $b\!\to\!\sigma(b)$ cross on {\it at most one} edge of any {\it reduced} graph (if the edge is removable), it is clear that graphs in $\partial^2$ are those obtained by removing a pair of edges. As such, the pair of edges can be removed in any order, proving that there are two paths from any graph to each graph in $\partial^2$.

(We should mention briefly that it remains an open and important problem to refine the definition of $\partial$ so that elements in $\partial(C)$ are decorated with signs $\pm1$ such that $\partial^2=0$ directly---not merely modulo 2.)

As mentioned above, an amazing feature of the positroid stratification is that the combinatorial structure of the inclusions induced by $\partial$ have the property that every positroid configuration defines an Eulerian poset---a combinatorial polytope. Because of this, we can loosely view each positroid configuration as a region of $G(k,n)$ with essentially the topology of an open ball---even though such a picture is only strictly known to be valid for relatively simple cases such as $G(2,n)$.

In the case of the positroid $G_+(2,4)$, the polytope is relatively easy to visualize. The four-dimensional top-cell has four, three-dimensional boundary configurations; and the boundaries of these cells collectively involve ten  two-dimensional configurations, etc. Starting with the generic configuration in $G_+(2,4)$, we find the boundaries defined by $\partial$ given as follows \cite{Williams:2003a}:
\vspace{.0cm}\eq{\hspace{-3.cm}\raisebox{-50pt}{\includegraphics[scale=.9]{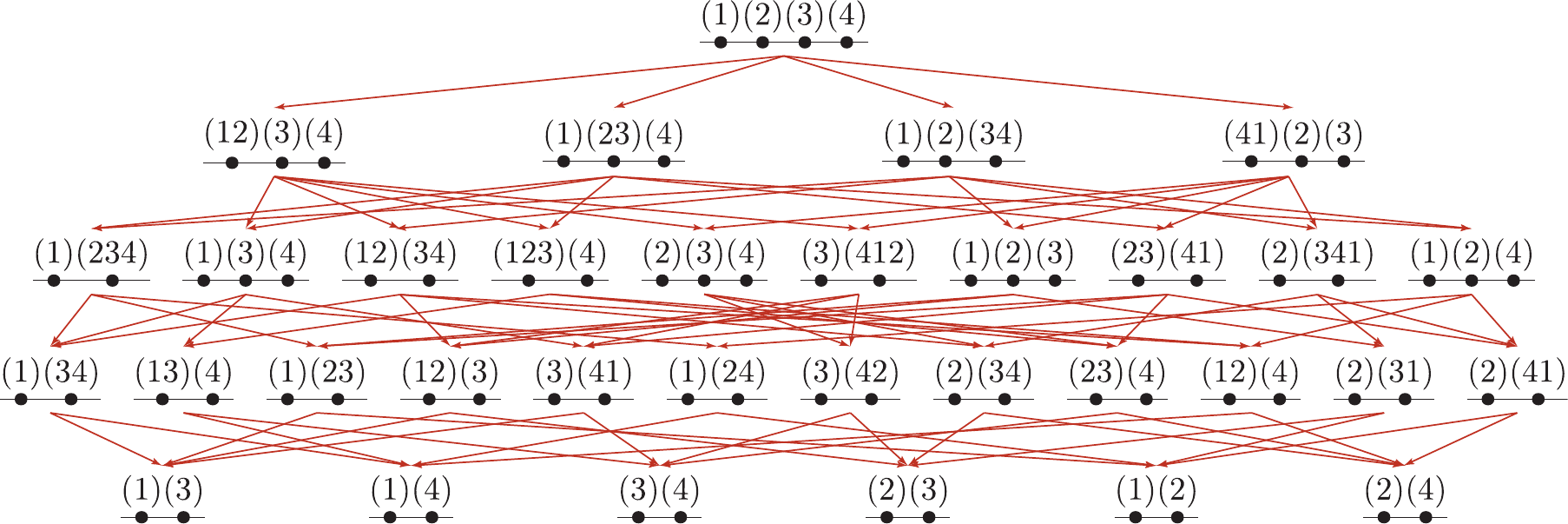}}\hspace{-3cm}\nonumber\vspace{.2cm}}

Although it is hard to draw the complete four-dimensional polytope, its four three-dimensional faces each define square-pyramidal regions of $G(2,4)$. For example, the polytope corresponding to the configuration \raisebox{-3.5pt}{{\footnotesize$\begin{array}{@{}c@{}c@{}c@{}}\\[-31pt]\\(1)&(2\,3)&(4)\\[-5pt]\bullet&\bullet&\bullet\\[-6.5pt]\hline\\[-6.25pt]\end{array}$}}\vspace{-4pt} of $G(2,4)$ labeled by the permutation ${\color{perm}\{4,3,5,6\}}$ is arranged as follows:
\vspace{-.1cm}\eq{\hspace{-3.354cm}\raisebox{-40pt}{\includegraphics[scale=.9]{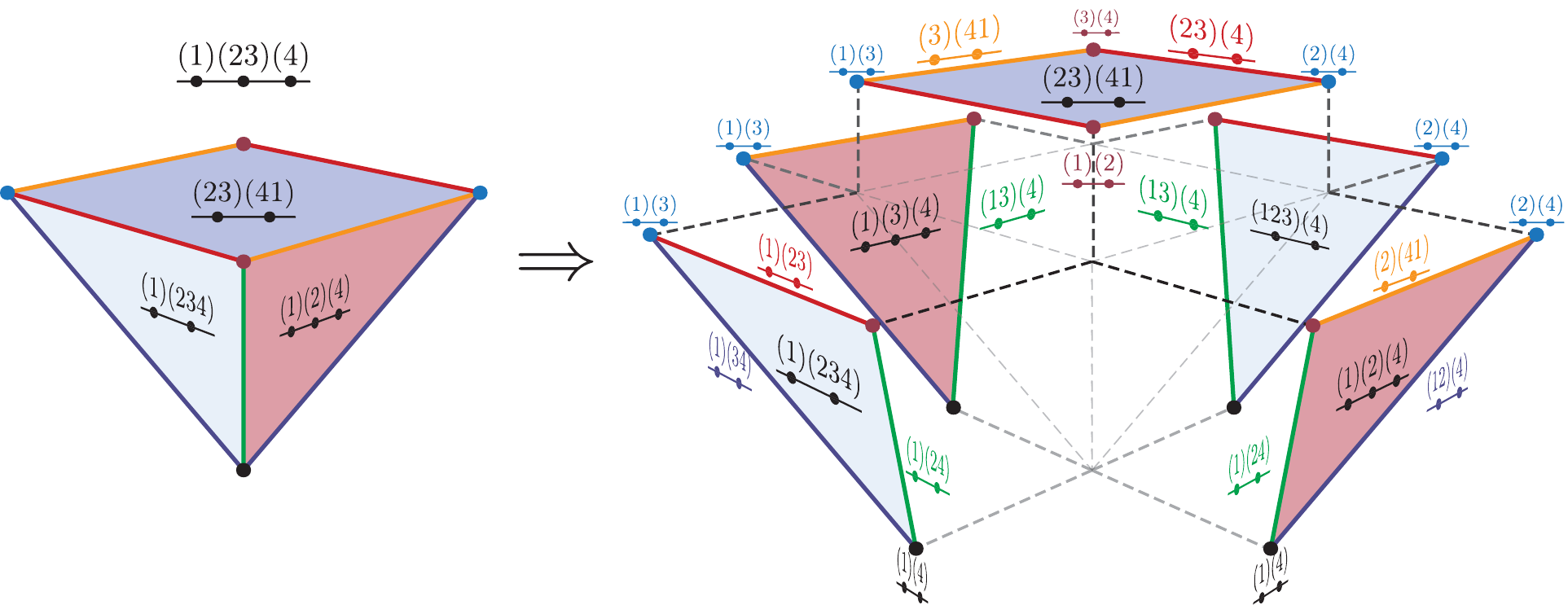}}\hspace{-3cm}\nonumber\vspace{-.5cm}}

\newpage
\subsection{Approaching Boundaries in Canonical Coordinates}\label{boundaries_in_canonical_coordinates}
Recall that the singularities of an on-shell differential form associated with an on-shell diagram are simply the residues of its poles. When written in terms of canonical coordinates on the Grassmannian as described above (see equation (\ref{physical_functions_in_canonical_coords_one})), it is tempting to identify the manifestly-logarithmic singularities in the measure with configurations in the `boundary'. But there are two important points which make such a correspondence a bit more delicate than it may appear at first-glance:
\begin{enumerate}
\vspace{-.2cm}\item the coordinate chart $\vec{\alpha}$ used to cover $C_\sigma$ may degenerate when some \mbox{$\alpha_i\!\to\!0$} ---such a degeneration would be signaled by the appearance of additional singularities in the Jacobian arising from the $\delta$-functions in (\ref{physical_functions_in_canonical_coords_one});
\vspace{-.2cm}\item no {\it single} coordinate chart $\vec{\alpha}$ covers {\it all} of the boundaries of $C_\sigma$.
\end{enumerate}
We can illustrate both points by considering a simple example. Recall from equation (\ref{bcfw_coordinates_for_g36_example}) the BCFW-bridge coordinates generated for the graph labeled by ${\color{perm}\{4,6,5,7,8,9\}}$:
\vspace{-0.45cm}\eq{\mbox{\hspace{-0.0cm}$\begin{array}{c}\\[-10pt]\raisebox{-58pt}{\includegraphics[scale=1]{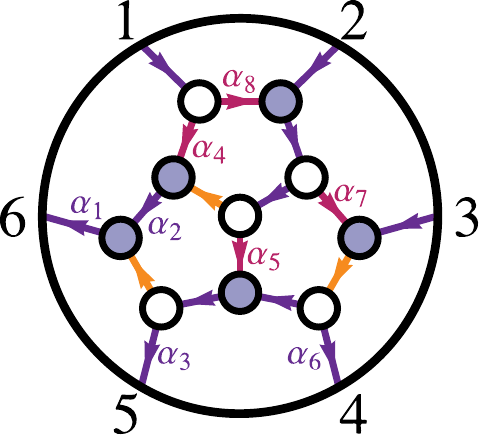}}\end{array}$\hspace{-0.0cm}}\label{g36_example_highlighting_boundary_in_coordinates}\vspace{-0.25cm}}
Because the BCFW coordinates $\vec{\alpha}$ correspond to edge-variables, sending any $\alpha_i\!\to\!0$ will have the effect of deleting the corresponding edge from the graph. The first subtlety mentioned above is reflected in the fact that some edge-variables---here, $\{\alpha_1,\alpha_2,\alpha_3,\alpha_6\}$---are attached to {\it irremovable} edges; the second subtlety is reflected in the fact that three of the seven removable edges---colored orange in the figure---are {\it not} dressed with edge-variables. Of course, if we introduce additional $GL(1)$-redundancies at each vertex as we did in \mbox{section \ref{boundary_measurements_section}}, every removable edge could be dressed by a variable whose vanishing would give the corresponding boundary; this would make all the boundaries accessible, but at the cost of introducing vast redundancy.

A surprising fact---not very difficult to prove---is that {\it all} the boundaries of any cell $C\!\in\!G_+(k,n)$ can be found at the zero-locus of single-coordinates in at least one chart from an atlas composed only of those charts generated by the BCFW-bridge construction (see \mbox{section \ref{BCFW_bridge_decomposition_subsection}}) in all its $n$ cyclic manifestations (taking each of the $n$ labels as the cyclic `starting-point' for the decomposition). To be clear, this claim only applies for the {\it specific scheme} described in \mbox{section \ref{BCFW_bridge_decomposition_subsection}} used to decompose a permutation into adjacent transpositions---no other scheme is known to have this remarkable property.

\newpage

\section{The Invariant Top-Form and the Positroid Stratification}\label{invariant_top_form_section}
We have seen that, associated with any $d$-dimensional cell of the positive Grassmannian, there is a natural associated form. In any of our natural coordinate charts, this $d$-form is just the ``$d\!\log$'' measure,
\vspace{-.2cm}\eq{\frac{d\alpha_1}{\alpha_1}\wedge\cdots\wedge\frac{d\alpha_d}{\alpha_d},\vspace{-.3cm}}
which is a special case of a more general cluster volume discussed in \mbox{section \ref{cluster_coordinates_section}}. This form makes it obvious that boundary configurations are associated with residues for some $\alpha_i=0$. It is also clear that we can view all cells $C\!\in\!G_+(k,n)$ as iterated residues of the {\it top-form} $\Omega^{{\rm top}}$ on a {\it generic} configuration \mbox{$C\!\in\!G_+(k,n)$}.

A natural question is  whether this top-form $\Omega^{{\rm top}}$ can be written directly in terms of the `matrix-coordinates' $c_{\alpha\,a}$ of $C$.
In terms of matrix-coordinates $C\equiv c_{\alpha\,a}$, the desired measure $G(k,n)$ would have the form,
\vspace{-.2cm}\eq{\Omega = \frac{d^{k\times n}C}{\mathrm{vol}(GL(k))}\frac{1}{f(C)},\vspace{-.2cm}}
where  $f(C)$ must be a function of the {\it minors} of $C$, and must scale uniformly as $f(tC)=t^{k\times n} f(C)$. Moreover, because the top-cell $G_+(k,n)$ {\it always} has precisely $n$ co-dimension one boundaries---corresponding to any $k$ consecutive columns becoming linearly-dependent---it is clear that $f(C)$ must have {\it at least} the $n$ cyclic-minors as factors:
\vspace{-.2cm}\eq{f(C)=(1\cdots k)\cdots(n\cdots k\mi1) f'(C).\vspace{-.2cm}}
Because the product of the cyclic minors scale as $f(C)$ must, $f'(C)$ must be scale-invariant: $f'(tC)=f'(C)$. And so, $f'(C)$ can at most involve ratios of minors. However, any non-consecutive minors appearing as factors in $f'(C)$ would generate new, unwanted singularities for the top-cell--poles corresponding to co-dimension one boundaries not in the positroid stratification---and any consecutive minors in $f'(C)$ would make a double-pole, spoiling the logarithmic singularities corresponding to one of the {\it necessary} boundary configurations. Therefore, we are forced to conclude that the only choice is to take $f'(C)\!\to\!1$. This means that the {\it only} viable ansatz  for a measure on $G(k,n)$ with the desired properties is:
\vspace{-.25cm}\eq{\Omega = \frac{d^{k\times n}C}{\mathrm{vol}(GL(k))}\frac{1}{(1\cdots k)\cdots(n\cdots k\mi1)}.\vspace{-.24cm}\label{cyclic_minor_top_form}}

This strikingly-simple form was first encountered in connection with ``leading singularities'' in \mbox{reference \cite{ArkaniHamed:2009dn}}.

It is not hard to see the plausibility of a guess that $\Omega = \Omega^{{\rm top}}$. We have just established that the poles of $\Omega$ and $\Omega^{{\rm top}}$ are the same, and furthermore $\Omega$ does not have any zeroes on the Grassmannian. Thus $\Omega^{{\rm top}}/\Omega$ is a function of the Grassmannian with no poles, and any such function must be a constant. So, we have
\vspace{-.3cm}\eq{\frac{d^{k\times n}C}{\mathrm{vol}(GL(k))}\frac{1}{(1\cdots k)\cdots(n\cdots k\mi1)}=\frac{d\alpha_1}{\alpha_1}\wedge\cdots\wedge\frac{d\alpha_{k(n-k)}}{\alpha_{k(n-k)}}.\vspace{-.1cm}}
This representation of the top form will be crucial for most transparently seeing the dual conformal symmetry and Yangian invariance of the theory.

We will momentarily prove that $\Omega = \Omega^{{\rm top}}$  by direct computation as well, but let us first step-back and observe some remarkable properties of $\Omega$. It is rather surprising that a form as simple as (\ref{cyclic_minor_top_form})---which has only $n$ poles!---should be able to capture all of the intricate and beautiful structure of the positive Grassmannian in its iterated singularities. The reason why this isn't obviously impossible is that each of these $n$ factors are generally $k^{\mathrm{th}}$-degree polynomials in the variables $c_{\alpha\,a}$, and whenever one such minor vanishes, other minors typically factorize, exposing further singularities and more structure below.

Let us consider an example which illustrates how the iterated factorizations of the consecutive minors exposes all the cells in the positroid stratification. Consider the top-cell of $G(3,6)$,
\vspace{-0.3cm}\eq{\raisebox{-42.5pt}{\includegraphics[scale=.85]{g36_configurations_9d}}\text{{\LARGE$\;\;\Leftrightarrow\;\;$}}\frac{d^{3\times 6}C}{\mathrm{vol}(GL(3))}\frac{1}{(123)(234)(345)(456)(561)(612)}.\vspace{-0.2cm}\label{g36_d_eq_9_configuration_v2}}
Upon restricting this form to the residue  where $(234)\!\to\!0$, the configuration becomes:
\vspace{-0.75cm}\eq{\raisebox{-42.5pt}{\includegraphics[scale=.85]{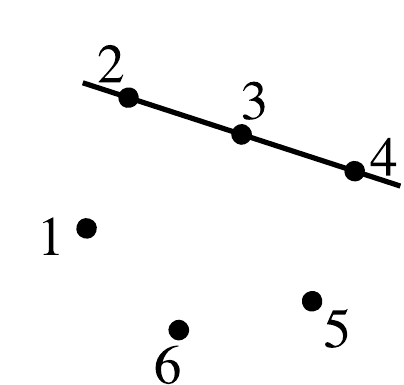}}}
Now, as described in \mbox{section \ref{configuration_of_vectors_section}}, this configuration contains $7$ boundary configurations. How are we to see {\it seven} logarithmic singularities arising from the {\it five} remaining cyclic minors of (\ref{g36_d_eq_9_configuration_v2})? The answer is simple: let us parameterize the pole $(234)\!\to\!0$ by sending $c_3\!\to\!\beta_{3\,2}c_2+\beta_{3\,4}c_4$, under which the minors $(123)$ and $(345)$ each factorize:
\vspace{-.1cm}\eq{\frac{1}{(123)(234)(345)(456)(561)(612)}\xrightarrow[\substack{\\[-5pt]\mathrm{via}\\c_3=\beta_{3\,2}c_2+\beta_{3\,4}c_4}]{(234)\to0}\frac{1}{\underbrace{\beta_{3\,4}(124)}_{(123)}\underbrace{\beta_{3\,2}(245)}_{(345)}(456)(561)(612)},\vspace{-.2cm}\nonumber}
exposing all seven of the boundary configurations! To further illustrate this point, let us now take a residue of this measure about the configuration setting $(561)\!\to\!0$, by setting $c_6\!\mapsto\!\beta_{6\,5}c_5+\beta_{6\,1}c_1$; as before, this leads to the factorization of minors $(456)$ and $(612)$, leaving us with,
\vspace{-.1cm}\eq{\frac{1}{\beta_{3\,4}(124)\beta_{3\,2}(245)(456)(561)(612)}\xrightarrow[\substack{\\[-5pt]\mathrm{via}\\c_6=\beta_{6\,5}c_5+\beta_{6\,1}c_1}]{(561)\to0}\frac{1}{\beta_{3\,4}(124)\beta_{3\,2}(245)\underbrace{\beta_{6\,1}(461)}_{(456)}\underbrace{\beta_{6\,5}(512)}_{(612)}},\vspace{-.2cm}\nonumber}
which shows that this configuration has {\it eight} further boundary configurations. Proceeding in this way we can reconstruct all the cells of $G_+(3,6)$.

\subsection{Proving Equivalence with the Canonical Positroid Measure}
In section \ref{polygon_coordinates_section} we showed that we can construct canonical coordinates for the top-cell of $G_+(k,n)$ recursively by first introducing coordinates
\vspace{-.2cm}\eq{C^{(1,n)}\equiv\left(\begin{array}{ccccc}{\color[rgb]{0,0,0.9}\beta_{1,1}}&{\color[rgb]{0,0,0.8}\beta_{1,2}}&\cdots&{\color[rgb]{0,0,0.7}\beta_{1,n-1}}&{\color[rgb]{0,0,0.6}\beta_{1,n}}\end{array}\right)\,,\vspace{-.2cm}\label{g1n_polygon_coordinates_v2}}
for $G(1,n)$, and then building-up coordinates for any $G(k,n)$ recursively via:
\vspace{-.2cm}\eq{C^{(k,n)}\equiv\left(\begin{array}{ccccc}{\color[rgb]{0.9,0,0}\beta_{k,1}}{\color[rgb]{0,0,0.7}\hat{c}^{\,(k,n)}_1}&\cdots&{\color[rgb]{0.7,0,0}\beta_{k,n-1}}{\color[rgb]{0,0,0.7}\hat{c}^{\,(k,n)}_{n-1}}&0\\{\color[rgb]{0.9,0,0}\beta_{k,1}}&\cdots&{\color[rgb]{0.7,0,0}\beta_{k,n-1}}&{\color[rgb]{0.6,0,0}\beta_{k,n}}\end{array}\right)\quad\mathrm{with}\quad {\color[rgb]{0,0,0.75}\hat{c}^{\,(k,n)}_a}\equiv\sum_{j=(a+1)}^{n}c^{(k-1,n)}_j\!.\vspace{-.2cm}}
Recall that these coordinates match those obtained by the BCFW bridge construction upon the trivial relabeling:
\vspace{-.75cm}\eq{\hspace{-1.25cm}\raisebox{-2.5pt}{$\left.\rule{0pt}{34pt}\right.$}\begin{array}{@{}|rccccccccl|@{}}\multicolumn{10}{c}{~}\\\hline{\color[rgb]{0.0,0.4,0.3}\alpha_d}&{\color[rgb]{0.8,0.1,0.2}\alpha_{d-2}}&\cdots&\cdots&\cdots&\cdots&{\color[rgb]{0.2,0.1,0.8}\alpha_{\ell}}&\cdots&\cdots&\alpha_{k(k-1)/2+1}\\[-5pt]
\raisebox{3pt}{${\color[rgb]{0.8,0.1,0.2}\alpha_{d-1}}$}&\iddots&\iddots&\iddots&\iddots&\raisebox{3pt}{${\color[rgb]{0.2,0.1,0.8}\alpha_{\ell+1}}$}&\iddots&\iddots&\iddots&\vdots\\[-8pt]
\vdots&\iddots&\iddots&\iddots&{\color[rgb]{0.2,0.1,0.8}\iddots}&\iddots&\iddots&\iddots&\iddots&\raisebox{2pt}{${\color[rgb]{0.8,0.1,0.2}\alpha_2}$}\\[-5pt]
\alpha_{d-k(k-1)/2}&\cdots&\cdots&{\color[rgb]{0.2,0.1,0.8}\alpha_{\ell+k-1}}&\cdots&\cdots&\cdots&\cdots&{\color[rgb]{0.8,0.1,0.2}\alpha_3}&{\color[rgb]{0.0,0.4,0.3}\alpha_1}\\\hline\multicolumn{1}{c}{~}\\
\end{array}\raisebox{-2.5pt}{$\left.\rule{0pt}{34pt}\right.\!\!\text{{\LARGE$\Leftrightarrow$}}$}
\begin{array}{@{}|ccccc|@{}}\multicolumn{5}{c}{\phantom{n-k}}\\\hline\beta_{1,k+1}&\beta_{1,k+2}&\cdots&\beta_{1,n-1}&\beta_{1,n}\\[-3.5pt]
\beta_{2,k+1}&\beta_{2,k+2}&\cdots&\beta_{2,n-1}&\beta_{2,n}\\[-3pt]
\vdots&\vdots&\ddots&\vdots&\vdots\\[-5pt]
\beta_{k,k+1}&\beta_{k,k+2}&\cdots&\beta_{k,n-1}&\beta_{k,n}\\\hline\multicolumn{1}{c}{~}\\
\end{array}\label{general_form_of_bcfw_variables_second}\hspace{-.75cm}\nonumber\vspace{-.7cm}}
and the gauge-choice of setting the first $k$ column-vectors to the identity matrix. The motivation for relabeling the coordinates in this way is that the BCFW-coordinates give rise a gauge-fixed parameterization of $C(\beta_{\alpha,a})$ of the form,
\vspace{-.2cm}\eq{\hspace{-0.75cm}\begin{array}{c}~\\1\\[-7pt]\phantom{\vdots}2\phantom{\vdots}\\[-6pt]\vdots\\[-3pt]k\end{array}\!\!\raisebox{-7.5pt}{$\left(\rule{0pt}{34pt}\right.$}\begin{array}{@{}cc@{$\!$}cc|cccc@{}}
1&2&\cdots&\multicolumn{1}{c}{k}&\,k\pl1\,&\cdots&\cdots&n\\\hline1&0&\cdots&0&({\color[rgb]{0.2,0.1,0.8}\beta_{1,k+1}}\!\cdots\beta_{k,k+1})\!+\!\cdots&({\color[rgb]{0.2,0.1,0.8}\beta_{1,k+2}}\!\cdots\beta_{k,k+2})\!+\!\cdots&\cdots&({\color[rgb]{0.2,0.1,0.8}\beta_{1,n}}\!\cdots\beta_{k,n})\!+\!\cdots\!\\[-7pt]
0&1&\ddots&\vdots&({\color[rgb]{0.2,0.1,0.8}\beta_{2,k+1}}\!\cdots\beta_{k,k+1})\!+\!\cdots&({\color[rgb]{0.2,0.1,0.8}\beta_{2,k+2}}\!\cdots\beta_{k,k+2})\!+\!\cdots&\cdots&({\color[rgb]{0.2,0.1,0.8}\beta_{2,n}}\!\cdots\beta_{k,n})\!+\!\cdots\!\\[-6pt]
\vdots&\ddots&\ddots&0&\vdots&\vdots&\ddots&\vdots\\[-3pt]
0&\cdots&0&1&{\color[rgb]{0.2,0.1,0.8}\beta_{k,k+1}}&{\color[rgb]{0.2,0.1,0.8}\beta_{k,k+2}}&\cdots&{\color[rgb]{0.2,0.1,0.8}\beta_{k,n}}
\end{array}\raisebox{-7.5pt}{$\left.\rule{0pt}{34pt}\right)$}\raisebox{-20pt}{}\label{beta_form_of_top_form_matrix}\hspace{-1cm}\vspace{-.2cm}}
Here, we have used color to highlight the fact that $c_{\alpha,a}\propto{\color[rgb]{0.2,0.1,0.8}\beta_{\alpha,a}}(\beta_{\alpha+1,a}\cdots\beta_{k,a})\!+\ldots$, and that only this factor contributes to the Jacobian in going from coordinates $c_{\alpha,a}$ to coordinates $\beta_{\alpha,a}$. In particular, it is easy to see that the entire Jacobian from this change of variables is simply,
\vspace{-.2cm}\eq{J\equiv\left|\frac{d c_{\alpha,a}}{d \beta_{\alpha,a}}\right|=\prod_{\alpha,a}\big(\beta_{\alpha,a}\big)^{\alpha-1}.\vspace{-.2cm}\label{jacobian_of_polygon_coordinates}}
Somewhat less obviously, the cyclic minors are all simply expressed in these coordinates: each is the product of all the {\it highlighted} ${\color[rgb]{0.2,0.1,0.8}\beta_{\alpha,a}}$ in the lower-right triangle of the corresponding sub-matrix of (\ref{beta_form_of_top_form_matrix}):
\vspace{-.4cm}\eq{(\ell\cdots\ell\pl k\mi1)=\prod_{\alpha=1}^{k}\left(\prod_{a=1}^{\alpha}{\color[rgb]{0.2,0.1,0.8}\beta_{\alpha,(k+\ell-a)}}\right)\raisebox{-3pt}{\text{{\LARGE$\;\Leftrightarrow\;$}}}\left|\begin{array}{cccc}{\color{deemph}\beta_{1,\ell}}&{\color{deemph}\cdots}&{\color{deemph}\cdots}&{\color[rgb]{0.2,0.1,0.8}\beta_{1,\ell+k-1}}\\[-3pt]{\color{deemph}\vdots}&{\color{deemph}\iddots}&{\color[rgb]{0.2,0.1,0.8}\beta_{2,\ell+k-2}}&{\color[rgb]{0.2,0.1,0.8}\beta_{2,\ell+k-1}}\\[-3pt]{\color{deemph}\vdots}&{\color[rgb]{0.2,0.1,0.8}\iddots}&{\color[rgb]{0.2,0.1,0.8}\vdots}&{\color[rgb]{0.2,0.1,0.8}\vdots}\\[-3pt]{\color[rgb]{0.2,0.1,0.8}\beta_{k,\ell}}&{\color[rgb]{0.2,0.1,0.8}\cdots}&{\color[rgb]{0.2,0.1,0.8}\beta_{k,\ell+k-2}}&{\color[rgb]{0.2,0.1,0.8}\beta_{k,\ell+k-1}}\end{array}\right|,\vspace{-.1cm}}
where the product of $\beta$'s only ranges over relevant columns: $k\pl1\leq(k\pl\,\ell\,\mi\,a)\leq n$. And so, the product of all the consecutive minors is simply,
\vspace{-.0cm}\eq{(1\cdots k)(2\cdots k\pl1)\cdots(n\cdots k\mi1)=\prod_{\alpha,a}\big(\beta_{\alpha,a}\big)^{\alpha}.\vspace{-.3cm}}

Therefore, combining the product of all the cyclic minors with the necessary Jacobian given in (\ref{jacobian_of_polygon_coordinates}) we have:
\vspace{-.2cm}\eq{\frac{d^{k\times n}c_{\alpha,a}}{\mathrm{vol}(GL(k))}\frac{1}{(1\cdots k)\cdots(n\cdots k\mi1)}=\left(\prod_{\alpha,a}d\beta_{\alpha,a}\right)\frac{J}{\prod_{\alpha,a}\big(\beta_{\alpha,a}\big)^{\alpha}}=\prod_{\alpha,a}\frac{d\beta_{\alpha,a}}{\beta_{\alpha,a}} \vspace{-.2cm}}
as desired.

Let us briefly consider one concrete example of this equivalence. Consider the top-cell of $G(3,6)$, where the BCFW-bridge construction gives the matrix-representative,
{\small\eqs{\hspace{-1cm}C(\alpha)&=\left(
\begin{array}{@{}cccccc@{}}
 1 & 0 & 0 & {\color[rgb]{0.2,0.1,0.8}\alpha _9} \alpha _8 \alpha _6 &{\color[rgb]{0.2,0.1,0.8}\alpha _7}\alpha_5\alpha _3{\color{deemph}\pl\,\alpha_3\alpha_9(\alpha_5\pl\,\alpha_8)}
 &{\color[rgb]{0.2,0.1,0.8}\alpha _4}\alpha_2\alpha _1{\color{deemph}\pl\,\alpha_1\left(\alpha_7(\alpha_2\pl\,\alpha_5)\pl\,\alpha_9(\alpha_2\pl\,\alpha_5\pl\,\alpha_8)\right)}\\
 0 & 1 & 0 & \mi\,{\color[rgb]{0.2,0.1,0.8}\alpha _8} \alpha _6 &\mi\,{\color[rgb]{0.2,0.1,0.8}\alpha_5}\alpha_3{\color{deemph}\mi\,\alpha _3 \alpha _8}&\mi\,{\color[rgb]{0.2,0.1,0.8}\alpha _2}\alpha _1{\color{deemph}\mi\,\alpha_1\left(\alpha _5+\alpha _8\right)}\\
 0 & 0 & 1 & {\color[rgb]{0.2,0.1,0.8}\alpha _6} & {\color[rgb]{0.2,0.1,0.8}\alpha _3} &{\color[rgb]{0.2,0.1,0.8}\alpha _1}
\end{array}
\right)\,,\nonumber}}which, upon relabeling the variables according to,
\vspace{-.2cm}\eq{\begin{array}{|ccc|}\hline{\color[rgb]{0,0.4,0.3}\alpha_9}&{\color[rgb]{0.8,0.1,0.2}\alpha_7}&{\color[rgb]{.2,.1,.8}\alpha_4}\\{\color[rgb]{0.8,0.1,0.2}\alpha_8}&{\color[rgb]{.2,.1,.8}\alpha_5}&{\color[rgb]{0.8,0.1,0.2}\alpha_2}\\{\color[rgb]{.2,.1,.8}\alpha_6}&{\color[rgb]{0.8,0.1,0.2}\alpha_3}&{\color[rgb]{0,.4,.3}\alpha_1}\\\hline\end{array}\raisebox{-3.0pt}{\text{{\LARGE$\;\Rightarrow\;$}}}\begin{array}{|ccc|}\hline{\color[rgb]{0,0.4,0.3}\beta_{1,4}}&{\color[rgb]{0.8,0.1,0.2}\beta_{1,5}}&{\color[rgb]{.2,.1,.8}\beta_{1,6}}\\{\color[rgb]{0.8,0.1,0.2}\beta_{2,4}}&{\color[rgb]{.2,.1,.8}\beta_{2,5}}&{\color[rgb]{0.8,0.1,0.2}\beta_{2,6}}\\{\color[rgb]{.2,.1,.8}\beta_{3,4}}&{\color[rgb]{0.8,0.1,0.2}\beta_{3,5}}&{\color[rgb]{0,0.4,0.3}\beta_{3,6}}\\\hline\end{array}\;,\vspace{-.4cm}}
becomes,
{\small\eqs{\hspace{-1cm}C(\beta)&=\left(
\begin{array}{@{}cccccc@{}}
 1 & 0 & 0 & {\color[rgb]{0.2,0.1,0.8}\beta_{1,4}}\beta_{2,4}\beta_{3,4}&{\color[rgb]{0.2,0.1,0.8}\beta_{1,5}}\beta_{2,5}\beta_{3,5}{\color{deemph}\pl\,\ldots}
 &{\color[rgb]{0.2,0.1,0.8}\beta_{1,6}}\beta_{2,6}\beta_{3,6}{\color{deemph}\pl\ldots}\\
 0 & 1 & 0 & \mi\,{\color[rgb]{0.2,0.1,0.8}\beta_{2,4}} \beta_{3,4}&\mi\,{\color[rgb]{0.2,0.1,0.8}\beta_{2,5}}\beta_{3,5}{\color{deemph}\,\mi\ldots}&\mi\,{\color[rgb]{0.2,0.1,0.8}\beta_{2,6}}\beta_{3,6}{\color{deemph}\,\mi\ldots}\\
 0 & 0 & 1 & {\color[rgb]{0.2,0.1,0.8}\beta_{3,4}} & {\color[rgb]{0.2,0.1,0.8}\beta_{3,5}} &{\color[rgb]{0.2,0.1,0.8}\beta_{3,6}}
\end{array}
\right)\,.}}
It is easy to see that the cyclic minors are given by,
\eq{\begin{array}{l@{$\,=\,$}c@{$\qquad$}l@{$=$}c}(1\,2\,3)&1&(4\,5\,6)&\beta_{1,6}\,\beta_{2,5}\,\beta_{2,6}\,\beta_{3,4}\,\beta_{3,5}\,\beta_{3,6}\\(2\,3\,4)&\beta_{1,4}\,\beta_{2,4}\,\beta_{3,4}&(5\,6\,1)&\beta_{2,6}\,\beta_{3,5}\,\beta_{3,6}\\(3\,4\,5)&\beta_{1,5}\,\beta_{2,4}\,\beta_{2,5}\,\beta_{3,4}\,\beta_{3,5}&(6\,1\,2)&\beta_{3,6}\end{array}}
so that their product gives,
\eq{(1\,2\,3)\cdots(6\,1\,2)=\big(\beta_{1,4}\,\beta_{1,5}\,\beta_{1,6}\big)^1\;\big(\beta_{2,4}\,\beta_{2,5}\,\beta_{2,6}\big)^2\;\big(\beta_{3,4}\,\beta_{3,5}\,\beta_{3,6}\big)^3;}
and the Jacobian of going from $c_{\alpha,a}$ to $\beta_{\alpha,a}$ is easily seen to be,
\eq{J\equiv\left|\frac{dc_{\alpha,a}}{d\beta_{\alpha,a}}\right|=\big(\beta_{1,4}\,\beta_{1,5}\,\beta_{1,6}\big)^0\;\big(\beta_{2,4}\,\beta_{2,5}\,\beta_{2,6}\big)^1\;\big(\beta_{3,4}\,\beta_{3,5}\,\beta_{3,6}\big)^2,}
so that
\eq{\frac{d^{3\times 6}C}{\mathrm{vol}(GL(3))}\frac{1}{(1\,2\,3)(2\,3\,4)(3\,4\,5)(4\,5\,6)(5\,6\,1)(6\,1\,2)}=\prod_{\alpha,a}\frac{d\beta_{\alpha,a}}{\beta_{\alpha,a}}.}

\section{\mbox{(Super) Conformal and Dual Conformal Invariance}}\label{superconformal_and_dual_superconformal_invariance}

In this section, we will describe how the Grassmannian formulation of on-shell diagrams makes all the symmetries of the theory---both the super-conformal {\it and} {\it dual} super-conformal symmetries---completely manifest. Along the way, we will find it useful to recast the on-shell differential form's dependence on external kinematical data in a way which more transparently reflects the geometry of momentum-conservation; doing so, we will discover a correspondence between (some) cells $C\!\in\!G(k,n)$ with cells \mbox{$\hat{C}\!\in\!G(k\,\mi\,2,n)$}.

\subsection{The Grassmannian Geometry of Momentum Conservation}\label{grassmannian_geometry_of_momentum_conservation_subsection}

Consider an arbitrary on-shell graph associated with the cell $\Gamma_\sigma\!\in\!G(k,n)$ labeled by the permutation $\sigma$ associated with an on-shell differential form $f^{(k)}_\sigma(1,\ldots,n)$. Using any of the canonical coordinates for the cell $C(\alpha_1,\ldots,\alpha_d)\subset\Gamma_\sigma\!\in\!G(k,n)$, this form is given by:
\vspace{-.2cm}\eq{\hspace{-.75cm}f_\sigma^{(k)}=\int\!\!\frac{d\alpha_1}{\alpha_1}\wedge\cdots\wedge\frac{d\alpha_d}{\alpha_d}\;\delta^{k\times4}\big(C\!\cdot\!\widetilde{\eta}\big)\delta^{k\times2}\big(C\!\cdot\!\widetilde{\lambda}\big)\delta^{2\times(n-k)}\big(\lambda\!\cdot\!C^\perp\!\big).\vspace{-.2cm}\label{general_on_shell_function_canonical_coordinates}}
As we saw in \mbox{section \ref{invariant_top_form_section}}, this can also be written as a residue of the top-form,
\vspace{-.2cm}\eq{\hspace{-0.75cm}f_\sigma^{(k)}\!=\!\oint\limits_{C\subset\Gamma_\sigma}\!\!\!\frac{d^{k \times n} C}{\mathrm{vol}(GL(k))}\;\frac{\delta^{k\times4}\big(C\!\cdot\!\widetilde{\eta}\big)}{(1\cdots k)\cdots(n\cdots k\mi1)}\delta^{k\times2}\big(C\!\cdot\!\widetilde{\lambda}\big)\delta^{2\times(n-k)}\big(\lambda\!\cdot\!C^\perp\!\big).\hspace{-1.3cm}\vspace{-.2cm}\label{general_on_shell_function}}

Recall from \mbox{section \ref{intro_to_grassmannian_section}}, the (ordinary) $\delta$-functions in (\ref{general_on_shell_function}) have the geometric interpretation of constraining the $k$-plane $C$ to be {\it orthogonal to} the 2-plane $\widetilde{\lambda}$ and to {\it contain} the $2$-plane $\lambda$, \cite{ArkaniHamed:2009dn}:
\vspace{-.2cm}\eq{\raisebox{-50pt}{\includegraphics[scale=1]{momentum_space_constraints}}\label{momentum_space_constraints}\vspace{-.2cm}}
Because $\widetilde{\lambda}\subset\lambda^\perp$, 4 of the $2n(=2(n\mi\,k)\pl\,2k)$ constraints always represent momentum-conservation, leaving $(2n\,\mi\,4)$ constraints imposed on $C$ in general. Therefore, cells of $G(k,n)$ with precisely $(2n\,\mi\,4)$ degrees of freedom can be {\it fully-localized} by these constraints, and become {\it ordinary} super-functions of the external momenta; cells of lower dimension become functions with $\delta$-function support, and cells of higher dimension represent  {\it integration measures} on auxiliary, {\it internal} degrees of freedom (which may represent, for example, the degrees of freedom of internal loop-momenta).

The simplest example illustrating this localization is for $k=2$. Here the $2$-plane $C$ is just identified with the $\lambda$-plane, and equation (\ref{general_on_shell_function}) directly becomes the familiar Parke-Taylor formula for tree-level MHV super-amplitudes, \cite{Parke:1986gb,Nair:1988bq}:
\vspace{-.2cm}\eq{\begin{array}{ll}\mathcal{A}_n^{(2)}&=\displaystyle\!\int\!\!\!\frac{d^{2\times n} C}{\mathrm{vol}(GL(2))}\;\frac{\delta^{2\times 4}\big(C\!\cdot\!\widetilde{\eta}\big)}{(12)(23)\cdots(n1)}\delta^{2\times 2}\big(C\!\cdot\!\widetilde{\lambda}\big)\delta^{2\times(n-2)}\big(\lambda\!\cdot\!C^\perp\!\big),\\[15pt]&\displaystyle=\frac{\delta^{2\times 4}\big(\lambda\!\cdot\!\widetilde\eta\big)}{\ab{1\,2}\ab{2\,3}\cdots\ab{n\,1}} \delta^{2\times 2}\big(\lambda\!\cdot\!\widetilde{\lambda}\big).\end{array}\vspace{-.2cm}\label{parke_taylor_tree_from_grassmannian}}

Let us look at a less trivial example of how this localization works for $k>2$. One of the on-shell diagrams contributing to the $6$-particle $k=3$ tree-amplitude is (see \mbox{section \ref{on_shell_scattering_amplitudes_section}}), \vspace{-.3cm}\eq{\raisebox{-46pt}{\includegraphics[scale=1]{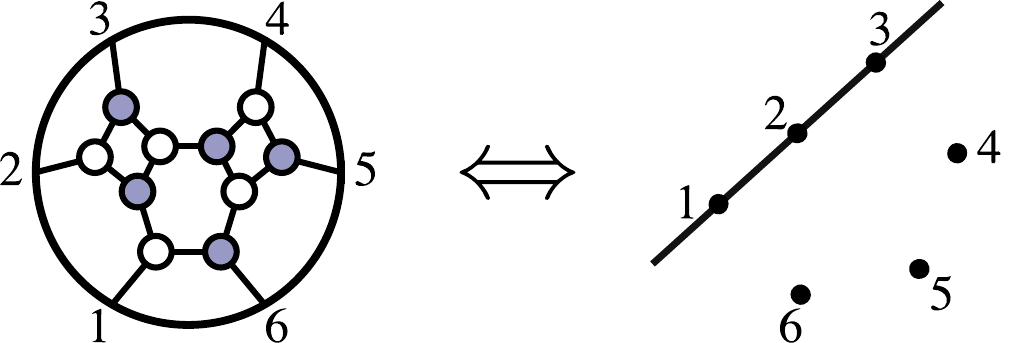}},\vspace{-.1cm}\label{g36_example_graph_and_configuration}}
which is labeled by the permutation ${\color{perm}\{3, 5, 6, 7, 8, 10\}}$. It is easy to see that (a $GL(3)$-representative of) the point $C^*$ in this positroid cell which satisfies the kinematical constraints is:
\vspace{-.2cm}\eq{C^*=\left(\begin{array}{@{}cccccc@{}}\lambda_1^1&\lambda_2^1&\lambda_3^1&\lambda_4^1&\lambda_5^1&\lambda_6^1\\ \lambda_1^2&\lambda_2^2&\lambda_3^2&\lambda_4^2&\lambda_5^2&\lambda_6^2\\\,\,\,0\,\,\,&\,\,\,0\,\,\,&\,\,\,0\,\,\,&\sb{5\,6}&\sb{6\,4}&\sb{4\,5}\end{array}\right)\vspace{-.2cm}\label{g36_configuration_cMatrix},}
where
$\sb{a\,b}\equiv\det\{\widetilde{\lambda}_a,\widetilde{\lambda}_b\}$ is a minor of the matrix $\widetilde{\lambda}$. (Notice that $C^*\!\!\cdot\!\widetilde{\lambda}=0$ because $\lambda\!\cdot\!\widetilde{\lambda}=0$, and the third-row dotted-into $\widetilde{\lambda}$ gives an instance of (\ref{Cramers_rule}).) Supported at this point, (\ref{general_on_shell_function}) generates the on-shell super-function,
\vspace{-.2cm}\eq{\hspace{-2.95cm}f^{(3)}_{{\color{perm}\{3, 5, 6, 7, 8, 10\}}}\!\!=\!\!\frac{\delta^{3\times4}(C^*\!\!\cdot\widetilde \eta)\,\delta^{2\times2}\big(\lambda\!\cdot\!\widetilde{\lambda}\big)}{\underbrace{\ab{2\,3}\sb{5\,6}}_{\phantom{|_{C^*}}(234)|_{C^*}}\!\underbrace{(\ab{3\,4}\sb{6\,4}\pl\ab{5\,3}\sb{5\,6})}_{\phantom{|_{C^*}}(345)|_{C^*}}\!\!\underbrace{\phantom{\ab{}}s_{4\,5\,6}\phantom{\ab{}}}_{\phantom{|_{C^*}}(456)|_{C^*}}\!\!\underbrace{(\ab{6\,1}\sb{6\,4}\pl\ab{1\,5}\sb{4\,5})}_{\phantom{|_{C^*}}(561)|_{C^*}}\!\underbrace{\ab{1\,2}\sb{4\,5}}_{\phantom{|_{C^*}}(612)|_{C^*}}},\hspace{-2cm}\vspace{-.2cm}\label{g36_on_shell_function}}
where
\eq{s_{456}\equiv(p_4+p_5+p_6)^2=\ab{4\,5}\sb{4\,5}+\ab{4\,6}\sb{4\,6}+\ab{5\,6}\sb{5\,6}.\vspace{-.1cm}\nonumber}
The particular $GL(3)$-representative of $C^*$ given in (\ref{g36_configuration_cMatrix}) was chosen so that the Jacobian from all the $\delta$-functions is 1, making the residue of (\ref{general_on_shell_function}) about the pole $(123)=0$ easy to read-off from $C^*$. Let us briefly mention that (\ref{g36_on_shell_function}) makes {\it super} momentum-conservation manifest: in addition to the obvious $\delta^{2\times2}\big(\lambda\!\cdot\!\widetilde\lambda\big)$ in (\ref{g36_on_shell_function}), the (fermionic) $\delta$-functions $\delta^{3\times4}\!\big(C^*\!\cdot\widetilde\eta\big)$ includes the factor $\delta^{2\times4}\big(\lambda\cdot\widetilde\eta\big)$---the supersymmetric-extension of ordinary momentum conservation.

\subsection{Twistor Space and the Super-Conformal Invariance of On-Shell Forms}\label{twistor_space_and_super_conformal_invariance_subsection}

In order to see the conformal symmetry of any theory, it is often wise to use {\it twistor} variables, \cite{Penrose:1967wn,Penrose:1972ia,Penrose:1968me,Kronheimer:1967mq,Penrose:1999cw}. Not surprisingly then, it is {\it twistor} space---not momentum-space---which gives us the simplest basis in which to describe scattering amplitudes conformally. Formally, we go to twistor space by assuming that $\lambda, \widetilde \lambda$ are independent, real variables, and then Fourier-transform with respect to either the $\lambda$ or $\widetilde \lambda$ variables, \cite{Witten:2003nn}. It is not hard to see how this Fourier transform makes the action of conformal transformations particularly transparent. Working with spinor-helicity variables, the generators of translations, $P_{\alpha\dot{\beta}}$, Lorentz transformations, $J_{\alpha\beta}$ and $\bar{J}_{\dot{\alpha}\dot{\beta}}$, dilatations $D$, and special conformal transformations, $K_{\alpha \dot{\beta}}$, all look very different:
\vspace{-.3cm}\eq{P_{\alpha\dot\beta}=\lambda_\alpha\widetilde\lambda_{\dot\beta},\qquad J_{\alpha\beta}=\frac{i}{2}\left(\lambda_{\alpha}\frac{\partial}{\partial \lambda^\beta}+\lambda_{\beta}\frac{\partial}{\partial \lambda^\alpha}\right),\quad\mathrm{and}\quad K_{\alpha\dot{\beta}}=\frac{\partial^2}{\partial\lambda^{\alpha}\partial\widetilde\lambda^{\dot{\beta}}}\,.\vspace{-.3cm}\label{conformal_generators_in_terms_of_spinors}}
($\bar{J}$ is defined analogously to $J$.) However, if we Fourier-transforms with respect to each of the $\lambda$'s, say, using $\int\!\!d^{2\times n}\!\lambda\,e^{i\lambda\cdot\widetilde\mu}$, denoting the $(2\!\times\!n)$-matrix of conjugate variables by $\widetilde{\mu}$, the generators (\ref{conformal_generators_in_terms_of_spinors}) become, (see \cite{Witten:2003nn} for a detailed discussion):
\vspace{-.3cm}\eq{P_{\dot\alpha\dot\beta}=i\widetilde\lambda_{\dot\alpha}\frac{\partial}{\partial\widetilde\mu^{\dot{\beta}}},\qquad J_{\dot\alpha\dot\beta}=\frac{i}{2}\left(\widetilde\mu_{\dot\alpha}\frac{\partial}{\partial\widetilde\mu^{\dot\beta}}+\widetilde\mu_{\dot\beta}\frac{\partial}{\partial\widetilde\mu^{\dot\alpha}}\right),\quad\mathrm{and}\quad K_{\dot\alpha\dot\beta}=i\widetilde\mu_{\dot\alpha}\frac{\partial}{\partial\widetilde\lambda^{\dot\beta}}\,.\vspace{-.3cm}\label{conformal_generators_in_terms_of_ft_spinors}}

These are easy to recognize as the generators of $SL(4)$-transformations on {\it twistor} variables, denoted $w_a$, which combine $\widetilde\lambda$ and $\widetilde\mu$ according to:
\vspace{-.2cm}\eq{w_a\equiv\left(\begin{array}{@{}c@{}}\widetilde\mu_a\\\widetilde\lambda_a\end{array}\right).\vspace{-.2cm}}
Very nicely, under the action of the little group, the $\widetilde\mu$'s transform oppositely to the $\lambda$'s so that the twistors transform uniformly like the $\widetilde\lambda$'s: $w_a\sim t_a^{-1} w_a$. Thus, we should view each $w_a$ projectively as a point in $\mathbb{P}^3$. Furthermore, we can combine these ordinary variables $w_a$ with the anti-commuting $\widetilde\eta$'s to form {\it super}-twistors $\mathcal{W}_a$, \cite{Ferber:1977qx},
\vspace{-.2cm}\eq{\mathcal{W}_a\equiv\left(\begin{array}{@{}c@{}}w_a\\\widetilde\eta_a\end{array}\right),\vspace{-.2cm}}
for which the generators of the super-conformal group are simply those of $SL(4|4)$ ---acting in the obvious way as super-linear transformations on the $\mathcal{W}$'s.

Now, given any of our on-shell forms, the Fourier-transform with respect to the $\lambda$ variables is straightforward as the only dependence on $\lambda$ is in the term $\delta^{2\times(n-k)}\big(\lambda\cdot C^\perp\!\big)$. It will be useful to re-write this to more directly reflect its geometric origin: the requirement that the plane $C$ {\it contains} $\lambda$. This means that there should exist a linear combination of the $k$ row-vectors of $C$ which {\it exactly} match $\lambda$. In other words, if we parameterize such a linear combination by a $(2\!\times\!k)$-matrix $\rho$, we should be able to find a $\rho$ for which $\rho\cdot C=\lambda$. Re-written in terms of this auxiliary matrix $\rho$, the constraint that $C$ contains $\lambda$ becomes,
\vspace{-.2cm}\eq{\delta^{2\times(n-k)}\big(\lambda\!\cdot\!C^\perp\!\big)=\int\!\!d^{2\times k}\!\rho\,\,\delta^{2\times n}\big(\rho\!\cdot\!C-\lambda\big),\vspace{-.2cm}} which makes it trivial to Fourier-transform to twistor space: \vspace{-.2cm}\eq{\int\!\!d^{2\times n}\!\lambda\,\,e^{i\lambda \cdot\widetilde\mu}\int\!\!d^{2\times k}\!\rho\,\,\delta^{2\times n}\big(\rho\!\cdot\!C-\lambda\big)=\int\!\!d^{2\times k}\!\rho\,\,e^{i(\rho\cdot C)\cdot\widetilde\mu}=\delta^{k\times2}\big(C\!\cdot\!\widetilde\mu\big).\vspace{-.2cm}}
Therefore, in twistor space the constraints $\delta^{k\times2}\big(C\!\cdot\!\widetilde\lambda)$ and $\delta^{2\times(n-k)}\big(\lambda\!\cdot\!C^\perp\!\big)$ together with the fermionic $\delta^{k\times4}\big(C\!\cdot\widetilde\eta\big)$ combine into the extremely elegant,
\vspace{-.2cm}\eq{\delta^{k\times4}\big(C\!\cdot\!\widetilde\eta\big)\delta^{k\times2}\big(C\!\cdot\!\widetilde\lambda\big)\delta^{k\times2}\big(C\!\cdot\!\widetilde\mu\big)\Rightarrow\delta^{4k|4k}\big(C\!\cdot\!\mathcal{W}\big),\vspace{-.2cm}}
which makes the $SL(4|4)$-invariance of on-shell forms completely manifest. And so, in twistor space, the general on-shell form, (\ref{general_on_shell_function}), is simply,
\vspace{-.2cm}\eq{\hspace{-0.0cm}f_\sigma^{(k)}\!=\!\oint\limits_{C\subset\Gamma_\sigma}\!\!\!\frac{d^{k \times n} C}{\mathrm{vol}(GL(k))}\frac{\delta^{4k|4k}\big(C\!\cdot\!\mathcal{W}\big)}{(1\cdots k)\cdots(n\cdots k\mi1)}.\hspace{-1.3cm}\vspace{-.2cm}\label{general_on_shell_function_in_twistor_space}}

Note that our brief passage to twistor space was done mostly for formal reasons: in order to make the super-conformal symmetry of on-shell forms manifest. One disadvantage of this formalism, however, is that---at first glance---it appears that the integral over $C\!\in\!\Gamma_\sigma$ could be localized by all $4k$ (ordinary) $\delta$-function constraints, while we know that on-shell forms associated with non-vanishing functions for generic (momentum-conserving) kinematical data correspond to $(2n\,\mi\,4)$-dimensional cells $\Gamma_\sigma\!\in\!G(k,n)$. The mismatch is due to the fact that Fourier-transforming to twistor space does not produce functions which are non-vanishing for a {\it generic} set of twistors. Instead, we get distributions on twistor space, imposing constraints on the twistor variables. Indeed, only $(2n\,\mi\,4)$ of the $4k$ \mbox{$\delta$-functions} in (\ref{general_on_shell_function_in_twistor_space}) can be used to localize the Grassmannian integral while the remaining impose constraints on the configuration of external twistors.

\subsection{Momentum-Twistors and  Dual Super-Conformal Invariance}\label{momentum_twistors_subsection}
In this subsection, we will review the arguments presented in \cite{ArkaniHamed:2009vw} in order to discover that on-shell forms are quite surprisingly {\it also} invariant under an additional super-conformal symmetry. This new symmetry, called {\it dual} super-conformal invariance, combines with ordinary super-conformal symmetry to generate an infinite-dimensional symmetry algebra of on-shell forms known as {\it the Yangian}, \cite{Drummond:2009fd,Drummond:2010uq,Drummond:2010qh,Beisert:2010gn}. (Dual super-conformal invariance was first noticed in multi-loop perturbative calculations, \cite{Drummond:2006rz}, and then at strong coupling, \cite{Alday:2007hr}; this led to a remarkable connection between null-polygonal Wilson loops and scattering amplitudes---see e.g.\ \cite{Alday:2007hr,Alday:2007he,Brandhuber:2007yx,Drummond:2007aua,Drummond:2007cf,Drummond:2007bm,Drummond:2008aq,Bern:2008ap,Alday:2009yn}.)

Let us start by reconsidering the condition that the plane $C$ contains the plane $\lambda$. Because this constraint is ubiquitous for on-shell forms, it is natural to sharpen our focus to the $(k\,\mi\,2)\equiv\hat{k}$-plane---denoted $\hat{C}$---which is the {\it projection of} $C$ onto the orthogonal-complement of $\lambda$. To be a bit more precise, suppose we have an operator \mbox{$Q\!:\!\mathbb{C}^n\!\to\!\mathbb{C}^n$} with $\mathrm{ker}(Q)=\lambda$ so that, \vspace{-.3cm}\eq{Q\!\cdot\!\lambda=0.\vspace{-.2cm}} With such an operator, we may define $\hat{C}\equiv C\!\cdot\!Q$ so that $\hat{C}\!\cdot\!\lambda=0$ trivially.

Now, super momentum-conservation is of course the statement that the planes $\widetilde\lambda$ and $\widetilde\eta$ are both in $\lambda^\perp$---which is the image of $Q$. And so we may use $Q$ to express $\widetilde\lambda$ and $\widetilde\eta$ in terms of some new, {\it generic} variables $\mu$ and $\eta$ according to:
\vspace{-.2cm}\eq{\widetilde\lambda\equiv\mu\!\cdot\!Q\quad\mathrm{and}\quad\widetilde\eta\equiv\eta\!\cdot\!Q\,.\vspace{-.2cm}}
Defined in this way, any {\it unconstrained} planes $\mu$ and $\eta$ will {\it automatically} define super momentum-conserving planes $\widetilde\lambda$ and $\widetilde\eta$.

Let us now consider the constraint that $C$ be orthogonal to the plane $\widetilde\lambda$. If $Q$ were symmetric, then $C\!\cdot\!\widetilde\lambda=\hat{C}\!\cdot\!\mu$; and similarly, $C\!\cdot\!\widetilde\eta=\hat{C}\!\cdot\!\eta$. Putting all this together, the constraints imposed on the image $\hat{k}$-plane $\hat{C}$ would become simply,
\vspace{-.2cm}\eq{\delta^{\hat{k}\times2}\big(\hat{C}\!\cdot\!\lambda\big)\delta^{\hat{k}\times2}\big(\hat{C}\!\cdot\!\mu\big)\delta^{\hat{k}\times4}\big(\hat{C}\!\cdot\!\eta\big)\Rightarrow\delta^{4\hat{k}|4\hat{k}}\big(\hat{C}\!\cdot\!\mathcal{Z}\big),\vspace{-.2cm}}
where we have introduced the super {\it momentum-twistors} $\mathcal{Z}$, \cite{Hodges:2009hk}, according to: \vspace{-.2cm}\eq{\mathcal{Z}_a\equiv\left(\begin{array}{@{}c@{}}z_a\\\eta_a\end{array}\right)\quad\mathrm{with}\quad z_a\equiv\left(\begin{array}{@{}c@{}}\lambda_a\\\mu_a\end{array}\right).\vspace{-.2cm}}

Geometrically, the $\delta$-functions $\delta^{k\times4}\big(\hat{C}\!\cdot\!Z\big)$ enforce that the plane $\hat{C}$ be orthogonal to the 4-plane $Z$:
\vspace{-.1cm}\eq{\raisebox{-42pt}{\includegraphics[scale=.85]{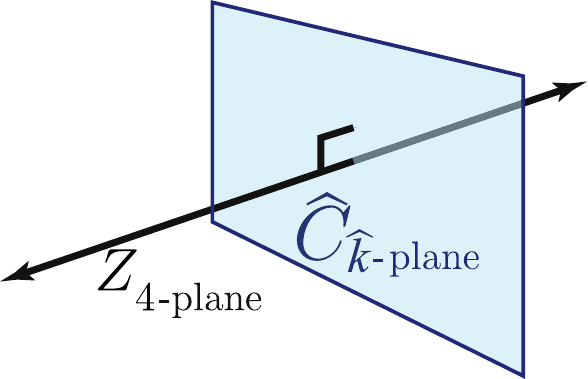}}\label{momentum_twistor_constraints}\vspace{-.1cm}}
Notice that these $\delta$-functions are invariant under a {\it new} $SL(4|4)$ symmetry, and thus it appears that we have uncovered a new super-conformal symmetry---one acting on the super-twistor variables ${\cal Z}_a$. However there is one small catch: the measure of integration over the $k$-plane $C$ does not necessarily descend to anything simple over the $\hat{k}$-plane $\hat{C}$. Indeed, depending on the choice of the projection operator $Q$, this resulting measure may have a complicated $\lambda$-dependence arising from the Jacobian of the change of variables from $(\widetilde\lambda,\widetilde\eta)$ to $(\mu,\eta)$, and this dependence on $\lambda$ may break the $SL(4)$ conformal symmetry.

But it turns out that for what is perhaps the most natural choice of a projection operator $Q$, everything works like magic. To better understand the scope of choices we could make in specifying $Q$, observe that such a projector can always be constructed via the Cramer's rule identities---the unique (up to rescaling) $(k\pl1)$-term identity satisfied by generic $k$-vectors. For a $2$-plane $\lambda$, Cramer's rule encodes the identities:
\vspace{-.3cm}\eq{\lambda_a\ab{b\,c}+\lambda_b\ab{c\,a}+\lambda_c\ab{a\,b}=0,\vspace{-.3cm}}
or equivalently, (if we prefer the identity to transform under the little group like $\widetilde\lambda_b$),
\vspace{-.4cm}\eq{\lambda_a\frac{1}{\ab{a\,b}}+\lambda_b\frac{\ab{c\,a}}{\ab{a\,b}\ab{b\,c}}+\lambda_c\frac{1}{\ab{b\,c}}=0.\vspace{-.1cm}\label{2_vector_cramers_rule_for_Q}}
If we combine any such $n$ cyclically-related identities, we will obtain a rank-$(n\,\mi\,2)$-matrix $Q$ which projects onto $\lambda^\perp$. In order for $Q$ to be {\it symmetric} as a matrix (which was necessary for $C\!\cdot\!\widetilde\lambda$ to be identified with $\hat{C}\!\cdot\!\mu$), we must have $\lambda_a$ and $\lambda_c$ equally-spaced about $\lambda_b$ in (\ref{2_vector_cramers_rule_for_Q}). Of course, the most obvious and natural choice (and the only one which generates the magic we seek) would be to use the {\it consecutive} 3-term identities:
\vspace{-.2cm}\eq{Q_{ab}\equiv\frac{\delta_{a-1\,b}\ab{a\,a\pl1}+\delta_{a\,b}\ab{a\pl1\,a\mi1}+\delta_{a+1\,b}\ab{a\mi1\,a}}{\ab{a\mi1\,a}\ab{a\,a\pl1}}.\vspace{-.2cm}\label{definition_of_Q}}

For this choice of $Q$, it turns out that for any plane $C$ containing $\lambda$, the plane $\hat{C}\equiv C\!\cdot\!Q$ will have the property that for {\it any} consecutive chain of columns \mbox{$\{c_a,\ldots,c_b\}$}, \mbox{$\mathrm{span}\{\hat{c}_a,\ldots,\hat{c}_b\}\subset(\mathrm{span}\{c_{a-1},\ldots,c_{b+1}\})$}. That is, $Q$ maps consecutive chains of columns onto consecutive chains of columns! An immediate consequence of this fact is that consecutive minors of $C$ and $\hat{C}$ are proportional to one another:
\vspace{-0.2cm}\eq{\left.(1\,2\cdots k\mi1\,k)\right|_{C}=\ab{1\,2}\ab{2\,3}\cdots\ab{k\mi1\,k}\left.(2\,3\cdots k\mi2\,k\mi1)\right|_{\hat{C}}\,.\vspace{-.2cm}\label{proportionality_of_cyclic_minors}}
Thus, for this choice of $Q$---up to an overall $\lambda$-dependent factor (which combines with the Jacobian arising from changing variables $(\widetilde\lambda,\widetilde\eta)$ to $(\mu,\eta)$)---the top-form measure on $C\!\in\!G(k,n)$ given as the product of its consecutive minors, is mapped to the top-form on $\hat{C}\!\in\!G(\hat{k},n)$ of precisely the same form. And so, $Q$ maps positroid cells in $G(k,n)$  (which contain a generic $2$-plane $\lambda$) to positroid cells in $G(\hat{k},n)$!

Conveniently, it turns out that the image of any cell $C\!\in\!G(k,n)$ in $G(\hat{k},n)$ is very easy to identify by its permutation label. Because $\mathrm{span}\{\hat{c}_a,\ldots,\hat{c}_b\}\subset(\mathrm{span}\{c_{a-1},\ldots,c_{b+1}\})$, we have that \mbox{$\hat{r}[a;b]=r[a\mi1;b\pl1]\mi 2$}; and so, the entire table of ranks, (\ref{deligne_table}), is preserved in going from $C$ to $\hat{C}$---merely shifted downward and to the right:
\vspace{-0.1cm}\eq{\hspace{-.75cm}\qquad\begin{array}{|c|c|}\hline~&\\[-10pt]\rnk{a}{\;\,\sigma(a)\,\;}&\rnk{a\pl1}{\;\,\sigma(a)\,\;}\rule{0pt}{0pt}\\[5pt]\hline~&\\[-10pt]\rnk{a}{\sigma(a)\mi1}&\rnk{a\pl1}{\sigma(a)\mi1}\\[5pt]\hline\end{array}\hspace{-2.835cm}\raisebox{-32.5pt}{\includegraphics[scale=1]{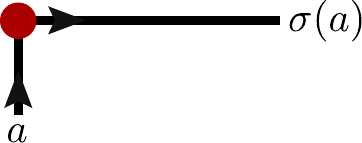}}\text{{\Large$\;\rightsquigarrow\;$}}\begin{array}{|c|c|}\hline~&\\[-10pt]\,\rnkhat{a\pl1}{\sigma(a)\mi1}\mi2\,&\,\rnkhat{a\pl2}{\sigma(a)\mi1}\mi2\,\rule{0pt}{0pt}\\[5pt]\hline~&\\[-10pt]\,\rnkhat{a\pl1}{\sigma(a)\mi2}\mi2\,&\,\rnkhat{a\pl2}{\sigma(a)\mi2}\mi2\\[5pt]\hline\end{array}\hspace{-3.56cm}\raisebox{-32.5pt}{\includegraphics[scale=1]{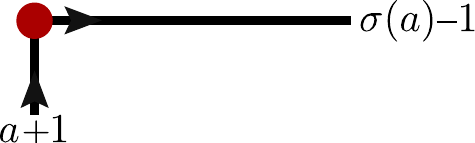}}\vspace{-0.2cm}\nonumber}
And so, a configuration $C_\sigma\!\in\!G(k,n)$ labeled by the permutation $\sigma$ will be mapped to a configuration $\hat{C}_{\hat{\sigma}}\!\in\!G(\hat{k},n)$ labeled by the permutation,
\vspace{-.2cm}\eq{\hat{\sigma}(a)\equiv\sigma(a-1)-1.\vspace{-.2cm}\label{permutation_map_to_momentum_twistors}}

One last remarkable aspect of this change of variables is that the combination of all the $\lambda$-dependent factors arising from (\ref{proportionality_of_cyclic_minors}) when mapping the cyclic minors of $G(k,n)$ to cyclic minors of $G(\hat{k},n)$ with the Jacobian of the change of variables from $(\widetilde\lambda,\widetilde\eta)$ to $(\mu,\eta)$ turns out to be nothing but the Parke-Taylor (MHV) tree-amplitude, (\ref{parke_taylor_tree_from_grassmannian})! And so,
\vspace{-.2cm}\eq{f_{\sigma}^{(k)}(\lambda,\widetilde \lambda,\widetilde \eta) =\frac{\delta^{2\times4}\big(\lambda\!\cdot\!\widetilde{\eta}\big)\delta^{2\times2}\big(\lambda\!\cdot\!\widetilde{\lambda}\big)}{\ab{1\,2}\ab{2\,3}\cdots\ab{n\,1}} \times f_{\hat{\sigma}}^{(\hat{k})}({\cal Z})\,,\vspace{-.2cm}\label{parke_taylor_jacobian}}
where,
\vspace{-.0cm}\eq{f_{\hat{\sigma}}^{(\hat{k})}({\cal Z})=\!\oint\limits_{\hat{C}\subset\Gamma_{\hat{\sigma}}}\!\!\!\frac{d^{\hat{k} \times n} \hat{C}}{\mathrm{vol}(GL(\hat{k}))}\frac{\delta^{4\hat{k}|4\hat{k}}\big(\hat{C}\!\cdot\!\mathcal{Z}\big)}{(1\cdots \hat{k})\cdots(n\cdots \hat{k}\mi1)}.\vspace{-.2cm}\label{general_on_shell_function_momentum_twistors}}
This should not be too surprising, as the Parke-Taylor amplitude can be thought of as the most concise differential form consistent with super momentum conservation---and we know that any generic set of super-momentum-twistors $\mathcal{Z}$ give rise to data $(\widetilde\lambda,\widetilde\eta)$ which {\it manifestly} conserve super-momentum (This Grassmannian formula in terms of momentum twistor was introduced in \cite{Mason:2009qx}).

Let us briefly see how the dimensionality of cells $C_\sigma\!\in\!G(k,n)$ and their images $\hat{C}_{\hat{\sigma}}\!\in\!G(\hat{k},n)$ are related. Because the rank of each chain $\hat{r}[a\pl1;\hat{\sigma}(a\pl1)]$ is lowered by $2$ relative to $r[a;\sigma(a)]$, recalling the way dimensionality is encoded by the permutation, (\ref{dimension_by_ranks}) we see that:\\[-7pt]
\vspace{-.1cm}\eqs{\dim(\hat{C}_{\hat{\sigma}})&=\dim(C_{\sigma})-2n+k^2-(k-2)^2,\\&=\dim(C_{\sigma})-(2n-4)+4\hat{k};\\\therefore\;\dim(\hat{C}_{\hat{\sigma}})-4\hat{k}\hspace{-0.95cm}&\hspace{0.95cm}=\dim(C_{\sigma})-(2n-4).\vspace{-.2cm}}
This is precisely as it should be: {\it generic} super momentum-twistors $\mathcal{Z}$ give rise to {\it generic} super-momentum conserving spinor-helicity data $\lambda, \widetilde \lambda, \widetilde\eta$. Thus, the degree of the form $f_{\hat{\sigma}}$ should be $\dim(\hat{C}_{\hat{\sigma}})$ minus the $4\hat{k}$ ordinary $\delta$-functions which enforce that $\hat{C}$ be orthogonal to the generic $4$-plane $Z$.

We should make one small point regarding the (existence of the) map between $G(k,n)\!\to\!G(\hat{k},n)$: it is only well-defined for cells $C_{\sigma}$ {\it which contain a generic $2$-plane $\lambda$} (a point which is completely obvious from the geometry involved in the map's construction). In terms of the permutation $\sigma$ which labels $C\!\in\!G(k,n)$, the criterion that $C$ can contain a generic $2$-plane $\lambda$ translates into the statement that $\sigma(a)\,\mi\,a\geq 2$ for all $a$. This guarantees that the permutation $\hat{\sigma}$ is well-defined as an {\it affine} permutation, that is, that $\hat{\sigma}(a) \geq a$. Suppose that instead we had $\sigma(a)=a\pl1$, then $c_{a}\in\mathrm{span}\{c_{a+1}\}$, and so $\lambda\!\subset\!C$ would require that $\ab{a\,a\pl1}=0$. This all makes perfect sense, of course, because $\ab{a\,a\pl1}\!\to\!0$ precisely corresponds to a singularity of the Parke-Taylor amplitude; and the Parke-Taylor amplitude being the {\it Jacobian} of the transformation to momentum-twistor space, any such singularity indicates that the change of variables is singular.

Let us conclude our discussion by illustrating the map to the `momentum-twistor Grassmannian' for the example discussed above, (\ref{g36_on_shell_function}), of the on-shell form associated with the cell in $G(3,6)$ labeled by the permutation ${\color{perm}\{3,5,6,7,8,10\}}$, (\ref{g36_example_graph_and_configuration}). The image of this cell in the momentum-twistor Grassmannian $G(1,6)$ is labeled by $\hat{\sigma}={\color{perm}\{3,2,4,5,6,7\}}$. Since $\hat{\sigma}(2) = 2$, we have that $\hat{c}_2=0$. A $GL(1)$-representative of the point $\hat{C}^*$ which is orthogonal to the $Z$-plane in this cell is,
\vspace{-.2cm}\eq{\hat{C}^*\equiv\left(\begin{array}{@{}cccccc@{}}\ab{3\,4\,5\,6}&0&\ab{4\,5\,6\,1}&\ab{5\,6\,1\,3}&\ab{6\,1\,3\,4}&\ab{1\,3\,4\,5}\end{array}\right),\vspace{-.2cm}}
where $\ab{a\,b\,c\,d}\equiv\mathrm{det}\{z_a,z_b,z_c,z_d\}$ is a minor of the matrix $Z$, and $\hat{C}^*\!\cdot Z=0$ because of the $4$-vector manifestation of Cramer's rule, (\ref{Cramers_rule}). Supported on this point, (\ref{general_on_shell_function_momentum_twistors}) generates the momentum-twistor super-function,
\vspace{-.2cm}\eq{\hspace{0.0cm}f^{(1)}_{{\color{perm}\{3, 2, 4,5, 6,7\}}}=\frac{\delta^{1\times4}\big(\hat{C}^*\!\!\cdot\!\eta\big)}{\underbrace{\ab{3\,4\,5\,6}}_{\phantom{|_{\hat{C}^*}}(1)|_{\hat{C}^*}}\!\underbrace{\ab{4\,5\,6\,1}}_{\phantom{|_{\hat{C}^*}}(3)|_{\hat{C}^*}}\!\underbrace{\ab{5\,6\,1\,3}}_{\phantom{|_{\hat{C}^*}}(4)|_{\hat{C}^*}}\!\underbrace{\ab{6\,1\,3\,4}}_{\phantom{|_{\hat{C}^*}}(5)|_{\hat{C}^*}}\!\underbrace{\ab{1\,3\,4\,5}}_{\phantom{|_{\hat{C}^*}}(6)|_{\hat{C}^*}}}.\vspace{-.0cm}\label{g16_momentum_twistor_function}}
And so, including the Parke-Taylor Jacobian, (\ref{parke_taylor_jacobian}), we have:
\eq{\hspace{-0.45cm}f^{(3)}_{{\color{perm}\{3, 5, 6, 7, 8, 10\}}}\!=\!\frac{\delta^{2\times4}\big(\lambda\!\cdot\!\widetilde{\eta}\big)\delta^{2\times2}\big(\lambda\!\cdot\!\widetilde{\lambda}\big)}{\ab{1\,2}\ab{2\,3}\ab{3\,4}\ab{4\,5}\ab{5\,6}\ab{6\,1}}\frac{\delta^{1\times4}\big(\hat{C}^*\!\!\cdot\!\eta\big)}{\ab{3\,4\,5\,6}\ab{4\,5\,6\,1}\ab{5\,6\,1\,3}\ab{6\,1\,3\,4}\ab{1\,3\,4\,5}}.\hspace{-3cm}}

\newpage
\section{Positive Diffeomorphisms and Yangian Invariance}\label{positive_diffeos_and_the_yangian_section}

We have seen that the map from twistor space to momentum-twistor space has a natural origin, providing an obvious geometric basis for dual conformal invariance. Let us now consider another obvious symmetry of the positive Grassmannian---namely, diffeomorphisms of Grassmannian coordinates which preserve the structure of the positroid stratification (equivalently, diffeomorphisms which leave measure on $G_+(k,n)$ invariant). Preserving the positive structure of the Grassmannian, we call this subset of diffeomorphisms {\it positive diffeomorphisms}. In this section, we illustrate the remarkable fact that the leading generators of infinitesimal positive diffeomorphisms directly match the level-one generators of the Yangian as described in \cite{Drummond:2010uq} (see also \cite{Drummond:2009fd,Drummond:2010qh,Korchemsky:2010ut,Beisert:2010gn}).

Let us begin by broadly characterizing the infinitesimal diffeomorphisms in which we are interested. Consider any infinitesimal variation $\delta C$ of $C\!\in\!G_+(k,n)$ which we may expand qualitatively as a power-series,
\vspace{-.2cm}\eq{\delta C  \sim C + C C + C C C + \cdots\;.\label{general_c_diffeo}\vspace{-.2cm}} We view a general infinitesimal diffeomorphism of $C$ in terms of the variations $\delta c_{\alpha\,a}$ for each matrix component of $C$. Because {\it positive} diffeomorphisms must preserve {\it all} positroid configurations, $\delta c_{\alpha\,a}$ must vanish whenever $c_a$ does; this restricts the class of diffeomorphisms to those of the form,
\vspace{-.2cm}\eq{\phantom{\text{(no summation on $a$),}}\quad\delta c_{\alpha\,a}=\big(\Omega_a[C]\big)_{\alpha}^{\beta}c_{\beta\,a}\quad\text{(no summation on $a$),}\vspace{-.2cm}\label{general_positive_diffeomorphism}}
where
each $\Omega_a[C]$ is itself expanded as a power-series in the components of $C$. Considering $\Omega_a[C]$ as a $(k\!\times\!k)$-matrix, we may simplify our notation by writing:
\vspace{-.2cm}\eq{\delta c_{a}=\big(\Omega_a[C]\big)\!\cdot\!c_{a}.\vspace{-.2cm}\label{general_positive_diffeomorphism_v2}}

Note that any variation where $\Omega$ is proportional to the identity matrix is just an un-interesting ($C$-dependent) little group transformation. Note also that this variation takes the form of a different $GL(k)$ transformation on each column. We can always use the global $GL(k)$-symmetry to bring the variation of any {\it one} column, say $c_1$, to zero: \vspace{-.2cm}\eq{\delta c_1=0.\vspace{-.2cm}} (And without loss of generality, we can always take  $c_1$ to be a non-vanishing column.)

Let us now determine what conditions must be imposed on $\Omega_a[C]$ in order to ensure that the variations $\delta c_a$ preserve {\it all} positroid configurations. We will now demonstrate that there are no {\it non-trivial} variations to leading order in $C$, and that the first non-trivial positive diffeomorphisms---those quadratic in $C$---precisely correspond to the level-one generators of the Yangian as described in reference \cite{Drummond:2010uq}.

To leading order, each $\Omega_a$ is a $C$-independent $(k\!\times\!k)$-matrix. Consider any configuration for which $c_1\!\propto\!c_2$, and let us use the $GL(k)$-symmetry to fix the variation of $c_1$ to zero. It is not hard to see that the only variation of $c_2$ which preserves the configuration in question would be the rescaling $\delta c_2=t\,c_2$. This variation can be fully compensated by a little group rescaling, allowing us to conclude that {\it no} {\it non-trivial} variation of $c_2$ is positive. Repeating this argument by starting with $c_2$ instead of $c_1$, and so on, we therefore see that the only {\it positive} leading-order diffeomorphisms are overall $GL(k)$-transformations and little group rescalings.

Non-trivial positive diffeomorphisms first arise at quadratic-order---when $\Omega_a[C]$ is linear in the components of $C$. Let us again consider any configuration for which $c_1\!\propto\!c_2$, and use the $GL(k)$-symmetry to fix the variation of $c_1$ to zero. Because positive diffeomorphisms must preserve $r[a;b]\equiv\mathrm{rank}\{c_a,\ldots,c_b\}$ generally---and $r[1;2]$ in particular---it is clear that the only allowed variations would be of the form,
\vspace{-.2cm}\eq{\delta c_2=(c_1\omega_1^\beta)c_{\beta\,2}\equiv c_1(\omega_1\!\cdot\!c_2).\label{transformation_of_c2}\vspace{-.2cm}}
We ignore any variation quadratic in $c_2$ as it represents  a little group rescaling. Here, $\omega_1^\beta$ is an arbitrary $k$-vector parameterizing the variation. Notice that (\ref{transformation_of_c2}) is just a simple $GL(k)$-transformation of column $c_2$ by the matrix $M_{\alpha}^{\beta}\equiv(c_{\alpha\,1}\omega_1^{\beta})$. Applying the inverse of this transformation to {\it all} columns would of course trivialize $\delta c_2\to 0$, allowing us to repeat the same logic to fix the most general form of $\delta c_3$, and so on. Continuing in this manner and then undoing each step's $GL(k)$-transformation so that we restore $\delta c_1=0$, the most general quadratic, positive diffeomorphism consistent with positivity would be of the form:
\vspace{-.2cm}\eqs{\\[-20pt]\delta c_1&=0;\\\delta c_2&=c_1(\omega_1\!\cdot\! c_2);\\\delta c_3&=c_1(\omega_1\!\cdot\! c_3)+c_2(\omega_2\!\cdot\! c_3);\\[-8pt]&\phantom{\,\,}\vdots\\[-5pt]\delta c_{n}&=c_1(\omega_1\!\cdot c_n)+\cdots+c_{n-1}(\omega_{n-1}\!\cdot\!c_n);\\[-25pt]\vspace{-.2cm}\label{most_general_quad_positive_diff_long_form}}
which we may summarize: \vspace{-.0cm}\eq{\delta c_a=\sum_{b<a}c_b(\omega_b^{\beta}c_{\beta\,a}).\vspace{-.2cm}\label{most_general_quad_positive_diff}}
We fixed the form of this transformation by demanding that the cells where $c_1 \propto c_2$ are left invariant, but quite nicely, we can see that this transformation leaves {\it all} cells invariant! Note that $r[1;a]\equiv\mathrm{rank}\{c_1,\ldots,c_a\}$ is unchanged for all $a$, as the variations in (\ref{most_general_quad_positive_diff_long_form}) transform each $c_{a}$ by factors proportional to columns which are always (trivially) spanned by the un-deformed chains. And so, (\ref{most_general_quad_positive_diff}) preserves all $r[1;b]$---the entire first column of the table (\ref{deligne_table}).

In order for the diffeomorphisms (\ref{most_general_quad_positive_diff}) to be {\it positive}, however, they must preserve the ranks $r[a;b]$ for {\it all} chains of columns; and so, we must find the subset which are independent of our choice to single-out $\delta c_1$. These can be found by continuing the sequence of successive variations in (\ref{most_general_quad_positive_diff_long_form}) back to $\delta c_1$, and requiring that this be consistent with our choice to fix $\delta c_1=0$:
\vspace{-.2cm}\eq{\delta c_1=c_1(\omega_1\!\cdot c_1)+\cdots+c_{n}(\omega_{n}\!\cdot\!c_1)=\Big(\sum_{b=1}^nc_b\omega_b^{\beta}\Big)c_{\beta\,1}=0.\vspace{-.2cm}}
Because this must be satisfied for {\it all} configurations in $G_+(k,n)$, this must be independent of $c_{\beta\,1}$. And so, the condition that ensures that (\ref{most_general_quad_positive_diff}) is positive is that,
\vspace{-.2cm}\eq{\sum_{b=1}^nc_b\omega_b^{\beta}=0.\vspace{-.2cm}\label{constraint_on_variations}}
This is simply the geometric statement that $\omega_a^\beta\!\subset\! C^\perp$ (for each index $\beta$ separately). We have therefore constructed the most general set of infinitesimal, quadratic diffeomorphisms which preserve all cells in the positroid stratification of $G(k,n)$.

Recall that kinematical data---specified, say, in terms of super-twistor variables $\mathcal{W}$---is communicated to the Grassmannian via the constraint $\delta^{4k|4k}\!\big(C\!\cdot\!\mathcal{W}\big)$. This means that any symmetry-transformation acting on the $\mathcal{W}$'s can be recast as a transformation on the configuration $C$. In reference\cite{Drummond:2010uq}, it was shown that the level-one generators of the Yangian can be translated in this way to become symmetry generators acting on the matrix $C$ by the operator:
\vspace{-.2cm}\eq{\mathcal{Q}\equiv\sum_{a=1}^{n}\mathcal{Q}_{a}\quad\mathrm{with}\quad \mathcal{Q}_{a}\equiv\Big(\sum_{b<a}c_{\alpha\,b}\mathcal{W}_b^{I}(\xi^{\beta}_{I}c_{\beta\,a})\Big)\frac{\partial}{\partial c_{\alpha\,a}}\vspace{-.2cm},}
which is easily seen to generate diffeomorphisms  of the form,
\vspace{-.2cm}\eq{\delta c_a=\sum_{b<a}c_bw_b^{I}(\xi^{\beta}_Ic_{\beta\,a}),\vspace{-.2cm}}
which we immediately recognize as nothing but the leading positive diffeomorphisms (\ref{most_general_quad_positive_diff}), where $w_b^{\beta}$ has been re-written as
\vspace{-.2cm}\eq{w_b^{\beta}\equiv w_b^I\xi_I^{\beta},\vspace{-.2cm}}
for some (arbitrary) $(4\!\times\!k)$-matrix $\xi_I^{\beta}$. Moreover, the condition on admissible variations, (\ref{constraint_on_variations}), is immediately seen to be {\it precisely} what is enforced by the constraint $\delta^{4k|4k}\big(C\!\cdot\!\mathcal{W}\big)$---which is imposed for all on-shell differential forms.

\newpage
\section{Combinatorics of Kinematical Support for On-Shell Forms}\label{kinematical_support_section}
On-shell graphs with the right number of degrees of freedom to be completely localized for {\it generic}, (super-)momentum conserving kinematical data are obviously of particular interest. In momentum space, this requires that a configuration $C$ associated with an on-shell graph admits solutions to both the constraint that it contains a generic 2-plane \mbox{$\lambda\!\in\!G(2,n)$}, and is contained within the geometric-dual of another 2-plane \mbox{$\widetilde{\lambda}\!\in\!G(2,n)$} satisfying $\lambda\!\cdot\!\widetilde\lambda=0$. In terms of the permutation $\sigma$ associated with an on-shell graph, these constraints minimally require that for any $a$, \mbox{$(a\pl\,2)\leq\sigma(a)\leq(a\pl\,n\,\mi\,2)$}. (Recall that the condition that $\sigma(a)\geq(a\pl\,2)$ is {\it necessary} for a configuration in $G(k\pl\,2,n)$ to even have a momentum-twistor dual in $G(k,n)$.) However, not all configurations which meet these conditions admit solutions to the combined constraints.

In this section, we will describe a purely-combinatorial solution to the question of whether or not an on-shell graph vanishes for generic kinematical data; and if so, how many solutions to the kinematical constraints exist. This turns out to be much simpler to do for the momentum-twistor Grassmannian rather than for configurations directly associated with on-shell graphs. This is partly because the kinematical constraints are much simpler for momentum-twistors than for the $\lambda$'s and $\widetilde\lambda$'s.

Recall that when kinematical data is specified by momentum-twistors, \mbox{$Z\!\in\!G(4,n)$}, the configuration $\bar{C}_{\bar{\sigma}}\!\in\!G(k\pl\,2,n)$ {\it directly} associated with an N$^k$MHV {\it on-shell graph} is mapped to its momentum-twistor image $\bar{C}_{\bar{\sigma}}\mapsto C_{\sigma}\!\in\!G(k,n)$, and the kinematical constraints become the simpler condition that $C\!\cdot\!Z=0$. This imposes $4k$ constraints in general, and so we are most interested in $4k$-dimensional cells of $G(k,n)$, as these can be completely isolated by generic kinematical data. In terms of the orthogonal complement $Z^\perp$ of the twistors $Z$, the number of solutions to $C\!\cdot\!Z=0$ is counted by the number of isolated points in $C\newcap Z^\perp$.

As with any intersection-number problem in algebraic geometry, the solution can be found by decomposing both $C$ and $Z^{\perp}$ into a homological basis for which the intersection numbers are known, such as Schubert cycles whose intersection numbers are given by the Littlewood-Richardson rule (see \cite{GriffithsHarris}). The decomposition of (the closure of) an arbitrary positroid cell into Schubert cycles was recently presented in ref. \cite{KLS}, and this provides us with a purely-combinatorial answer to the `number of intersections' question in which we are interested. And it turns out that for the special case of {\it generic} kinematical data, the machinery of \cite{KLS} simplifies considerably. (We are thankful to Thomas Lam and David Speyer for helpful discussions regarding this specialization of the general case.)

A complete discussion of this story would require more space than warranted here; but let us briefly describe the ultimate, combinatorial solution to the question of kinematical support. The first step is to generalize our discussion slightly, and consider kinematical data specified for any number of dimensions:\\[-7pt]

\noindent{\bf Definition:} For any $(m\!\times\!k)$-dimensional cell $C\!\in\!G_+(k,n)$, let $\Gamma^m(C^{})$ denote the number of isolated points in $C\newcap Z^{\perp}$ for a {\it generic} $m$-plane $Z\!\in\!G(m,n)$. \\[-7pt]

The basic strategy is to define a {\it distinguished subset} $[\partial^k](C)$ of $k^{\mathrm{th}}$-degree boundary elements of $C$ which contain non-overlapping subsets of the intersection points as projected to these boundaries, such that each element $C'\in[\partial^k](C)$ contains precisely $\Gamma^{m-1}(C')$ points. If this can be done, then $\Gamma^m(C)$ will be determined recursively by,
\vspace{-.2cm}\eq{\Gamma^m(C)=\sum_{C'\in[\partial^k](C^{})}\Gamma^{m-1}(C')\qquad\mathrm{with}\qquad\Gamma^0(C)\equiv1.\label{intersection_number_recursion}\vspace{-.2cm}}
The magic, then, is entirely in the definition of the distinguished boundary elements $[\partial^k](C)$. Before we describe these in general, however, it may be helpful to build some intuition with two (very) simple cases for which (\ref{intersection_number_recursion}) is easy to understand.

\subsection{Kinematical Support of NMHV Yangian-Invariants}
Although perhaps a bit trivial, it is worth noting that $\Gamma^m(C)=1$ for {\it all} $m$-dimensional configurations in $G(1,n)$---those relevant to NMHV amplitudes: given any generic $m$-plane $Z$, there is a unique configuration $C^*\!\in\!C\newcap Z^{\perp}$ supplied by  Cramer's rule, (\ref{Cramers_rule})---the unique $(m\pl1)$-term identity satisfied by generic $m$-vectors. This is of course obvious; but let us see what it suggests about how we may define the {\it distinguished} boundary elements $[\partial^1](C)$ which we seek to understand.

Just as $\Gamma^m(C)=1$ for any $m$-dimensional configuration in $G(1,n)$, $\Gamma^{m-1}(C)=1$ for any $(m\,\mi\,1)$-dimensional configuration. Therefore, in order for the recursive formula (\ref{intersection_number_recursion}) to give us the right answer, we need only define $[\partial^1](C)$ to systematically choose any {\it one} element of the boundary of $C$. One natural choice would be the configuration in $\partial C$ which deletes the {\it first} non-vanishing column of $C$---that is, the boundary for which the maximum image of the configuration's permutation is `raised.'

\subsection{Kinematical Support for One-Dimensional Kinematics}
Let us now consider the slightly less trivial case of one-dimensional kinematics, where $Z\!\in\!G(1,n)$ and we are interested in finding $\Gamma^1(C)$ for $k$-dimensional configurations in $G(k,n)$. Unlike the situation for $k=1$, it is no longer the case that every $k$-dimensional configuration admits solutions to $C\!\cdot\!Z=0$. The simplest example of a configuration for which $\Gamma^1(C)=0$ occurs for the $2$-dimensional configuration in $G(2,4)$ labeled by the permutation ${\color{perm}\{2,3,5,8\}}$: \vspace{-.2cm}\eq{C(\alpha)=\left(\begin{array}{@{}cccc@{}}1&\alpha_1&\alpha_2&0\\0&0&0&1\end{array}\right).\vspace{-.2cm}}
Notice that $C\!\cdot\!Z=0$ implies that $z_4=0$, which is obviously not satisfied by a generic set of ($1$-dimensional) momentum-twistors. In contrast, consider the configuration labeled by the permutation ${\color{perm}\{2,5,4,7\}}$ represented by,
\vspace{-.2cm}\eq{C(\alpha)=\left(\begin{array}{@{}cccc@{}}1&\alpha_1&0&0\\0&0&1&\alpha_2\end{array}\right),\quad\text{for which}\quad C^*\equiv\left(\begin{array}{@{}cccc@{}}z_2&\mi\,z_1\,\,&0&0\\0&0&z_4&\mi\,z_3\,\,\end{array}\right)\vspace{-.3cm}}
is the unique solution to $C\!\cdot\!Z=0$.

We can understand that a solution exists in the second case because each row of the matrix-representative of $C$ has one degree of freedom---reducing each row to the simple case of $k=1$ described above. In the first example, however, no solution exists because its second row has no degrees of freedom---which can itself be viewed as a zero-dimensional configuration in $G(1,n)$. Heuristically, then, in order for any solutions to $C\!\cdot\!Z=0$ to exist, there must exist at least one degree of freedom in every row of any matrix-representative of $C$.

In terms of the permutation, the existence of a row without any degrees of freedom is indicated by any column $a$ such that $\sigma(a)=a\,\pl\,n$. And so, a $k$-dimensional cell $C\!\in\!G(k,n)$ admits solutions to $C\!\cdot\!Z=0$ for a generic $1$-plane $Z$ if and only if $\sigma(a)\neq a\,\pl\,n$ for all $a$.

\subsection{General Combinatorial Test of Kinematical Support}
Combining the lessons learned from the two simple cases above, it is clear that solutions to $C\cdot Z=0$ exist only if there are in some sense $m$ degrees of freedom in each row of any matrix-representative of $C$. A systematic way to test this combinatorially would be to find boundary elements of $C$ which successively remove one degree of freedom from each row of $C$. Let us now describe how such boundary configurations can be found.

Recall that the lexicographically-first non-vanishing minor $A(\sigma)\equiv(a_1,\ldots,a_k)$ of any configuration $C_\sigma$ is given simply by the images of $\sigma$ which extend beyond $n$ (see \mbox{section \ref{geometry_of_the_positroid_stratification_subsection}}). Because of this, we can always give a matrix-representative of $C$ in the following, gauge-fixed form:
\vspace{-.4cm}\eq{\raisebox{5pt}{$\displaystyle\begin{array}{c}~\\1\\2\\[-0.17cm]\vdots\\[-0.05cm]k\end{array}\raisebox{-0.25cm}{$\left(\raisebox{1.1cm}{$\,$}\right.$}\begin{array}{@{}cccccccccccccc@{}}&\cdots&\cdots&{\color{red}a_1}&\cdots&\cdots&{\color{red}a_2}&\cdots&\cdots&{\color{red}a_k}&\cdots&\cdots&\\\hline
{\color{deemph}0}&{\color{deemph}\cdots}&{\color{deemph}0}&{\color{red}1}&{\color{blue}*}&{\color{blue}\cdots}&{\color{blue}*}&{\color{blue}\cdots}&{\color{blue}\cdots}&{\color{blue}\cdots}&{\color{blue}\cdots}&{\color{blue}\cdots}&{\color{blue}*}\\
{\color{deemph}0}&{\color{deemph}\cdots}&{\color{deemph}\cdots}&{\color{deemph}\cdots}&{\color{deemph}\cdots}&{\color{deemph}0}&{\color{red}1}&{\color{blue}*}&{\color{blue}\cdots}&{\color{blue}\cdots}&{\color{blue}\cdots}&{\color{blue}\cdots}&{\color{blue}*}\\[-0.17cm]
{\color{deemph}\vdots}&~&{\color{deemph}\ddots}&{\color{deemph}\ddots}&{\color{deemph}\ddots}&{\color{deemph}\ddots}&{\color{deemph}\ddots}&~&{\color{blue}\ddots}&{\color{blue}\ddots}&{\color{blue}\ddots}&&{\color{blue}\vdots}\\[-0.05cm]
{\color{deemph}0}&{\color{deemph}\cdots}&{\color{deemph}\cdots}&{\color{deemph}\cdots}&{\color{deemph}\cdots}&{\color{deemph}\cdots}&{\color{deemph}\cdots}&{\color{deemph}\cdots}&{\color{deemph}0}&{\color{red}1}&{\color{blue}*}&{\color{blue}\cdots}&{\color{blue}*}\end{array}\raisebox{-0.25cm}{$\left.\raisebox{1.1cm}{$\,$}\right).$}$}\vspace{-.2cm}}
A boundary-element which removes one degree of freedom from the $k^{\mathrm{th}}$-row of $C$, for example, would be any which `raises' $a_k$---that is, if $\sigma(c_k)=a_k$, then a boundary for which $\sigma'(c_k)=a'_k>a_k$. After this, a degree of freedom can be removed from the $(k\,\mi\,1)^{\mathrm{th}}$ row, and so on. Notice, however, that at each stage, $A(\sigma)$ must be {\it raised}: if $A(\sigma)$ remained unchanged, then it would not indicate that a degree of freedom from any particular row had been removed, as we desire.

With this picture serving as motivation, we define the distinguished, $k^{\mathrm{th}}$-degree boundary-elements of $C$, $[\partial^k](C)$, as follows. Let $\sigma$ be the permutation labeling the configuration $C$, and let us define $A(\sigma)\equiv(a_1,\ldots,a_k)$---with $a_1<a_2<\cdots<a_k$---to be the images of $\sigma$ which extend beyond $n$ (the necklace $A^{(1)}(\sigma)$). Then $[\partial^k](C)$ is the set of all $k^{\mathrm{th}}$-degree boundaries of $C$ obtained by a sequence of boundaries labeled by permutations $\sigma\xrightarrow{\partial}\sigma^{(1)}\xrightarrow{\partial}\sigma^{(2)}\xrightarrow{\partial}\cdots\xrightarrow{\partial}\sigma^{(k)}$ such that, {\it lexicographically},\vspace{-0.2cm}\eq{A(\sigma)<A(\sigma^{(1)})<A(\sigma^{(2)})<\cdots<A(\sigma^{(k)}).\vspace{-0.2cm}}
That is, if the configuration labeled by $\sigma^{(\ell-1)}$ with \mbox{$A(\sigma^{(\ell-1)})=(a_1,\ldots,a_{k-\ell},\ldots,a_k)$} is found at the $\ell^{\mathrm{th}}$ successive boundary, we take all $\sigma^{(\ell)}$ in its boundary for which \mbox{$A(\sigma^{(\ell)})=(a_1,\ldots,a_{k-\ell}',\ldots,a_k)$} with $a_{k-\ell}<a_{k-\ell}'$.

In general, there can be many such elements of $[\partial^k](C)$, and each can contribute to $\Gamma^m(C)$. Putting all these contributions together, we find the recursive formula given above:
\vspace{-.2cm}\eq{\Gamma^m(C)=\sum_{C'\in[\partial^k](C^{})}\Gamma^{m-1}(C')\qquad\mathrm{with}\qquad\Gamma^0(C)\equiv1.\label{intersection_number_recursion_second}\vspace{-.2cm}}

The utility of this combinatorial test is hard to overstate, as the number of $4k$-dimensional cells in $G(k,n)$ with {\it non-vanishing} support become increasingly rare with large $k$ and $n$. Cells with $\Gamma^4(C)=0$---for which $C\newcap Z^\perp=\{\}$---represent generally-vanishing functions which do not contribute to identities, for example. Many of these fail the simple test of \mbox{$(a\pl\,2)\leq\sigma(a)\leq(n\pl\,a\pl\,2)$}, but with increasing frequency, configurations fail to have kinematical support for much more subtle reasons---demonstrating the value of having a more robust yet simple combinatorial test available. For example, neither of the following on-shell graphs---in $G(4,8)$ and $G(5,10)$, respectively---have kinematical support:
\vspace{-.2cm}\eq{\hspace{0.3cm}\begin{array}{c}\\[-20pt]\text{{\small$\Gamma^4(C)=0$}}\\[0.085cm]\includegraphics[scale=.8]{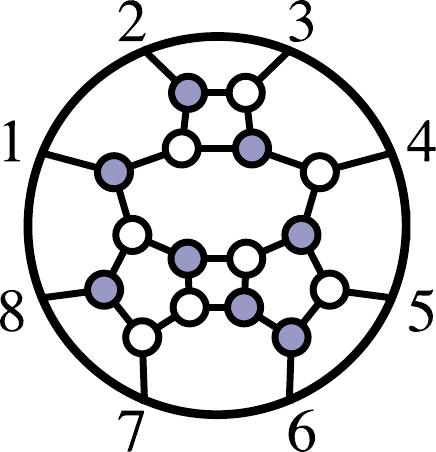}\\[.00cm]\text{{\footnotesize{\color{perm}$\{6,\!4,\!9,\!7,\!8,\!10,\!11,\!13\}$}}}\end{array}\qquad\mathrm{and}\qquad\begin{array}{c}\\[-20pt]\text{{\small$\Gamma^4(C)=0$}}\\\includegraphics[scale=.8]{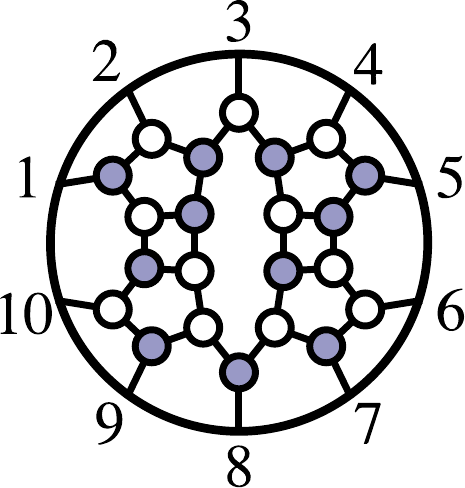}\\[-0.24cm]\text{{\footnotesize{\color{perm}$\!\!\!\!\{8,\!9,\!6,\!7,\!11,\!10,\!14,\!15,\!12,\!13\}\!\!\!\!$}}}\end{array}\nonumber\vspace{-.2cm}}

Configurations for which $\Gamma^4(C)=1$ correspond to manifestly {\it rational} functions of the kinematical data. More generally, however, when $\Gamma^4(C)>1$ the isolation of internal degrees of freedom via $\delta^{k\times4}\big(C\!\cdot\!Z\big)$ results in a (generally) {\it algebraic} function of the external twistors for each isolated solution $C^*\!\in\!C\newcap Z^\perp$---each point giving us a Yangian-invariant which is individually of some physical interest.  However, a highly non-trivial but general result is that the function obtained by summing-over all isolated solutions to $C\cdot Z=0$ is {\it always} {\it rational}. Throughout the rest of this paper, whenever we speak of `the' function associated with a graph for which $\Gamma^4>1$---for example, when appearing in a identity (see \mbox{section \ref{geometric_origin_of_identities_section}})---we always implicitly mean the rational function obtained by summing-over all particular solutions to $C\!\cdot\!Z=0$.

On-shell graphs which admit multiple solutions to the kinematical constraints are comparatively rare. The first on-shell graph for which more than one solution exists occurs for $G(4,8)$ and is well-known to physicists as the `four-mass-box':
\vspace{-.2cm}\eq{\hspace{0.0cm}\begin{array}{c}\includegraphics[scale=.8]{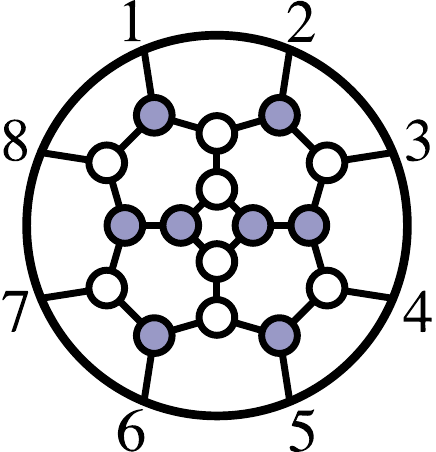}\end{array}\label{four_mass_box}\vspace{-0.3cm}}
The image of this configuration in the momentum-twistor Grassmannian is labeled by ${\color{perm}\{2,5,4,7,6,9,8,11\}}$, for which we calculate $\Gamma^4(C)=2$ recursively as follows:\\[-20pt]

\indent\scalebox{0.85}{\setlength{\unitlength}{2.5mm}\hspace{0.25cm}\begin{minipage}[h]{\textwidth}\eq{\hspace{-2.45cm}\begin{array}{c}\begin{array}{c}\text{{\footnotesize$\Gamma^4(C)=2$}}\\\text{{\footnotesize$\left(\begin{array}{@{}cccccccc@{}}{\color{red}1}&{\color{blue}*}&{\color{blue}*}&{\color{blue}*}&{\color{blue}*}&{\color{blue}*}&{\color{deemph}0}&{\color{deemph}0}\\{\color{deemph}0}&{\color{deemph}0}&{\color{red}1}&{\color{blue}*}&{\color{blue}*}&{\color{blue}*}&{\color{blue}*}&{\color{blue}*}\end{array}\right)$}}\end{array}\\\text{{\small ${\color{perm}\{2,\!5,\!4,\!7,\!6,}{\color{red}\mathbf{9}}{\color{perm},\!8,}{\color{red} \mathbf{11}}{\color{perm}\}}$}}\\~\\~\\~\\[0.15cm]~\end{array}\;\;
\raisebox{0.2cm}{\begin{picture}(2,4)\put(-1,5.3){$[\partial^2]$}\put(-1,4.3){\vector(1,0){3}}\end{picture}}\raisebox{1.15cm}{$\left\{\rule{0pt}{65pt}\right.\!\!\!$}
\begin{array}{c}\begin{array}{c}\text{{\footnotesize$\Gamma^3(C)=1$}}
\\ \text{{\footnotesize$\left(\begin{array}{@{}cccccccc@{}}{\color{deemph}0}&{\color{red}1}&{\color{deemph}0}&{\color{blue}*}&{\color{blue}*}&{\color{blue}*}&{\color{deemph}0}&{\color{deemph}0}\\{\color{deemph}0}&{\color{deemph}0}&{\color{deemph}0}&{\color{red}1}&{\color{blue}*}&{\color{blue}*}&{\color{blue}*}&{\color{blue}*}\end{array}\right)$}}\end{array}\\
\text{{\small${\color{perm}
\{1,\!5,\!3,\!7,\!6,}{\color{red}\mathbf{10}}{\color{perm},\!8,}{\color{red}\mathbf{12}}{\color{perm}\}}$}}\\~\\\begin{array}{c}\text{{\footnotesize$\Gamma^3(C)=1$}}
\\ \text{{\footnotesize$\left(\begin{array}{@{}cccccccc@{}}{\color{deemph}0}&{\color{red}1}&{\color{blue}*}&{\color{blue}*}&{\color{blue}*}&{\color{blue}*}&{\color{deemph}0}&{\color{deemph}0}\\{\color{deemph}0}&{\color{deemph}0}&{\color{deemph}0}&{\color{deemph}0}&{\color{red}1}&{\color{blue}*}&{\color{blue}*}&{\color{blue}*}\end{array}\right)$}}\end{array}\\
\text{{\small ${\color{perm}\{1,\!3,\!4,\!7,\!6,}{\color{red}
\mathbf{10}}{\color{perm},\!8,}{\color{red}
\mathbf{13}}{\color{perm}\}}$}}
\\~\\~\\~\\[0.35cm]~\\\end{array}\;\;
\raisebox{0.3cm}{\begin{picture}(2,4)\put(-1,10.7){$[\partial^2]$}\put(-1,9.7){\vector(1,0){3}}\put(-0.8,-1.6){$[\partial^2]$}\put(-1.5,-0.6){\vector(1,-2){3.}}\end{picture}}\raisebox{-1.45cm}{$\left\{\rule{0pt}{65pt}\right.\!\!\!\!$}
\begin{array}{c}\begin{array}{c}\text{{\footnotesize$\Gamma^2(C)=1$}}
\\ \text{{\footnotesize$\left(\begin{array}{@{}cccccccc@{}}{\color{deemph}0}&{\color{deemph}0}&{\color{deemph}0}&{\color{red}1}&{\color{blue}*}&{\color{blue}*}&{\color{deemph}0}&{\color{deemph}0}\\{\color{deemph}0}&{\color{deemph}0}&{\color{deemph}0}&{\color{deemph}0}&{\color{red}1}&{\color{blue}*}&{\color{blue}*}&{\color{blue}*}\end{array}\right)$}}\end{array}\\
\text{{\small
${\color{perm}\{1,\!2,\!3,\!7,\!6,}{\color{red}\mathbf{12}}{\color{perm},\!8,}{\color{red}\mathbf{13}}{\color{perm}\}}$}}
\\~\\\begin{array}{c}\text{{\footnotesize$\Gamma^2(C)=1$}}
\\ \text{{\footnotesize$\left(\begin{array}{@{}cccccccc@{}}{\color{deemph}0}&{\color{deemph}0}&{\color{red}1}&{\color{blue}*}&{\color{deemph}0}&{\color{blue}*}&{\color{deemph}0}&{\color{deemph}0}\\{\color{deemph}0}&{\color{deemph}0}&{\color{deemph}0}&{\color{deemph}0}&{\color{deemph}0}&{\color{red}1}&{\color{blue}*}&{\color{blue}*}\end{array}\right)$}}\end{array}\\
\text{{\small
${\color{perm}\{1,\!2,\!4,\!7,\!5,}{\color{red}\mathbf{11}}{\color{perm},\!8,}{\color{red}\mathbf{14}}{\color{perm}\}}$}}
\\~\\\begin{array}{c}\text{{\footnotesize$\Gamma^2(C)=0$}}
\\ \text{{\footnotesize$\left(\begin{array}{@{}cccccccc@{}}{\color{deemph}0}&{\color{deemph}0}&{\color{red}1}&{\color{blue}*}&{\color{blue}*}&{\color{blue}*}&{\color{deemph}0}&{\color{deemph}0}\\{\color{deemph}0}&{\color{deemph}0}&{\color{deemph}0}&{\color{deemph}0}&{\color{deemph}0}&{\color{deemph}0}&{\color{red}1}&{\color{blue}*}\end{array}\right)$}}\\
\text{{\small ${\color{perm}\{1,\!2,\!4,\!5,\!6,}{\color{red}
\mathbf{11}}{\color{perm},\!8,}{\color{red}\mathbf{15}}{\color{perm}\}}$}}
\end{array}\end{array}\;\;
\raisebox{-0.8cm}{\begin{picture}(2,4)\put(-1,15.15){$[\partial^2]$}\put(-1,14.15){\vector(1,0){3}}\put(-1,4.7){$[\partial^2]$}\put(-1,3.7){\vector(1,0){3}}
\put(-1,-5.7){$[\partial^2]$}\put(-1,-6.7){\vector(1,0){3}}\end{picture}}
\begin{array}{c}\begin{array}{c}\text{{\footnotesize$\Gamma^1(C)=1$}}
\\ \text{{\footnotesize$\left(\begin{array}{@{}cccccccc@{}}{\color{deemph}0}&{\color{deemph}0}&{\color{deemph}0}&{\color{deemph}0}&{\color{red}1}&{\color{blue}*}&{\color{deemph}0}&{\color{deemph}0}\\{\color{deemph}0}&{\color{deemph}0}&{\color{deemph}0}&{\color{deemph}0}&{\color{deemph}0}&{\color{deemph}0}&{\color{red}1}&{\color{blue}*}\end{array}\right)$}}\end{array}\\
\text{{\small${\color{perm}
\{1,\!2,\!3,\!4,\!6,}{\color{red}\mathbf{13}}{\color{perm},\!8,}{\color{red}\mathbf{15}}{\color{perm}\}}$}}
\\~\\\begin{array}{c}\text{{\footnotesize$\Gamma^1(C)=1$}}
\\ \text{{\footnotesize$\left(\begin{array}{@{}cccccccc@{}}{\color{deemph}0}&{\color{deemph}0}&{\color{deemph}0}&{\color{red}1}&{\color{deemph}0}&{\color{blue}*}&{\color{deemph}0}&{\color{deemph}0}\\{\color{deemph}0}&{\color{deemph}0}&{\color{deemph}0}&{\color{deemph}0}&{\color{deemph}0}&{\color{deemph}0}&{\color{red}1}&{\color{blue}*}\end{array}\right)$}}\end{array}\\
\text{{\small${\color{perm}\{1,\!2,\!3,\!6,\!5,}{\color{red}\mathbf{12}}{\color{perm},\!8,}{\color{red}\mathbf{15}}{\color{perm}\}}$}}
\\~\\\begin{array}{c}\text{{\footnotesize$\Gamma^1(C)=0$}}
\\ \text{{\footnotesize$\left(\begin{array}{@{}cccccccc@{}}{\color{deemph}0}&{\color{deemph}0}&{\color{deemph}0}&{\color{red}1}&{\color{blue}*}&{\color{blue}*}&{\color{deemph}0}&{\color{deemph}0}\\{\color{deemph}0}&{\color{deemph}0}&{\color{deemph}0}&{\color{deemph}0}&{\color{deemph}0}&{\color{deemph}0}&{\color{deemph}0}&{\color{red}1}\end{array}\right)$}}\\
\text{{\small
${\color{perm}\{1,\!2,\!3,\!5,\!6,}{\color{red}\mathbf{12}}{\color{perm},\!7,}{\color{red}\mathbf{16}}{\color{perm}\}}$}}
\end{array}\end{array}\;\;
\raisebox{0.6cm}{\begin{picture}(2,4)\put(0,8.25){$[\partial^2]$}\put(-1,8.7){\vector(1,-1){3}}\put(0,-1.5){$[\partial^2]$}\put(-1,-1.2){\vector(1,1){3.}}\put(-1,-11.3){$[\partial^2]$}\put(-1,-12.3){\vector(1,0){3}}\put(2.5,-12.7){$\{\}$}\end{picture}}
\begin{array}{c}\begin{array}{c}\text{{\footnotesize$\Gamma^0(C)=1$}}
\\ \text{{\footnotesize$\left(\begin{array}{@{}cccccccc@{}}{\color{deemph}0}&{\color{deemph}0}&{\color{deemph}0}&{\color{deemph}0}&{\color{deemph}0}&{\color{red}1}&{\color{deemph}0}&{\color{deemph}0}\\{\color{deemph}0}&{\color{deemph}0}&{\color{deemph}0}&{\color{deemph}0}&{\color{deemph}0}&{\color{deemph}0}&{\color{deemph}0}&{\color{red}1}\end{array}\right)$}}\\
\text{{\small${\color{perm}
\{1,\!2,\!3,\!4,\!5,}{\color{red}\mathbf{14}}{\color{perm},\!7,}{\color{red}\mathbf{16}}{\color{perm}\}}$}}
\end{array}\\~\\~\\~\\[0.55cm]~\end{array}\nonumber}\end{minipage}}\\[-4pt]

Configurations admitting more than two solutions to $C\cdot Z=0$ are even rarer---and their rarity increases dramatically with increasing $\Gamma^4$. Indeed, almost no examples of Yangian-invariant functions for which $\Gamma^4(C)>2$ were even known before the advent of the tools described in this section. But having the combinatorial test available allows us to systematically find and classify them. Three striking examples of on-shell graphs which admit many solutions to the kinematical constraints---for $G(6,12)$, $G(8,16)$, and $G(10,20)$, respectively---are:
\vspace{-.2cm}\eq{\hspace{-0.65cm}\begin{array}{c}\Gamma^4(C)=4\\\includegraphics[scale=.9]{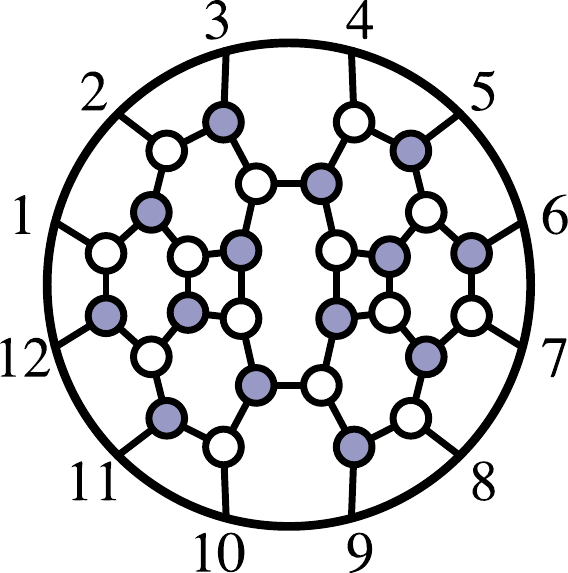}\\[-5pt]\text{{\footnotesize{\color{perm}$\{10,\!8,\!12,\!7,\!11,\!9,\!16,\!14,\!18,\!13,\!17,\!15\}$}}}\\[-50pt]
\end{array}\;\begin{array}{c}\Gamma^4(C)=4\\\includegraphics[scale=.9]{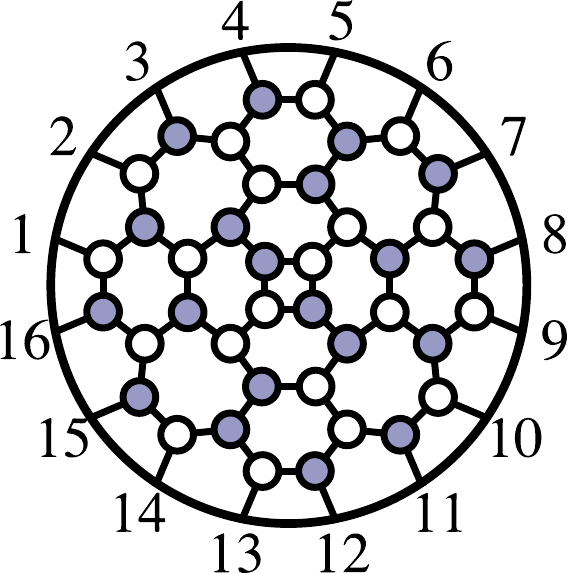}\\[-5pt]\text{{\footnotesize{\color{perm}$\begin{array}{r@{}c@{,}c@{,}c@{,}c@{,}c@{,}c@{,}c@{,}c@{}l@{}}\{&11&5&16&10&15&9&20&14,\\&19&13&24&18&23&17&28&22&\}\end{array}$}}}\\[-50pt]\end{array}\;\begin{array}{c}\Gamma^4(C)=34\\\includegraphics[scale=.9]{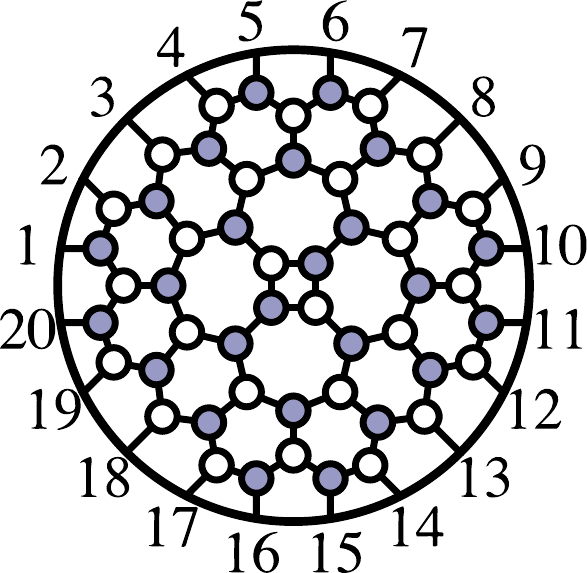}\\[-5pt]\text{{\footnotesize{\color{perm}$\begin{array}{r@{}r@{,}c@{,}c@{,}c@{,}c@{,}c@{,}c@{,}c@{,}c@{,}c@{}l@{}}\{&15&14&8&7&21&20&19&13&12&26,\\&25&24&18&17&31&30&29&23&22&36&\}\end{array}$}}}\\[-50pt]\end{array}\nonumber\vspace{-0.1cm}}

\newpage

\newpage
\section{The Geometric Origin of Identities Among Yangian-Invariants}\label{geometric_origin_of_identities_section}

In this section, we will focus primarily on on-shell differential forms for which the integral over auxiliary Grassmannian degrees of freedom is fully localized by the \mbox{$\delta$-function} constraints, without imposing any conditions on the external kinematical data other than momentum conservation. These are on-shell diagrams with $(2n\,\mi\,4)$ degrees of freedom or their momentum-twistor images with $4k$ degrees of freedom, and for which $\Gamma^4(C)>0$; we will refer to such on-shell forms as {\it Yangian-invariants}, and frequently refer to them (improperly) as `functions' of the kinematical variables.

One of the most remarkable and important properties about Yangian-invariants is that they satisfy many, intricate functional identities. Examples of such identities have long been known, and are crucial for our understanding of many important physical properties of scattering amplitudes. Perhaps the simplest and most familiar examples of such identities come from equating the various implementations of the BCFW recursion relations, (\ref{all_loop_recursion}); for example, for the $6$-particle NMHV tree-level scattering amplitude, the BCFW recursion can alternatively lead to two distinct formulae depending on which pair of adjacent legs are singled-out by the recursion:
\vspace{-0.45cm}\eq{\hspace{-1.cm}\begin{array}{ccccccc}\raisebox{-40pt}{\includegraphics[scale=1]{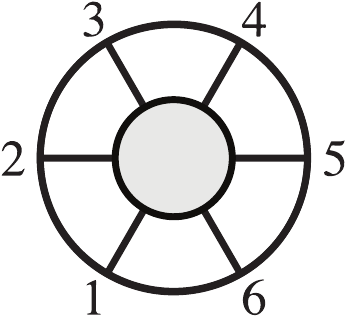}}&\text{{\Large$=\!\!$}}&\begin{array}{c}\raisebox{-10pt}{\includegraphics[scale=1]{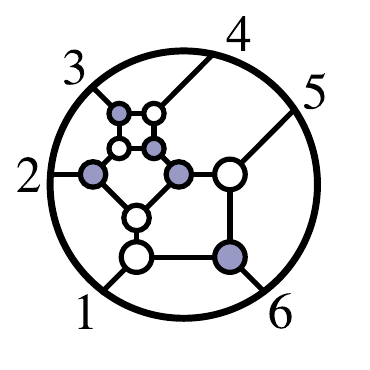}}\\[-10pt]{\color{perm}\{4,5,6,8,7,9\}}\end{array}&\text{{\Large$\!\!\!\!\!+\!\!\!$}}&\begin{array}{c}\raisebox{-10pt}{\includegraphics[scale=1]{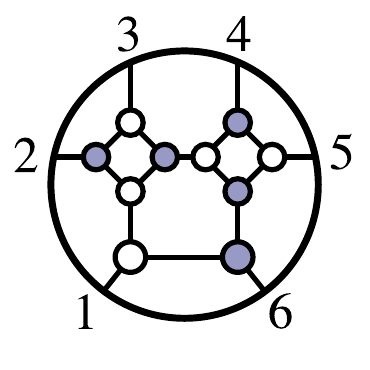}}\\[-10pt]{\color{perm}\{3,5,6,7,8,10\}}\end{array}&\text{{\Large$\!\!\!+\!\!\!\!\!$}}&\begin{array}{c}\raisebox{-10pt}{\includegraphics[scale=1]{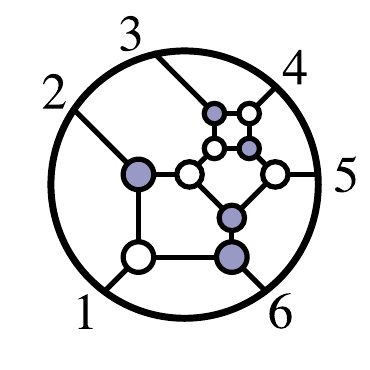}}\\[-10pt]{\color{perm}\{4,6,5,7,8,9\}}\end{array}\\[-6pt]\text{{\normalsize$\rule{0.75pt}{11.5pt}\rule{3pt}{0pt}\rule{0.75pt}{11.5pt}$}}\\[-6pt]\raisebox{-40pt}{\includegraphics[scale=1]{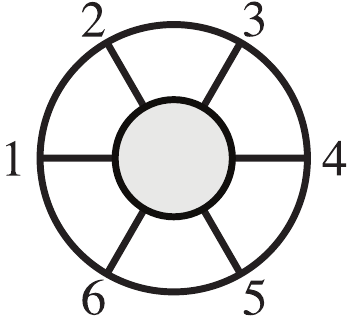}}&\text{{\Large$=\!\!$}}&\begin{array}{c}\raisebox{-10pt}{\includegraphics[scale=1]{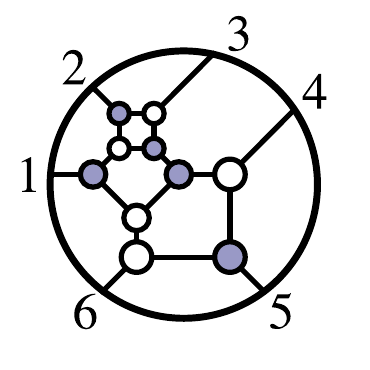}}\\[-10pt]{\color{perm}\{4,5,7,6,8,9\}}\end{array}&\text{{\Large$\!\!\!\!\!+\!\!\!$}}&\begin{array}{c}\raisebox{-10pt}{\includegraphics[scale=1]{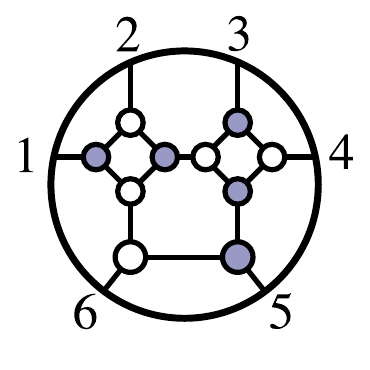}}\\[-10pt]{\color{perm}\{4,5,6,7,9,8\}}\end{array}&\text{{\Large$\!\!\!+\!\!\!\!\!$}}&\begin{array}{c}\raisebox{-10pt}{\includegraphics[scale=1]{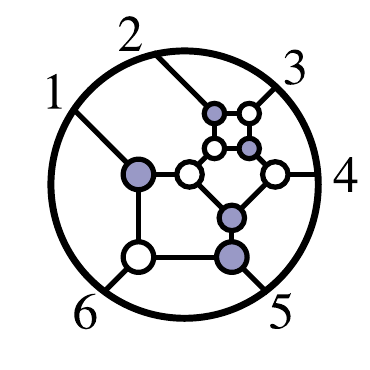}}\\[-10pt]{\color{perm}\{5,4,6,7,8,9\}}\end{array}\end{array}\vspace{-0.2cm}\hspace{-1cm}\nonumber}
This identity is not easy to prove directly if each term is viewed as a multivariate, rational `function' of the kinematical data. However, its veracity is crucial to our understanding of many important properties of the complete amplitude. For example, although the BCFW-recursion breaks cyclicity by the choice of legs to deform, the entire amplitude---being cyclically-invariant---must be independent of this choice.

A wide variety of such identities can be generated simply by equating all the myriad BCFW `formulae' obtained by recursing the left- and right-amplitudes appearing across the BCFW-bridge in all possible ways (at each stage of the recursion). For example, for the $8$-particle N$^2$MHV tree amplitude, there are many hundreds of ways to follow the recursion all the way down to a sum of 20 trivalent, on-shell diagrams; this multitude of BCFW `formulae' involves a total of $176$ distinct Yangian-invariants in $G(4,8)$, and equating every pair leads to 74 linearly-independent, 40-term identities satisfied among them.

Other than the equality of different BCFW formulae, however, few identities among Yangian-invariants were known until the Grassmannian formulation---the contour integral ``$\mathcal{L}_{n,k}$'' was discovered, \cite{ArkaniHamed:2009dn}. But a complete understanding of the range of possible Yangian-invariants, and a systematic understanding of the relations they satisfy remained to be understood. In the remainder of this section, we will describe how {\it all} such identities arise {\it homologically} in the Grassmannian, and can be understood in purely geometric (even combinatorial) terms. In \mbox{section \ref{classification_section}}, we will illustrate the power of the combinatorial tools at our disposal, by giving an explicit classification of  all Yangian-invariants and their relations through N$^4$MHV.

\subsection{Homological Identities in the Grassmannian}\label{homological_identities_subsection}
The six-term identity described above which equates the two possible representations of the $6$-particle N$^{(k=1)}$MHV tree-amplitude turns out to generate {\it all} the identities among NMHV Yangian-invariants. In order to see how this can be, let us first descend to the somewhat simpler situation which arises in the momentum-twistor Grassmannian, where NMHV Yangian-invariants correspond to $4$-dimensional cells of $G(1,n)$.

All NMHV Yangian-invariants are essentially equivalent, as any $4$-dimensional configuration in $C\!\in\!G(1,n)$ involves precisely 5 non-vanishing `columns'; and so, such configurations differ only in which of the 5 columns are involved. In terms of canonical coordinates, such a configuration would be represented by,
\vspace{-.1cm}\eq{C(\alpha)\equiv\big(\begin{array}{cccccccccccccccccccccc}&&a&&&&b&&&&c&&&&d&&&&e\\\hline{\color{deemph}\!\cdots\!}&{\color{deemph}0}&{\color{black}1}&{\color{deemph}0}&{\color{deemph}\!\cdots\!}&{\color{deemph}0}&{\color[rgb]{0.1333,0,0.666}\alpha_1}&{\color{deemph}0}&{\color{deemph}\!\cdots\!}&{\color{deemph}0}&{\color[rgb]{0.155555,0,0.77777}\alpha_2}&{\color{deemph}0}&{\color{deemph}\!\cdots\!}&{\color{deemph}0}&{\color[rgb]{0.177777,0,0.8888}\alpha_3}&{\color{deemph}0}&{\color{deemph}\!\cdots\!}&{\color{deemph}0}&{\color[rgb]{0.2,0,1}\alpha_4}&{\color{deemph}0}&{\color{deemph}\!\cdots\!}\\&\end{array}\big),\nonumber\vspace{-0.5cm}}
and would be labeled by a permutation,
\vspace{-0.2cm}\eq{\sigma\equiv\left(\begin{array}{@{$\!$}ccccc@{$\!$}}a&b&c&d&e\\[-0.2cm]\,\,\downarrow\,\,&\,\,\downarrow\,\,&\,\,\downarrow\,\,&\,\,\downarrow\,\,&\,\,\downarrow\,\,\\[-0.1cm]b&c&d&e&a\end{array}\right),\vspace{-0.2cm}}
with $\sigma(j)=j$ for all other columns. Instead of labeling the configuration by its permutation, it is tempting to label it instead by its 5 non-vanishing columns---as a {\it 5-bracket}, `$[a\,b\,c\,d\,e]$'. Given any generic momentum-twistors $Z\!\in\!G(4,n)$, there is a unique point $C^*\!\in\!C\newcap Z^\perp$, which can be represented by the matrix,
\vspace{-.4cm}\eq{C^*\equiv\big(\begin{array}{cccccccccccccccccccccc}&&a&&&&b&&&&c&&&&d&&&&e\\\hline{\color{deemph}\!\cdots\!}&{\color{deemph}0}&\ab{b\,c\,d\,e}&{\color{deemph}0}&{\color{deemph}\!\cdots\!}&{\color{deemph}0}&\ab{c\,d\,e\,a}&{\color{deemph}0}&{\color{deemph}\!\cdots\!}&{\color{deemph}0}&\ab{d\,e\,a\,b}&{\color{deemph}0}&{\color{deemph}\!\cdots\!}&{\color{deemph}0}&\ab{e\,a\,b\,c}&{\color{deemph}0}&{\color{deemph}\!\cdots\!}&{\color{deemph}0}&\ab{a\,b\,c\,d}&{\color{deemph}0}&{\color{deemph}\!\cdots\!}\\&\end{array}\big),\nonumber\vspace{-0.5cm}}
for which it is easy to see that $C^*\!\!\cdot\!Z=0$ as an instance of Cramer's rule, (\ref{Cramers_rule}). This leads to the general form of the essentially-unique NMHV Yangian-invariant,
\vspace{-.2cm}\eq{\hspace{-0.35cm}\begin{array}{c}[a\,b\,c\,d\,e]\\\text{{\small${\color{perm}\!\!\{b,\!c,\!d,\!e,\!a\}}\!\!$}}\end{array}\raisebox{-3pt}{\text{{\LARGE$\,\Leftrightarrow\,$}}}\frac{\delta^{1\times4}\big(\eta_a\ab{b\,c\,d\,e}+\eta_b\ab{c\,d\,e\,a}+\eta_c\ab{d\,e\,a\,b}+\eta_d\ab{e\,a\,b\,c}+\eta_e\ab{a\,b\,c\,d}\big)}{\ab{b\,c\,d\,e}\ab{c\,d\,e\,a}\ab{d\,e\,a\,b}\ab{e\,a\,b\,c}\ab{a\,b\,c\,d}}.\label{5bracket_not_consecutive}\hspace{-1cm}\vspace{-.2cm}}
(Notice that the 5-bracket $[a\,b\,c\,d\,e]$ as we have defined it is antisymmetric with respect to its arguments. This reflects the fact that the measure \mbox{$d\!\log(\alpha_1)\wedge\cdots\wedge d\!\log(\alpha_4)$} is {\it oriented}. This 5-bracket is the simplest dual super-conformal invariant and was first found in the literature in \cite{Drummond:2008vq} in momentum space).

If we considered instead a $5$-dimensional configuration in $G(1,n)$, then the constraint $\delta^{1\times4}\big(C\!\cdot\!Z\big)$ would fix only four of the internal degrees of freedom, leaving us with a 1-dimensional integral over $G(1,n)$. In this case, Cauchy's theorem informs us that the sum of all the residues of this one-form will vanish. As each of these residues is itself a 4-dimensional configuration of the form above (\ref{5bracket_not_consecutive}), this gives rise to an identity among $5$-brackets. Motivated by the notation used above, let us denote a generic $5$-dimensional configuration in $G(1,n)$ by the {\it 6-bracket} $[a\,b\,c\,d\,e\,f]$; then we find,
\vspace{-.1cm}\eq{\hspace{-0.5cm}\left.\begin{array}{c}\partial[a\,b\,c\,d\,e\,f]\phantom{\partial}\\\text{{\footnotesize${\color{perm}\!\{b,\!c,\!d,\!e,\!f,\!a\}\!}$}}\end{array}\right.\!\!\!\!\!\!\begin{array}{c}=\\~\end{array}\!\!\begin{array}{c}[a\,b\,c\,d\,e]\\\text{{\footnotesize${\color{perm}\!\{b,\!c,\!d,\!e,}{\color{red}\textbf{\textit{a}}}{\color{perm},}{\color{red}\textbf{\textit{f}}}{\color{perm}\}\!}$}}\end{array}\!\!\begin{array}{c}-\\~\end{array}\!\!\begin{array}{c}[a\,b\,c\,d\,f]\\\text{{\footnotesize${\color{perm}\!\{b,\!c,\!d,}{\color{red}\textbf{\textit{f}}}{\color{perm},}{\color{red}\textbf{\textit{e}}}{\color{perm},\!a\}\!}$}}\end{array}\!\!\begin{array}{c}+\\~\end{array}\!\!\begin{array}{c}[a\,b\,c\,e\,f]\\\text{{\footnotesize${\color{perm}\!\{b,\!c,}{\color{red}\textbf{\textit{e}}}{\color{perm},}{\color{red}\textbf{\textit{d}}}{\color{perm},\!f,\!a\}\!}$}}\end{array}\!\!\begin{array}{c}-\\~\end{array}\!\!\begin{array}{c}[a\,b\,d\,e\,f]\\\text{{\footnotesize${\color{perm}\!\{b,}{\color{red}\textbf{\textit{d}}}{\color{perm},}{\color{red}\textbf{\textit{c}}}{\color{perm},\!e,\!f,\!a\}\!}$}}\end{array}\!\!\begin{array}{c}+\\~\end{array}\!\!\begin{array}{c}[a\,c\,d\,e\,f]\\\text{{\footnotesize${\color{perm}\!\{}{\color{red}\textbf{\textit{c}}}{\color{perm},}{\color{red}\textbf{\textit{b}}}{\color{perm},\!d,\!e,\!f,\!a\}\!}$}}\end{array}\!\!\begin{array}{c}-\\~\end{array}\!\!\begin{array}{c}[b\,c\,d\,e\,f]\\\text{{\footnotesize${\color{perm}\!\{}{\color{red}\textbf{\textit{a}}}{\color{perm},\!c,\!d,\!e,\!f,}{\color{red}\textbf{\textit{b}}}{\color{perm}\}\!}$}}\end{array}\!\!\begin{array}{c}=0.\\~\end{array}\label{nmhv_identity_with_general_labels}\nonumber\vspace{-.1cm}}
(Here, the signs are important: they reflect the fact that our formula for the \mbox{$5$-bracket} (\ref{5bracket_not_consecutive}) corresponds to a {\it particular orientation} of the $4$-dimensional cells; and so, when taking the boundary of $[a\,b\,c\,d\,e\,f]$ we must re-order the coordinates for each boundary cell accordingly---at the cost of introducing signs. Notice that the alternating signs here {\it precisely} capture the equality between two three-term, all-plus formulae as generated by equating BCFW formulae as described above.)

Notice that this 6-term identity precisely reproduces the identity among $6$-particle NMHV Yangian invariants generated by equating BCFW recursion schemes. More importantly, however, because we understand that all NMHV Yangian-invariants are of the same basic form, the identity given above captures {\it all} the identities satisfied among NMHV Yangian-invariants.

The essential point in the example above is that if we consider a configuration $C\!\in\!G(k,n)$ {\it whose boundary} includes those associated with Yangian-invariant `functions', then the $\delta$-function constraints will localize the Grassmannian integral to a 1-dimensional integral, allowing us to use Cauchy's theorem to conclude that the sum of all the residues in the boundary will vanish; equivalently, that the combination of Yangian-invariants along any boundary $\partial(C)$ add to zero. This turns out to generate {\it all} the functional identities satisfied by Yangian-invariants, including many of impressive complexity.

Recall that the boundary of an on-shell diagram is the collection of diagrams obtained by deleting its {\it removable} edges. And so we can find identities among N$^k$MHV on-shell differential forms by taking the boundary of any $(2n\,\mi\,3)$-dimensional cell in $G(k\pl2,n)$ for ordinary kinematical data, or any $(4k\pl1)$-dimensional cell of $G(k,n)$ for momentum-twistor kinematical data. One example of an identity found in this way generates an identity among $8$-particle N$^2$MHV Yangian-invariants which is independent of all those identities found by equating various BCFW formulae, and can be understood as a way to represent the `four-mass box' (which generally involves quadratic roots, as $\Gamma^4(C)=2$) as a sum of purely-rational Yangian-invariants:
\noindent\scalebox{0.7}{\begin{minipage}[h]{\textwidth}\eq{
\hspace{.075cm}\raisebox{-53.5pt}{\raisebox{58.5pt}{\text{\Huge$\partial$}$\begin{array}{c}~\\\includegraphics[scale=1]{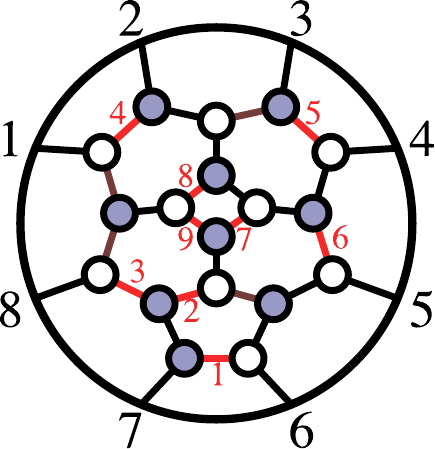}\\[-0pt]\text{{\large${\color{perm}\{4,\!7,\!6,\!9,\!8,\!10,\!11,\!13\}}$}}\end{array}$}\raisebox{60.5pt}{\text{\Large$=$}}}\left(\begin{array}{@{}cccccc@{}}&\begin{array}{c}\\[-15pt]\includegraphics[scale=1]{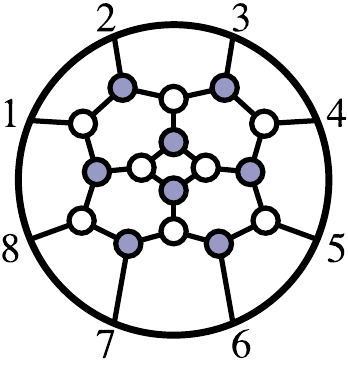}\\[-8pt]
\text{{\normalsize$\!\!{\color{perm}\{4,\!7,\!6,\!9,\!8,}{\color{red}\mathbf{11}}{\color{perm},}{\color{red}\mathbf{10}}{\color{perm},\!13\}}\!\!$}}\end{array}&\raisebox{6.5pt}{\text{{\Large$-$}}}&\begin{array}{c}\\[-15pt]\includegraphics[scale=1]{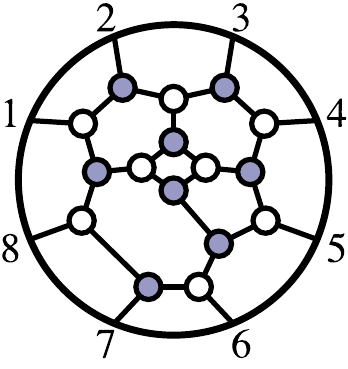}\\[-8pt]\text{{\normalsize$\!\!{\color{perm}\{4,\!7,\!6,\!9,}{\color{red}\mathbf{10}}{\color{perm},}{\color{red}\mathbf{8}},{\color{perm}\!11,\!13\}}\!\!$}}
\end{array}&\raisebox{6.5pt}{\text{{\Large$+$}}}&\begin{array}{c}\\[-15pt]\includegraphics[scale=1]{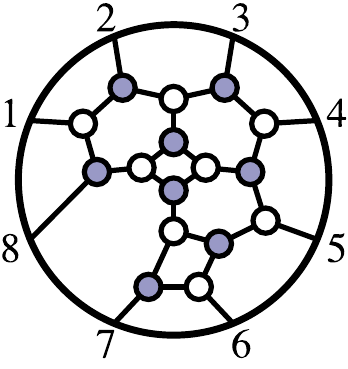}\\[-8pt]\text{{\normalsize$\!\!{\color{perm}\{4,}{\color{red}\mathbf{8}}{\color{perm},\!6,\!9,}{\color{red}\mathbf{7}}{\color{perm},\!10,\!11,\!13\}}\!\!$}}\end{array}\\[-15pt]\raisebox{0pt}{\text{{\Large$-$}}}&\begin{array}{c}\\[-0pt]\includegraphics[scale=1]{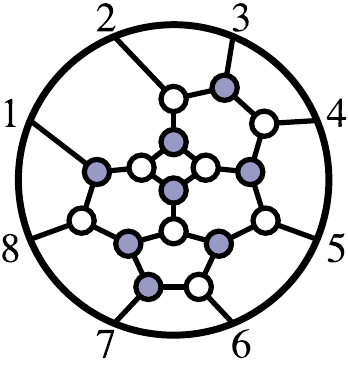}\\[-8pt]\text{{\normalsize\!\!${\color{perm}\{}{\color{red}\mathbf{7}}{\color{perm},}{\color{red}\mathbf{4}}{\color{perm},\!6,\!9,\!8,\!10,\!11,\!13\}}\!\!$}}\end{array}&\raisebox{0pt}{\text{{\Large$+$}}}&\begin{array}{c}\\[-0pt]\includegraphics[scale=1]{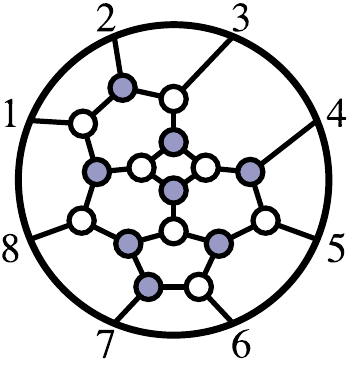}\\[-8pt]\text{{\normalsize$\!\!{\color{perm}\{}{\color{red}\mathbf{3}}{\color{perm},\!7,\!6,\!9,\!8,\!10,}{\color{red}\mathbf{12}}{\color{perm},\!13\}}\!\!$}}\end{array}&\raisebox{0pt}{\text{{\Large$-$}}}&\begin{array}{c}\\[-0pt]\includegraphics[scale=1]{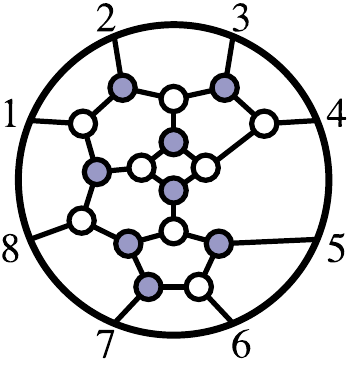}\\[-8pt]\text{{\normalsize$\!\!{\color{perm}\{4,\!7,\!6,\!9,\!8,\!10,}{\color{red}\mathbf{13}}{\color{perm},}{\color{red}\mathbf{11}}{\color{perm}\}}\!\!$}}\end{array}\\[-15pt]\raisebox{0pt}{\text{{\Large$+$}}}&\begin{array}{c}\\[-0pt]\includegraphics[scale=1]{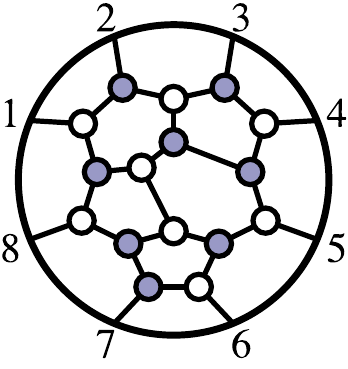}\\[-8pt]\text{{\normalsize$\!\!{\color{perm}\{4,\!7,\!6,}{\color{red}\mathbf{10}}{\color{perm},\!8,}{\color{red}\mathbf{9}}{\color{perm},\!11,\!13\}}\!\!$}}\end{array}&\raisebox{0pt}{\text{{\Large$-$}}}&\begin{array}{c}\\[-0pt]\includegraphics[scale=1]{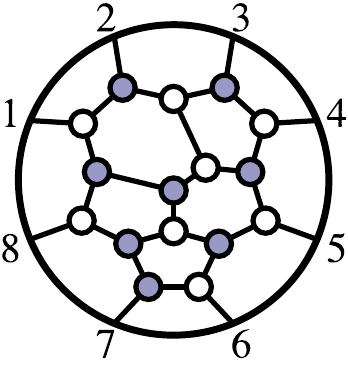}\\[-8pt]\text{{\!\!\normalsize${\color{perm}\{4,\!7,}{\color{red}\mathbf{5}}{\color{perm},\!9,\!8,\!10,\!11,}{\color{red}\mathbf{14}}{\color{perm}\}}$\!\!}}\end{array}&\raisebox{0pt}{\text{{\Large$+$}}}&\begin{array}{c}\\[-0pt]\includegraphics[scale=1]{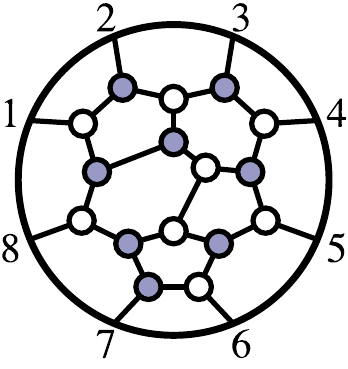}\\[-8pt]\text{{\normalsize$\!\!{\color{perm}\{4,\!7,}{\color{red}\mathbf{9}}{\color{perm},}{\color{red}\mathbf{6}}{\color{perm},\!8,\!10,\!11,\!13\}}\!\!$}}\end{array}\\
\end{array}\right)\raisebox{5pt}{${\text{\Large$=\!0.\;\;$}}$}\nonumber\hspace{-1cm}}\end{minipage}}\\[0.0cm]

\noindent It is worth noting that we have only included the {\it non-vanishing} contributions to this identity---those graphs for which $\Gamma^4>0$; in addition to the nine graphs above, the boundary of {\small${\color{perm}\{4,\!7,\!6,\!9,\!8,\!10,\!11,\!13\}}$} also includes the graphs,
\vspace{-.0cm}\eq{\begin{array}{c}\includegraphics[scale=.75]{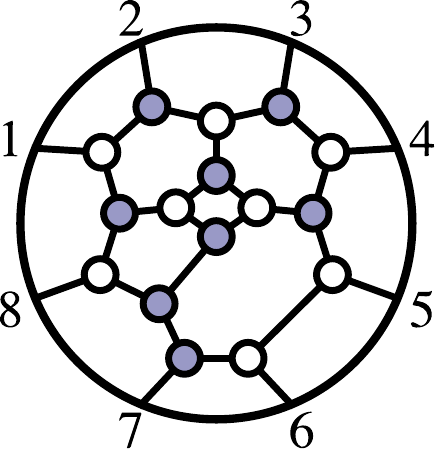}\\\text{{\small${\color{perm}\{4,\!7,}{\color{red}\mathbf{8}}{\color{perm},\!9,}{\color{red}\mathbf{6}}{\color{perm},\!10,\!11,\!13\}}$}}\end{array},\begin{array}{c}\includegraphics[scale=.75]{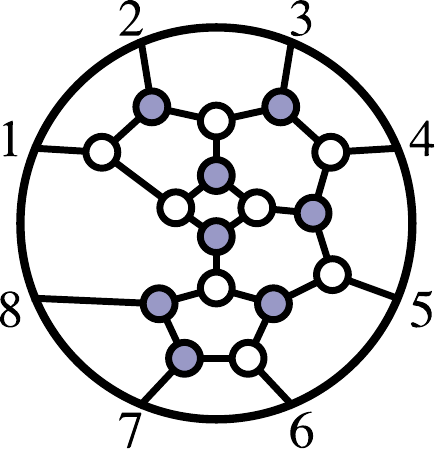}\\\text{{\small${\color{perm}\{4,}{\color{red}\mathbf{5}}{\color{perm},\!6,\!9,\!8,\!10,\!11,}{\color{red}\mathbf{15}}{\color{perm}\}}$}}\end{array},\begin{array}{c}\includegraphics[scale=.75]{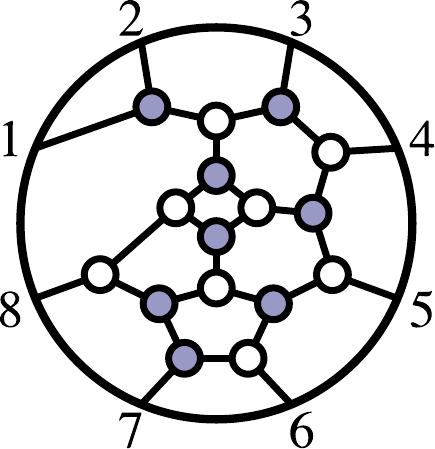}\\\text{{\small${\color{perm}\{4,}{\color{red}\mathbf{9}}{\color{perm},\!6,}{\color{red}\mathbf{7}}{\color{perm},\!8,\!10,\!11,\!13\}}$}}\end{array},\begin{array}{c}\includegraphics[scale=.75]{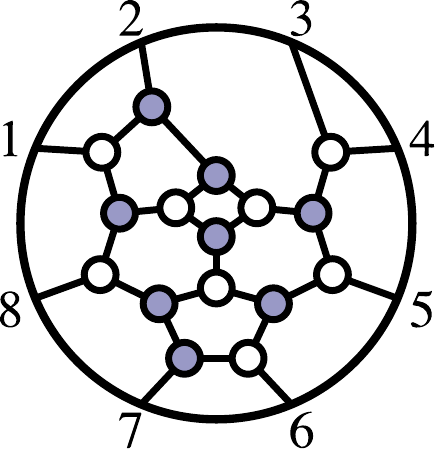}\\\text{{\small${\color{perm}\{}{\color{red}\mathbf{6}}{\color{perm},\!7,}{\color{red}\mathbf{4}}{\color{perm},\!9,\!8,\!10,\!11,\!13\}}$}}\end{array},\vspace{-.1cm}\nonumber}
which all have $\Gamma^4=0$, and so lead to generally-vanishing functions of the external, kinematical data and therefore do not contribute to the identity.

Another particularly impressive example of an identity generated in this way is a 24-term identity among $15$-particle N$^4$MHV Yangian-invariants, generated by the boundary of the 27-dimensional cell,
\vspace{-.5cm}\eq{~\hspace{-.35cm}\begin{array}{c}\\[-8pt]\raisebox{-00.5pt}{\includegraphics[scale=.65]{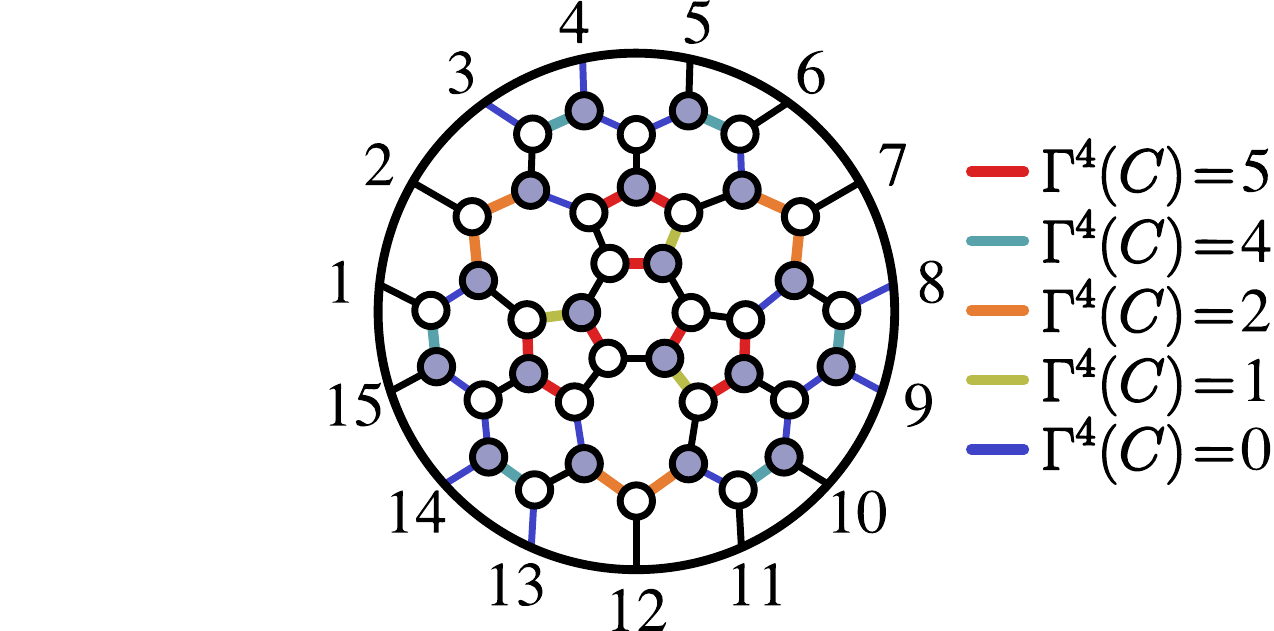}}\\[-5pt]\text{{\footnotesize${\color{perm}\,\{9,\!7,\!6,\!15,\!8,\!14,\!12,\!11,\!20,\!13,\!19,\!17,\!16,\!25,\!18\}}$}}\\[-29pt]\end{array}\nonumber\vspace{-0.1cm}}
which includes 8 cyclic classes of Yangian-invariants---three of which are quintic ($\Gamma^4(C)=5$), two quartic, two quadratic, and one of which is rational:
\vspace{-.2cm}\eq{\hspace{-1.5cm}\includegraphics[scale=0.45]{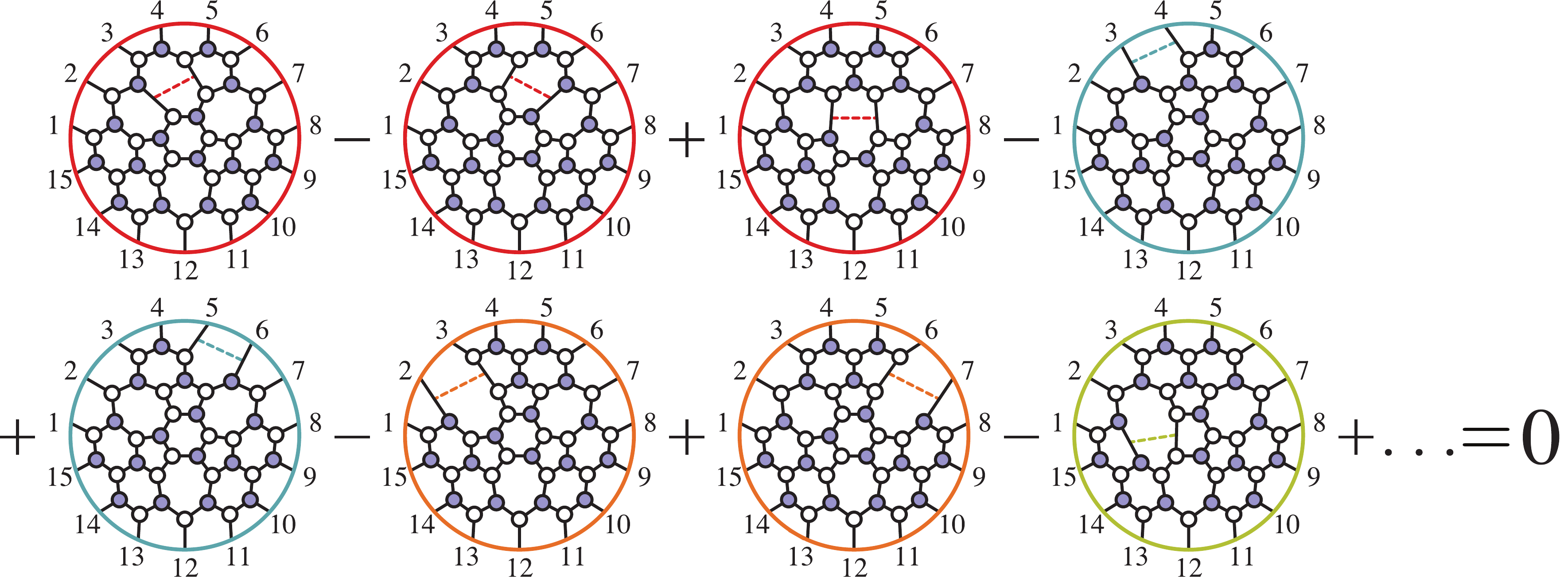}\hspace{-1.5cm}\nonumber\vspace{-.2cm}}
where `$\cdots$' indicates a sum over all cyclic classes.

\newpage
\section{Classification of Yangian-Invariants and Their Relations}\label{classification_section}

As we have seen, N$^{k}$MHV Yangian-invariant functions---expressed in terms of momentum-twistors---are determined by $4k$-dimensional configurations $C\!\in\!G(k,n)$ with non-vanishing kinematical support. As such, they are trivially classified by the permutations which label such cells. Importantly, because the dimensionality of these cells is independent of $n$, it turns out that for any fixed $k$, there {\it only} are only a {\it finite number of N$^k$MHV Yangian-invariant functions} with non-trivial dependence on the external momentum-twistors.

This is easy to see: consider any cell $C\!\in\! G(k,n)$ involving $\nu(C)$ non-vanishing columns; because $(\nu(C)\,\mi\,k)$ of the columns must each represent at least one degree of freedom, $\dim(C)\geq \nu(C)\,\mi\,k$; and so, for $4k$-dimensional cells, $\nu(C)\leq 5k$. Because any vanishing column $c_a$ of $C$ implies that $C\newcap Z^\perp$ is geometrically-independent of the column $z_a$ of $Z$, the Yangian-invariant associated with $C$ will be functionally-independent of the momentum-twistor $z_a$. And so, we immediately see that for any fixed $k$, the number of configurations $C\!\in\!G(k,n)$ with only non-trivial dependence on external momentum-twistors is strictly finite.

In addition to the finiteness of Yangian-invariant functions for fixed $k$, because {\it all} relations among them are generated as boundaries of $(4k\pl\,1)$-dimensional cells, their {\it relations} are also finite in number and can be similarly classified (for early work on classification of Yangian invariants, see\cite{ArkaniHamed:2009dn, ArkaniHamed:2009dg,ArkaniHamed:2009sx,Bullimore:2009cb, Ashok:2010ie}). Given the tools described in \mbox{section \ref{kinematical_support_section}} to distinguish the permutations corresponding to cells with non-vanishing support, such a classification can be carried-out relatively efficiently.

As described in \mbox{section \ref{homological_identities_subsection}}, N$^{(k=1)}$MHV Yangian-invariant functions correspond to cells $C\!\in\!G(1,n)$ with precisely 5 non-vanishing `columns'; because of this, there is only one NMHV Yangian-invariant function---up to trivial dependence on further twistors which is commonly denoted by the {\it 5-bracket} symbol $[1\,2\,3\,4\,5]$:
\eq{\hspace{-.0cm}\begin{array}{c}[1\,2\,3\,4\,5]\\\text{{\small${\color{perm}\!\!\{2,\!3,\!4,\!5,\!6\}}\!\!$}}\end{array}\raisebox{-1.5pt}{{\Large$\,\Leftrightarrow\,$}}\frac{\delta^{1\times4}\big(\eta_1\ab{2\,3\,4\,5}\pl\,\eta_2\ab{3\,4\,5\,2}\pl\,\eta_3\ab{4\,5\,1\,2}\pl\,\eta_4\ab{5\,1\,2\,3}\pl\,\eta_5\ab{1\,2\,3\,4}\big)}{\ab{2\,3\,4\,5}\ab{3\,4\,5\,1}\ab{4\,5\,1\,2}\ab{5\,1\,2\,3}\ab{1\,2\,3\,4}}.\hspace{-0.8cm}\label{r-invariant}}
All other NMHV Yangian-invariants are equivalent to this object, but with the 5 non-vanishing columns arbitrary, `$[a\,b\,c\,d\,e]$', (\ref{5bracket_not_consecutive}). (A form very similar to this one was first introduced in \cite{Hodges:2009hk} and used to develop a polytope description of NMHV tree-level amplitudes.)

Similarly, there is a unique $5$-dimensional cell in $G(1,6)$ involving 6 columns, which we may denote $[1\,2\,3\,4\,5\,6]$. Its boundary generates the unique identity satisfied by $5$-brackets:
\vspace{-.1cm}\eq{\hspace{-0.05cm}\left.\begin{array}{c}\partial[1\,2\,3\,4\,5\,6]\phantom{\partial}\\\text{{\footnotesize${\color{perm}\!\{2,\!3,\!4,\!5,\!6,\!7\}\!}$}}\end{array}\right.\!\!\!\!\!\!\begin{array}{c}=\!\\~\end{array}\!\!\begin{array}{c}[1\,2\,3\,4\,5]\\\text{{\footnotesize${\color{perm}\!\{2,\!3,\!4,\!5,}{\color{red}\mathbf{7}}{\color{perm},}{\color{red}\mathbf{6}}{\color{perm}\}\!}$}}\end{array}\!\!\begin{array}{c}\!-\!\\~\end{array}\!\!\begin{array}{c}[1\,2\,3\,4\,6]\\\text{{\footnotesize${\color{perm}\!\{2,\!3,\!4,}{\color{red}\mathbf{6}}{\color{perm},}{\color{red}\mathbf{5}}{\color{perm},\!7\}\!}$}}\end{array}\!\!\begin{array}{c}\!+\!\\~\end{array}\!\!\begin{array}{c}[1\,2\,3\,5\,6]\\\text{{\footnotesize${\color{perm}\!\{2,\!3,}{\color{red}\mathbf{5}}{\color{perm},}{\color{red}\mathbf{4}}{\color{perm},\!6,\!7\}\!}$}}\end{array}\!\!\begin{array}{c}\!-\!\\~\end{array}\!\!\begin{array}{c}[1\,2\,4\,5\,6]\\\text{{\footnotesize${\color{perm}\!\{2,}{\color{red}\mathbf{4}}{\color{perm},}{\color{red}\mathbf{3}}{\color{perm},\!5,\!6,\!7\}\!}$}}\end{array}\!\!\begin{array}{c}\!+\!\\~\end{array}\!\!\begin{array}{c}[1\,3\,4\,5\,6]\\\text{{\footnotesize${\color{perm}\!\{}{\color{red}\mathbf{3}}{\color{perm},}{\color{red}\mathbf{2}}{\color{perm},\!4,\!5,\!6,\!7\}\!}$}}\end{array}\!\!\begin{array}{c}\!-\!\\~\end{array}\!\!\begin{array}{c}[2\,3\,4\,5\,6]\\\text{{\footnotesize${\color{perm}\!\{}{\color{red}\mathbf{1}}{\color{perm},\!3,\!4,\!5,\!6,}{\color{red}\mathbf{8}}{\color{perm}\}\!}$}}\end{array}\!\!\begin{array}{c}\!=0.\\~\end{array}\label{nmhv_identity}\nonumber\vspace{-.1cm}}

For N$^2$MHV, it is not hard to find that there are precisely 14 cyclically-distinct Yangian-invariant functions (again, ignoring trivial dependence on further twistors). These are listed in \mbox{Table \ref{g2n_yangian_invariants}}, where we have also given canonical coordinates for each, and also expressed each as a product of $5$-brackets---a notation somewhat more familiar to physicists, but which is generally cumbersome, as they frequently depend on geometrically-defined momentum-twistors such as ``$(12)\newcap(345)$'' which means ``the momentum-twistor supported at the intersection of $\mathrm{span}\{z_1,z_2\}$ with $\mathrm{span}\{z_3,z_4,z_5\}$'', found as a trivial application of Cramer's rule, (\ref{Cramers_rule}):\vspace{-.2cm}\eq{\begin{array}{ll}(12)\newcap(345)&\equiv z_1\ab{2\,3\,4\,5}+z_2\ab{3\,4\,5\,1};\\&\propto z_3\ab{4\,5\,1\,2}+z_4\ab{5\,1\,2\,3}+z_5\ab{1\,2\,3\,4}.\end{array}\vspace{-.4cm}}
\begin{table}[t*]\vspace{-1.5cm}\caption{The complete classification of N$^2$MHV Yangian-invariant functions\label{g2n_yangian_invariants}}
\vspace{-0.2cm}\scalebox{0.85}{\begin{minipage}[h]{\textwidth}\eq{\hspace{-1.22cm}\begin{array}{|c|c|l|}\hline\text{\hspace{0.85cm}Configuration\hspace{0.85cm}}&\text{Canonical coordinates for $C\!\in\! G_+(2,n)$}&\multicolumn{1}{c|}{\text{Written in terms of $5$-brackets}}\\\hline\begin{array}{@{}c@{}c@{}c@{}c@{}c@{}c@{}}\\[-31pt]\\(1)&(2)&(3)&(4)&(5)&(6)\\[-5pt]\bullet&\bullet&\bullet&\bullet&\bullet&\bullet\\[-7.9pt]\hline\\[-4.25pt]\\[-24pt]\multicolumn{6}{c}{${\color{perm}\{3,4,5,6,7,8\}}$}\end{array}&\text{{\footnotesize$\left(\begin{array}{@{}cccccc@{}}1&(\alpha_{1}\pl\alpha_{2}\pl\alpha_{3}\pl\alpha_{4})&(\alpha_{2}\pl\alpha_{3}\pl\alpha_{4})\,\alpha_{5}&(\alpha_{3}\pl\alpha_{4})\,\alpha_{6}&\alpha_{4}\,\alpha_{7}&0\\0&1&\alpha_{5}&\alpha_{6}&\alpha_{7}&\alpha_{8}\end{array}\right)$}}&\text{{\small$\begin{array}{l}\phantom{\times\hspace{-0.035cm}}\left[1,2,(23)\newcap(456),(234)\newcap(56),6\right]\\\times\left[2,3,4,5,6\right]\end{array}$}}\\\hline\begin{array}{@{}c@{}c@{}c@{}c@{}c@{}}\\[-31pt]\\(1\,2)&(3\,4)&(5)&(6)&(7)\\[-5pt]\bullet&\bullet&\bullet&\bullet&\bullet\\[-7.9pt]\hline\\[-4.25pt]\\[-24pt]\multicolumn{5}{c}{${\color{perm}\{2,5,4,6,7,8,10\}}$}\end{array}&\text{{\footnotesize$\left(\begin{array}{@{}ccccccc@{}}1&\alpha_{1}&(\alpha_{2}\pl\alpha_{3}\pl\alpha_{4})&(\alpha_{2}\pl\alpha_{3}\pl\alpha_{4})\,\alpha_{5}&(\alpha_{3}\pl\alpha_{4})\,\alpha_{6}&\alpha_{4}\,\alpha_{7}&0\\0&0&1&\alpha_{5}&\alpha_{6}&\alpha_{7}&\alpha_{8}\end{array}\right)$}}&\text{{\small$\begin{array}{l}\phantom{\times\hspace{-0.035cm}}\left[1,2,(34)\newcap(567),(345)\newcap(67),7\right]\\\times\left[3,4,5,6,7\right]\end{array}$}}\\\hline\begin{array}{@{}c@{}c@{}c@{}c@{}c@{}}\\[-31pt]\\(1\,2)&(3)&(4\,5)&(6)&(7)\\[-5pt]\bullet&\bullet&\bullet&\bullet&\bullet\\[-7.9pt]\hline\\[-4.25pt]\\[-24pt]\multicolumn{5}{c}{${\color{perm}\{2,4,6,5,7,8,10\}}$}\end{array}&\text{{\footnotesize$\left(\begin{array}{@{}ccccccc@{}}1&\alpha_{1}&(\alpha_{2}\pl\alpha_{3}\pl\alpha_{4})&(\alpha_{3}\pl\alpha_{4})\,\alpha_{5}&(\alpha_{3}\pl\alpha_{4})\,\alpha_{6}&\alpha_{4}\,\alpha_{7}&0\\0&0&1&\alpha_{5}&\alpha_{6}&\alpha_{7}&\alpha_{8}\end{array}\right)$}}&\text{{\small$\begin{array}{l}\phantom{\times\hspace{-0.035cm}}\left[1,2,3,(345)\newcap(67),7\right]\\\times\left[3,4,5,6,7\right]\end{array}$}}\\\hline\begin{array}{@{}c@{}c@{}c@{}c@{}}\\[-31pt]\\(1\,2\,3)&(4\,5\,6)&(7)&(8)\\[-5pt]\bullet&\bullet&\bullet&\bullet\\[-7.9pt]\hline\\[-4.25pt]\\[-24pt]\multicolumn{4}{c}{${\color{perm}\{2,3,7,5,6,8,9,12\}}$}\end{array}&\text{{\footnotesize$\left(\begin{array}{@{}cccccccc@{}}1&\alpha_{1}&\alpha_{2}&(\alpha_{3}\pl\alpha_{4})&(\alpha_{3}\pl\alpha_{4})\,\alpha_{5}&(\alpha_{3}\pl\alpha_{4})\,\alpha_{6}&\alpha_{4}\,\alpha_{7}&0\\0&0&0&1&\alpha_{5}&\alpha_{6}&\alpha_{7}&\alpha_{8}\end{array}\right)$}}&\text{{\small$\begin{array}{l}\phantom{\times\hspace{-0.035cm}}\left[1,2,3,(456)\newcap(78),8\right]\\\times\left[4,5,6,7,8\right]\end{array}$}}\\\hline\begin{array}{@{}c@{}c@{}c@{}c@{}}\\[-31pt]\\(1\,2\,3)&(4)&(5\,6\,7)&(8)\\[-5pt]\bullet&\bullet&\bullet&\bullet\\[-7.9pt]\hline\\[-4.25pt]\\[-24pt]\multicolumn{4}{c}{${\color{perm}\{2,3,5,8,6,7,9,12\}}$}\end{array}&\text{{\footnotesize$\left(\begin{array}{@{}cccccccc@{}}1&\alpha_{1}&\alpha_{2}&(\alpha_{3}\pl\alpha_{4})&\alpha_{4}\,\alpha_{5}&\alpha_{4}\,\alpha_{6}&\alpha_{4}\,\alpha_{7}&0\\0&0&0&1&\alpha_{5}&\alpha_{6}&\alpha_{7}&\alpha_{8}\end{array}\right)$}}&\text{{\small$\begin{array}{l}\phantom{\times}\left[1,2,3,4,8\right]\\\times\left[4,5,6,7,8\right]\end{array}$}}\\\hline\begin{array}{@{}c@{}c@{}c@{}c@{}}\\[-31pt]\\(1\,2\,3)&(4\,5)&(6\,7)&(8)\\[-5pt]\bullet&\bullet&\bullet&\bullet\\[-7.9pt]\hline\\[-4.25pt]\\[-24pt]\multicolumn{4}{c}{${\color{perm}\{2,3,6,5,8,7,9,12\}}$}\end{array}&\text{{\footnotesize$\left(\begin{array}{@{}cccccccc@{}}1&\alpha_{1}&\alpha_{2}&(\alpha_{3}\pl\alpha_{4})&(\alpha_{3}\pl\alpha_{4})\,\alpha_{5}&\alpha_{4}\,\alpha_{6}&\alpha_{4}\,\alpha_{7}&0\\0&0&0&1&\alpha_{5}&\alpha_{6}&\alpha_{7}&\alpha_{8}\end{array}\right)$}}&\text{{\small$\begin{array}{l}\phantom{\times\hspace{-0.035cm}}\left[1,2,3,(45)\newcap(678),8\right]\\\times\left[4,5,6,7,8\right]\end{array}$}}\\\hline\begin{array}{@{}c@{}c@{}c@{}c@{}}\\[-31pt]\\(1\,2\,3)&(4\,5)&(6)&(7\,8)\\[-5pt]\bullet&\bullet&\bullet&\bullet\\[-7.9pt]\hline\\[-4.25pt]\\[-24pt]\multicolumn{4}{c}{${\color{perm}\{2,3,6,5,7,9,8,12\}}$}\end{array}&\text{{\footnotesize$\left(\begin{array}{@{}cccccccc@{}}1&\alpha_{1}&\alpha_{2}&(\alpha_{3}\pl\alpha_{4})&(\alpha_{3}\pl\alpha_{4})\,\alpha_{5}&\alpha_{4}\,\alpha_{6}&0&0\\0&0&0&1&\alpha_{5}&\alpha_{6}&\alpha_{7}&\alpha_{8}\end{array}\right)$}}&\text{{\small$\begin{array}{l}\phantom{\times\hspace{-0.035cm}}\left[1,2,3,(45)\newcap(678),(456)\newcap(78)\right]\\\times\left[4,5,6,7,8\right]\end{array}$}}\\\hline\begin{array}{@{}c@{}c@{}c@{}c@{}}\\[-31pt]\\(1\,2\,3)&(4)&(5\,6)&(7\,8)\\[-5pt]\bullet&\bullet&\bullet&\bullet\\[-7.9pt]\hline\\[-4.25pt]\\[-24pt]\multicolumn{4}{c}{${\color{perm}\{2,3,5,7,6,9,8,12\}}$}\end{array}&\text{{\footnotesize$\left(\begin{array}{@{}cccccccc@{}}1&\alpha_{1}&\alpha_{2}&(\alpha_{3}\pl\alpha_{4})&\alpha_{4}\,\alpha_{5}&\alpha_{4}\,\alpha_{6}&0&0\\0&0&0&1&\alpha_{5}&\alpha_{6}&\alpha_{7}&\alpha_{8}\end{array}\right)$}}&\text{{\small$\begin{array}{l}\phantom{\times\hspace{-0.035cm}}\left[1,2,3,4,(456)\newcap(78)\right]\\\times\left[4,5,6,7,8\right]\end{array}$}}\\\hline\!\!\text{{\Large$*$}}\;\begin{array}{@{}c@{}c@{}c@{}c@{}}\\[-31pt]\\(1\,2)&(3\,4)&(5\,6)&(7\,8)\\[-5pt]\bullet&\bullet&\bullet&\bullet\\[-7.9pt]\hline\\[-4.25pt]\\[-24pt]\multicolumn{4}{c}{${\color{perm}\{2,5,4,7,6,9,8,11\}}$}\end{array}\phantom{\!\!\text{{\Large$*$}}\;}&\text{{\footnotesize$\left(\begin{array}{@{}cccccccc@{}}1&\alpha_{1}&(\alpha_{2}\pl\alpha_{3})&(\alpha_{2}\pl\alpha_{3})\,\alpha_{4}&\alpha_{3}\,\alpha_{5}&\alpha_{3}\,\alpha_{6}&0&0\\0&0&1&\alpha_{4}&\alpha_{5}&\alpha_{6}&\alpha_{7}&\alpha_{8}\end{array}\right)$}}&\text{{\small$\begin{array}{l}\psi\,\left[A,1,2,3,4\right]\\\times\left[B,5,6,7,8\right]\end{array}$}}
\\\hline\begin{array}{@{}c@{}c@{}c@{}}\\[-31pt]\\(1\,2\,3\,4)&(5\,6\,7\,8)&(9)\\[-5pt]\bullet&\bullet&\bullet\\[-7.9pt]\hline\\[-4.25pt]\\[-24pt]\multicolumn{3}{c}{${\color{perm}\!\!\!\{2,3,4,9,6,7,8,10,14\}\!\!\!}$}\end{array}&\text{{\footnotesize$\left(\begin{array}{@{}ccccccccc@{}}1&\alpha_{1}&\alpha_{2}&\alpha_{3}&\alpha_{4}&\alpha_{4}\,\alpha_{5}&\alpha_{4}\,\alpha_{6}&\alpha_{4}\,\alpha_{7}&0\\0&0&0&0&1&\alpha_{5}&\alpha_{6}&\alpha_{7}&\alpha_{8}\end{array}\right)$}}&\text{{\small$\begin{array}{l}\phantom{\times}\left[1,2,3,4,9\right]\\\times\left[5,6,7,8,9\right]\end{array}$}}\\\hline\begin{array}{@{}c@{}c@{}c@{}}\\[-31pt]\\(1\,2\,3\,4)&(5\,6\,7)&(8\,9)\\[-5pt]\bullet&\bullet&\bullet\\[-7.9pt]\hline\\[-4.25pt]\\[-24pt]\multicolumn{3}{c}{${\color{perm}\!\!\!\{2,3,4,8,6,7,10,9,14\}\!\!\!}$}\end{array}&\text{{\footnotesize$\left(\begin{array}{@{}ccccccccc@{}}1&\alpha_{1}&\alpha_{2}&\alpha_{3}&\alpha_{4}&\alpha_{4}\,\alpha_{5}&\alpha_{4}\,\alpha_{6}&0&0\\0&0&0&0&1&\alpha_{5}&\alpha_{6}&\alpha_{7}&\alpha_{8}\end{array}\right)$}}&\text{{\small$\begin{array}{l}\phantom{\times\hspace{-0.035cm}}\left[1,2,3,4,(567)\newcap(89)\right]\\\times\left[5,6,7,8,9\right]\end{array}$}}\\\hline\begin{array}{@{}c@{}c@{}c@{}}\\[-31pt]\\(1\,2\,3\,4)&(5\,6)&(7\,8\,9)\\[-5pt]\bullet&\bullet&\bullet\\[-7.9pt]\hline\\[-4.25pt]\\[-24pt]\multicolumn{3}{c}{${\color{perm}\!\!\!\{2,3,4,7,6,10,8,9,14\}\!\!\!}$}\end{array}&\text{{\footnotesize$\left(\begin{array}{@{}ccccccccc@{}}1&\alpha_{1}&\alpha_{2}&\alpha_{3}&\alpha_{4}&\alpha_{4}\,\alpha_{5}&0&0&0\\0&0&0&0&1&\alpha_{5}&\alpha_{6}&\alpha_{7}&\alpha_{8}\end{array}\right)$}}&\text{{\small$\begin{array}{l}\phantom{\times\hspace{-0.035cm}}\left[1,2,3,4,(56)\newcap(789)\right]\\\times\left[5,6,7,8,9\right]\end{array}$}}\\\hline\begin{array}{@{}c@{}c@{}c@{}}\\[-31pt]\\(1\,2\,3)&(4\,5\,6)&(7\,8\,9)\\[-5pt]\bullet&\bullet&\bullet\\[-7.9pt]\hline\\[-4.25pt]\\[-24pt]\multicolumn{3}{c}{${\color{perm}\!\!\!\{2,3,7,5,6,10,8,9,13\}\!\!\!}$}\end{array}&\text{{\footnotesize$\left(\begin{array}{@{}ccccccccc@{}}1&\alpha_{1}&\alpha_{2}&\alpha_{3}&\alpha_{3}\,\alpha_{4}&\alpha_{3}\,\alpha_{5}&0&0&0\\0&0&0&1&\alpha_{4}&\alpha_{5}&\alpha_{6}&\alpha_{7}&\alpha_{8}\end{array}\right)$}}&\text{{\small$\begin{array}{l}\varphi\hspace{0.05cm}\left[1,2,3,(45)\newcap(789),(46)\newcap(789)\right]\\\times\left[(45)\newcap(123),(46)\newcap(123),7,8,9\right]\end{array}$}}\\\hline\begin{array}{@{}c@{}c@{}}\\[-31pt]\\(1\,2\,3\,4\,5)&(6\,7\,8\,9\,10)\\[-5pt]\bullet&\bullet\\[-7.9pt]\hline\\[-4.25pt]\\[-24pt]\multicolumn{2}{c}{${\color{perm}\!\!\!\!\!\{2,3,4,5,11,7,8,9,10,16\}\!\!\!\!\!}$}\end{array}&\text{{\footnotesize$\left(\begin{array}{@{}cccccccccc@{}}1&\alpha_{1}&\alpha_{2}&\alpha_{3}&\alpha_{4}&0&0&0&0&0\\0&0&0&0&0&1&\alpha_{5}&\alpha_{6}&\alpha_{7}&\alpha_{8}\end{array}\right)$}}&\text{{\small$\begin{array}{l}\phantom{\times}\left[1,2,3,4,5\right]\\\times\left[6,7,8,9,10\right]\end{array}$}}\\\hline\end{array}\nonumber}\vspace{-0.4cm}{\footnotesize\eq{\hspace{-0.225cm}\mathrm{with}\quad\varphi\equiv\frac{\ab{4\,5\,(123)\newcap(789)}\ab{4\,6\,(123)\newcap(789)}}{\ab{1\,2\,3\,4}\ab{4\,7\,8\,9}\ab{5\,6\,(123)\newcap(789)}},\quad\mathrm{and}\;\;\;\psi\equiv\left(1-\frac{\ab{A\,4\,5\,6}\ab{B\,8\,1\,2}}{\ab{A\,4\,1\,2}\ab{B\,8\,5\,6}}\right)^{-1}\!\!\;\;\mathrm{where}\;\left\{\begin{array}{@{}lr@{}}A\equiv&(7\,8)\newcap(5\,6\,B)\\B\equiv&(3\,4)\newcap(1\,2\,A)\end{array}\right\}.\nonumber}}\end{minipage}}\vspace{-0.5cm}
\end{table}}

The configuration in \mbox{Table \ref{g2n_yangian_invariants}} marked with a `$*$', when lifted to an on-shell graph in $G(4,8)$, corresponds to the four-mass box, (\ref{four_mass_box}). It is the unique N$^2$MHV function which admits more than one solution to the kinematical constraints. The two solutions to $C(\vec{\alpha})\!\cdot\!Z=0$ in this case are implicitly given in the $5$-bracket form given, due to the quadratic-relation which defines the auxiliary twistors ``$A$'' and ``$B$''; to be explicit, these auxiliary twistors are found as the solutions to:
\vspace{-.2cm}\eq{\hspace{.15cm}\left\{\begin{array}{@{}c@{}}A=z_7+\alpha z_8\\B=z_3+\beta z_4\end{array}\right\},\;\mathrm{with}\;\text{{\small$\alpha=\frac{\ab{5\,6\,3\,7}+\beta\ab{5\,6\,4\,7}}{\ab{8\,5\,6\,3}+\beta\ab{8\,5\,6\,4}}\;\mathrm{and}\;\beta=\frac{\ab{1\,2\,7\,3}+\alpha\ab{1\,2\,8\,3}}{\ab{4\,1\,2\,7}+\alpha\ab{4\,1\,2\,8}}$}}.\hspace{-.5cm}\vspace{-.2cm}\label{four_mass_explicit_solution}}

Similarly, we may easily classify all the identities satisfied by N$^2$MHV Yangian-invariant functions, by classifying the $9$-dimensional configurations of $G(2,n)$ whose boundaries generate such relations. It turns out that there are 24 cyclic-classes of generators of identities---those listed in \mbox{Table \ref{g2n_identity_seeds}}.
\begin{table}[b*]\vspace{-0.5cm}
\scalebox{0.9}{\begin{minipage}[h]{\textwidth}\eq{\hspace{-4cm}\begin{array}{|c|c|c|c|}\hline\begin{array}{@{}c@{}c@{}c@{}c@{}c@{}c@{}}\\[-31pt]\\(1\,2)&(3)&(4)&(5)&(6)&(7)\\[-5pt]\bullet&\bullet&\bullet&\bullet&\bullet&\bullet\\[-7.9pt]\hline\\[-4.25pt]\\[-24pt]\multicolumn{6}{c}{\mbox{$\!\!\!\!{\color{perm}\{2,\!4,\!5,\!6,\!7,\!8,\!10\}}\!\!\!\!$}}\end{array}&\begin{array}{@{}c@{}c@{}c@{}c@{}c@{}}\\[-31pt]\\(1\,2\,3)&(4\,5)&(6)&(7)&(8)\\[-5pt]\bullet&\bullet&\bullet&\bullet&\bullet\\[-7.9pt]\hline\\[-4.25pt]\\[-24pt]\multicolumn{5}{c}{\mbox{$\!\!\!\!{\color{perm}\{2,\!3,\!6,\!5,\!7,\!8,\!9,\!12\}}\!\!\!\!$}}\end{array}&\begin{array}{@{}c@{}c@{}c@{}c@{}c@{}}\\[-31pt]\\(1\,2\,3)&(4)&(5\,6)&(7)&(8)\\[-5pt]\bullet&\bullet&\bullet&\bullet&\bullet\\[-7.9pt]\hline\\[-4.25pt]\\[-24pt]\multicolumn{5}{c}{\mbox{$\!\!\!\!{\color{perm}\{2,\!3,\!5,\!7,\!6,\!8,\!9,\!12\}}\!\!\!\!$}}\end{array}&\begin{array}{@{}c@{}c@{}c@{}c@{}c@{}}\\[-31pt]\\(1\,2\,3)&(4)&(5)&(6\,7)&(8)\\[-5pt]\bullet&\bullet&\bullet&\bullet&\bullet\\[-7.9pt]\hline\\[-4.25pt]\\[-24pt]\multicolumn{5}{c}{\mbox{$\!\!\!\!{\color{perm}\{2,\!3,\!5,\!6,\!8,\!7,\!9,\!12\}}\!\!\!\!$}}\end{array}\\\hline\begin{array}{@{}c@{}c@{}c@{}c@{}c@{}}\\[-31pt]\\(1\,2\,3)&(4)&(5)&(6)&(7\,8)\\[-5pt]\bullet&\bullet&\bullet&\bullet&\bullet\\[-7.9pt]\hline\\[-4.25pt]\\[-24pt]\multicolumn{5}{c}{\mbox{$\!\!\!\!{\color{perm}\{2,\!3,\!5,\!6,\!7,\!9,\!8,\!12\}}\!\!\!\!$}}\end{array}&\begin{array}{@{}c@{}c@{}c@{}c@{}c@{}}\\[-31pt]\\(1\,2)&(3\,4)&(5\,6)&(7)&(8)\\[-5pt]\bullet&\bullet&\bullet&\bullet&\bullet\\[-7.9pt]\hline\\[-4.25pt]\\[-24pt]\multicolumn{5}{c}{\mbox{$\!\!\!\!{\color{perm}\{2,\!5,\!4,\!7,\!6,\!8,\!9,\!11\}}\!\!\!\!$}}\end{array}&\begin{array}{@{}c@{}c@{}c@{}c@{}c@{}}\\[-31pt]\\(1\,2)&(3\,4)&(5)&(6\,7)&(8)\\[-5pt]\bullet&\bullet&\bullet&\bullet&\bullet\\[-7.9pt]\hline\\[-4.25pt]\\[-24pt]\multicolumn{5}{c}{\mbox{$\!\!\!\!{\color{perm}\{2,\!5,\!4,\!6,\!8,\!7,\!9,\!11\}}\!\!\!\!$}}\end{array}&\begin{array}{@{}c@{}c@{}c@{}c@{}}\\[-31pt]\\(1\,2\,3\,4)&(5\,6\,7)&(8)&(9)\\[-5pt]\bullet&\bullet&\bullet&\bullet\\[-7.9pt]\hline\\[-4.25pt]\\[-24pt]\multicolumn{4}{c}{\mbox{$\!\!\!\!{\color{perm}\{2,\!3,\!4,\!8,\!6,\!7,\!9,\!10,\!14\}}\!\!\!\!$}}\end{array}\\\hline\begin{array}{@{}c@{}c@{}c@{}c@{}}\\[-31pt]\\(1\,2\,3\,4)&(5)&(6\,7\,8)&(9)\\[-5pt]\bullet&\bullet&\bullet&\bullet\\[-7.9pt]\hline\\[-4.25pt]\\[-24pt]\multicolumn{4}{c}{\mbox{$\!\!\!\!{\color{perm}\{2,\!3,\!4,\!6,\!9,\!7,\!8,\!10,\!14\}}\!\!\!\!$}}\end{array}&\begin{array}{@{}c@{}c@{}c@{}c@{}}\\[-31pt]\\(1\,2\,3\,4)&(5)&(6)&(7\,8\,9)\\[-5pt]\bullet&\bullet&\bullet&\bullet\\[-7.9pt]\hline\\[-4.25pt]\\[-24pt]\multicolumn{4}{c}{\mbox{$\!\!\!\!{\color{perm}\{2,\!3,\!4,\!6,\!7,\!10,\!8,\!9,\!14\}}\!\!\!\!$}}\end{array}&\begin{array}{@{}c@{}c@{}c@{}c@{}}\\[-31pt]\\(1\,2\,3\,4)&(5\,6)&(7\,8)&(9)\\[-5pt]\bullet&\bullet&\bullet&\bullet\\[-7.9pt]\hline\\[-4.25pt]\\[-24pt]\multicolumn{4}{c}{\mbox{$\!\!\!\!{\color{perm}\{2,\!3,\!4,\!7,\!6,\!9,\!8,\!10,\!14\}}\!\!\!\!$}}\end{array}&\begin{array}{@{}c@{}c@{}c@{}c@{}}\\[-31pt]\\(1\,2\,3\,4)&(5\,6)&(7)&(8\,9)\\[-5pt]\bullet&\bullet&\bullet&\bullet\\[-7.9pt]\hline\\[-4.25pt]\\[-24pt]\multicolumn{4}{c}{\mbox{$\!\!\!\!{\color{perm}\{2,\!3,\!4,\!7,\!6,\!8,\!10,\!9,\!14\}}\!\!\!\!$}}\end{array}\\\hline\begin{array}{@{}c@{}c@{}c@{}c@{}}\\[-31pt]\\(1\,2\,3\,4)&(5)&(6\,7)&(8\,9)\\[-5pt]\bullet&\bullet&\bullet&\bullet\\[-7.9pt]\hline\\[-4.25pt]\\[-24pt]\multicolumn{4}{c}{\mbox{$\!\!\!\!{\color{perm}\{2,\!3,\!4,\!6,\!8,\!7,\!10,\!9,\!14\}}\!\!\!\!$}}\end{array}&\begin{array}{@{}c@{}c@{}c@{}c@{}}\\[-31pt]\\(1\,2\,3)&(4\,5\,6)&(7\,8)&(9)\\[-5pt]\bullet&\bullet&\bullet&\bullet\\[-7.9pt]\hline\\[-4.25pt]\\[-24pt]\multicolumn{4}{c}{\mbox{$\!\!\!\!{\color{perm}\{2,\!3,\!7,\!5,\!6,\!9,\!8,\!10,\!13\}}\!\!\!\!$}}\end{array}&\begin{array}{@{}c@{}c@{}c@{}c@{}}\\[-31pt]\\(1\,2\,3)&(4\,5\,6)&(7)&(8\,9)\\[-5pt]\bullet&\bullet&\bullet&\bullet\\[-7.9pt]\hline\\[-4.25pt]\\[-24pt]\multicolumn{4}{c}{\mbox{$\!\!\!\!{\color{perm}\{2,\!3,\!7,\!5,\!6,\!8,\!10,\!9,\!13\}}\!\!\!\!$}}\end{array}&\begin{array}{@{}c@{}c@{}c@{}c@{}}\\[-31pt]\\(1\,2\,3)&(4\,5)&(6\,7\,8)&(9)\\[-5pt]\bullet&\bullet&\bullet&\bullet\\[-7.9pt]\hline\\[-4.25pt]\\[-24pt]\multicolumn{4}{c}{\mbox{$\!\!\!\!{\color{perm}\{2,\!3,\!6,\!5,\!9,\!7,\!8,\!10,\!13\}}\!\!\!\!$}}\end{array}\\\hline\begin{array}{@{}c@{}c@{}c@{}c@{}}\\[-31pt]\\(1\,2\,3)&(4\,5)&(6\,7)&(8\,9)\\[-5pt]\bullet&\bullet&\bullet&\bullet\\[-7.9pt]\hline\\[-4.25pt]\\[-24pt]\multicolumn{4}{c}{\mbox{$\!\!\!\!{\color{perm}\{2,\!3,\!6,\!5,\!8,\!7,\!10,\!9,\!13\}}\!\!\!\!$}}\end{array}&\begin{array}{@{}c@{}c@{}c@{}}\\[-31pt]\\(1\,2\,3\,4\,5)&(6\,7\,8\,9)&(10)\\[-5pt]\bullet&\bullet&\bullet\\[-7.9pt]\hline\\[-4.25pt]\\[-24pt]\multicolumn{3}{c}{\mbox{$\!\!\!\!{\color{perm}\{2,\!3,\!4,\!5,\!10,\!7,\!8,\!9,\!11,\!16\}}\!\!\!\!$}}\end{array}&\begin{array}{@{}c@{}c@{}c@{}}\\[-31pt]\\(1\,2\,3\,4\,5)&(6)&(7\,8\,9\,10)\\[-5pt]\bullet&\bullet&\bullet\\[-7.9pt]\hline\\[-4.25pt]\\[-24pt]\multicolumn{3}{c}{\mbox{$\!\!\!\!{\color{perm}\{2,\!3,\!4,\!5,\!7,\!11,\!8,\!9,\!10,\!16\}}\!\!\!\!$}}\end{array}&\begin{array}{@{}c@{}c@{}c@{}}\\[-31pt]\\(1\,2\,3\,4\,5)&(6\,7\,8)&(9\,10)\\[-5pt]\bullet&\bullet&\bullet\\[-7.9pt]\hline\\[-4.25pt]\\[-24pt]\multicolumn{3}{c}{\mbox{$\!\!\!\!{\color{perm}\{2,\!3,\!4,\!5,\!9,\!7,\!8,\!11,\!10,\!16\}}\!\!\!\!$}}\end{array}\\\hline\begin{array}{@{}c@{}c@{}c@{}}\\[-31pt]\\(1\,2\,3\,4\,5)&(6\,7)&(8\,9\,10)\\[-5pt]\bullet&\bullet&\bullet\\[-7.9pt]\hline\\[-4.25pt]\\[-24pt]\multicolumn{3}{c}{\mbox{$\!\!\!\!{\color{perm}\{2,\!3,\!4,\!5,\!8,\!7,\!11,\!9,\!10,\!16\}}\!\!\!\!$}}\end{array}&\begin{array}{@{}c@{}c@{}c@{}}\\[-31pt]\\(1\,2\,3\,4)&(5\,6\,7\,8)&(9\,10)\\[-5pt]\bullet&\bullet&\bullet\\[-7.9pt]\hline\\[-4.25pt]\\[-24pt]\multicolumn{3}{c}{\mbox{$\!\!\!\!{\color{perm}\{2,\!3,\!4,\!9,\!6,\!7,\!8,\!11,\!10,\!15\}}\!\!\!\!$}}\end{array}&\begin{array}{@{}c@{}c@{}c@{}}\\[-31pt]\\(1\,2\,3\,4)&(5\,6\,7)&(8\,9\,10)\\[-5pt]\bullet&\bullet&\bullet\\[-7.9pt]\hline\\[-4.25pt]\\[-24pt]\multicolumn{3}{c}{\mbox{$\!\!\!\!{\color{perm}\{2,\!3,\!4,\!8,\!6,\!7,\!11,\!9,\!10,\!15\}}\!\!\!\!$}}\end{array}&\begin{array}{@{}c@{}c@{}}\\[-31pt]\\(1\,2\,3\,4\,5\,6)&(7\,8\,9\,10\,11)\\[-5pt]\bullet&\bullet\\[-7.9pt]\hline\\[-4.25pt]\\[-24pt]\multicolumn{2}{c}{\mbox{$\!\!\!\!{\color{perm}\{2,\!3,\!4,\!5,\!6,\!12,\!8,\!9,\!10,\!11,\!18\}}\!\!\!\!$}}\end{array}\\\hline\multicolumn{1}{c}{\rule{130pt}{0pt}}&\multicolumn{1}{c}{\rule{130pt}{0pt}}&\multicolumn{1}{c}{\rule{130pt}{0pt}}&\multicolumn{1}{c}{\rule{130pt}{0pt}}\end{array}\hspace{-5.65cm}\nonumber}\end{minipage}}\vspace{-0.75cm}\caption{Representatives of all generators of identities among N$^2$MHV Yangian-invariants\label{g2n_identity_seeds}\vspace{-10pt}}
\end{table}
One interesting example of these is:\\[-5pt]
\noindent\scalebox{0.85}{\begin{minipage}[h]{\textwidth}\eq{\hspace{-0.1cm}\raisebox{0.05cm}{{\text{\Large$\partial\!$}}}\begin{array}{c}\\[-12.5pt]\left[\begin{array}{@{}c@{}c@{}c@{}c@{}c@{}}(1\,2)&(3\,4)&(5\,6)&(7)&(8)\\[-5pt]\bullet&\bullet&\bullet&\bullet&\bullet\\[-7.9pt]\hline\\[-10pt]\end{array}\right]\\[-0pt]\text{{\normalsize${\color{perm}\{2,\!5,\!4,\!7,\!6,\!8,\!9,\!11\}}$}}\end{array}\!\raisebox{0.05cm}{{\text{\Large$=$}}}\!\left(\begin{array}{@{}ccccccc@{}}\raisebox{-0.0cm}{{\text{\Large$~$}}}&\begin{array}{@{}c@{}c@{}c@{}c@{}}\\[-30pt]\\(1\,2)&(3\,4)&(5\,6)&(7\,8)\\[-5pt]\bullet&\bullet&\bullet&\bullet\\[-7.9pt]\hline\\[-4.25pt]\\[-22pt]\multicolumn{4}{c}{\text{{\normalsize$\!\!{\color{perm}\{2,\!5,\!4,\!7,\!6,}{\color{red}\mathbf{9}}{\color{perm},}{\color{red}\mathbf{8}}{\color{perm},\!11\!\}}\!\!$}}}\end{array}&\raisebox{-0.05cm}{{\text{\Large$-$}}}&\begin{array}{@{}c@{}c@{}c@{}c@{}c@{}}\\[-30pt]\\(1\,2)&(3\,4)&(5)&(7)&(8)\\[-5pt]\bullet&\bullet&\bullet&\bullet&\bullet\\[-7.9pt]\hline\\[-4.25pt]\\[-22pt]\multicolumn{5}{c}{\text{{\normalsize$\!\!{\color{perm}\{2,\!5,\!4,\!7,}{\color{red}\mathbf{8}}{\color{perm},}{\color{red}\mathbf{6}},{\color{perm}\!9,\!11\!\}}\!\!$}}}\end{array}&\raisebox{-0.05cm}{{\text{\Large$+$}}}&\begin{array}{@{}c@{}c@{}c@{}c@{}c@{}}\\[-30pt]\\(1\,2)&(3\,4)&(6)&(7)&(8)\\[-5pt]\bullet&\bullet&\bullet&\bullet&\bullet\\[-7.9pt]\hline\\[-4.25pt]\\[-22pt]\multicolumn{5}{c}{\text{{\normalsize$\!\!{\color{perm}\{2,}{\color{red}\mathbf{6}}{\color{perm},\!4,\!7,}{\color{red}\mathbf{5}}{\color{perm},\!8,\!9,\!11\!\}}\!\!$}}}\end{array}\\[-5pt]\raisebox{-0.25cm}{{\text{\Large$-$}}}&\begin{array}{@{}c@{}c@{}c@{}c@{}c@{}}\\[-16pt]\\(1)&(3\,4)&(5\,6)&(7)&(8)\\[-5pt]\bullet&\bullet&\bullet&\bullet&\bullet\\[-7.9pt]\hline\\[-4.25pt]\\[-22pt]\multicolumn{5}{c}{\text{{\normalsize\!\!${\color{perm}\{}{\color{red}\mathbf{5}}{\color{perm},}{\color{red}\mathbf{2}}{\color{perm},\!4,\!7,\!6,\!8,\!9,\!11\!\}}\!\!$}}}\end{array}&\raisebox{-0.25cm}{{\text{\Large$+$}}}&\begin{array}{@{}c@{}c@{}c@{}c@{}c@{}}\\[-16pt]\\(2)&(3\,4)&(5\,6)&(7)&(8)\\[-5pt]\bullet&\bullet&\bullet&\bullet&\bullet\\[-7.9pt]\hline\\[-4.25pt]\\[-22pt]\multicolumn{5}{c}{\text{{\normalsize$\!\!{\color{perm}\{}{\color{red}\mathbf{1}}{\color{perm},\!5,\!4,\!7,\!6,\!8,}{\color{red}\mathbf{10}}{\color{perm},\!11\!\}}\!\!$}}}\end{array}&\raisebox{-0.25cm}{{\text{\Large$-$}}}&\begin{array}{@{}c@{}c@{}c@{}c@{}}\\[-16pt]\\(8\,1\,2)&(3\,4)&(5\,6)&(7)\\[-5pt]\bullet&\bullet&\bullet&\bullet\\[-7.9pt]\hline\\[-4.25pt]\\[-22pt]\multicolumn{4}{c}{\text{{\normalsize$\!\!{\color{perm}\{2,\!5,\!4,\!7,\!6,\!8,}{\color{red}\mathbf{11}}{\color{perm},}{\color{red}\mathbf{9}}{\color{perm}\}}\!\!$}}}\end{array}\\[-5pt]\raisebox{-0.25cm}{{\text{\Large$+$}}}&\begin{array}{@{}c@{}c@{}c@{}c@{}}\\[-16pt]\\(1\,2)&(3\,4)&(5\,6\,7)&(8)\\[-5pt]\bullet&\bullet&\bullet&\bullet\\[-7.9pt]\hline\\[-4.25pt]\\[-22pt]\multicolumn{4}{c}{\text{{\normalsize$\!\!{\color{perm}\{2,\!5,\!4,}{\color{red}\mathbf{8}}{\color{perm},\!6,}{\color{red}\mathbf{7}}{\color{perm},\!9,\!11\!\}}\!\!$}}}\end{array}&\raisebox{-0.25cm}{{\text{\Large$-$}}}&\begin{array}{@{}c@{}c@{}c@{}c@{}c@{}}\\[-16pt]\\(1\,2)&(4)&(5\,6)&(7)&(8)\\[-5pt]\bullet&\bullet&\bullet&\bullet&\bullet\\[-7.9pt]\hline\\[-4.25pt]\\[-22pt]\multicolumn{5}{c}{\text{{\!\!\normalsize${\color{perm}\{2,\!5,}{\color{red}\mathbf{3}}{\color{perm},\!7,\!6,\!8,\!9,}{\color{red}\mathbf{12}}{\color{perm}\}}$\!\!}}}\end{array}&\raisebox{-0.25cm}{{\text{\Large$+$}}}&\begin{array}{@{}c@{}c@{}c@{}c@{}c@{}}\\[-16pt]\\(1\,2)&(3)&(5\,6)&(7)&(8)\\[-5pt]\bullet&\bullet&\bullet&\bullet&\bullet\\[-7.9pt]\hline\\[-4.25pt]\\[-22pt]\multicolumn{5}{c}{\text{{\normalsize$\!\!{\color{perm}\{2,\!5,}{\color{red}\mathbf{7}}{\color{perm},}{\color{red}\mathbf{4}}{\color{perm},\!6,\!8,\!9,\!11\!\}}\!\!$}}}\end{array}\end{array}\right)\raisebox{-0.05cm}{{\text{\Large$\!=\!0.$}}}\nonumber}\end{minipage}}\\[.2cm]
\noindent Here, we have only listed those boundary-configurations which have non-vanishing kinematical support. This identity lifts to the identity among on-shell graphs given in \mbox{section \ref{geometric_origin_of_identities_section}} which can be understood as expressing the (sum-over particular solutions to $C\!\cdot\!Z=0$ for the) four-mass box function as a sum of $8$ purely-rational, N$^2$MHV Yangian-invariants.\\
\begin{table}[t]\caption{Numbers of cyclic-classes of N$^k$MHV Yangian-invariant functions nontrivially involving $n$ particles, and the number of cyclic-classes of identities among them.\label{statistics_of_yangian_invariants}}
\vspace{-0.2cm}$\hspace{-1.15cm}\begin{array}{|r|l|r|r|r|r|r|r|r|r|r|r|r|r|r|r|r|r|r|r|}
\multicolumn{2}{c}{~}&\multicolumn{16}{c}{n}\\\cline{3-18}\multicolumn{1}{c}{k}&\multicolumn{1}{c|}{~}&\multicolumn{1}{c|}{5}&\multicolumn{1}{c|}{6}&\multicolumn{1}{c|}{7}&\multicolumn{1}{c|}{8}&\multicolumn{1}{c|}{9}&\multicolumn{1}{c|}{10}&\multicolumn{1}{c|}{11}&\multicolumn{1}{c|}{12}&\multicolumn{1}{c|}{13}&\multicolumn{1}{c|}{14}&\multicolumn{1}{c|}{15}&\multicolumn{1}{c|}{16}&\multicolumn{1}{c|}{17}&\multicolumn{1}{c|}{18}&\multicolumn{1}{c|}{19}&\multicolumn{1}{c|}{20}&\multicolumn{1}{c}{\text{{\bf Total}}}&\multicolumn{1}{c}{\text{{\bf IDs}}}\\\hline\,1\,&\text{total:}&1&{\color{deemph}0}&{\color{deemph}0}&{\color{deemph}0}&{\color{deemph}0}&{\color{deemph}0}&{\color{deemph}0}&{\color{deemph}0}&{\color{deemph}0}&{\color{deemph}0}&{\color{deemph}0}&{\color{deemph}0}&{\color{deemph}0}&{\color{deemph}0}&{\color{deemph}0}&{\color{deemph}0}&\mathbf{1}&\mathbf{1}\\\hline
\multirow{3}{*}{\raisebox{2.5pt}{$\,2\,$}}&\text{{\footnotesize${\color[rgb]{0.8,0.1,0.2}\Gamma=2\!:}$}}&\text{{\footnotesize${\color{deemph}0}$}}&\text{{\footnotesize${\color{deemph}0}$}}&\text{{\footnotesize${\color{deemph}0}$}}&\text{{\footnotesize${\color[rgb]{0.8,0.1,0.2}1}$}}&\text{{\footnotesize${\color{deemph}0}$}}&\text{{\footnotesize${\color{deemph}0}$}}&\text{{\footnotesize${\color{deemph}0}$}}&\text{{\footnotesize${\color{deemph}0}$}}&\text{{\footnotesize${\color{deemph}0}$}}&\text{{\footnotesize${\color{deemph}0}$}}&\text{{\footnotesize${\color{deemph}0}$}}&\text{{\footnotesize${\color{deemph}0}$}}&\text{{\footnotesize${\color{deemph}0}$}}&\text{{\footnotesize${\color{deemph}0}$}}&\text{{\footnotesize${\color{deemph}0}$}}&\text{{\footnotesize${\color{deemph}0}$}}&&\\[-5pt]&\text{{\footnotesize${\color[rgb]{0.2,0.1,0.8}\Gamma=1\!:}$}}&\text{{\footnotesize${\color{deemph}0}$}}&\text{{\footnotesize${\color[rgb]{0.2,0.1,0.8}1}$}}&\text{{\footnotesize${\color[rgb]{0.2,0.1,0.8}2}$}}&\text{{\footnotesize${\color[rgb]{0.2,0.1,0.8}5}$}}&\text{{\footnotesize${\color[rgb]{0.2,0.1,0.8}4}$}}&\text{{\footnotesize${\color[rgb]{0.2,0.1,0.8}1}$}}&\text{{\footnotesize${\color{deemph}0}$}}&\text{{\footnotesize${\color{deemph}0}$}}&\text{{\footnotesize${\color{deemph}0}$}}&\text{{\footnotesize${\color{deemph}0}$}}&\text{{\footnotesize${\color{deemph}0}$}}&\text{{\footnotesize${\color{deemph}0}$}}&\text{{\footnotesize${\color{deemph}0}$}}&\text{{\footnotesize${\color{deemph}0}$}}&\text{{\footnotesize${\color{deemph}0}$}}&\text{{\footnotesize${\color{deemph}0}$}}&&\\\cline{2-18}&\text{total:}&{\color{deemph}0}&1&2&6&4&1&{\color{deemph}0}&{\color{deemph}0}&{\color{deemph}0}&{\color{deemph}0}&{\color{deemph}0}&{\color{deemph}0}&{\color{deemph}0}&{\color{deemph}0}&{\color{deemph}0}&{\color{deemph}0}&\mathbf{14}&\mathbf{24}\\\hline
\multirow{4}{*}{\raisebox{5pt}{$\,3\,$}}&\text{{\footnotesize${\color[rgb]{0.0,0.4,0.3}\Gamma=3\!:}$}}&\text{{\footnotesize${\color{deemph}0}$}}&\text{{\footnotesize${\color{deemph}0}$}}&\text{{\footnotesize${\color{deemph}0}$}}&\text{{\footnotesize${\color{deemph}0}$}}&\text{{\footnotesize${\color{deemph}0}$}}&\text{{\footnotesize${\color{deemph}0}$}}&\text{{\footnotesize${\color[rgb]{0.0,0.4,0.3}1}$}}&\text{{\footnotesize${\color{deemph}0}$}}&\text{{\footnotesize${\color{deemph}0}$}}&\text{{\footnotesize${\color{deemph}0}$}}&\text{{\footnotesize${\color{deemph}0}$}}&\text{{\footnotesize${\color{deemph}0}$}}&\text{{\footnotesize${\color{deemph}0}$}}&\text{{\footnotesize${\color{deemph}0}$}}&\text{{\footnotesize${\color{deemph}0}$}}&\text{{\footnotesize${\color{deemph}0}$}}&&\\[-5pt]&\text{{\footnotesize${\color[rgb]{0.8,0.1,0.2}\Gamma=2\!:}$}}&\text{{\footnotesize${\color{deemph}0}$}}&\text{{\footnotesize${\color{deemph}0}$}}&\text{{\footnotesize${\color{deemph}0}$}}&\text{{\footnotesize${\color{deemph}0}$}}&\text{{\footnotesize${\color[rgb]{0.8,0.1,0.2}2}$}}&\text{{\footnotesize${\color[rgb]{0.8,0.1,0.2}15}$}}&\text{{\footnotesize${\color[rgb]{0.8,0.1,0.2}27}$}}&\text{{\footnotesize${\color[rgb]{0.8,0.1,0.2}19}$}}&\text{{\footnotesize${\color[rgb]{0.8,0.1,0.2}2}$}}&\text{{\footnotesize${\color{deemph}0}$}}&\text{{\footnotesize${\color{deemph}0}$}}&\text{{\footnotesize${\color{deemph}0}$}}&\text{{\footnotesize${\color{deemph}0}$}}&\text{{\footnotesize${\color{deemph}0}$}}&\text{{\footnotesize${\color{deemph}0}$}}&\text{{\footnotesize${\color{deemph}0}$}}&&\\[-5pt]&\text{{\footnotesize${\color[rgb]{0.2,0.1,0.8}\Gamma=1\!:}$}}&\text{{\footnotesize${\color{deemph}0}$}}&\text{{\footnotesize${\color{deemph}0}$}}&\text{{\footnotesize${\color[rgb]{0.2,0.1,0.8}1}$}}&\text{{\footnotesize${\color[rgb]{0.2,0.1,0.8}6}$}}&\text{{\footnotesize${\color[rgb]{0.2,0.1,0.8}54}$}}&\text{{\footnotesize${\color[rgb]{0.2,0.1,0.8}177}$}}&\text{{\footnotesize${\color[rgb]{0.2,0.1,0.8}298}$}}&\text{{\footnotesize${\color[rgb]{0.2,0.1,0.8}274}$}}&\text{{\footnotesize${\color[rgb]{0.2,0.1,0.8}134}$}}&\text{{\footnotesize${\color[rgb]{0.2,0.1,0.8}30}$}}&\text{{\footnotesize${\color[rgb]{0.2,0.1,0.8}1}$}}&\text{{\footnotesize${\color{deemph}0}$}}&\text{{\footnotesize${\color{deemph}0}$}}&\text{{\footnotesize${\color{deemph}0}$}}&\text{{\footnotesize${\color{deemph}0}$}}&\text{{\footnotesize${\color{deemph}0}$}}&&\\\cline{2-18}&\text{total:}&{\color{deemph}0}&{\color{deemph}0}&1&6&56&192&326&293&136&30&1&{\color{deemph}0}&{\color{deemph}0}&{\color{deemph}0}&{\color{deemph}0}&{\color{deemph}0}&\mathbf{1041}&\mathbf{3669}\\\hline
\multirow{6}{*}{\raisebox{15pt}{$\,4\,$}}&\text{{\footnotesize${\color[rgb]{0.5,0.1,0.2}\Gamma=5\!:}$}}&\text{{\footnotesize${\color{deemph}0}$}}&\text{{\footnotesize${\color{deemph}0}$}}&\text{{\footnotesize${\color{deemph}0}$}}&\text{{\footnotesize${\color{deemph}0}$}}&\text{{\footnotesize${\color{deemph}0}$}}&\text{{\footnotesize${\color{deemph}0}$}}&\text{{\footnotesize${\color{deemph}0}$}}&\text{{\footnotesize${\color{deemph}0}$}}&\text{{\footnotesize${\color{deemph}0}$}}&\text{{\footnotesize${\color[rgb]{0.5,0.1,0.2}9}$}}&\text{{\footnotesize${\color{deemph}0}$}}&\text{{\footnotesize${\color{deemph}0}$}}&\text{{\footnotesize${\color{deemph}0}$}}&\text{{\footnotesize${\color{deemph}0}$}}&\text{{\footnotesize${\color{deemph}0}$}}&\text{{\footnotesize${\color{deemph}0}$}}&&\\[-5pt]&\text{{\footnotesize${\color[rgb]{0.3,0.1,0.5}\Gamma=4\!:}$}}&\text{{\footnotesize${\color{deemph}0}$}}&\text{{\footnotesize${\color{deemph}0}$}}&\text{{\footnotesize${\color{deemph}0}$}}&\text{{\footnotesize${\color{deemph}0}$}}&\text{{\footnotesize${\color{deemph}0}$}}&\text{{\footnotesize${\color{deemph}0}$}}&\text{{\footnotesize${\color{deemph}0}$}}&\text{{\footnotesize${\color[rgb]{0.3,0.1,0.5}6}$}}&\text{{\footnotesize${\color[rgb]{0.3,0.1,0.5}40}$}}&\text{{\footnotesize${\color[rgb]{0.3,0.1,0.5}60}$}}&\text{{\footnotesize${\color[rgb]{0.3,0.1,0.5}30}$}}&\text{{\footnotesize${\color[rgb]{0.3,0.1,0.5}3}$}}&\text{{\footnotesize${\color{deemph}0}$}}&\text{{\footnotesize${\color{deemph}0}$}}&\text{{\footnotesize${\color{deemph}0}$}}&\text{{\footnotesize${\color{deemph}0}$}}&&\\[-5pt]&\text{{\footnotesize${\color[rgb]{0.0,0.4,0.3}\Gamma=3\!:}$}}&\text{{\footnotesize${\color{deemph}0}$}}&\text{{\footnotesize${\color{deemph}0}$}}&\text{{\footnotesize${\color{deemph}0}$}}&\text{{\footnotesize${\color{deemph}0}$}}&\text{{\footnotesize${\color{deemph}0}$}}&\text{{\footnotesize${\color{deemph}0}$}}&\text{{\footnotesize${\color[rgb]{0.0,0.4,0.3}1}$}}&\text{{\footnotesize${\color[rgb]{0.0,0.4,0.3}47}$}}&\text{{\footnotesize${\color[rgb]{0.0,0.4,0.3}179}$}}&\text{{\footnotesize${\color[rgb]{0.0,0.4,0.3}232}$}}&\text{{\footnotesize${\color[rgb]{0.0,0.4,0.3}125}$}}&\text{{\footnotesize${\color[rgb]{0.0,0.4,0.3}12}$}}&\text{{\footnotesize${\color{deemph}0}$}}&\text{{\footnotesize${\color{deemph}0}$}}&\text{{\footnotesize${\color{deemph}0}$}}&\text{{\footnotesize${\color{deemph}0}$}}&&\\[-5pt]&\text{{\footnotesize${\color[rgb]{0.8,0.1,0.2}\Gamma=2\!:}$}}&\text{{\footnotesize${\color{deemph}0}$}}&\text{{\footnotesize${\color{deemph}0}$}}&\text{{\footnotesize${\color{deemph}0}$}}&\text{{\footnotesize${\color{deemph}0}$}}&\text{{\footnotesize${\color{deemph}0}$}}&\text{{\footnotesize${\color[rgb]{0.8,0.1,0.2}10}$}}&\text{{\footnotesize${\color[rgb]{0.8,0.1,0.2}177}$}}&\text{{\footnotesize${\color[rgb]{0.8,0.1,0.2}1008}$}}&\text{{\footnotesize${\color[rgb]{0.8,0.1,0.2}2630}$}}&\text{{\footnotesize${\color[rgb]{0.8,0.1,0.2}3829}$}}&\text{{\footnotesize${\color[rgb]{0.8,0.1,0.2}3158}$}}&\text{{\footnotesize${\color[rgb]{0.8,0.1,0.2}1413}$}}&\text{{\footnotesize${\color[rgb]{0.8,0.1,0.2}272}$}}&\text{{\footnotesize${\color[rgb]{0.8,0.1,0.2}18}$}}&\text{{\footnotesize${\color{deemph}0}$}}&\text{{\footnotesize${\color{deemph}0}$}}&&\\[-5pt]&\text{{\footnotesize${\color[rgb]{0.2,0.1,0.8}\Gamma=1\!:}$}}&\text{{\footnotesize${\color{deemph}0}$}}&\text{{\footnotesize${\color{deemph}0}$}}&\text{{\footnotesize${\color{deemph}0}$}}&\text{{\footnotesize${\color[rgb]{0.2,0.1,0.8}1}$}}&\text{{\footnotesize${\color[rgb]{0.2,0.1,0.8}13}$}}&\text{{\footnotesize${\color[rgb]{0.2,0.1,0.8}263}$}}&\text{{\footnotesize${\color[rgb]{0.2,0.1,0.8}1988}$}}&\text{{\footnotesize${\color[rgb]{0.2,0.1,0.8}7862}$}}&\text{{\footnotesize${\color[rgb]{0.2,0.1,0.8}18532}$}}&\text{{\footnotesize${\color[rgb]{0.2,0.1,0.8}28204}$}}&\text{{\footnotesize${\color[rgb]{0.2,0.1,0.8}28377}$}}&\text{{\footnotesize${\color[rgb]{0.2,0.1,0.8}18925}$}}&\text{{\footnotesize${\color[rgb]{0.2,0.1,0.8}8034}$}}&\text{{\footnotesize${\color[rgb]{0.2,0.1,0.8}2047}$}}&\text{{\footnotesize${\color[rgb]{0.2,0.1,0.8}270}$}}&\text{{\footnotesize${\color[rgb]{0.2,0.1,0.8}1}$}}&&\\\cline{2-18}&\text{total:}&{\color{deemph}0}&{\color{deemph}0}&{\color{deemph}0}&1&13&273&2166&8923&21381&32334&31690&20353&8306&2065&270&1&\mathbf{127776}&\mathbf{603121}\\\hline
\end{array}$\vspace{-0.05cm}
\end{table}
\\[-14pt]
\indent Although it is not difficult to continue such a classification to higher and higher $k$, the space required to tabulate all objects soon grows overwhelming. Being of some interest, however, we summarize the statistical structure of the classification of Yangian-invariants and their identities through N$^{4}$MHV in \mbox{Table \ref{statistics_of_yangian_invariants}}.

This classification makes some interesting structures visible. For example, notice the extreme rarity of configurations with large $\Gamma^4$. Curiously we see that configurations with highest $\Gamma^4$ appear to be found when for $n=3k\pl\,2$ (this trend is known to continue well-beyond $k\!=\!4$).

To illustrate the Yangian-invariants with exceptionally-large $\Gamma^4$, below we give the on-shell graph which represents the unique `cubic' N$^3$MHV Yangian-invariant, and one of those which represents a `quintic' N$^4$MHV Yangian-invariant:
\vspace{-0.5cm}\eq{\hspace{0.2cm}\begin{array}{c}\\[10pt]\Gamma^4(C)=3\\[-0.075cm]\includegraphics[scale=1]{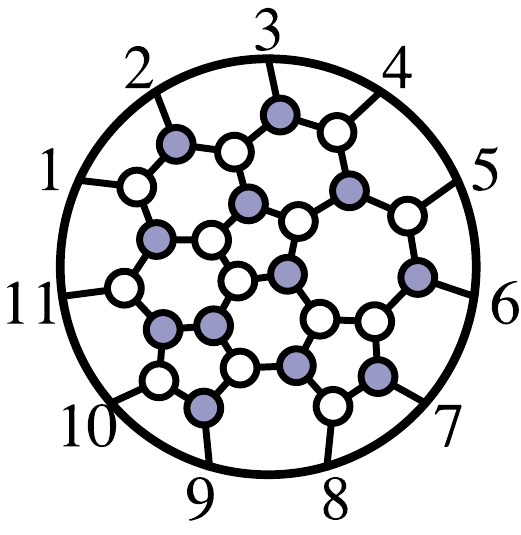}\\[-0.2cm]\text{{\small{\color{perm}$\phantom{12345}\{4,\!9,\!6,\!11,\!8,\!14,\!10,\!13,\!12,\!18,\!16\}\phantom{12345}$}}}\\[-30pt]\end{array}\qquad\begin{array}{c}\\[10pt]\Gamma^4(C)=5\\[-0.075cm]\includegraphics[scale=1]{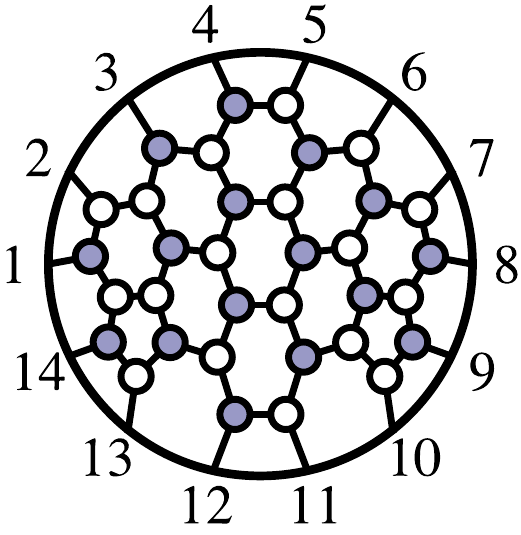}\\[-0.2cm]\text{{\small{\color{perm}$\{6,\!5,\!13,\!12,\!11,\!9,\!10,\!18,\!17,\!15,\!14,\!22,\!16,\!21\}$}}}\\[-30pt]\end{array}\vspace{-0.2cm}\nonumber}

~\newpage
\section{\mbox{The Yang-Baxter Relation and ABJM Theories}}\label{yang_baxter_and_abjm_section}
We began our discussion linking permutations and scattering amplitudes in \mbox{section \ref{combinatorics_of_scattering_amplitudes_section}} by recalling the story of scattering in $(1\pl1)$-dimensional integrable theories (for a review see \cite{Staudacher:2010jz}). In this section, we will see that this familiar story is actually contained as a special case of our new picture linking permutations to on-shell diagrams. And there is another special case which will turn out to give a theory of on-shell graphs for the ABJM theory \cite{Aharony:2008ug} (see also \cite{Bianchi:2011aa,Bianchi:2011dg,Bianchi:2011fc,Bianchi:2012cq}) defined in $(2\pl1)$ dimensions! Although both these stories are physically very rich on their own, we will content ourselves here by briefly sketching-out the main points involved, leaving  more detailed exposition and exploration to future work.

\subsection{The On-Shell Avatar of the Yang-Baxter Relation}
Recall the basic structure of the $(1\pl1)$-dimensional amplitudes from our discussion in \mbox{section \ref{combinatorics_of_scattering_amplitudes_section}}, for which the fundamental interactions involved are $4$-particle vertices. In order to relate these to our story, we must find a way to recast each $4$-particle interaction (each carrying only one degree of freedom) in terms of an on-shell diagram with only trivalent vertices. The simplest way of doing this is to `blow-up' each $4$-point vertex according to:
\vspace{-.3cm}\eq{\raisebox{-45pt}{\includegraphics[scale=1]{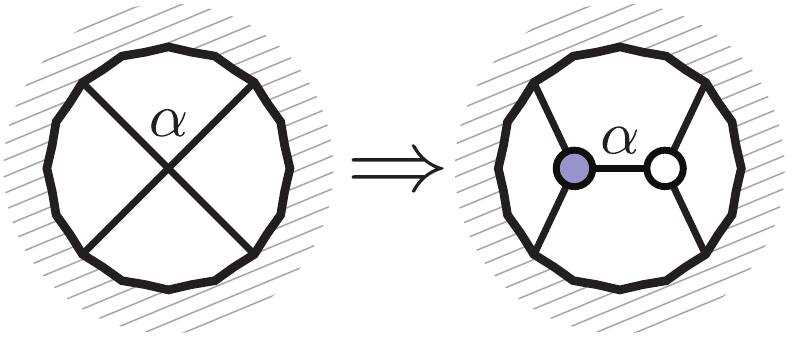}}\vspace{-.3cm}}
where only edges from blown-up vertices have weight different from unity.

Notice that the left-right path permutation moving from the bottom to the top of the graph agrees with the `$(1\pl1)$-permutation', while the left-right path permutations from top to bottom are trivial:
\vspace{-.3cm}\eq{\hspace{-1cm}\raisebox{-45pt}{\includegraphics[scale=1]{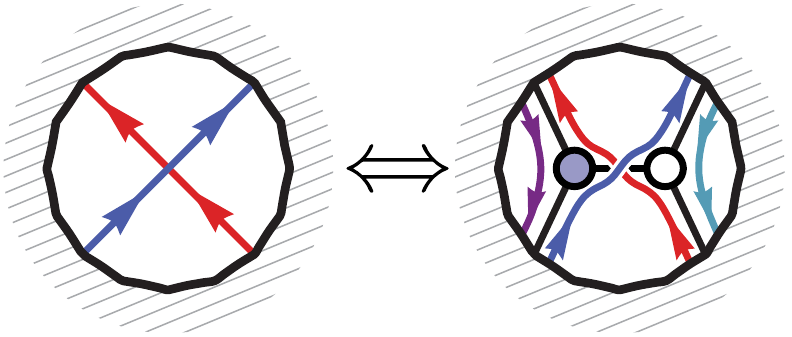}}\hspace{-1cm}\vspace{-.3cm}}

Consider for example the $4$-point vertex by itself,
\vspace{-.3cm}\eq{\hspace{-1cm}\raisebox{-34pt}{\includegraphics[scale=1]{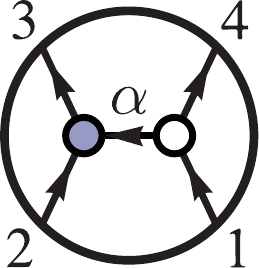}}\quad\raisebox{-2pt}{\text{{\LARGE$\;\Leftrightarrow\;$}}}\quad C=\left(\begin{array}{@{}cc@{$\!$}|@{$\,\,$}cc@{}} 1 & \;0\;\,\, & \!\mi\alpha\,\, & \!\mi1\,\, \\ 0 & 1 & \!\mi1\,\, & 0 \end{array} \right) \equiv \left(1_{2 \times 2}|  R_{12} \right),\hspace{-1cm}\vspace{-.3cm}}
where we have given the point in the Grassmannian $C$ obtained using edge variables and the perfect-orientation indicated in the figure (see \mbox{section \ref{boundary_measurements_section}}), from which we can read-off the $2\!\to\!2$ ``scattering-matrix'' which we have denoted $R_{12}$. In general, blowing-up each 4-particle vertex allows us to translate any $(1\pl1)$-scattering diagram into a trivalent, on-shell diagram from which we can identify an $(n\!\times\!n)$ scattering-matrix in the same way---identifying the point $C$ in the Grassmannian in the gauge-fixed form,
\vspace{-.2cm}\eq{C_{n \times 2n} = \left(1_{n \times n}|R_{n \times n} \right)\vspace{-.2cm}}

As an example, let us look at the familiar configuration,
\vspace{-.2cm}\eq{\raisebox{-45pt}{\includegraphics[scale=1]{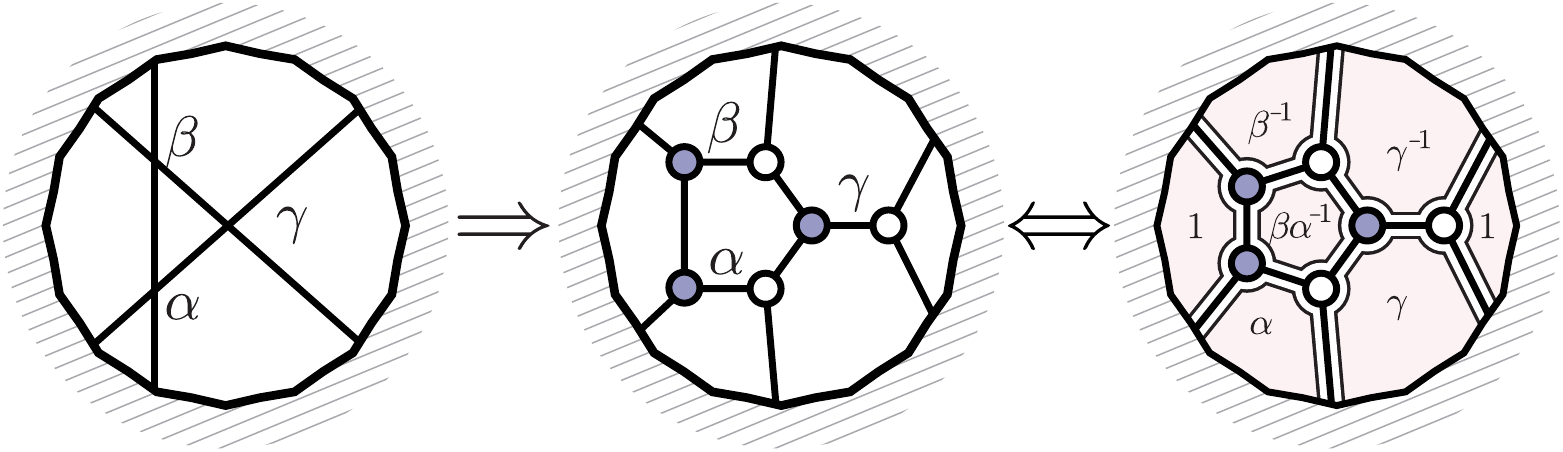}}\vspace{-.2cm}\nonumber}
where on the right, we have recast the edge-variables into corresponding face-variables. Notice that because we are only putting non-trivial edge-weights on the ``bridges'' in the diagram, there are relations between the face variables.

Now, quite beautifully, we can see that the Yang-Baxter relation follows as a consequence of the more elementary actions of the merge- and square-moves! We can see this explicitly through the following sequence of moves, observing the effects induced on the face variables  (see \mbox{section \ref{coordinate_transformations_induced_by_moves}}):
\vspace{-.2cm}\eq{\raisebox{-45pt}{\includegraphics[scale=1]{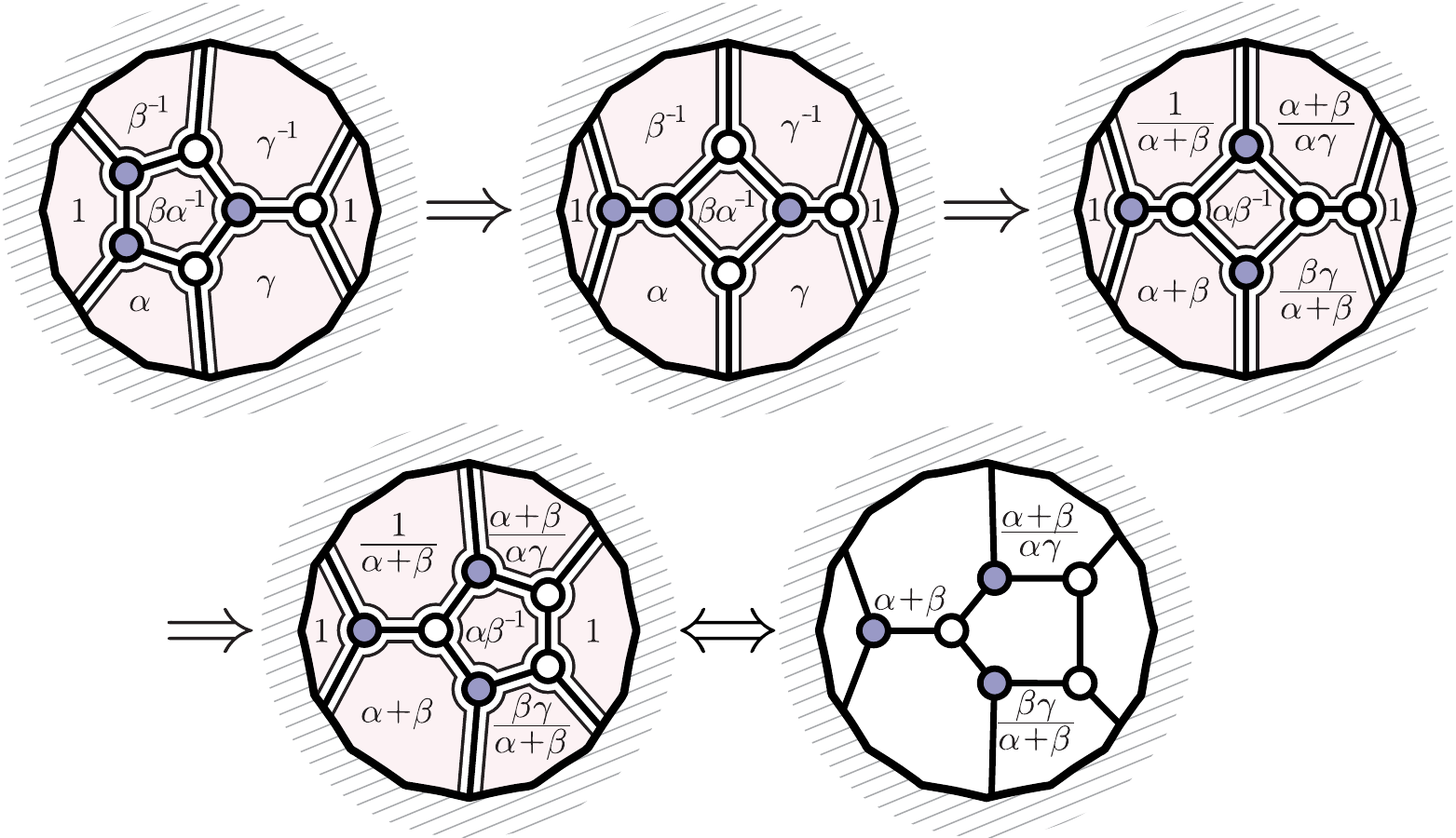}}\vspace{-.2cm}\nonumber}
From this, we may conclude that,
\vspace{-.2cm}\eq{\raisebox{-45pt}{\includegraphics[scale=1]{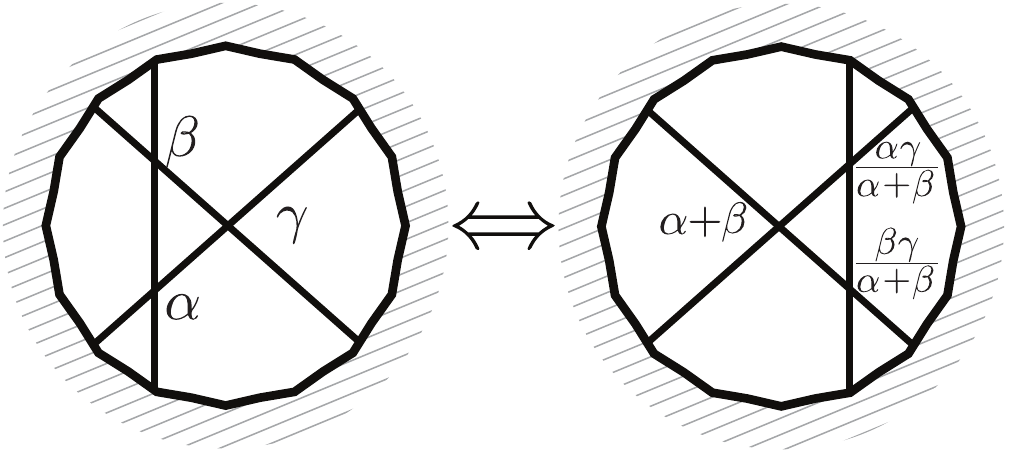}}\vspace{-.2cm}\nonumber}
This equivalence can be interpreted as a generalized Yang-Baxter relation for the $R$-matrices:
\vspace{-.2cm}\eq{R_{12}(\beta) R_{13}(\gamma) R_{23}(\alpha) \, = \, R_{23}\left(\frac{\alpha\gamma}{\alpha + \beta}\right) R_{13}(\alpha + \beta) R_{12}\left(\frac{\beta\gamma}{\alpha + \beta}\right).\vspace{-.2cm}}
In particular, if we set $\alpha + \beta = \gamma$, we recover the familiar Yang-Baxter equation:
\vspace{-.2cm}\eq{R_{12}(\beta) R_{13}(\alpha + \beta) R_{23}(\alpha) = R_{23}(\alpha) R_{13}(\alpha + \beta) R_{12}(\beta).\vspace{-.2cm}}

\subsection{ABJM Theories}
There is yet another natural way to associate a permutation with a scattering process. Suppose we have an even number, $2k$, of particle labels. We can divide them into two sets, $A$ and $B$, of $k$ elements each and draw arrows between them. Such a permutation takes some $a\!\to\!b$ and back via $b\!\to\!a$. We can then represent such a permutation graphically, with all labels on the boundary, as in the following:
\vspace{-.3cm}\eq{\hspace{2.45cm}\raisebox{-62pt}{\includegraphics[scale=.85]{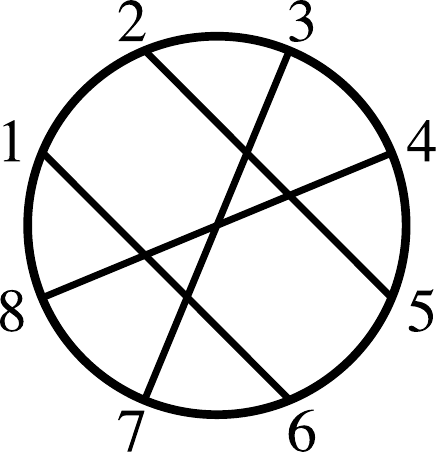}}\vspace{-.1cm} \qquad\begin{array}{c}A\phantom{\longleftrightarrow}B\\\hline 1 \longleftrightarrow 6 \\ 2 \longleftrightarrow 5 \\ 3 \longleftrightarrow 7 \\ 4 \longleftrightarrow 8\\ \end{array}}

We can then interpret this as an on-shell scattering process in a theory where each interaction is fundamentally a 4-particle vertex; and we can ``blow-up'' each four-particle vertex into an element of $G(2,4)$, preserving the symmetrical nature of the permutation according to:
\vspace{-.1cm}\eq{\raisebox{-45pt}{\includegraphics[scale=1]{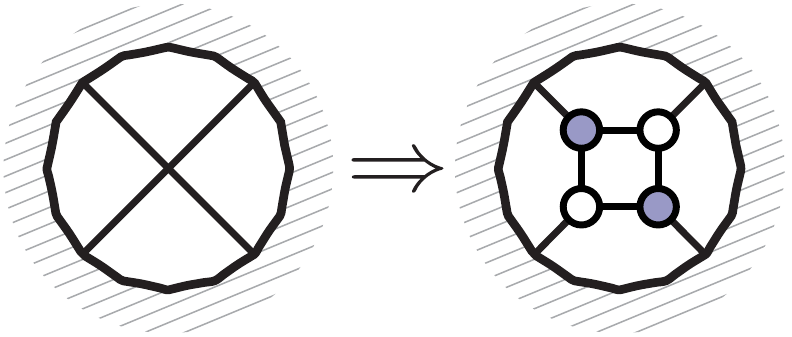}}\vspace{-.3cm}}
This structure was also recently considered in \cite{GK}.
As in the $(1\pl1)$-dimensional example, it is natural to try and associated each vertex with a single degree of freedom. Unlike the $(1\pl1)$-dimensional example, however, this restriction should keep us within the top cell of $G(2,4)$. A very simple way of doing this would be to impose the restriction that the 2-plane is {\it null}; that is,
\vspace{-.2cm}\eq{C\!\cdot\!C=0.\vspace{-.2cm}}
Notice that because the constraint $C\!\cdot\!C=0$ is {\it symmetric}, it represents $k(k\pl1)/2$ constraints in general; for $C\!\in\!G(2,4)$, this imposes only {\it three} constraints, leaving us with a single degree of freedom. In a canonical-gauge, we can write:
\vspace{-.2cm}\eq{C = \left(\begin{array}{@{}cc|cc@{}} 1 & 0 & \,\,\,is &\mi\,ic\!\\ 0 & 1 &\,\,\,ic\, &\,\,\,\,is \end{array} \right),\vspace{-.2cm}\label{null_plane_theta_form}}
where $c \equiv \cos(\theta)$ and $s \equiv \sin(\theta)$ for some angle $\theta$.

Exactly this Grassmannian structure has been found to represent scattering amplitudes for the $(2\pl1)$-dimensional ABJM theory, \cite{Bargheer:2010hn,Lee:2010du,Huang:2010qy,Gang:2010gy}. As in $(3\pl1)$ dimensions, we can motivate the appearance of the Grassmannian  by first looking at the geometry of external data. In $(2\pl1)$ dimensions, the momenta are grouped into a {\it symmetric} $(2\!\times\!2)$-matrix according to,
\vspace{-.4cm}\eq{p^{\alpha\beta} = \left(\begin{array}{@{}cc@{}} p^ 0\pl\,p^2 & p^1 \\ p^1 & p^0\,\mi\,p^2 \end{array} \right),\vspace{-.3cm}}
so that null momenta are given by,
\vspace{-.2cm}\eq{p_a^{\alpha\beta} = \lambda_a^{\alpha} \lambda_a^{\beta},\vspace{-.2cm}}
without any need for conjugate $\widetilde \lambda$'s. The Lorentz group acts as a single copy of $SL(2)$, so the $\lambda_a$ are still represented by a $2$-plane in $n$ dimensions. However, momentum-conservation,
\vspace{-.2cm}\eq{\sum \lambda_a^{\alpha} \lambda_a^{\beta}= 0,\vspace{-.2cm}}
is now the statement that the $\lambda$ plane is orthogonal to itself. Thus, the external data is given not by a general point in $G(2,n)$, but by a point in the  {\it null} Grassmannian of 2-planes in $n$ dimensions. It is therefore not surprising to find the null Grassmannian playing a role in ABJM theory.

ABJM theories have ${\cal N} = 6$ supersymmetries; if we diagonalize half of the supercharges, then the corresponding Grassmann coherent states are labeled by $\eta^I$ for $I = 1,\ldots, 3$. Thus, the on-shell data can be collected into,
\vspace{-.3cm}\eq{\Lambda_a = \left(\begin{array}{@{}c@{}} \lambda_a \\[-1pt] \eta_a \end{array} \right).\vspace{-.3cm}}
The ABJM amplitudes are not cyclically-invariant, but {\it are} invariant under a cyclic shift by {\it two}. Notice that since we only have $\lambda$'s, there is not the same little group rescaling symmetry as we had in three dimensions; rather, we have only the symmetry of sending
$\lambda_a\!\to\mi\, \lambda_a$, under which on-shell differential forms transform according to $f(\mi\,\Lambda_a) = (\mi\,1)^a f(\Lambda_a)$.

Let us now return to the basic $4$-point vertex, and determine the natural measure on the space of null $2$-planes in $4$ dimensions. This space is easily seen to be equivalent to $G(1,2)\simeq\mathbb{P}^1$: the two rows of a $(2\!\times\!4)$-matrix can be viewed as four-vectors $p_1, p_2$ which are null and mutually orthogonal; we can therefore write,
\vspace{-.2cm}\eq{p_1=\lambda\,\widetilde\lambda_1,\qquad p_2=\lambda\,\widetilde\lambda_2,\vspace{-.2cm}}
and use the $GL(2)$-freedom to write $\widetilde\lambda_1\equiv(1\,0)$, $\widetilde\lambda_2\equiv(0\,1)$, and $\lambda\equiv(1\,z)$. This demonstrates the equivalence of the null Grassmannian $C\!\subset\!G(2,4)$ with $\mathbb{P}^1$, and also provides us with a natural measure: $d\!\log(z)$. Using this identification, we can write the null-plane $C\!\subset\!G(2,4)$ in terms of $z$ according to:
\vspace{-.2cm}\eq{\left(\begin{array}{@{}cccc@{}}\mi\,i&\mi\,iz&\,\,z&1\\\mi\,z&\,\,1&\mi\,i&iz\end{array}\right).\vspace{-.2cm}}
Performing a $GL(2)$-transformation to recast this matrix-representative of $C$ in a canonical-gauge brings it to the form given above in (\ref{null_plane_theta_form}), with the identification:
\vspace{-.4cm}\eq{s=\frac{2z}{z^2+1},\quad\mathrm{and}\quad c=\frac{z^2-1}{z^2+1}.\vspace{-.1cm}}

In terms of the natural measure $d\!\log(z)$ on the null subspace, the fundamental $4$-point interaction in the ABJM theory can then be represented by,
\vspace{-.2cm}\eq{{\cal A}_4 = \int\!\!\frac{dz}{z}\,\,\delta^{4|6}(C(z)\!\cdot\!\Lambda);\vspace{-.1cm}}
equivalently, we may view this as having been obtained from a measure defined on all of $G(2,4)$, but restricted to the null subspace by the constraint $\delta^{3}(C\!\cdot\!C)$:
\vspace{-.2cm}\eq{{\cal A}_4 = \int\!\!\frac{d^{2 \times 4} C}{\mathrm{vol}(GL(2))}\,\,\frac{1}{(12)(23)} \delta^3(C\!\cdot\!C) \delta^{4|6}(C\!\cdot\!\Lambda).\vspace{-.1cm}\label{abjm_four_point}}

With this, we can define on-shell diagrams for the ABJM theory just as for $\mathcal{N}\!=\!4$ by gluing together these basic $4$-point vertices. Note that unlike for $\mathcal{N}\!=\!4$, $n$ and $k$ are not independent for ABJM: we always have $n\!=\!2k$.

It is easy to see that the on-shell representation of a BCFW shift is simply,
\vspace{-0.2cm}\eq{\mbox{\hspace{-1.5cm}\raisebox{-51pt}{\includegraphics[scale=.75]{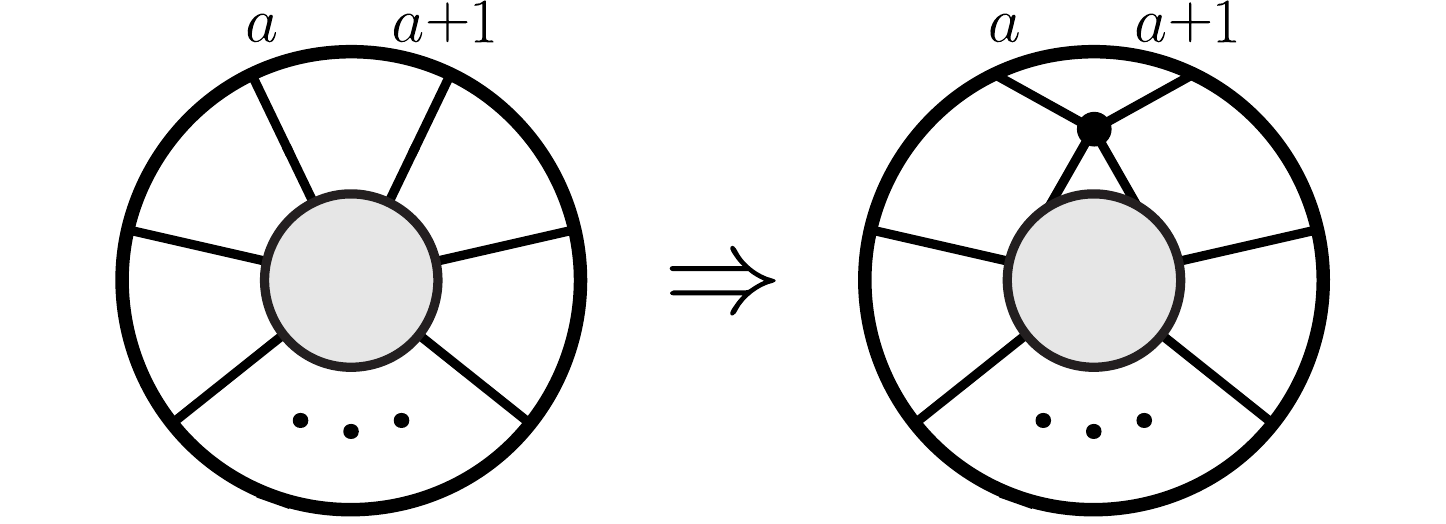}}\hspace{-1.5cm}}\label{adding_a_BCFW_bridge_in_abjm}\vspace{-.2cm}}
The action on the column-vectors is simply a rotation between $c_a$ and $c_{a+1}$:
\vspace{-.2cm}\eq{c_a \mapsto c\,c_a- s\,c_{a+1}, \quad c_{a+1} \mapsto s\,c_{a}+c\,c_{a+1}\;.\vspace{-.2cm}}

\noindent And the all-loop integrand can be given in terms of on-shell diagrams just as before:
\vspace{-0.6cm}\eq{\hspace{-3.cm}\raisebox{-57.5pt}{\includegraphics[scale=.9]{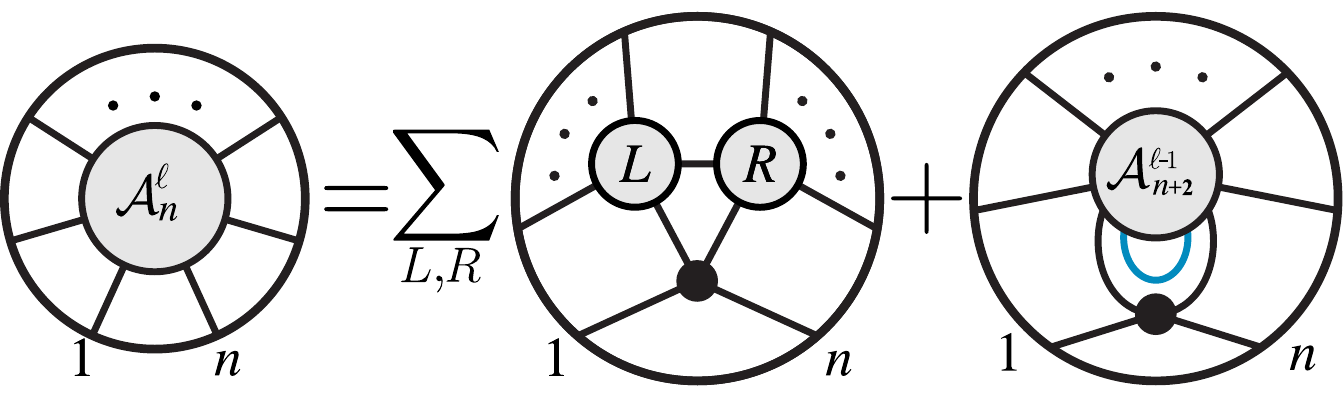}}\hspace{-2cm}\label{bcfw_all_loop_recursion_abjm}\vspace{-0.2cm}}
(For recent computations at one and two loops see \cite{Agarwal:2008pu,Bargheer:2012cp,Brandhuber:2012un,Brandhuber:2012wy}.)

The rules for amalgamation are essentially identical to the ${\cal N}\!=\!4$ case---the only difference being some factors of $i$ that must be included. In $(2\pl1)$ dimensions, because we write momenta as $p_a= \lambda_a \lambda_a$, switching $p_a\!\mapsto\!\mi\,p_a$ corresponds to taking $\lambda_a\!\mapsto\!i \lambda$. And so, when identifying two legs for the ``projection'' operation, instead of projecting relative to $(c_A\,\mi\,c_B)$, we must project relative to $(c_A\,\mi\,ic_B)$. The result is that minors of $C\!\in\!G(k,n)$ are related to those of the pre-image $\hat{C}\!\in\! G(k\pl1,n\pl2)$ via:
\vspace{-.4cm}\eq{\left.(a_1 \cdots a_k)\right|_{C} =\left.(A a_1 \cdots a_k)\right|_{\hat{C}} +\!i\left. (B a_1 \cdots a_k)\right|_{\hat{C}}.\vspace{-.2cm}}

It is very easy to see that, starting with elementary $4$-point vertices in the null Grassmannian, amalgamation preserves  this property; translated in terms of minors, this is the statement that for all $a$,
\vspace{-.3cm}\eq{(c_1 \cdots c_{k-1} a) (d_1 \cdots d_{k-1} a) = 0.\vspace{-.3cm}}
This is trivial for the direct product. For projection, we easily verify that\\[-12pt]
\eqs{\hspace{-1cm}(c_1\!\cdots\!c_{k-1} a) (d_1\!\cdots\!d_{k-1} a) &= [(A c_1\!\cdots\!c_{k-1} a)\!+\!i (B\!c_1\!\cdots\!c_{k-1}a)][(A d_1 \cdots d_{k-1} a)\!+\!i (B d_1\!\cdots\!d_{k-1}a)]\\&
= (A c_1\!\cdots\!c_{k-1} B) (A d_1\!\cdots\!d_{k-1} B)\!-\!(B c_1\!\cdots\!c_{k-1} A)(Ad_1\!\cdots\!d_{k-1} B)\\&=0.\\[-24pt]&\nonumber}
Thus, amalgamation of many little null $G(2,4)$'s produces a point in the null Grassmannian $G(k,2k)$, together with the measure,
\vspace{-.3cm}\eq{\prod_{{\rm vertices~}v} \!\!\!d\!\log(z_v).\vspace{-.3cm}}
Notice that an important difference between this and the case of ${\cal N}\!=\!4$ is that the fundamental variables are associated with the {\it vertices} of an on-shell graph, rather than its faces.

The measure on the top-cell can be given in terms of the $C$ matrix via \cite{Lee:2010du}
\vspace{-.2cm}\eq{\frac{d^{k\times2k}C}{\mathrm{vol}(GL(k))}\frac{\delta^{k(k+1)/2}\big(C\!\cdot\!C\big)}{(1\,2\cdots k)\cdots(k\,k\pl1\cdots 2k\mi1)}.\vspace{-.2cm}}

It is also straightforward to find the analog of boundary measurements by summing over all the paths joining sources to sinks in a perfectly oriented graphs. We can orient each vertex with two incoming and two outgoing lines. Traversing any internal line contributes a factor of $i$, and at each vertex we get a $is$, $ic$ or $-ic$ according to:
\vspace{-.2cm}\eq{\hspace{-2.0cm}\mbox{\raisebox{-45pt}{\includegraphics[scale=1]{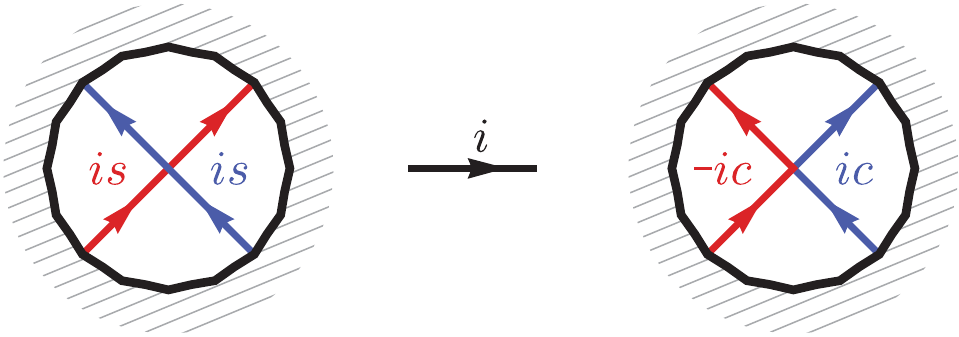}}}\hspace{-2.0cm}\vspace{-.5cm}}

As an example, consider the following on-shell diagram involving 6 particles,
\vspace{-.4cm}\eq{\hspace{-2.0cm}\mbox{\raisebox{-43pt}{\includegraphics[scale=1]{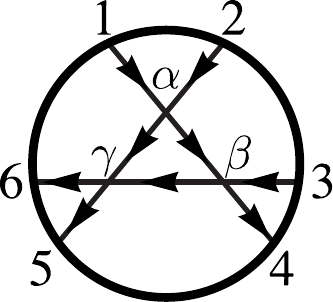}}}\hspace{-2.0cm}\vspace{-.3cm}}
We find that this diagram is associated with a configuration $C$ in the Grassmannian represented by,
\vspace{-.5cm}\eq{C = -i \left(\begin{array}{@{}ccc|ccc@{}} i & 0 & 0 & s_{\alpha} s_{\beta} & s_{\alpha} c_{\beta} c_{\gamma}\pl\,c_{\alpha} s_{\gamma} & c_{\alpha} c_{\gamma}\,\mi\,s_{\alpha} c_{\beta} s_{\gamma} \\ 0 & i & 0 &\,\mi\,c_{\alpha} s_{\beta}\;& s_{\alpha} s_{\gamma}\,\mi\,c_{\alpha} c_{\beta} c_{\gamma} & s_{\alpha} c_{\gamma}\pl\,c_{\alpha} c_{\beta} s_{\gamma} \\ 0 & 0 & i & c_{\beta} &\,\mi\,s_{\beta} c_{\gamma} & s_{\beta} s_{\gamma} \end{array} \right).\vspace{-.2cm}}
In general, we can write the ($k\!\times\!2k$)-matrix representative $C\!\in\!G(k,2k)$ associated with any such graph in the form,
\vspace{-.3cm}\eq{C = -i \left( i1_{k \times k}|R_{k \times k} \right),\vspace{-.2cm}}
where $R$ is an $SO(k)$-rotation matrix. This gives us a pretty interpretation for amalgamation. The basic $4$-point vertex is just a rotation in two dimensions. Amalgamation provides a way of building general rotations in higher dimensions by a composing many rotations in two-dimensional subspaces. The example above for $6$ particles corresponds to a canonical way of representing three-dimensional rotations using Euler angles. The analog of the square move in ABJM looks much like the Yang-Baxter move, and represents the equality of two different Euler-angle representations of the same three-dimensional rotation.

Just as with $\mathcal{N}\!=\!4$ SYM, the invariant content of any reduced on-shell diagram is read-off from its associated permutation.  We also have an analog of reduction, looking at the $4$-point bubble diagram connecting two $4$-particle vertices with parameters $\alpha$ and $\beta$:
\vspace{-.2cm}\eq{\hspace{-2.0cm}\mbox{\raisebox{-43pt}{\includegraphics[scale=1]{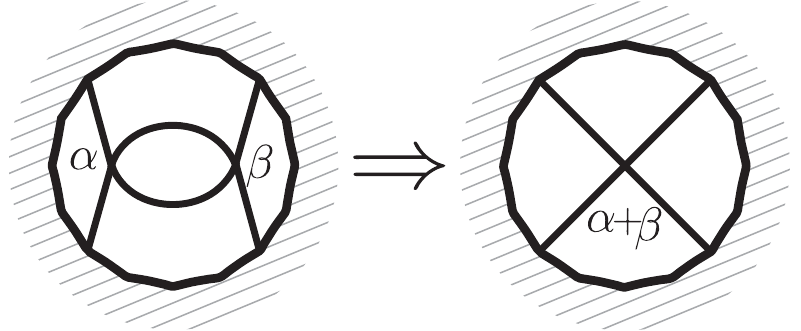}}}\hspace{-2.0cm}\vspace{-.2cm}}
Finally, we can take a boundary, lowering the dimension by one, by deleting a vertex, and re-connecting the lines according to:
\vspace{-.2cm}\eq{\hspace{-2.0cm}\mbox{\raisebox{-43pt}{\includegraphics[scale=1]{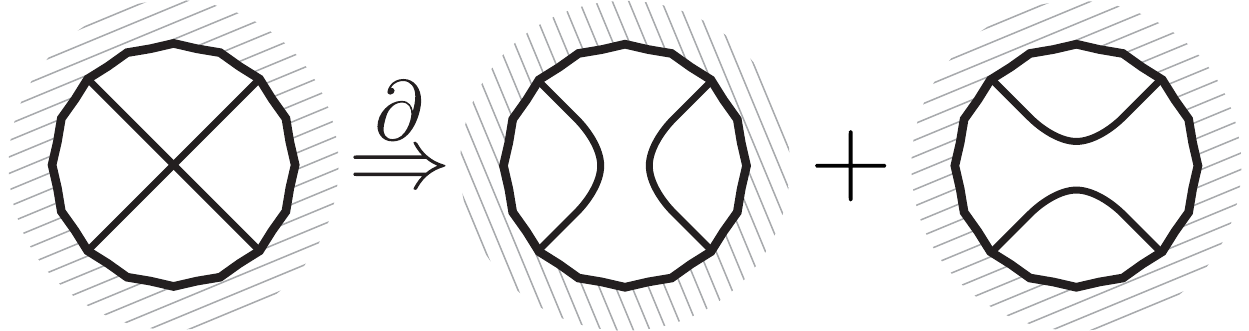}}}\hspace{-2.0cm}\vspace{-.6cm}}

~\newpage

\section{On-Shell Diagrams with $\mathcal{N}<$ 4 Supersymmetries}\label{less_supersymmetries_section}

On-shell diagrams can be defined for any theory with fundamental trivalent vertices, and in particular for gauge theories with any number, $\mathcal{N}$, of supersymmetries. There is obviously a rich structure to be unearthed here; in this short section we will content ourselves with setting-up some of the basic formalism and highlighting the central new mathematical object that makes an appearance---reflecting the physics of ultraviolet singularities which are present in theories with less supersymmetry.

Let us begin our discussion by focusing on non-supersymmetric theories, those of ``${\cal N}\!=\!0$''. It is useful to represent the helicities involved in each basic $3$-particle vertex by giving each of the edges an orientation:
\vspace{-.5cm}\eq{\raisebox{-50pt}{\includegraphics[scale=1]{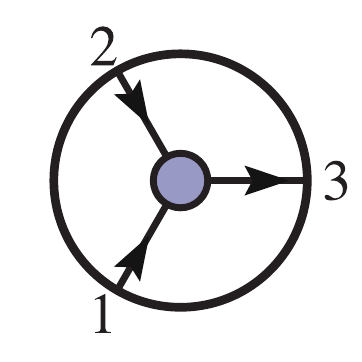}}\qquad\mathrm{and}\qquad\raisebox{-50pt}{\includegraphics[scale=1]{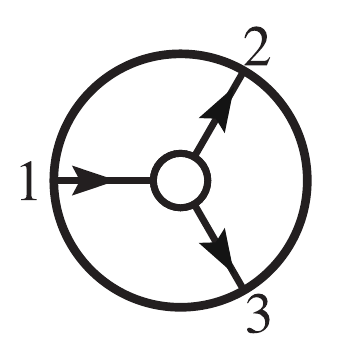}}\vspace{-.5cm}}
We can then glue these vertices together to build-up more complex on-shell diagrams as before---leading to, for example:
\vspace{-.2cm}\eq{\raisebox{-57pt}{\includegraphics[scale=1]{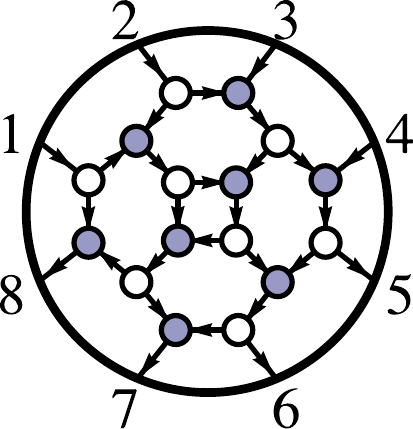}}\vspace{-.2cm}}
In such decorated on-shell diagrams, the arrows are useful because they automatically encode the helicities of the internal particles involved. In general, we consider the particles as Grassmann coherent states labeled by $\widetilde \eta^I$ for $I = 1,\ldots, {\cal N}$. In theories with ${\cal N}<$ 4 supersymmetry, we have ``$+$'' and ``$-$'' multiplets, which include gluons of helicity $\pm 1$ as their top components, respectively; thus, on-shell diagrams must be labeled in exactly the same way for any $\mathcal{N}< 4$.

The Grassmannian formalism is just as powerful in integrating over the phase space of the internal particles regardless of the amount of supersymmetry. However, when $\mathcal{N}<4$, the diagrams really are fundamentally oriented, whereas for $\mathcal{N}\!=\!4$ such an orientation merely encodes a convenient translation of the on-shell diagram into a particular gauge-fixed  matrix-representative $C\!\in\!G(k,n)$. If the $k$ incoming ``source'' indices are from a set  $A$ and the $(n\,\mi\,k)$ outgoing ``sink'' indices are from $a$, we find exactly the same linear relation between the external kinematical data:
\vspace{-.2cm}\eq{\prod_A\delta^2\big(\widetilde\lambda_A - c_{A\,a} \widetilde\lambda_a \big)\prod_A\delta^{\cal N} \big(\widetilde \eta_A-c_{A\,a} \widetilde \eta_a \big) \prod_a  \delta^2\big(\lambda_a + c_{A\,a}  \lambda_A \big),\vspace{-.2cm}}
where the $c_{A\,a}$ are exactly as in equation (\ref{general_link_in_terms_of_edges}), which we reproduce below:
\vspace{-.2cm}\eq{c_{A\,a}=-\sum_{\Gamma\in\{A\rightsquigarrow a\}}\prod_{e\in\Gamma}\alpha_e.\vspace{-.2cm}}
The only difference between general $\mathcal{N}$ and ${\cal N}\!=\!4$ is the measure on the Grassmannian which ultimately encodes the on-shell differential form in terms of the auxiliary, Grassmannian degrees of freedom. For ${\cal N}\!=\!4$, we didn't have to include any Jacobian resulting from the elimination of internal variables, because the fermionic $\delta$-functions always canceled the contributions between the internal bosons and internal fermions.  However, when ${\cal N}<4$, these two factors do not cancel, and leave a net Jacobian contribution to the measure which we may write as:
\vspace{-.2cm}\eq{\left(\prod_{\mathrm{vertices~}v}\!\!\frac{1}{\mathrm{vol}(GL(1)_v)}\right)\left(\prod_{\mathrm{edges~}e}\frac{d\alpha_e}{\alpha_e}\right) \times {\cal J}^{{\cal N} - 4}.\vspace{-.2cm}}
If the vertices of the graph are labeled $i,j$, then we define the {\it adjacency matrix} $A_{ij}$ of the graph by,
\vspace{-.3cm}\eq{A_{ij} = \text{the weight of the {\it directed} edge $i\!\to\!j$ (if any);}\vspace{-.3cm}}
then the Jacobian ${\cal J}$ is given by
\vspace{-.3cm}\eq{{\cal J} =\det(1-A).\vspace{-.3cm}}

We know that the edge variables can only occur in the $GL(1)$ gauge-invariant ``flux'' combinations associated with faces, and we can give a simple formula for ${\cal J}$ in terms of these face variables. In general, if we have a collection of closed, {\it orientated} orbits bounding faces $f_i$, with {\it disjoint} pairs $(f_i,f_j)$, {\it disjoint} triples $(f_i,f_j,f_k)$, and so on, then $\mathcal{J}$ is given by:
\vspace{-.2cm}\eq{{\cal J} = 1 + \sum_{{\rm faces}} f_i + \sum_{\substack{\text{disjoint}\\\text{pairs}\,i,j}} f_i f_j + \sum_{\substack{\text{disjoint}\\\text{triples}\,i,j,k}} f_i f_j f_k + \cdots\,.\vspace{-.2cm}}
(Here, each `face' is really a clockwise-oriented product of edge-variables around an orbit---and so may be the inverse of a face variable, or the product of face-variables which are bounded by a single orbit.)

For any oriented graph without any closed, oriented orbits, the spectrum is trivial, and ${\cal J}\!=\!1$; for any such oriented on-shell diagram, the maximally-supersymmetric and non-supersymmetric on-shell forms are identical. This is easy to understand because when an on-shell diagram is free of such oriented orbits, only gluons propagate internally. In contrast, when there are oriented orbits, the rest of the super-multiplet {\it can} propagate internally, differentiating theories with different amounts of supersymmetry.

When an oriented on-shell diagram has closed, oriented orbits, the Jacobian is nontrivial. The simplest example occurs for four particles, where we can have,
\vspace{-.0cm}\eq{\raisebox{-56pt}{\includegraphics[scale=1]{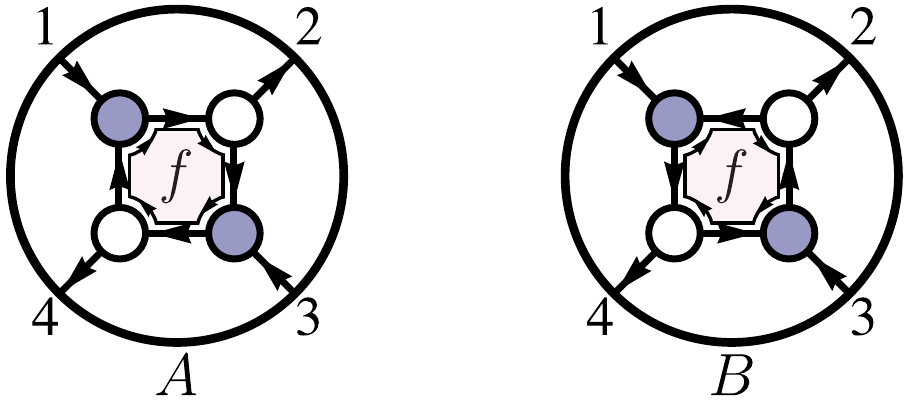}}\vspace{-.2cm}}
for which the corresponding Jacobian is,
\vspace{-.2cm}\eq{{\cal J}_A = 1 + f\qquad\mathrm{and}\qquad{\cal J}_B = 1 + f^{-1}.\vspace{-.2cm}}
In order to compute the full on-shell process for fixed external sources and sinks, we have to sum-over all the possible orientations of the internal graph. And so, in this case we would be obliged to sum-over both diagrams, giving us a final contribution to the measure of:
\vspace{-.2cm}\eq{{\cal J}_A^{{\cal N} - 4} + {\cal J}_B^{{\cal N} - 4} = (1 + f)^{{\cal N} - 4} + (1 + f^{-1})^{{\cal N} - 4}.\vspace{-.2cm}}
Notice that when $\mathcal{N}\!=\!3$, the complete contribution is simply:
\vspace{-.2cm}\eq{{\cal J}_A^{-1} + {\cal J}_B^{-1} = (1 + f)^{-1} + (1 + f^{-1})^{-1}=1.\vspace{-.2cm}}
This is good, because the ``$+$'' and ``$-$'' super-multiplets of $\mathcal{N}\!=\!3$ combine to give us a complete $\mathcal{N}\!=\!4$ super-multiplet. Of course, when $\mathcal{N}<3$, the sum is not unity, and the result differs from what we would have found for $\mathcal{N}\!=\!4$.

Let us consider a somewhat more interesting example:
\vspace{-.2cm}\eq{\raisebox{-76pt}{\includegraphics[scale=1]{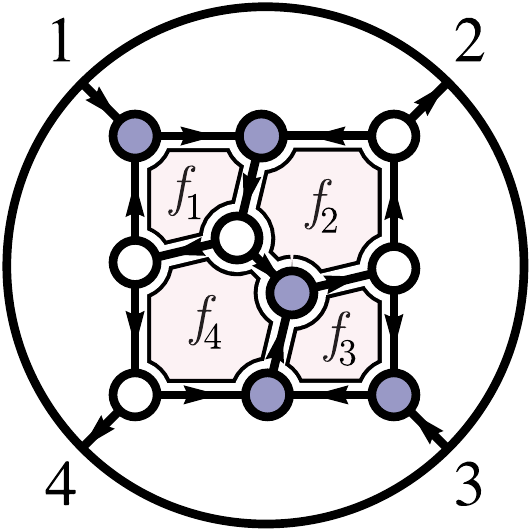}}\vspace{-.2cm}}
Here we have four closed orbits, and one disjoint pair of orbits: three of the orbits bound single faces, contributing $f_1,f_2^{-1},$ and $f_3$, and one orbit involving both $f_2$ and $f_4$---contributing $f_2^{-1}f_4^{-1}$; the disjoint pair are $f_1,f_3$. Putting everything together, we find that the complete Jacobian is:
\vspace{-.0cm}\eq{{\cal J} = 1 + (f_1 + f_2^{-1} + f_3+f_{2}^{-1}f_4^{-1}) + f_1 f_3.\vspace{-.0cm}}

We stress again that the point in the Grassmannian obtained from amalgamation is the same as it is for the maximally supersymmetric theory; the {\it only} difference between the theories is the presence of the Jacobian factor $\mathcal{J}$ in the measure. The merge/un-merge moves still leaves the point in the Grassmannian and the rest of the form invariant; but now, the square-move and bubble-reduction---while leaving the point in the Grassmannian fixed---{\it can} change the measure.

If we consider a reduced graph with the dimension required to completely localize all the auxiliary variables associated with the matrix $C\!\in\!G(k,n)$, then the net effect is not particularly interesting---as theories with $\mathcal{N}<4$ differ from those with maximal supersymmetry only by the prefactor of $\mathcal{J}$ in the measure, evaluated at this particular point in $G(k,n)$. However, the situation is considerably more interesting when we consider on-shell graphs for which some auxiliary variables are not fixed by the $\delta$-function constraints, leaving us with an integration measure over these internal degrees of freedom. Such graphs occur, for instance, in the forward-limits that generate loop integrands in the all-loop, on-shell BCFW recursion (\ref{all_loop_recursion}). In such cases, the factor of $\mathcal{J}$ can lead to a qualitatively-new set of singularities where poles are generated by $\mathcal{J}$.

As a simple example of such a situation, consider a ``wrong'' BCFW-bridge acting on the four-particle tree amplitude's on-shell graph:
\vspace{-.1cm}\eq{\raisebox{-48pt}{\includegraphics[scale=1]{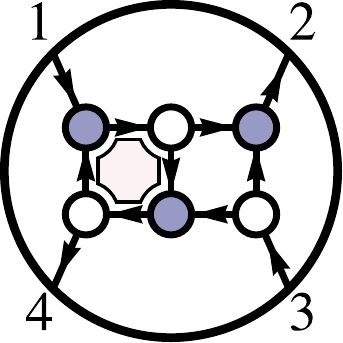}}\vspace{-.1cm}\label{g24_wrong_bcfw_shift}}
The shift is ``wrong'' in the familiar sense of BCFW deforming the ``wrong'' helicities, for which the deformed amplitudes don't vanish at infinity. This is reflected in the on-shell graph by the presence of a closed oriented loop (making the resulting on-shell differential form differ for theories with different amounts of supersymmetry). Because this graph's measure includes the a non-trivial Jacobian $\mathcal{J}$, the corresponding function does not vanish in the deep ultraviolet---taking the shift-parameter $\alpha\!\to\!\infty$. This ``pole at infinity'' is characterized by the residue about $\mathcal{J}\!\mapsto\!0$. Notice that this allows us to fully characterize the non-trivial, ultraviolet singularities present in theories with less than maximal supersymmetry. The presence of such poles indicate ``lower-transcendentality'' contributions to scattering amplitudes. For instance, the object above, (\ref{g24_wrong_bcfw_shift}) can be interpreted as the triple-cut of the one-loop four-particle amplitude, and the residue about $\mathcal{J}=0$ computes the coefficient of the ``triangle integral'' for the amplitude. The coefficients of ``bubbles'' can be exposed in similar ways.

One of the most fundamental consequences of space-time locality is that the ultraviolet and infrared singularities are completely independent. It is fascinating to see that this physical fact is sharply captured by the Grassmannian formalism, where IR and UV singularities are associated with  disparate contributions to the integration measure of the auxiliary Grassmannian: the positroid's ``$d\!\log$'' measure captures all the long-distance singularities---where internal particles go on-shell---and the prefactor ${\cal J}$ captures  ultraviolet singularities. This ultraviolet/infrared decoupling has an even more striking incarnation in the planar sector of the theory: it can be shown that $\mathcal{J}$ is completely regular everywhere in the positive-part of $G(k,n)$---literally {\it separating} the ultraviolet singularities of $\mathcal{J}$ from infrared singularities of the positroid, their boundaries being completely disjoint!

\newpage

\newtheorem{definition}{Definition}[section]

\section{Dual Graphs and Cluster Algebras}\label{cluster_coordinates_section}

So far in this paper, we have extensively studied planar on-shell diagrams. In \mbox{section \ref{amalgamation_subsection}}, we introduced two natural classes of operations: {\it amalgamation}, the operation which allows us to build-up very complex diagrams from very simple ones; and {\it mergers} and {\it square moves}, which allow us to connect very distinct on-shell diagrams which nevertheless encode the same physical information.

In this section we turn to the very obvious question that arises when dealing with planar diagrams of any sort: what are the corresponding dual graphs? what do they mean? and how are the operations we have found realized in terms of them?
Of course, being two-colored, on-shell diagrams carry more information than ordinary graphs, and whatever definition of a dual graph we introduce must encode this additional information. Luckily, the theory of dual graphs for bipartite planar graphs is both known and simple; in fact, the dual of a {\it bipartite} graph is a familiar object in the physics of ${\cal N}=1$ supersymmetric gauge theories: it is a quiver diagram! Indeed, the connection between bipartite graphs and quiver gauge theories is already an active research area in the physics community and has led to beautiful constructions such as those described in \cite{Xie:2012dw,Franco:2012mm,Xie:2012mr,Xie:2012jd,Heckman:2012jh,Franco:2012wv}. Bipartite graphs are also intimately related to dimer models, with the recent mathematical work \cite{GK} particularly closely related to our discussion.

\subsection{The `Dual' of an On-Shell Diagram}

Recall that the dual of an ordinary planar graph (one without colored vertices) is obtained by drawing a vertex for each face, and connected adjacent faces with edges. In our case, we have graphs on a {\it disc}, and so the faces of an on-shell diagram can be divided into two distinct classes: those in the {\it interior} of the graph, and those on the {\it exterior} (those adjacent to  the boundary of the disc).

As mentioned above, the dual of a {\it bipartite} graph turns out to be nothing but an {\it oriented} quiver diagram. Let us now describe how this dual ``quiver'' of a general bipartite graph on a disc is defined. Let $\Gamma$ denote a bipartite graph on a disc; we define a {\it flag} $F$ of $\Gamma$ to be the combination of one vertex of $\Gamma$ with one edge connected to it. (Here, the word ``flag'' can just be taken as merely a name used this construction; but---not surprisingly---this terminology stems from its more familiar use in algebraic geometry.) Each flag inherits a coloring according to the color of its vertex. We orient each black flag of $\Gamma$ to be `out of the vertex' and each white flag to be `toward the vertex' . (Notice that every {\it external} edge is a member of a single flag, while each {\it internal} edge is part of two flags---one black, and one white.)

Let us suppose that $\Gamma$ is {\it oriented} according to its internal flags---that is, we take each internal edge to be directed `black-to-white'. Choosing a clockwise-orientation for the boundary of each face $f$ of $\Gamma$, we define the adjacency matrix of for the dual graph $\widetilde\Gamma$ as follows: for each flag $F$, we define
\vspace{-.2cm}\eq{\nonumber\delta^F_{f_1,f_2}\equiv\left\{ \begin{array}{@{}l@{$\qquad$}l} \phantom{+}0 & \mbox{$F\notin\big(\partial f_1\newcap\partial f_2\big);$}\\ +\frac{1}{2}& \mbox{$F\in\big(\partial f_1\newcap\partial f_2\big)$ and $F$ is oriented according to $\partial f_1;$}\\-\frac{1}{2}& \mbox{$F\in\big(\partial f_1\newcap\partial f_2\big)$ and $F$ is oriented according to $\partial f_2.$}\end{array}\right.\label{123a}\vspace{-.2cm}}
The dual graph's adjacency matrix $\varepsilon_{f_1,f_2}$ is then obtained by summing over all flags:\\[-6pt]
\vspace{-.2cm}\eq{\varepsilon_{f_1,f_2} \equiv \sum_F\delta^F_{f_1, f_2}.\vspace{-.2cm}\label{123f}}

(The factors of ``$1/2$'' in the definition of $\delta^F_{f_1,f_2}$ is important for both mathematical and physical reasons, but will only show up in the final quiver for edges connecting external faces; to see this, recall that each internal edge is a member of  {\it two} flags. They are important for the consistency of the general procedure of ``amalgamation'' \cite{FG3}, to be described more fully below.)

In order to illustrate the preceding discussion, let us consider---as usual---the on-shell diagram which generates the four-particle tree-amplitude,
\vspace{-0.1cm}\eq{\raisebox{-47pt}{\includegraphics[scale=1]{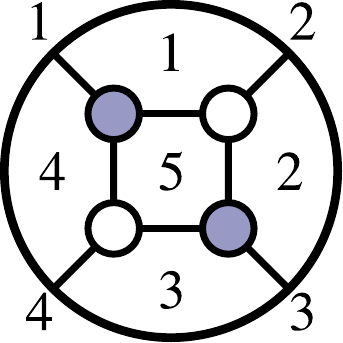}}\vspace{-0.1cm}\label{four_point_box_later}}
Labeling the faces of the graph as indicated above, we find the dual quiver to be,
\vspace{-.2cm}\eq{\raisebox{-47pt}{\includegraphics[scale=1]{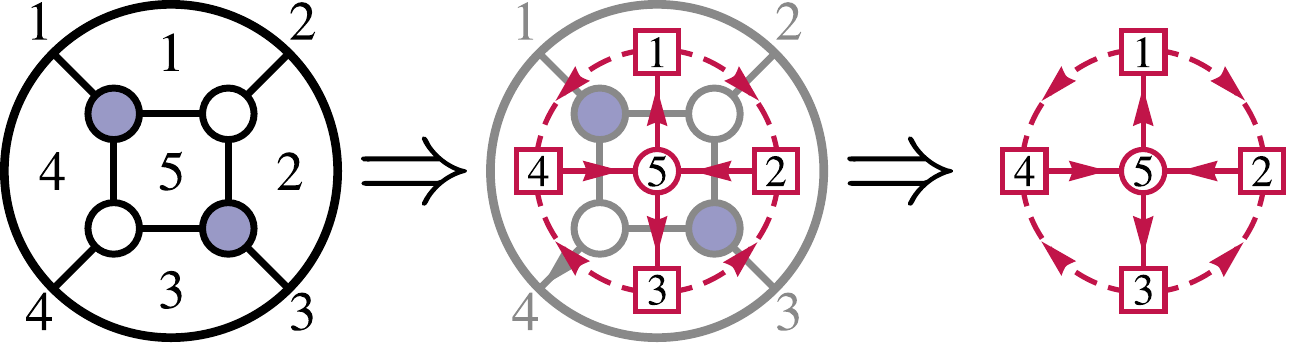}}\vspace{-.2cm}\label{g24_to_quiver_a}}
defined by the adjacency matrix $\varepsilon_{i,j}$:
\vspace{-.2cm}\eq{\varepsilon=\left(\begin{array}{@{}ccccc@{}}0&\frac{1}{2}&0&\frac{1}{2}&\mi1\phantom{\mi}\\\mi\frac{1}{2}\phantom{\mi}&0&\mi\frac{1}{2}\phantom{\mi}&0&1\\0&\frac{1}{2}&0&\frac{1}{2}&\mi1\phantom{\mi}\\\mi\frac{1}{2}\phantom{\mi}&0&-\frac{1}{2}&0&1\\1&\mi1\phantom{\mi}&1&\mi1\phantom{\mi}&0\end{array}\right).\vspace{-.2cm}}
Notice that in drawing (\ref{g24_to_quiver_a}), we have denoted each internal face by a circle and each external face by a square.

This quiver can  be given a  gauge theory interpretation. Let all the vertices represent $U(N)$ groups, with external ones being flavor groups and internal ones dynamical, gauge groups. The adjacency matrix $\varepsilon_{i,j}$ denotes the number of bi-fundamental fields charged under the groups $(i,j)$. Anomaly cancellation is the statement that all the rows of $\varepsilon$ add up to zero. Of course, having a $1/2$ bi-fundamental field is clearly not possible in a physical theory; but a physical realization can be always be obtained by including also super-potential terms (see e.g.\ \cite{Heckman:2012jh,Franco:2012wv}) which we will not need for our purposes of formal analogy.

Let's see how a square-move affects the resulting dual-quiver; starting from,
\vspace{-.2cm}\eq{\raisebox{-47pt}{\includegraphics[scale=1]{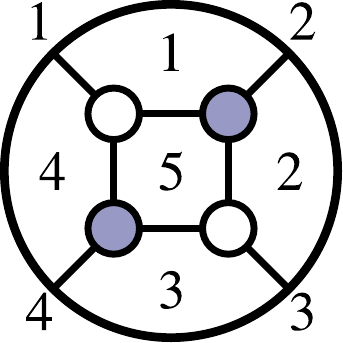}}\vspace{-.2cm}}
it is easy to find the corresponding quiver:
\vspace{-.2cm}\eq{\raisebox{-47pt}{\includegraphics[scale=1]{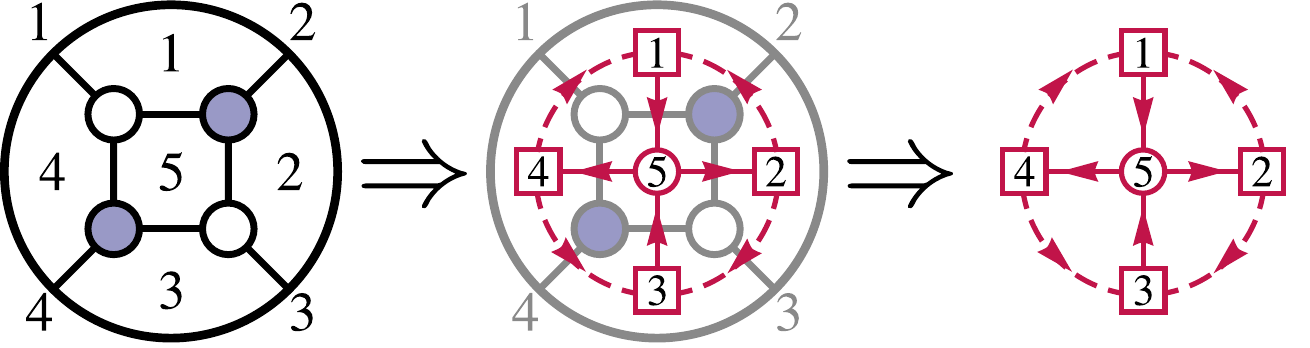}}\vspace{-.2cm}}
Very nicely, this new quiver corresponds to nothing but the {\it Seiberg-dual} of quiver (\ref{g24_to_quiver_a}) with respect to the internal node $(5)$, \cite{Seiberg:1994pq}. In other words, the arrows connected to node $(5)$ are reversed and new bi-fundamentals are created between two flavor nodes every time the bi-fundamentals connecting them to $(5)$ can pair-up. This happens when the arrows are in opposite directions. Whenever possible, the new bi-fundamentals pair-up with existing ones to get a mass and disappear from the theory in the infrared. (Here, we again must stretch the analogy a bit, declaring that, e.g.\ the new bi-fundamental going from $(1)$ to $(2)$ pairs up with the existing `half bi-fundamental' from $(2)$ to $(1)$ to leave behind a new  `half bi-fundamental' from $(1)$ to $(2)$.)

The factors of $1/2$ appearing in this discussion might seem like an unnecessary annoyance. In order to see the importance of these factors in the definition of $\delta^F_{f_1,f_2}$, let us consider the $5$-particle on-shell diagram (in bipartite form):
\vspace{-.2cm}\eq{\raisebox{-57pt}{\includegraphics[scale=1]{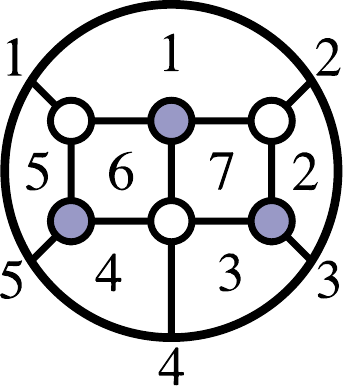}}\vspace{-.2cm}}
whose quiver is found to be given by:
\vspace{-.2cm}\eq{\raisebox{-57pt}{\includegraphics[scale=1]{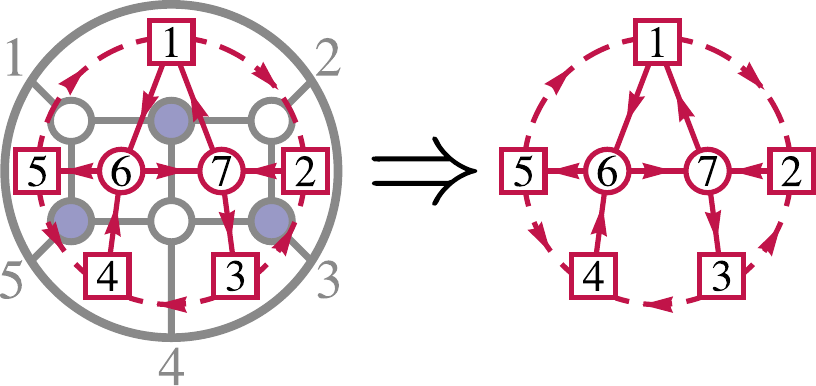}}\vspace{-.2cm}}
After performing a Seiberg duality on node $(6)$, we find the quiver:
\vspace{-.2cm}\eq{\raisebox{-57pt}{\includegraphics[scale=1]{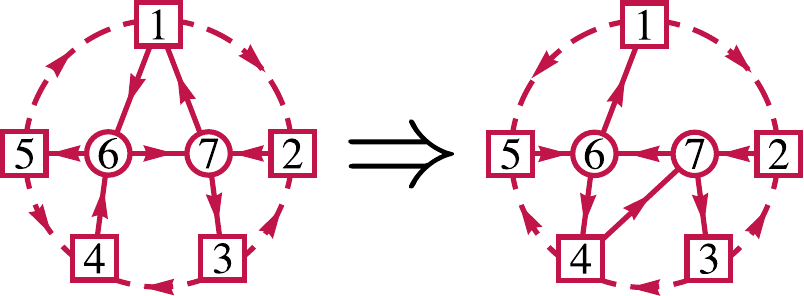}}\vspace{-.2cm}}
This can be seen to agree with the one obtained by expanding the 4-valent white vertex in the original on-shell diagram, applying a square-move and merging the two new adjacent white vertices to make a new bipartite graph. This case shows that the factors of $1/2$ are needed in order to make the square-move correspond to Seiberg duality when not all the four surrounding faces are external. As mentioned above, a fully physical alternative requires adding the natural super-potential terms associated with each closed loop in the quiver consistent with orientations (see e.g.\ \cite{Heckman:2012jh,Franco:2012wv}).

These examples make it clear that Seiberg duality on an internal nodes with {\it exactly} four edges corresponds to a square-move in the on-shell diagram. Of course, once we have the dual-quiver of a given on-shell diagram is determined, it is tempting to perform Seiberg duality on any internal node---not only those of valency four. The question is then: what happens when a Seiberg duality is taken for nodes of arbitrary valency?

Let us start by considering the case of valency two. A valency-two node arises in the dual quiver only for on-shell diagrams with `bubbles'. Recall that all nodes correspond to gauge groups $U(N)$; therefore, upon taking the Seiberg dual of a bivalent vertex, one finds that the rank goes from \mbox{$N\!\mapsto\!N'\!=\!N\,\mi\,N\!=\!0$}---that is, it `disappears' from the theory, leaving only the bi-fundamental created by the duality:\\[-6pt]
\vspace{-.0cm}\eq{\hspace{-3cm}\raisebox{-57pt}{\includegraphics[scale=1]{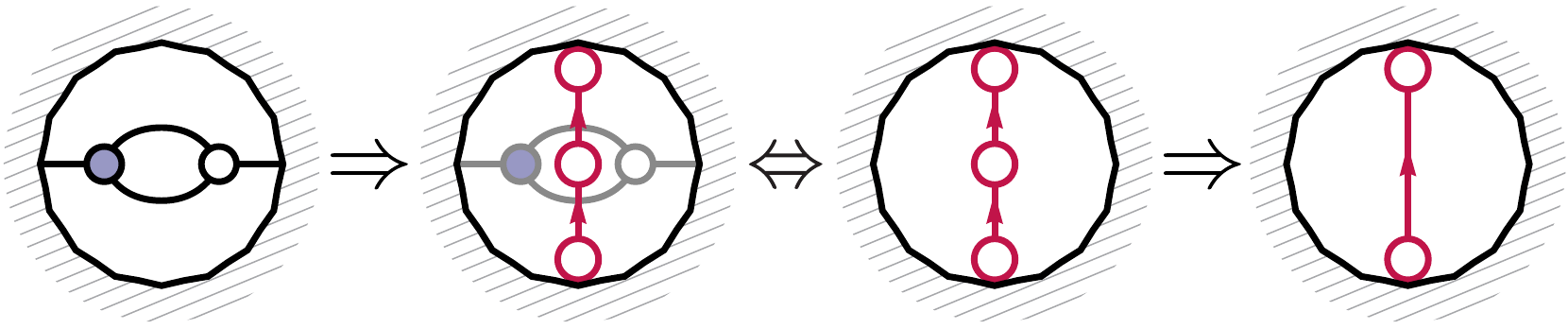}}\nonumber\hspace{-3cm}\vspace{-.1cm}}
The resulting quiver is precisely that which would have been obtained for the on-shell diagram after bubble-deletion.

Let us now see the effect of applying Seiberg duality to a vertex whose valency is greater than four. Consider for example the dual-quiver of the following diagram for the top-cell of $G(3,6)$:
\vspace{-.2cm}\eq{\raisebox{-57pt}{\includegraphics[scale=1.25]{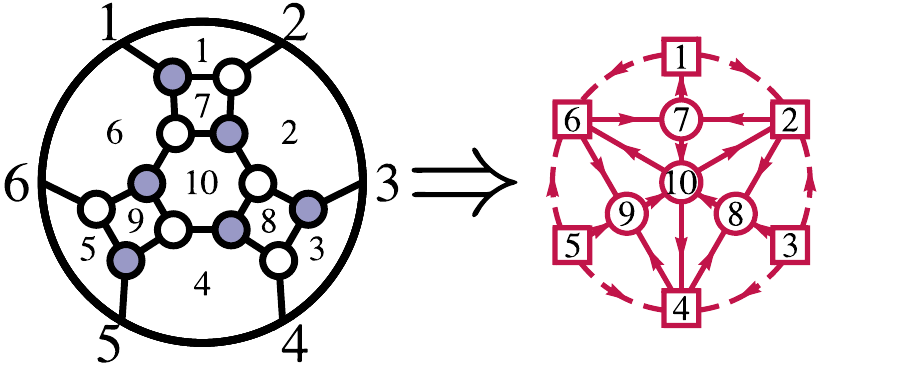}}\vspace{-.2cm}}
Applying Seiberg duality to any of the nodes $(7)$, $(8)$, or $(9)$, would correspond to a square-move as expected. However, when we perform a Seiberg duality on the $6$-valent vertex $(10)$, we find the following:
\vspace{-.2cm}\eq{\raisebox{-57pt}{\includegraphics[scale=1.5]{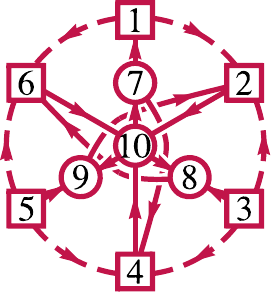}}\vspace{-.2cm}}
This new quiver is non-planar and therefore does not correspond to the dual of an on-shell diagram on the disc!

The presence of these new objects suggests that on-shell diagrams and operations like the square-move are part of a much larger mathematical structure. As it turns out, the general story of such transformations is a very active area of research in mathematics: the theory of {\it cluster algebras}.

Let us now turn to a (basic) summary of this rich, more general story. The reason for doing so, is that cluster algebras have recently made an appearance in the descriptions of seemingly unrelated developments in gauge theory---and we anticipate that a deeper understanding will lead to important connections between these areas of physics and the physics of scattering amplitudes.

\newpage
\subsection{Cluster Algebras: Seeds, Mutations, and Cluster Coordinates}\label{cluster_coordinates_and_mutations_subsection}

The definition of a cluster algebra \cite{FZ} begins with the notion of a {\it seed} and the various {\it mutations} that can take convert one seed into another. We will start with a general mathematical description of {\it seeds} and {\it mutations}, and try to relate these objects to on-shell diagrams as we proceed.\\[-7pt]

\noindent{\bf Seeds.} A {\it seed} is a set of combinatorial data $\mathbf{s}\!=\!\{S,S_0,\varepsilon\}$, where $S$ is a set, $S_0$ a distinguished subset of $S$ called the {\it frozen subset}, and $\varepsilon_{i,j}$ is a skew-symmetric matrix with $(i,j)\!\in\!S$ for which $\varepsilon_{i,j}\!\in\!\mathbb{Z}$ unless {\it both} of $(i,j)\!\in\!S_0$, in which case $\varepsilon_{i,j}\in\frac{1}{2}\mathbb{Z}$. \\[-7pt]

Notice that for the discussion about, the dual-quiver of an on-shell diagram is an example of a seed where $S$ is the set of faces, $S_0$ the subset of faces on the boundary of the disc, and $\varepsilon$ precisely as defined above. Notice that the distinction between $S$ and $S_0$ nicely matches the physical distinction between flavor and dynamical groups of the quiver theory.\\[-7pt]

\noindent{\bf Mutations.} Given a seed $\mathbf{s}\!=\!\{S,S_0,\varepsilon\}$ and any non-frozen element $k\!\in\!(S\raisebox{1pt}{\text{{\footnotesize$\backslash$}}} S_0)$, the {\it mutation} of $\mathbf{s}$ ``in the direction $k$'' is $\mu_k(\mathbf{s})\!\equiv\!\{S,S_0,\varepsilon'\}$ where the {\it mutated} matrix $\varepsilon'$ is given by the Fomin-Zelevinsky formula, \cite{FZ}:
\vspace{-.2cm}\eq{\varepsilon_{i,j}'\equiv\left\{\begin{array}{@{}l@{$\quad$}l}\mi\,\varepsilon_{i,j}&k\in\{i,j\};\\\phantom{\mi\,}\varepsilon_{i,j}&k\notin\{i,j\}\text{ and }\varepsilon_{i,k}\varepsilon_{k,j}\leq0;\\\phantom{\mi\,}\varepsilon_{i,j}+|\varepsilon_{i,k}|\!\cdot\!\varepsilon_{k,j}&k\notin\{i,j\}\text{ and }\varepsilon_{i,k}\varepsilon_{k,j}>0.\end{array}\right.\vspace{-.2cm}\label{5.11.03.6}}

Notice that this procedure is {\it involute}: the mutation of $\varepsilon'$ in the direction $k$ is the original matrix $\varepsilon$. Moreover, the rule for mutations {\it exactly} corresponds to the transformation of bi-fundamental matter fields occurring under a Seiberg duality of a dynamical gauge-group node (corresponding to an element of $(S\raisebox{1pt}{\text{{\footnotesize$\backslash$}}} S_0)$).\\[-7pt]

\noindent{\bf Cluster Coordinates.} Recall the description the {\it face variables} $\{f_i\}$ associated with any on-shell diagrams given in \mbox{section \ref{boundary_measurements_section}}; one of the most important features of the face variables is that if they are chosen to be {\it positive} (see \mbox{section \ref{polygon_coordinates_section}}) then they remain positive after a square-move. Also important is that face variables are {\it canonical} coordinates on the Grassmannian configuration $C$ associated with an on-shell diagram: that is, the positroid volume-form on the Grassmannian---when expressed in these coordinates---is simply,
\vspace{-.2cm}\eq{d\!\log(\hat{f}_1)\wedge d\!\log(\hat{f}_2)\wedge\cdots\wedge d\!\log(\hat{f}_d),\vspace{-.2cm}}
where we have defined the {\it rescaled} coordinates $\hat{f}_i\!\equiv\! f_i/f_0$. Importantly, this measure is invariant (up to a sign) under square-moves.

Now that we allow mutations that are {\it not} square moves and obtain quivers which are not related to planar on-shell diagrams, one could ask if there is any analog of `face variables' or the volume form on the Grassmannian. It turns out that in fact there are two sets of coordinates that can be defined in general, and one of them coincides with the face variables described in \mbox{section \ref{boundary_measurements_section}} when restricted to on-shell diagrams. These are known as the cluster ${\cal A}$-coordinates and the cluster ${\cal X}$-coordinates.

Given a seed ${\bf s}$ the two sets of coordinates are each parameterized by the set $S$---the set of `faces' in the case of planar on-shell diagrams. Let us denote the ${\cal X}$-coordinates by $\{X_i\}$ and the ${\cal A}$-coordinates by $\{A_i\}$. (This is a slight abuse of notation because we have suppressed their dependence on the seed $\mathbf{s}$.) {\it A priori}, one makes a proposal for what the coordinates are in terms of the data defining the problem at hand, and then the kind of variables will be determined by their behavior under mutations.

Given a mutation of seeds $\mu_k:\mathbf{s}\!\mapsto\!\mathbf{s}'$, the cluster coordinates assigned to these seeds are related as follows. If we denote the cluster coordinates related to
the seed ${\bf s}$ by $X_i$ and $A_i$, and the ones assigned to the seed ${\bf s}'$ by $X_i'$ and $A_i'$, then we have mutation formulae for ${\cal A}$-coordinates
\vspace{-.2cm}\eq{A_{k}A'_{k}\equiv \prod_{j| \varepsilon_{kj} >0} A_{j}^{\varepsilon_{kj}} + \prod_{j| \varepsilon_{kj} <0} A_{j}^{-\varepsilon_{kj}},\quad\mathrm{and}\quad A'_{i} =  A_{i}\text{ for } i\neq k;\vspace{-.2cm}\label{5.11.03.1a}}
and for ${\cal X}$-coordinates we have,
\vspace{-.2cm}\eq{X'_{i}\equiv\left\{\begin{array}{@{}l@{$\qquad$}l} X_k^{-1}&i=k; \\ X_i\left(1+X_k^{-{\rm sgn} (\varepsilon_{ik})}\right)^{-\varepsilon_{ik}} & i\neq k.\end{array} \right.\label{5.11.03.1x}\vspace{-.2cm}}
(In (\ref{5.11.03.1a}), if any of the sets $\{j| \varepsilon_{kj} >0\}$ or $\{j| \varepsilon_{kj} < 0\}$ is empty, the corresponding monomial is $1$.)

The set of transformations among cluster coordinates obtained by composing mutations are known as {\it cluster transformations}.

Cluster ${\cal A}$-coordinates and mutation formulae are the main ingredients of the definition of cluster algebras of Fomin and Zelevinsky, \cite{FZ}. So far in this paper these coordinates have not made an appearance but they can be nicely defined for on-shell diagrams using left-right paths as follows. For each left-right path take the end point label and write it on all faces to the left of the path. If the on-shell diagram is reduced and represents a cell of $G(k,n)$, this procedure provides each face with $k$-labels. The Pl\"{u}cker coordinates given by the sets of labels in each face provides a set of cluster ${\cal A}$-coordinates. These coordinates are generically non-vanishing and, under a square move,  indeed transform as ${\cal A}$-coordinates.

Cluster ${\cal X}$-coordinates and mutation formulae (\ref{5.11.03.1x}) describe a dual object, introduced in \cite{FG2} under the name {\it cluster ${\cal X}$-variety} or {\it cluster Poisson variety}. The transformation (\ref{5.11.03.1x}) is precisely the way face variables were found to transform in \mbox{section \ref{coordinate_transformations_induced_by_moves}}.  So, our face variables are an example of cluster ${\cal X}$-coordinates.

One of the key features of cluster ${\cal X}$-varieties is that they are endowed with a natural Poisson structure. We can define
\vspace{-.2cm}\eq{\{X_i, X_j\}\equiv \varepsilon_{ij}X_iX_j.\vspace{-.2cm}}
Of course, the crucial fact is that the cluster transformations (\ref{5.11.03.1x}) preserve the Poisson bracket. This fascinating structure has made an appearance in a number of physical settings, but has not yet  played a role in our understanding of scattering amplitudes.

Since the cluster transformations are given by subtraction free formulas,
they identify the sets of points with the real positive cluster coordinates
assigned to different seeds. Gluing these sets according to the
cluster transformations,
we arrive at the spaces ${\cal A}^+$ and respectively ${\cal X}^+$ on
which all cluster coordinates are perfectly well defined and take
positive real values.

The Grassmannian is an example of a cluster ${\cal X}$-variety, and
the corresponding space of positive points  ${\cal X}^+$ is nothing
else but the Positive Grassmannian.

It turns out that just as for face variables in on-shell diagrams, one can define canonical volume forms for both kind of coordinates \cite{FG4}. Given a seed ${\bf s}$, consider the volume forms
\vspace{-.2cm}\eq{{\rm vol}^{\bf s}_{\cal A}\equiv d\!\log A_1 \wedge \cdots \wedge d\!\log A_n,\qquad{\rm vol}^{\bf s}_{\cal X}\equiv d\!\log X_1 \wedge \cdots \wedge d\!\log X_n.\vspace{-.2cm}}
It is an easy but fundamental fact that the cluster transformations preserve the corresponding volume form up to a sign. Precisely, given a seed mutation ${\bf s}\!\mapsto\!{\bf s}'$, we have
\vspace{-.2cm}\eq{{\rm vol}^{\bf s'}_{\cal A} = -{\rm vol}^{\bf s}_{\cal A}, \qquad {\rm vol}^{\bf s'}_{\cal X} = - {\rm vol}^{\bf s}_{\cal X}.\vspace{-.2cm}}
To check the first identity, let us do a mutation at $k$. Then only the coordinate $A_k$ changes, and due to the exchange relation (\ref{5.11.03.1a}), one has
\vspace{-.2cm}\eq{d\!\log A'_k + d\!\log A_k =0 ~~\mbox{mod $dA_j$, where $j \not = k$}. \vspace{-.2cm}}
To check the second, notice that under a mutation at $k$, one has $d\!\log X_k'=\mi\, d\!\log X_k$, while $d\!\log X_j' = d\!\log X_j$ modulo $dX_k$. These forms are known as the ${\cal A}$- and ${\cal X}$-cluster volume forms. Of course, our top-form on the positive Grassmannian precisely coincides with the ${\cal X}$ volume form.\\[-7pt]

\noindent{\bf Singularities of the cluster volume-form and frozen variables.}\\Take a variety equipped with a cluster ${\cal A}$-coordinate system $\{A_i\}$. Let us assume that $k\!\in\!(S\raisebox{1pt}{\text{{\footnotesize$\backslash$}}} S_0)$ is non-frozen, and $\varepsilon_{kj}\not =0$ for some $j$. Then the residue of the cluster volume form ${\rm vol}_{\cal A}$ at the locus $A_k=0$ is zero:
\vspace{-.2cm}\eq{{\rm Res}_{A_k=0}({\rm vol}_{\cal A})=0.\vspace{-.2cm}}
Indeed, the residue  is given by ${\rm Res}_{A_k=0}({\rm vol}_{\cal A}) = \pm \bigwedge_{i\not = k}d\!\log A_i$. Since $k$ is non-frozen, there is an exchange relation (\ref{5.11.03.1a}). It implies a monomial relation on the locus $A_k=0$:
\vspace{-.2cm}\eq{\prod_{j} A_{j}^{\varepsilon_{kj}}= -1.\label{monrelA}\vspace{-.2cm}}
Since $\varepsilon_{kj}$ is not identically zero, the monomial in (\ref{monrelA}) is nontrivial. This implies that the form $\bigwedge_{i\not = k}d\!\log A_i$ vanishes at the $A_k=0$ locus.

This explains the role of frozen variables in  a cluster coordinate system $\{A_i\}$ on a space $M$. Indeed, a coordinate $A_k$, with $\varepsilon_{kj}\not =0$ for some $j$, can be declared non-frozen only if ${\rm Res}_{A_k=0}(d\!\log A_1 \wedge \cdots \wedge d\!\log A_n)=0$. This condition just means that the functions $A_1,\ldots, \widehat A_k, \ldots, A_n$ become dependent on every component of the $A_k=0$ locus.

\newpage 
\subsection{Cluster Amalgamation}
The ``atomic'' principle in which complicated objects and their properties can all be simply derived from constituent building blocks has played a fundamental role in understanding on-shell diagrams, and  is more generally making an appearance more and more often in both physics and mathematics. Given the potential importance of this phenomenon, let us now describe the more general procedure of amalgamation of cluster structures, of which our sense of amalgamation is a special case.  We find it convenient to use a different but equivalent description of seeds which is known as the geometric description. A definition of amalgamation via the combinatorial description of seeds is given in \cite{FG3}.

The following geometric definition is taken from \cite{FG2}:\\[-7pt]

\noindent{\bf Definition:} A {\it seed} is a set of combinatorial data $\mathbf{s}\!=\!\big\{\Lambda,\Lambda_0,\{e_i\},\varepsilon\big\}$, where $\Lambda$ is a free abelian group, $\Lambda_0$ a distinguished subgroup of $\Lambda$, $\{e_i\}$ is a basis of $\Lambda$ such that $\Lambda_0$ is generated by a subset of {\it frozen} basis vectors, and $\varepsilon_{i,j}\!\equiv\!\varepsilon(e_i,e_j)$ is a skew-symmetric bilinear form on $\Lambda$ such that $\varepsilon_{i,j}\!\in\!\mathbb{Z}$ unless {\it both} of $(e_i,e_j)\!\in\!\Lambda_0$, in which case $\varepsilon_{i,j}\in\frac{1}{2}\mathbb{Z}$. \\[-7pt]

To see that this definition of a seed is equivalent to the previous one, note that given a combinatorial data $\{S,S_0,\varepsilon\}$, the abelian group $\Lambda$ is the free abelian group generated by the set $S$, where the generator $e_i$ is the one assigned to an element $i\!\in\!S$.  The subgroup $\Lambda_0$ is then generated by the subset $S_0$, and the bilinear form $\varepsilon(\cdot,\cdot)$ is defined as above. Vice versa,
given a $\{\Lambda, \Lambda_0, \{e_i\}, \varepsilon\}$ data, the set $S$ is the set parameterizing the basis vectors, etc.

Given this geometric definition, this is a good point to mention that mutations can be interpreted as half-reflections. This bring us again closer to the known description of Seiberg duality in quiver gauge theories as Weyl reflections where coupling constants in the form $1/g_i^2$ transform as root vectors $e_i$. In full generality we have that the  seed ${\bf s}'$ obtained from ${\bf s}$ by the mutation in the direction $k$ is defined by changing the basis $\{e_i\}$ (the rest of the data stays the same). The new basis $\{e'_i\}$ is defined as a half-reflection of the one $\{e_i\}$ along the hyperplane $\varepsilon(e_k, \cdot)=0$:
\vspace{-.2cm}\eq{e'_i\equiv\left\{ \begin{array}{lll} e_i + [\varepsilon_{ik}]_+e_k& \mbox{ if } &  i\not = k\\-e_k& \mbox{ if } &  i = k.\end{array}\right.\label{12.12.04.2a}\vspace{-.2cm}}
Here we set  $[\alpha]_+\equiv\alpha$ if $\alpha\geq 0$ and $[\alpha]_+\equiv0$ otherwise. One can check that formula (\ref{12.12.04.2a}) amounts to formula (\ref{5.11.03.6}), telling how the $\varepsilon$-matrix changes under mutations.

By definition, the frozen/non-frozen basis vectors of the mutated seed are the images of the  frozen/non-frozen basis vectors of the original seed. The composition of two mutations in the same direction $k$ is no longer the identity, but rather an isomorphism of seeds.

We are now ready to turn to the amalgamation procedure.
Take a pair of seeds, where we emphasize the set of frozen basis vectors $\{f_i\}$:
\vspace{-.2cm}\eq{{\mathbf s}' = \Big\{\Lambda', \varepsilon', \{e'_{i}\}, \{f'_{j}\}\Big\},\qquad{\mathbf s}'' = \Big\{\Lambda'', \varepsilon'', \{e''_{i}\}, \{f''_{j}\}\Big\}. \vspace{-.2cm} \label{thepair}}
First, we define their direct product according to:
\vspace{-.2cm}\eq{{\mathbf s}'\!\otimes\!{\mathbf s}''\equiv \Big\{\Lambda,\varepsilon, \{e_{i}\}, \{f_{j}\}\Big\},\vspace{-.2cm}}
where $\Lambda\!\equiv\!\Lambda'\!\oplus\!\Lambda''$, and the form $\varepsilon$ is defined to be the direct product of the forms $\varepsilon'$ and $\varepsilon''$. The basis vectors and the frozen ones, are inherited in an obvious way.

Next, given a seed ${\mathbf s} = \Big\{\Lambda,\varepsilon, \{e_{i}\}, \{f_{j}\}\Big\}$, and a pair of frozen basis vectors $f_a$ and $f_b$, we define the {\it reduced seed}
\vspace{-.2cm}\eq{{\mathbf s}_{a \ast  b} = \Big\{\Lambda_{a \ast  b}, \varepsilon_{a \ast  b}, \{e_{s}\}, \{f_{t}\}\Big\}.\vspace{-.2cm}}
Here $\Lambda_{a \ast  b}$ is a co-rank one subgroup of $\Lambda$ whose basis vectors are $(f_a\pl\,f_b)$ and those of $\Lambda$ different from $f_{a}$ and $f_b$, and the vectors  $(f_a\pl\,f_b)$ and $f_j\notin\{f_a,f_b\}$ are the frozen ones; the form $\varepsilon_{a \ast  b}$ is the restriction to $\Lambda_{a \ast  b}$ of the form $\varepsilon$ on $\Lambda$.

Given a pair of seeds (\ref{thepair}), a pair of subsets $\{f'_a\}, a\in A$ and $\{f''_b\}, b\in B$ of the frozen basis vectors in ${\mathbf s}'$ and ${\mathbf s}''$, and an isomorphism of sets $\varphi: A\!\to\!B$, we define the {\it amalgamation ${\mathbf s}'\ast_{\varphi}{\mathbf s}''$} of the seeds ${\mathbf s}'$ and ${\mathbf s}''$ along $\varphi$. This is done in a few steps:\\[-25pt]
\begin{enumerate}
\item take the direct product ${\mathbf s}'\!\otimes\!{\mathbf s}''$;\\[-25pt]
\item perform the reduction along a pair of frozen vectors $f'_a$ and $f''_{\varphi(a)}$, for each $a\in A$;\\[-25pt]
\item if the restriction of the form $\varepsilon$ of the seed ${\mathbf s}'\ast_{\varphi}{\mathbf s}''$ to the basis vectors $f'_a\pl\,f''_{\varphi(a)}$ with $a\in A$ takes values in $\mathbb{Z}$, then {\it defrost} these basis vectors by declaring them to be {\it unfrozen} (meaning that we now allow mutations at these vectors).\\[-25pt]
\end{enumerate}
The first two steps amounts to taking the subgroup of $\Lambda'\oplus \Lambda''$ generated by the vectors $f'_a + f''_{\varphi(a)}$, $a\in A$, and the rest of the basis vectors, and inducing on it a seed structure. The amalgamation of seeds evidently commutes with the seed mutations.

The cluster coordinates $X_i$ on the set $S$ are related to the ones $X'_i$ and $X''_i$ by:
\vspace{-.2cm}\eq{X_i\equiv\left\{\begin{array}{@{}l@{$\qquad\;\;\;\,\,$}l}X_i'&i\!\in\!(S'\raisebox{1pt}{\text{{\footnotesize$\backslash$}}} A);\\X_i''&i\!\in\!(S''\raisebox{1pt}{\text{{\footnotesize$\backslash$}}} \varphi(A));\\X_a'X''_{\varphi(a)}&i=a\!\in\!A.\end{array}\right.\vspace{-.2cm}\label{restrx}}
It is easy to check that the amalgamation respects the Poisson structure. For the cluster ${\cal A}$-coordinates, we have
\vspace{-.2cm}\eq{A_i\equiv\left\{\begin{array}{@{}l@{$\quad$}l}A_i'&i\!\in\!(S'\raisebox{1pt}{\text{{\footnotesize$\backslash$}}} A);\\A_i''&i\!\in\!(S''\raisebox{1pt}{\text{{\footnotesize$\backslash$}}} \varphi(A));\\A_a'\big(=A''_{\varphi(a)}\big)&i=a\!\in\!A.\end{array}\right.\vspace{-.2cm}\label{restra}}

The algebra generated by the cluster ${\cal X}$-coordinates of the amalgamated seed embeds to the product of similar algebras assigned to the original seeds via formulae (\ref{restrx}).

Contrary to this, the algebra generated by the cluster ${\cal A}$-coordinates of the amalgamated seed is the quotient of the product of the similar algebras assigned to the original seeds: we impose the relations  $A'_a = A''_{\varphi(a)}$.

\newpage
\subsection{Brief Overview of the Appearance of Cluster Structures in Physics}

The theory of cluster algebras had it origins in a very unexpected area: the study of totally positive square matrices. This investigation began in the 1930's with Gantmacher and Krein, \cite{GKr}, and Schoenberg, \cite{Sch}, and had immediate applications to the theory of oscillators in classical mechanics.

The notion of total positivity was vastly generalized by Lusztig, \cite{L2}, to the case of arbitrary split real reductive groups $G$. Lusztig defined the positive part of group $G_{>0}$ by using the Chevalley generators. The study of total positivity, related parameterizations and canonical bases in simple Lie groups theory led to discovery of cluster algebras, \cite{FZ}.

A crucial feature of cluster Poisson varieties in connection with physics is that they provide a very general example of {\it non-perturbative} quantization, \cite{FG4}.

Recall that to quantize a Poisson space $(X, \{,\})$ means to deform its algebra of functions to a non-commutative algebra ${\cal O}_g({X})$, depending on a ``coupling'' constant $g>0$ (normally referred to in the literature as ``$\hbar$'') so that $\widehat a\widehat b\,\mi\,\widehat b \widehat a = g \{a,b\}\pl\ldots$, and represent the algebra ${\cal O}_g({X})$  by operators in a Hilbert space. The Heisenberg quantization does this for a flat space with canonical coordinates $(p_i, q_i)$. Kontsevich, \cite{Kontsevich}, proved that a perturbative version of the algebra ${\cal O}_g({X})$, where the dependence on the coupling $g$ is via formal power series always exists.

Any cluster Poisson variety ${\cal X}$ admits a  {non-perturbative quantization},  which is manifestly invariant under the ``S-duality'' $g\!\to\! g^{-1}$. It comes with a series of $\ast$-representations in  Hilbert spaces, modeled on in a single Hilbert space $L^2({\cal A}^+, \Omega_{\cal A})$, defined using the space ${\cal A}^+$ of positive real points of the dual cluster variety ${\cal A}$, and the canonical cluster volume form $\Omega_{\cal A}$ providing the Lebesgue measure there.

Many (if not most) interesting examples of cluster structures appear when one couples a reductive Lie group $G$ to a topological surface $S$, studying moduli spaces of flat $G$-bundles on a topological surface $S$ and related moduli spaces introduced and studies in \cite{FG1}. The corresponding spaces of positive real points are the Higher Teichm\"{u}ller spaces related to the pair $(G, S)$.

Let us give a broad-view description of how the non-perturbative cluster quantization, and in particular the  canonical cluster volume form play a crucial role in defining a Hilbert space that has made an appearance now several times in quantum field theory.

The staring point is a {\it decorated surface} $S$. Let $S$ be a surface with or without boundary, and a finite collection of points on the boundary, considered modulo isotopy. For example, a disc with $n$ points on the boundary is the topological background for the $n$-particle scattering amplitudes.

Given $S$ and a split reductive Lie group $G$, there are two moduli spaces defined in \cite{FG1} closely related to the moduli space of $G$-bundles with flat connections on $S$:\\[-6pt] 
\vspace{-.2cm}\eq{{\cal X}_{G, S}\quad\mbox{and}\quad {\cal A}_{G, S}.\label{dual}\vspace{-.2cm}} 
The first is equipped with an ${\cal X}$-cluster atlas, and the second with an ${\cal A}$-cluster atlas.

This immediately implies that the sets ${\cal X}^+_{G, S}$ and ${\cal A}^+_{G, S}$ of real positive points of these spaces are defined. This is the dual pair of higher Teichm\"{u}ller spaces assigned to $(G,S)$.

As described in \cite{FG2}, the existence of the cluster atlas on the space ${\cal X}_{G, S}$ implies that the algebra ${\cal O}({\cal X}_{G, S})$ of regular functions on this space admits a canonical non-commutative $q$-deformation to a $\ast$-algebra ${\cal O}_q({\cal X}_{G, S})$, where $q\!=\!\exp(\pi i g)$ (for $g>0$). It is invariant under the action of the mapping class group of $S$.

The {\it modular double} of the algebra ${\cal O}_q({\cal X}_{G, S})$ is the tensor product of the original $\ast$-algebra with the coupling $g$, and the $\ast$-algebra related to the Langlands dual group $G^L$ at the ``inverse'' $1/(d_G g)$ of the coupling ($d_G=1$ for simply-laced group $G$): \vspace{-.2cm}\eq{{\cal O}_q({\cal X}_{G, S}) \bigotimes {\cal O}_{q^\vee}({\cal X}_{G^L, S}),\vspace{-.2cm}} where $q = {\rm exp}(\pi i g)$ and $q^\vee = {\rm exp}(\pi i /d_G g)$. 

It follows from the general result on quantization of cluster varieties proved in \cite{FG4} that the modular double admits a series of $\ast$-representations in Hilbert spaces ${\cal H}_\chi$, depending on a parameter $\chi \in \mathbb{R}^m$.   Here $m$ is the dimension of the center of the Poisson bracket on the space ${\cal X}_{G, S}$, and $\chi$ is a unitary character of the center of the algebra ${\cal O}_q({\cal X}_{G, S})$. The Hilbert spaces ${\cal H}_\chi$ are expected to be the spaces of conformal blocks for higher Toda theories.

On the other hand, since the space ${\cal A}_{G, S}$ has a cluster ${\cal A}$-atlas, it carries the canonical volume form $\Omega_{\cal A}$. The latter restricts to a canonical volume form on the positive real space  ${\cal A}^+_{G, S}$. Therefore we arrive at a canonical Hilbert space assigned to the pair $(G, S)$: 
\vspace{-.2cm}\eq{L^2({\cal A}^+_{G, S}, \Omega_{\cal A}).\vspace{-.2cm}} 
The mapping class group of $S$ acts by its unitary symmetries. It was proved in \cite{FG4} that this Hilbert space is the integral of the spaces of operators acting in the Hilbert spaces ${\cal H}_\chi$: \vspace{-.2cm}\eq{L^2({\cal A}^+_{G, S}, \Omega_{\cal A}) = \int {\cal H}^*_\chi\otimes{\cal H}_\chi d\chi.\vspace{-.2cm}}

We conclude that the positive structure and the canonical cluster volume form on the space  ${\cal A}_{G, S}$ provide us with the Hilbert space $L^2({\cal A}^+_{G, S}, \Omega_{\cal A})$ describing the conformal blocks.

As we have seen in this paper, it is quite amazing that exactly the same data---the positive structure and the canonical cluster volume form---for the Grassmannian $G(k,n)$ provides us the measure for the integrand of the scattering amplitudes in $\mathcal{N}\!=\!4$ SYM. It is even more striking that there are structures crucial for each of these stories which have not made an appearance in the other: we need the quantized dual ${\cal X}$-moduli space in one, and the rich external kinematic data in the other. This strongly suggests that a deeper study is bound to reveal the roles the ``missing structures'' in each of the stories and lead to a beautiful unified picture.

\newpage

\section{On-Shell Representations of Scattering Amplitudes}\label{on_shell_scattering_amplitudes_section}
\vspace{-.2cm}
Although we have focused on understanding individual on-shell diagrams for most of the paper, let us return to a study of how these can combine to entire scattering amplitudes. As discussed in \mbox{section \ref{on_shell_diagrams_section}}, the defining property of the full amplitude is that it satisfies the ``differential equation'',
\vspace{-.3cm}\eq{\hspace{-4.1cm}\raisebox{-41pt}{\includegraphics[scale=1]{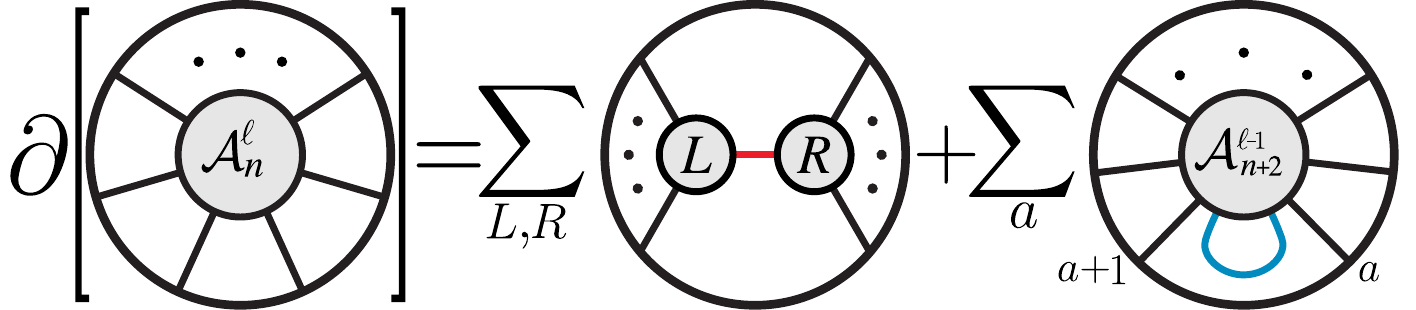}}\hspace{-3cm}\label{general_amplitude_boundary_with_discs}\vspace{-.3cm}}
which specifies the two kinds of singularities it can have---corresponding to ``factorization channels'' (red) and ``forward limits'' (blue), respectively. All known representations of scattering amplitudes can be thought of as particular ways of building objects with these---and {\it only} these---(co-dimension one) singularities.

The usual Feynman-diagrammatic expansion for scattering amplitudes makes these singularities (together with conformal invariance) manifest, but at the cost of introducing  unphysical, off-shell variables and gauge-redundancies which obscure the underlying Yangian-invariance of the theory. (The same can be said for the equivalent Wilson-loop representation---except that it is  the {\it dual} conformal symmetry which is made manifest.) By contrast, the BCFW recursion relations,
\vspace{-.5cm}\eq{\hspace{-3.7cm}\raisebox{-54pt}{\includegraphics[scale=1]{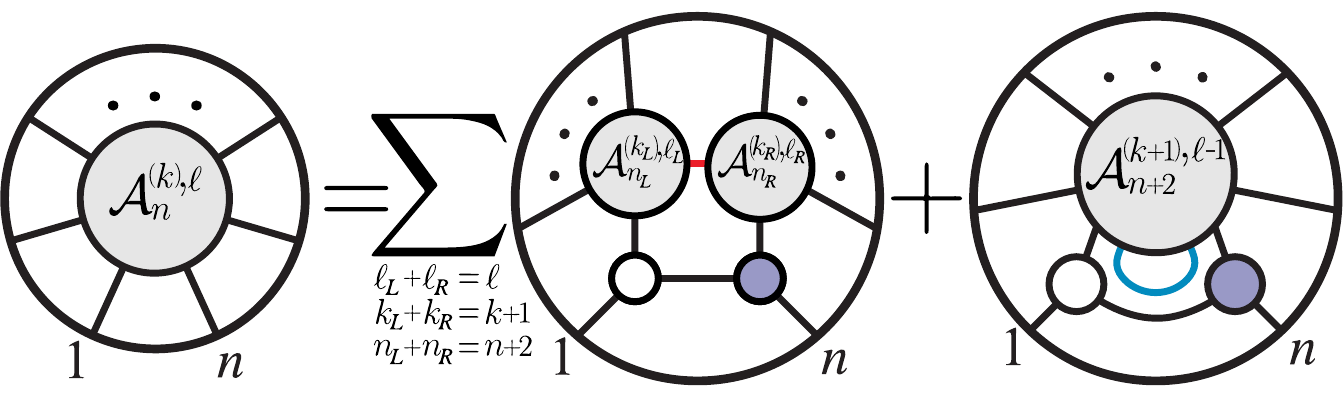}}\label{bcfw_all_loop_recursion_disc}\hspace{-3cm}\vspace{-.4cm}}
can be understood of as a {\it direct} integration of the defining equation (\ref{general_amplitude_boundary_with_discs}), and provides us with a representation of scattering amplitudes for which {\it every term} enjoys the full Yangian-invariance of the theory. However, the recursion requires that two legs be singled-out to play a special role---in (\ref{bcfw_all_loop_recursion_disc}), these are the legs $(n\,1)$. Although this choice is arbitrary, it breaks the cyclic-symmetry of the complete amplitude, and makes manifest only a rather small subset of the singularities required by (\ref{general_amplitude_boundary_with_discs}).

Of course, the BCFW recursion relations can be derived from field theory, starting either with the ``scattering amplitude'' \cite{ArkaniHamed:2010kv} or ``Wilson loop'' \cite{CaronHuot:2010ek,Mason:2010yk,Bullimore:2011ni} pictures (for the relation to light-like correlation functions, see \cite{Alday:2010zy,Eden:2010zz,Eden:2010ce,Eden:2011yp,Eden:2011ku}). We will however begin by showing how they can also be proven directly by induction. That is, we will show that the boundary of (\ref{bcfw_all_loop_recursion_disc}) includes {\it precisely} the singularities required by (\ref{general_amplitude_boundary_with_discs}); this proof will be entirely {\it diagrammatic}. In \mbox{section \ref{tree_amplitudes_and_basic_structures}} we will review some important features encountered in the tree-level $(\ell=0)$ version of the recursion relations, and in \mbox{section \ref{canonical_coordinates_for_loop_integrands}} we will see how the structure of tree amplitudes is reflected at loop-level, giving rise to a canonical---purely `$d\!\log$'---form for all loop-integrands.

\subsection{(Diagrammatic) Proof of the BCFW Recursion Relations}\label{proof_of_the_recursion_relations}
Let us take the BCFW recursion relations, (\ref{bcfw_all_loop_recursion_disc}) as an ansatz, and demonstrate inductively that its boundary includes {\it all} the correct factorization channels and forward limits, and no other singularities (for earlier work along these lines see \cite{Schuster:2008nh,He:2008nj}). Recall that the four-point, tree-amplitude, $\mathcal{A}^{(2),\ell=0}_{4}$, manifestly has all the correct factorization channels in its boundary,
\vspace{-.2cm}\eq{\hspace{-3.15cm}\begin{array}{r}\raisebox{-40pt}{\includegraphics[scale=.85]{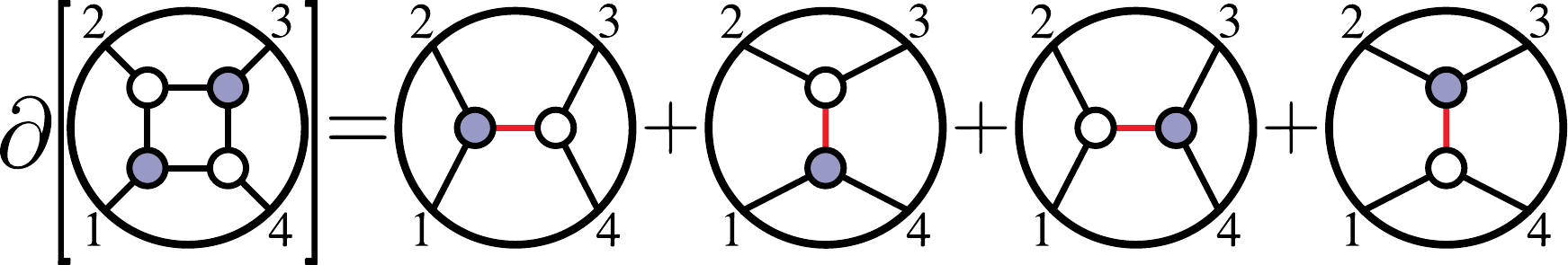}}\\[-4pt]\text{{\small${\color{perm}\{3,4,5,6\}}\hspace{1.5cm}{\color{perm}\{3,\,}{\color{red}\mathbf{5}}{\color{perm},\,}{\color{red}\mathbf{4}}{\color{perm},6\}}\hspace{1.25cm}{\color{perm}\{}{\color{red}\mathbf{4}}{\color{perm},\,}{\color{red}\mathbf{3}}{\color{perm},5,6\}}\hspace{1.25cm}{\color{perm}\{}{\color{red}\mathbf{2}}{\color{perm},4,5,\,}{\color{red}\mathbf{7}}{\color{perm}\}}\hspace{1.25cm}{\color{perm}\{3,4,\,}{\color{red}\mathbf{6}}{\color{perm},\,}{\color{red}\mathbf{5}}{\color{perm}\}}\hspace{.25cm}$}}\end{array}\vspace{-.3cm}\hspace{-3cm}\nonumber}

We may therefore suppose that the ansatz is correct for all amplitudes $\mathcal{A}_{\hat{n}}^{(\hat{k}),\hat{\ell}}\vspace{-.0cm}$ with $\vspace{-.0cm}\hat{n}<n$, $\vspace{-.0cm}\hat{k}\leq k$, and $\hat{\ell}\leq\ell$; we must show that this suffices to prove that it also holds for $\mathcal{A}^{(k),\ell}_{n}$. We may divide the argument into two parts: first, demonstrating that the boundary includes all the correct factorization channels; and then showing that it includes all the correct forward-limits.

Among the factorization channels, those for which particles $1$ and $n$ are on opposite sides are trivially present:
\vspace{-.3cm}\eq{\hspace{-3cm}\raisebox{-52pt}{\includegraphics[scale=0.825]{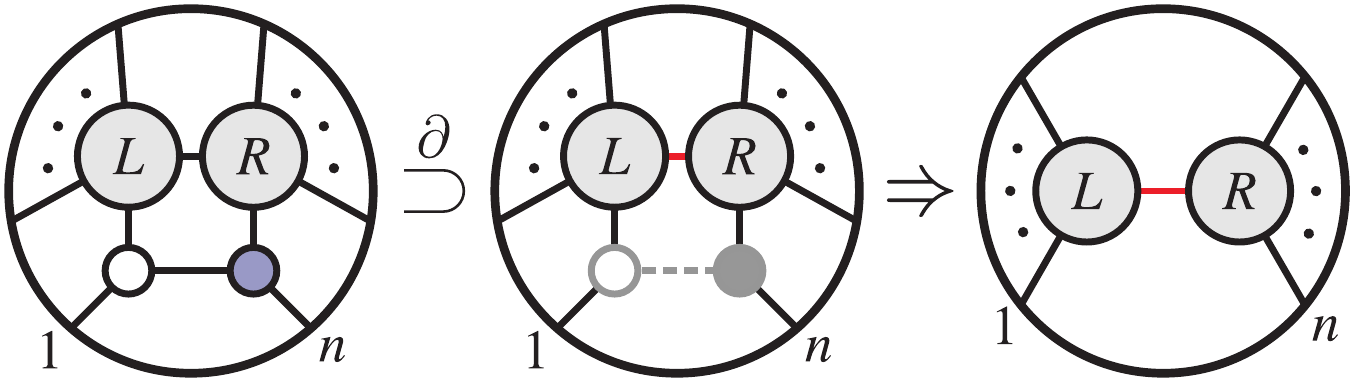}}\hspace{-3cm}\nonumber\vspace{-.2cm}}
What we first need to check is that the BCFW recursion formula also generates all those factorizations for which $1$ and $n$ are on the same side. Factorization channels for which legs $1$ and $n$ are not alone on one side arise from the factorizations of the bridged amplitudes. For example, the boundaries of the left-amplitudes include:
\vspace{-.4cm}\eq{\hspace{-2cm}\raisebox{-69pt}{\includegraphics[scale=0.825]{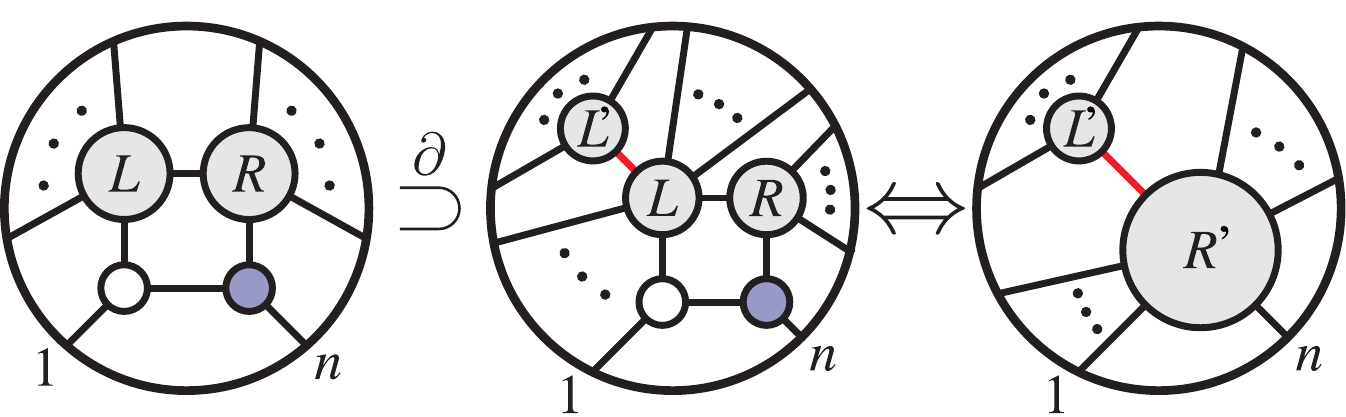}}\vspace{-.1cm}\nonumber\hspace{-2cm}}
where we have used our induction hypothesis to identify the terms appearing on the right-side of the factorization as a lower-point amplitude denoted $R$'. We also have the analogous diagrams arising from the right-amplitudes.

The case of a two-particle factorization involving {\it just} $1$ and $n$ together, however, arises somewhat differently. The factorization for which particles $1$ and $n$ are connected via a $\mathcal{A}_3^{(1)}$-vertex arises from the boundary,
\vspace{-.3cm}\eq{\hspace{-3cm}\raisebox{-57pt}{\includegraphics[scale=0.825]{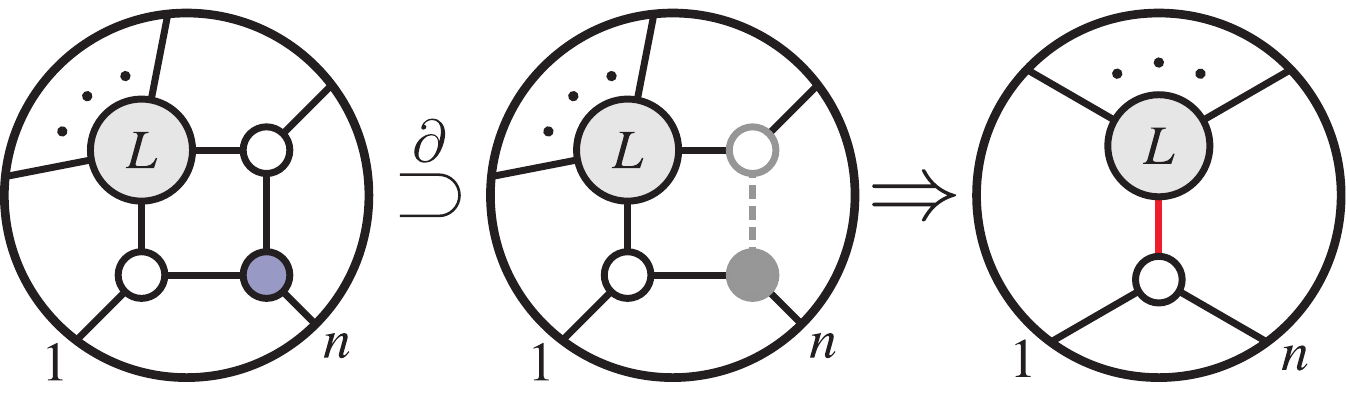}}\hspace{-3cm}\nonumber\vspace{-.4cm}}
Similarly, the case where particles $1$ and $n$ are connected via a $\mathcal{A}_3^{(2)}$-vertex arises from,
\vspace{-.3cm}\eq{\hspace{-3cm}\raisebox{-57pt}{\includegraphics[scale=0.825]{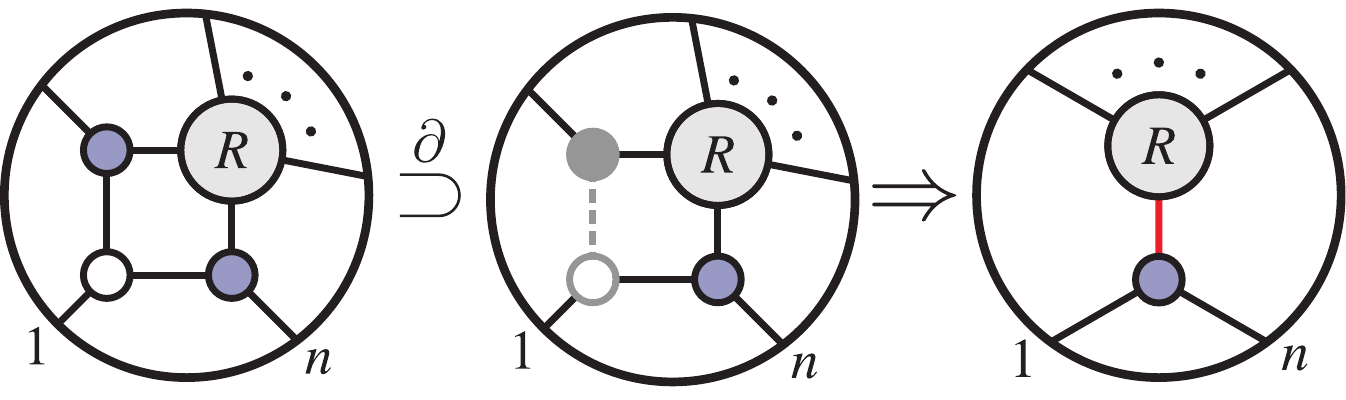}}\hspace{-3cm}\nonumber\vspace{-.3cm}}

We have therefore shown that all factorization channels are present in the boundary of the BCFW ansatz. However, we must also show that these are the {\it only} such boundaries. Our induction hypothesis would suggest that such `spurious' poles could arise from factorizations of separating $(\hat{1}\,I)$ on the left, or $(I\,\hat{n})$ on the right:
\vspace{-.3cm}\eq{\hspace{-3cm}\raisebox{-57pt}{\includegraphics[scale=0.825]{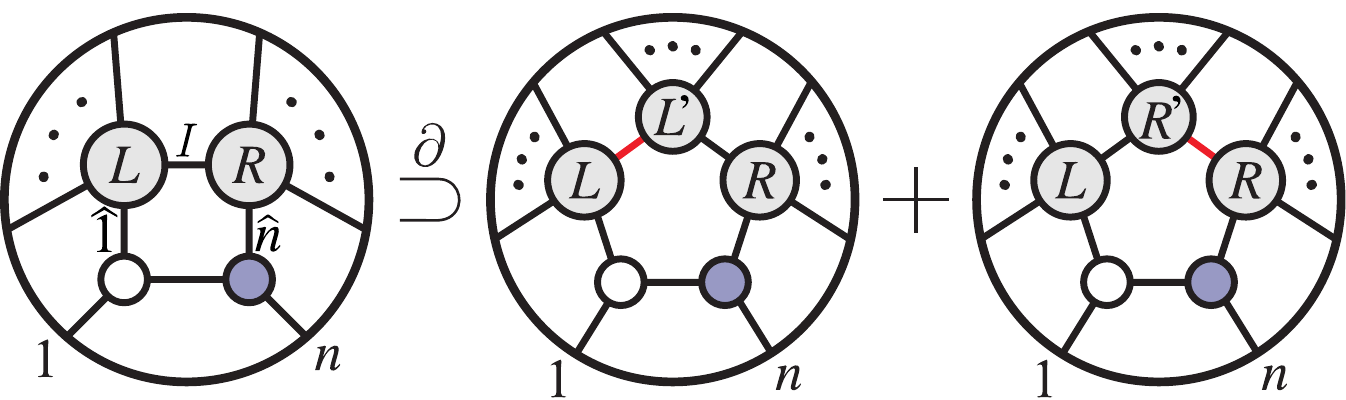}}\hspace{-3cm}\nonumber\vspace{-.3cm}}
Conveniently, such boundaries are {\it always} generated symmetrically from the left- and right-amplitudes, and cancel in the sum.

Let us now demonstrate that the BCFW recursion ansatz generates all the correct forward-limits as co-dimension one boundaries---and {\it only} these. As with the factorization channels, the BCFW recursion ansatz always makes one of the forward-limits manifest---those where the forward limit is taken between $1$ and $n$:
\vspace{-.2cm}\eq{\hspace{-3cm}\raisebox{-12pt}{\includegraphics[scale=0.825]{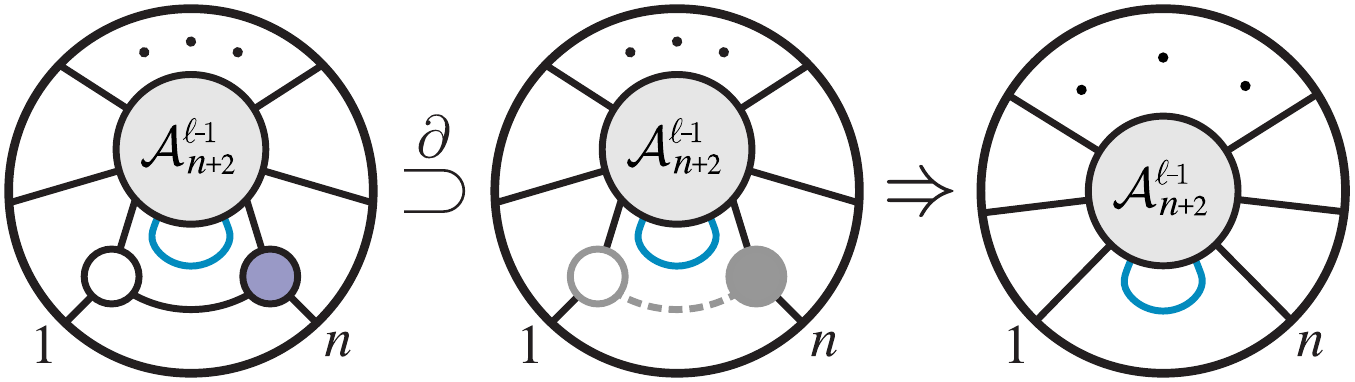}}\hspace{-3cm}\nonumber\vspace{-.2cm}}
When the identified legs are not between $(n\,1)$, but say $(a\,a\pl1)$, something more interesting happens. Some of these arise trivially from the boundary of `bridged' terms in the recursion,
\vspace{-.2cm}\eq{\hspace{-2.5cm}\raisebox{-40pt}{\includegraphics[scale=0.825]{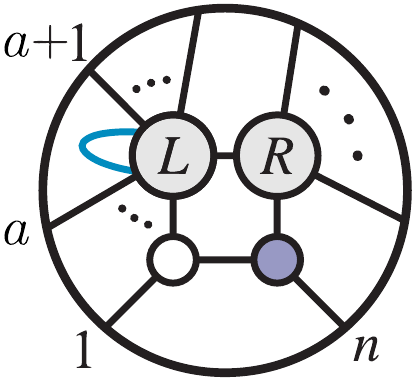}}\quad\mathrm{and}\quad\raisebox{-40pt}{\includegraphics[scale=0.825]{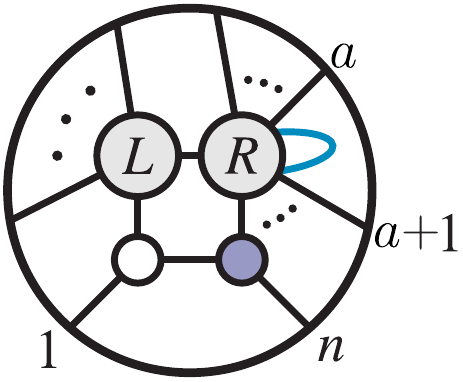}}\hspace{-3cm}\nonumber\vspace{-.1cm}}
but these terms alone do not represent the complete BCFW-representation of the lower-loop, higher-point amplitude including the identified legs: the problem is that we are missing both the terms where the identified legs (before the forward-limit) are separated across the BCFW-bridge, and also the terms for which they are identified in the `forward-limit' term. By our induction hypothesis, both of these terms arise from the boundary of the forward-limit term: as factorization and forward-limit boundaries of the forward-limit term, respectively:
\vspace{-.3cm}\eq{\hspace{-3cm}\raisebox{-57pt}{\includegraphics[scale=0.85]{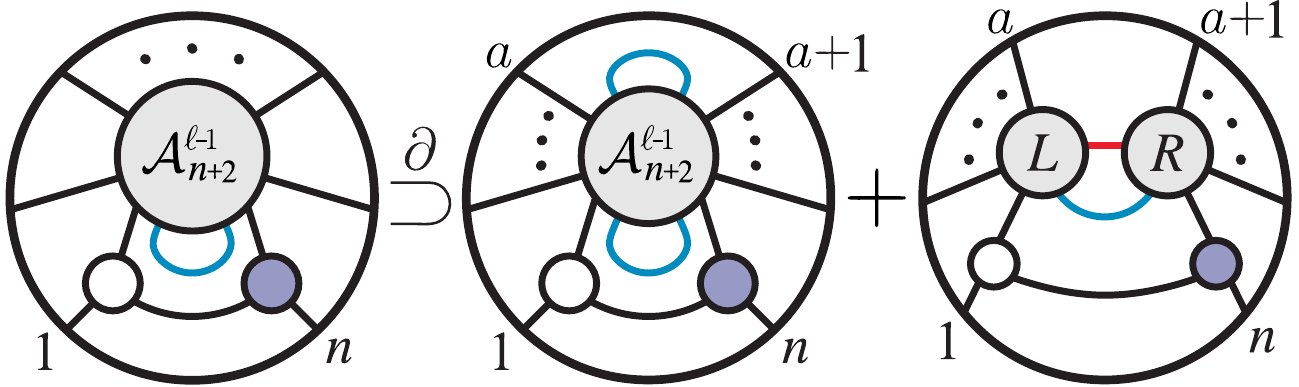}}\hspace{-3cm}\nonumber\vspace{-.2cm}}
The first of these is needed by `forward-limit'  term in the BCFW recursion ansatz, and the second term is needed to complete the `bridge' term of the recursion ansatz; to see this more clearly, notice that the second term can be redrawn more suggestively:
\vspace{-.5cm}\eq{\hspace{-3cm}\raisebox{-40pt}{\includegraphics[scale=0.85]{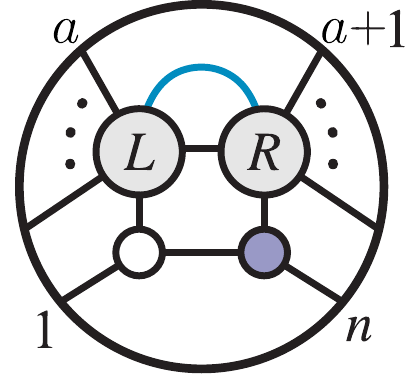}}\hspace{-3cm}\nonumber\vspace{-.1cm}}

And so, we have shown that the induction hypothesis ensures that all the necessary forward-limit terms are generated in the boundary of the BCFW recursion formula. But as with the factorization-channels studied earlier, we must show that no `spurious' forward-limit terms are generated. Such spurious forward-limit terms can be generated by the `bridge' term in the recursion---when the identified legs appear either between $(\hat{1}\,I)$ on the left, or between $(I\,\hat{n})$ on the right---or from the factorization-channels of the `forward-limit' term; these are always generated in pairs, and cancel accordingly; for example,
\vspace{-.2cm}\eq{\hspace{-3cm}\raisebox{-57pt}{\includegraphics[scale=0.85]{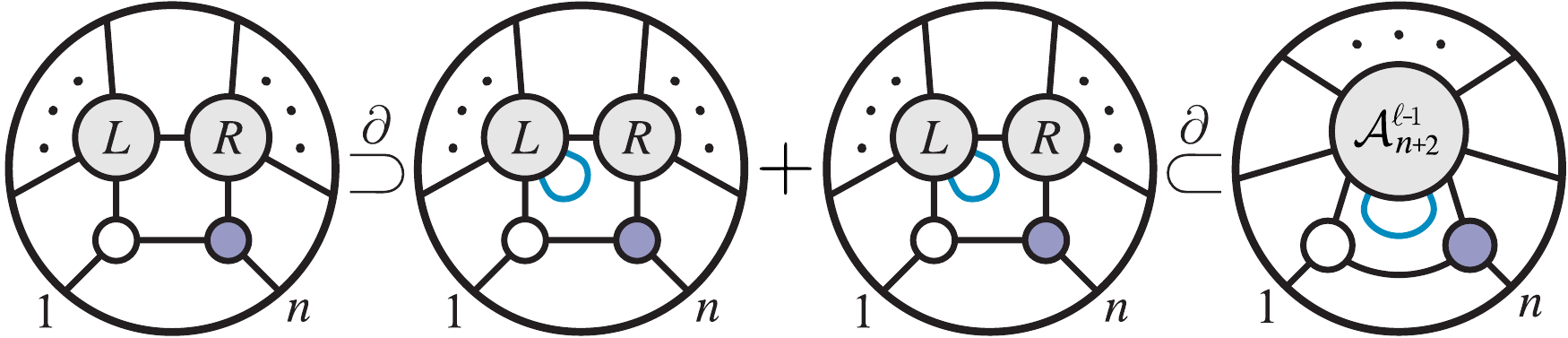}}\hspace{-3cm}\nonumber\vspace{-.1cm}}

\newpage
\subsection{The Structure of (Tree-)Amplitudes in the Grassmannian}\label{tree_amplitudes_and_basic_structures}
The BCFW recursion relations provide us with a powerful description of scattering amplitudes to all-loop orders. Although the tree-level recursion relations have been largely understood for nearly a decade (see e.g.\ \cite{Britto:2004ap,Britto:2005fq,Drummond:2008cr,Schuster:2008nh,Feng:2011np}), its extension to all-loop integrands remains relatively novel---and until now, has only been understood in terms of momentum-twistor variables (as described in \cite{ArkaniHamed:2010kv}). Because of this novelty, it is worthwhile to explore some of the features of the recursion and the structures that emerge. In this subsection, we will mostly review aspects of tree-amplitudes that are well known to most practitioners; this will provide us with the background necessary to
discuss some of the novelties that arise loop-level in \mbox{section \ref{canonical_coordinates_for_loop_integrands}}.

When restricted to tree-level, the recursion relations (\ref{bcfw_all_loop_recursion_disc}) become,
\vspace{-.2cm}\eq{\hspace{-3.cm}\raisebox{-54pt}{\includegraphics[scale=1]{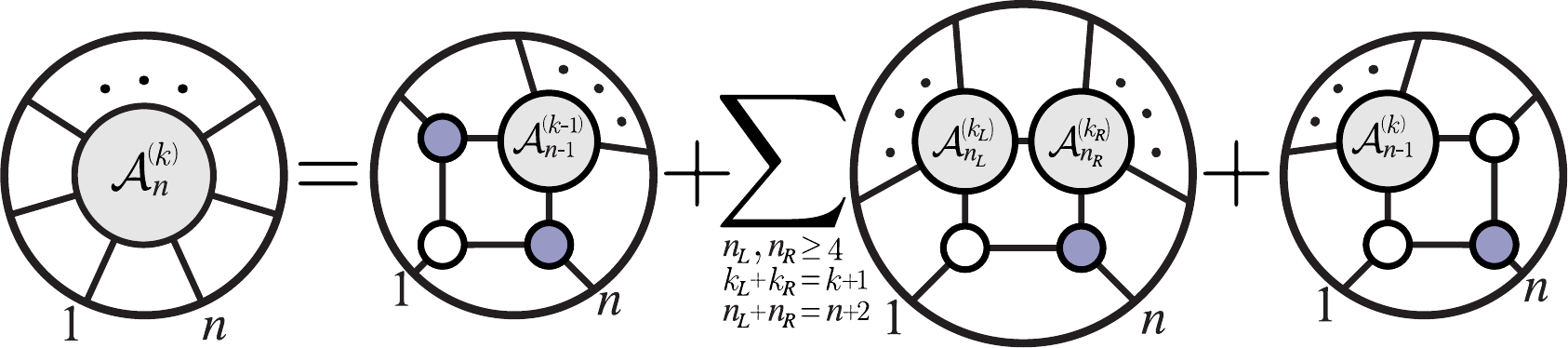}}\label{bcfw_tree_recursion_disc}\hspace{-3cm}\nonumber\vspace{-.2cm}}
Here, we have separated the terms in the recursion which involve a $3$-particle amplitude on either side of the bridge; this is because one of the $3$-particle amplitudes when bridged on either side will lead to an on-shell form with vanishing-support for generic kinematical data---for example, bridging $\mathcal{A}_3^{(1)}$ on the left would give,
\vspace{-.3cm}\eq{\hspace{-3cm}\raisebox{-38pt}{\includegraphics[scale=1]{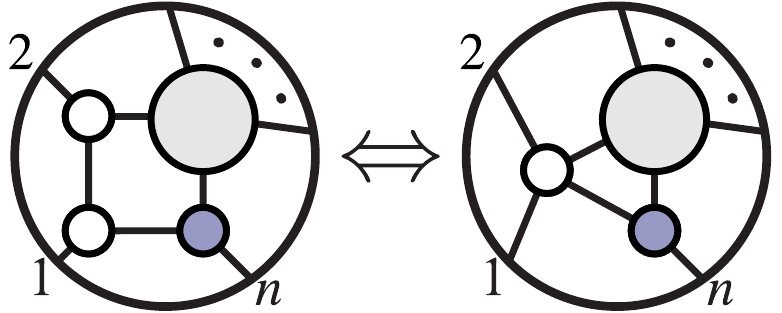}}\hspace{-3cm}\nonumber\vspace{-.3cm}}
which is only non-vanishing if $\lambda_1\propto\lambda_2$. (Moreover, it turns out that these graphs are always reducible, and so have less than the necessary $(2n\,\mi\,4)$ independent degrees of freedom required to solve the kinematical constraints.)

Let us begin to build intuition about the structure that arises from the recursion by considering the simplest examples. Recall that the $4$-particle amplitude is entirely given by the single on-shell graph, (\ref{four_point_box})---the familiar `box',
\vspace{-.2cm}\eq{\raisebox{8pt}{{\normalsize$\mathcal{A}_4^{(2)}\!=\mathcal{A}_3^{(2)}\!\!\otimes\!\mathcal{A}_3^{(1)}\!=$}}\begin{array}{c}\raisebox{-10pt}{\includegraphics[scale=1]{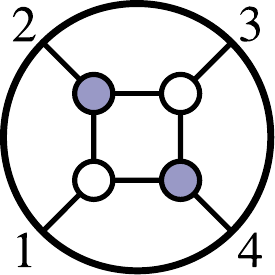}}\\[-0pt]\text{{\small$\phantom{\mathcal{A}_{3}^{(2)}\!\!\otimes\!\mathcal{A}_3^{(1)}}$}}\end{array}\phantom{\raisebox{8pt}{{\normalsize$\mathcal{A}_4^{(2)}\!=\mathcal{A}_3^{(2)}\!\!\otimes\!\mathcal{A}_3^{(1)}\!=$}}}\vspace{-.8cm}\nonumber}
This of course follows trivially from the recursion relations. But it is not the only amplitude which is so simple: for example, the two $5$-particle amplitudes are simply,\\[-6pt]
\vspace{-.2cm}\eq{\hspace{-0.75cm}\raisebox{4pt}{{\normalsize$\mathcal{A}_5^{(2)}\!=\mathcal{A}_4^{(2)}\!\!\otimes\!\mathcal{A}_3^{(1)}\!=$}}\begin{array}{c}\raisebox{-10pt}{\includegraphics[scale=1]{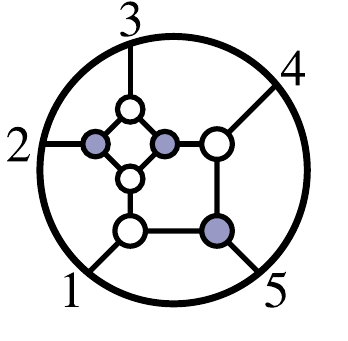}}\\[-8pt]\text{{\small$\phantom{\mathcal{A}_{4}^{(2)}\!\!\otimes\!\mathcal{A}_3^{(1)}}$}}\end{array}\phantom{\raisebox{8pt}{{\large$\mathcal{A}_4^{(2)}\!=\,$}}}\hspace{-1.15cm}\raisebox{4pt}{and}\quad\raisebox{4pt}{{\normalsize$\mathcal{A}_5^{(3)}\!=\mathcal{A}_3^{(2)}\!\!\otimes\!\mathcal{A}_4^{(2)}\!=\,$}}\hspace{-.3cm}\begin{array}{c}\raisebox{-10pt}{\includegraphics[scale=1]{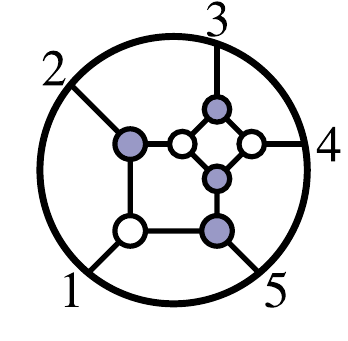}}\\[-8pt]\text{{\small$\phantom{\mathcal{A}_{3}^{(2)}\!\!\otimes\!\mathcal{A}_4^{(2)}}$}}\end{array}\phantom{\raisebox{8pt}{{\large$\mathcal{A}_4^{(2)}\!=\,$}}}\hspace{-2cm}\vspace{-.8cm}\nonumber}
This trend continues for all MHV and $\bar{\text{MHV}}$ amplitudes, $\mathcal{A}_{n}^{(2)}$ and $\mathcal{A}_n^{(n-2)}$, respectively. For $6$-particles, these amplitudes are:
\vspace{-.2cm}\eq{\hspace{-0.75cm}\raisebox{4pt}{{\normalsize$\mathcal{A}_6^{(2)}\!=\mathcal{A}_5^{(2)}\!\!\otimes\!\mathcal{A}_3^{(1)}\!=$}}\begin{array}{c}\raisebox{-10pt}{\includegraphics[scale=1]{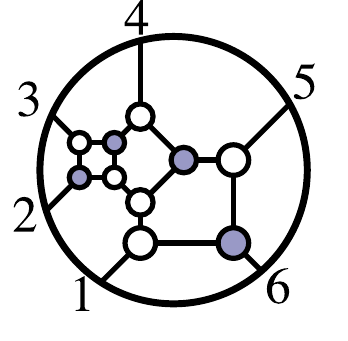}}\\[-8pt]\text{{\small$\phantom{\mathcal{A}_{4}^{(2)}\!\!\otimes\!\mathcal{A}_3^{(1)}}$}}\end{array}\phantom{\raisebox{8pt}{{\large$\mathcal{A}_4^{(2)}\!=\,$}}}\hspace{-1.15cm}\raisebox{4pt}{and}\quad\raisebox{4pt}{{\normalsize$\mathcal{A}_6^{(4)}\!=\mathcal{A}_3^{(2)}\!\!\otimes\!\mathcal{A}_5^{(3)}\!=\,$}}\hspace{-.3cm}\begin{array}{c}\raisebox{-10pt}{\includegraphics[scale=1]{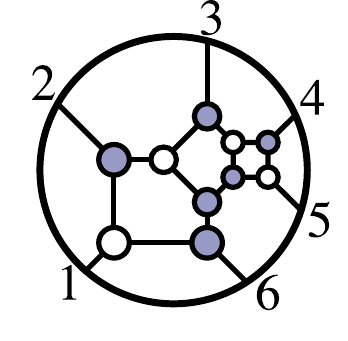}}\\[-8pt]\text{{\small$\phantom{\mathcal{A}_{3}^{(2)}\!\!\otimes\!\mathcal{A}_4^{(2)}}$}}\end{array}\phantom{\raisebox{8pt}{{\large$\mathcal{A}_4^{(2)}\!=\,$}}}\hspace{-2cm}\vspace{-.8cm}\nonumber}
Thus, the BCFW-recursion {\it directly} represents all MHV (and $\bar{\text{MHV}}$) amplitudes as single terms---directly giving the famous formula guessed by Parke and Taylor, (\ref{parke_taylor_tree_from_grassmannian}).

Although fairly trivial, notice that in obtaining these formulae, it is natural to view the act of attaching a 3-particle amplitude across the BCFW bridge as an operation which `adds a particle'. This operation is of course well-defined not just for the amplitude, but for any on-shell graph; thus, we have a way to add a particle in a way which `preserves $k$',  $(\bullet\otimes\!\mathcal{A}_3^{(1)}):G(k,n)\!\mapsto\! G(k,n\pl1)$, and in way which `increases $k$', $(\mathcal{A}_3^{(2)}\!\!\otimes\bullet):G(k,n)\!\mapsto\! G(k\pl1,n\pl1)$. These are called `inverse-soft factors'. As a reference, these operations correspond to:
\vspace{-.1cm}\eq{~\hspace{-1.5cm}\raisebox{-68pt}{\includegraphics[scale=1]{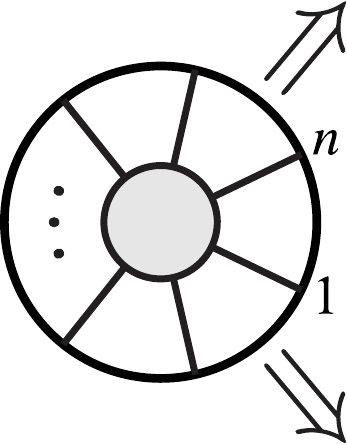}}\hspace{-1cm}\begin{array}{l}\begin{array}{ll}\raisebox{-51pt}{\includegraphics[scale=1]{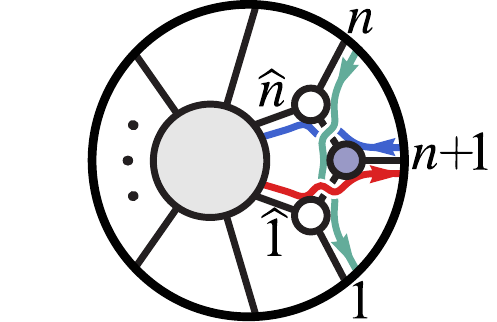}}&\hspace{-.2cm}\raisebox{-6pt}{$\left\{\rule{0pt}{57pt}\right.$}\text{{\footnotesize$\begin{array}{|l|l|}\multicolumn{2}{c}{}\\[-15pt]\multicolumn{2}{c}{\text{$k$-preserving or {\it holomorphic} inverse-soft factor}}\\\hline\multicolumn{1}{|c|}{\begin{array}{c}\text{momentum-space}\end{array}}&\multicolumn{1}{c|}{\begin{array}{c}\text{momentum-twistors}\end{array}}\\\hline\begin{array}{l}\lambda_{\hat{n}}=\lambda_n\\\widetilde\lambda_{\hat{n}}=\widetilde\lambda_n\!-\alpha_{(n\,n+1)}\widetilde\lambda_{n+1}\end{array}&\begin{array}{l}z_{\hat{n}}=z_n\\\phantom{\widetilde{\lambda}_{n}}\end{array}\\
\hline\begin{array}{l}\lambda_{\hat{1}}=\lambda_1\\\widetilde\lambda_{\hat{1}}=\widetilde\lambda_1\!-\alpha_{(1\,n+1)}\widetilde{\lambda}_{n+1}\end{array}&\begin{array}{l}z_{\hat{1}}=z_1\phantom{+\alpha_{(n1)}z_2}\\\phantom{\widetilde\lambda_1-\alpha_{(1n)}}\end{array}\\\hline
\begin{array}{l}f(\cdots\!, n,n\pl1,1,\cdots)\\\raisebox{-1.5pt}{{\large$\!\Rightarrow$}}\! f(\cdots\!, \hat{n},\hat{1},\cdots)\times\\\phantom{=}\,\text{{\footnotesize$\delta^2\big(\lambda_{n+1}\mi\,\alpha_{(n\,n+1)}\lambda_{n}\mi\,\alpha_{(1\,n+1)}\lambda_1\big)$}}\end{array}&\begin{array}{l}f(\cdots\!, n,n\pl1,1,\cdots)\\\raisebox{-1.5pt}{{\large$\!\Rightarrow$}}\! f(\cdots\!, n,1,\cdots)\\\phantom{=}\end{array}\\\hline
\end{array}$}}\end{array}\\~\\\begin{array}{ll}\raisebox{-51pt}{\includegraphics[scale=1]{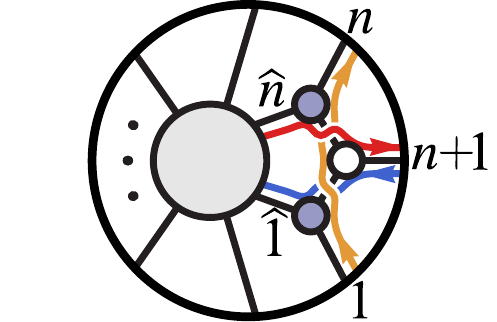}}&\hspace{-.2cm}\raisebox{-6pt}{$\left\{\rule{0pt}{57pt}\right.$}\text{{\footnotesize$\begin{array}{|l|l|}\multicolumn{2}{c}{}\\[-15pt]\multicolumn{2}{c}{\text{$k$-increasing or {\it anti-holomorphic} inverse-soft factor}}\\\hline\multicolumn{1}{|c|}{\begin{array}{c}\text{momentum-space}\end{array}}&\multicolumn{1}{c|}{\begin{array}{c}\text{momentum-twistors}\end{array}}\\\hline\begin{array}{l}\lambda_{\hat{n}}=\lambda_n+\alpha_{(n+1\,n)}\lambda_{n+1}\\\widetilde\lambda_{\hat{n}}=\widetilde\lambda_n\end{array}&\begin{array}{l}z_{\hat{n}}=z_n\!\pl\,\alpha_{(n+1\,n)}z_{n-1}\\\phantom{\widetilde{\lambda}_{n}}\end{array}\\
\hline\begin{array}{l}\lambda_{\hat{1}}=\lambda_1+\alpha_{(n+1\,1)}\lambda_{n+1}\\\widetilde\lambda_{\hat{1}}=\widetilde\lambda_1\phantom{-\alpha_{(1n)}\widetilde\lambda_n}\end{array}&\begin{array}{l}z_{\hat{1}}=z_1\pl\,\alpha_{(n+1\,1)}z_2\\\phantom{\widetilde\lambda_1-\alpha_{(1n)}}\end{array}\\\hline
\begin{array}{l}f(\cdots\!, n,n\pl1,1,\cdots)\\\raisebox{-1.5pt}{{\large$\!\Rightarrow$}}\! f(\cdots\!, \hat{n},\hat{1},\cdots)\times\\[-1pt]\phantom{=}\,\text{{\footnotesize$\delta^2\big(\widetilde\lambda_{n+1}\!\pl\,\alpha_{(n\,n+1)}\widetilde\lambda_{n}\!\pl\,\alpha_{(1\,n+1)}\widetilde\lambda_1\big)\!$}}\end{array}&\begin{array}{l}f(\cdots\!, n,n\pl1,1,\cdots)\\\raisebox{-1.5pt}{{\large$\!\Rightarrow$}}\! f(\cdots\!,\hat{n},\hat{1},\cdots)\\\phantom{=}\times\![n\mi1\,n\,n\pl1\,1\,2]\end{array}\\\hline
\end{array}$}}\end{array}\end{array}\hspace{-1cm}\vspace{-.2cm}\nonumber}
(Here, the $\widetilde\eta$'s transform identically to the $\widetilde\lambda$'s.) Each of these can be seen to follow from the action of two successive BCFW-bridges:
\vspace{-.3cm}\eq{~\hspace{-1.5cm}\raisebox{-54pt}{\includegraphics[scale=1]{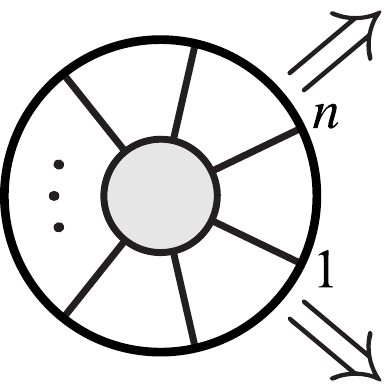}}\hspace{-0.45cm}\begin{array}{l}\begin{array}{ll}\raisebox{-51pt}{\includegraphics[scale=1]{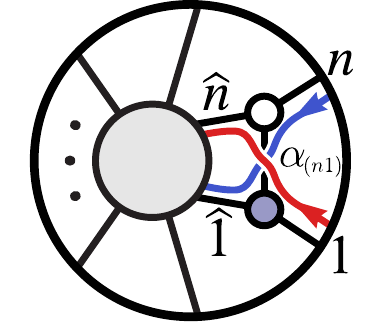}}&\hspace{-.2cm}\raisebox{-6pt}{$\left\{\rule{0pt}{37pt}\right.$}\text{{\footnotesize$\begin{array}{|l|l|}\multicolumn{2}{c}{}\\[-15pt]\multicolumn{2}{c}{\text{BCFW-bridge `$(n\,1)$'}}\\\hline\multicolumn{1}{|c|}{\begin{array}{c}\text{momentum-space}\end{array}}&\multicolumn{1}{c|}{\begin{array}{c}\text{momentum-twistors}\end{array}}\\\hline\begin{array}{l}\lambda_{\hat{n}}=\lambda_n\\\widetilde\lambda_{\hat{n}}=\widetilde\lambda_n\!-\alpha_{(n1)}\widetilde\lambda_1\end{array}&\begin{array}{l}z_{\hat{n}}=z_n\\\phantom{\widetilde{\lambda}_{n}}\end{array}\\
\hline\begin{array}{l}\lambda_{\hat{1}}=\lambda_1+\alpha_{(n1)}\lambda_n\\\widetilde\lambda_{\hat{1}}=\widetilde\lambda_1\end{array}&\begin{array}{l}z_{\hat{1}}=z_1+\alpha_{(n1)}z_2\\\phantom{\widetilde\lambda_1-\alpha_{(1n)}}\end{array}\\\hline
\end{array}$}}\end{array}\\~\\\begin{array}{ll}\raisebox{-51pt}{\includegraphics[scale=1]{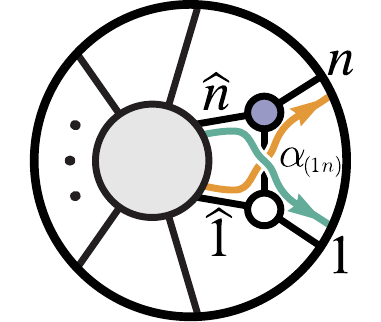}}&\hspace{-.2cm}\raisebox{-6pt}{$\left\{\rule{0pt}{37pt}\right.$}\text{{\footnotesize$\begin{array}{|l|l|}\multicolumn{2}{c}{}\\[-15pt]\multicolumn{2}{c}{\text{BCFW-bridge `$(1\,n)$'}}\\\hline\multicolumn{1}{|c|}{\begin{array}{c}\text{momentum-space}\end{array}}&\multicolumn{1}{c|}{\begin{array}{c}\text{momentum-twistors}\end{array}}\\\hline\begin{array}{l}\lambda_{\hat{n}}=\lambda_n+\alpha_{(1n)}\lambda_{1}\\\widetilde\lambda_{\hat{n}}=\widetilde\lambda_n\end{array}&\begin{array}{l}z_{\hat{n}}=z_n+\alpha_{(1n)}z_{n-1}\\\phantom{\widetilde{\lambda}_{n}}\end{array}\\
\hline\begin{array}{l}\lambda_{\hat{1}}=\lambda_1\\\widetilde\lambda_{\hat{1}}=\widetilde\lambda_1\!-\alpha_{(1n)}\widetilde\lambda_n\end{array}&\begin{array}{l}z_{\hat{1}}=z_1\phantom{\lambda_1}\\\phantom{\widetilde\lambda_1-\alpha_{(1n)}}\end{array}\\\hline
\end{array}$}}\end{array}\end{array}\hspace{-1cm}\vspace{-.2cm}\nonumber}
Notice that whenever an on-shell graph has a leg $a$ such that $\sigma(a\mi1)\!=\!a\pl1$ or $\sigma(a\pl1)\!=\!a\mi1$ we can view it as having been obtained by adding particle $a$ to a lower-point graph using a $k$-preserving or $k$-increasing inverse soft-factor, respectively. In such cases, $a$ is said to be an `inverse-soft factor'; and any on-shell graph which can be constructed by successively adding particles to a $3$-particle amplitude using inverse-soft factors is said to be `inverse-soft {\it constructible}'.

The notion of `inverse-soft constructibility' proves useful because the auxiliary variables associated with any inverse-soft factor can be completely fixed by the associated $\delta$-function constraint, making it very easy to recursively eliminate all the auxiliary, Grassmannian degrees of freedom. It turns out that for 13 or fewer legs, {\it all} on-shell forms generated by the tree-level recursion relations---{\it regardless} of how lower-point amplitudes are themselves recursed---are inverse-soft constructible. However, for 14 or more particles, some objects can be generated by the recursion relations which are {\it not} inverse-soft constructible, such as the following possible contribution to the $14$-particle N$^5$MHV tree-amplitude (labeled by {\footnotesize${\color{perm}\{4,\!7,\!6,\!10,\!16,\!17,\!14,\!15,\!12,\!13,\!19,\!23,\!22,\!25\}}$}):\\[-6pt]
\vspace{-.2cm}\eq{\hspace{-3.cm}\raisebox{-8pt}{$\begin{array}{c}\raisebox{-4.5pt}{\includegraphics[scale=0.95]{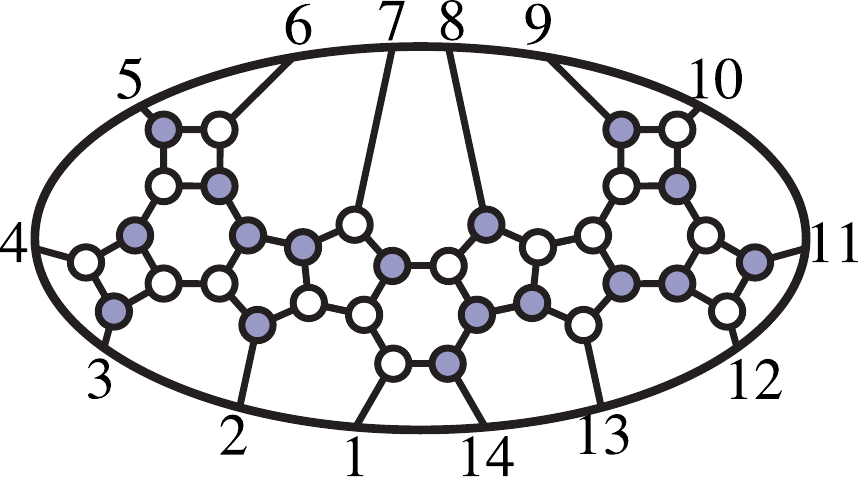}}\\[3pt]\text{{\footnotesize$\big(\big(\mathcal{A}_3^{(2)}\!\!\otimes\!\big(\mathcal{A}_4^{(2)}\!\!\otimes\!\mathcal{A}_4^{(2)}\big)\big)\!\otimes\!\mathcal{A}_3^{(1)}\big)\!\otimes\!\big(\mathcal{A}_3^{(2)}\!\!\otimes\big(\big(\mathcal{A}_4^{(2)}\!\!\otimes\!\mathcal{A}_4^{(2)}\big)\!\otimes\!\mathcal{A}_3^{(1)}\big)\big)$}}\\[2pt]\end{array}$}\label{g714_non_soft_constructible_tree_graph}\hspace{-3cm}\vspace{-.2cm}}
Notice that this graph was generated by always using {\it internal edges} to recurse the objects appearing across the BCFW-bridge---$(\hat{1}\,I)$ on the left and $(I\,\hat{n})$ on the right. (We should mention in passing that if one {\it always} recurses the lower-point amplitudes according to the marked legs as follows,
\vspace{-.4cm}\eq{\raisebox{-42pt}{\includegraphics[scale=.8]{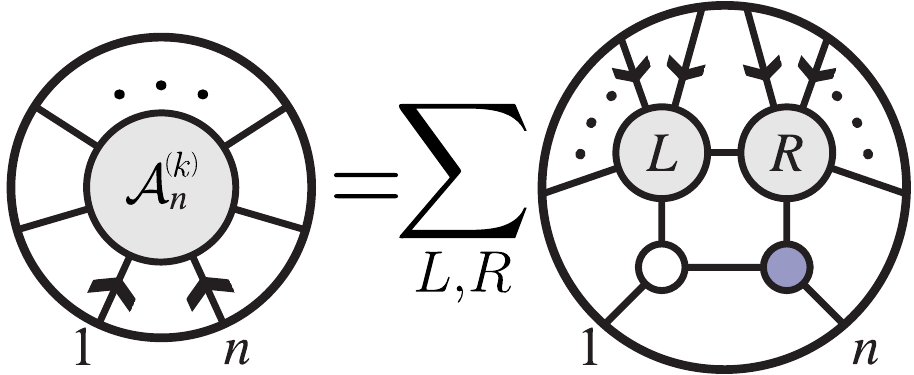}}\vspace{-.2cm}}
then {\it all} tree-amplitudes will be given in terms of only inverse-soft constructible graphs. This corresponds to the recursion `scheme' $\{\mi2,2,0\}$ of reference \cite{Bourjaily:2010wh}.)

As described in \mbox{section \ref{geometric_origin_of_identities_section}}, the first amplitude which is given as the combination of several on-shell graphs is $\mathcal{A}_6^{(3)}$, the $6$-particle NMHV tree-amplitude. This is given by three terms, \mbox{$\mathcal{A}_5^{(3)}\!\!\otimes\!\mathcal{A}_3^{(1)}$}, \mbox{$\mathcal{A}_4^{(2)}\!\!\otimes\!\mathcal{A}_4^{(2)}$}, and \mbox{$\mathcal{A}_3^{(2)}\!\!\otimes\!\mathcal{A}_5^{(2)}$}:
\vspace{-.6cm}\eq{\hspace{0.05cm}\raisebox{0pt}{{\large$\mathcal{A}_6^{(3)}=$}}\begin{array}{c}\\[-10pt]\includegraphics[scale=1]{g36_tree_a_term_1}\\[-15pt]\text{{\small${\color{perm}\{4,5,6,8,7,9\}}$}}\\[-00pt]\end{array}\raisebox{-2pt}{$\text{{\LARGE$+$}}$}\begin{array}{c}\\[-10pt]\includegraphics[scale=1]{g36_tree_a_term_2}\\[-15pt]\text{{\small${\color{perm}\{3,5,6,7,8,10\}}$}}\\[-00pt]\end{array}\raisebox{-2pt}{$\text{{\LARGE$+$}}$}\begin{array}{c}\\[-10pt]\includegraphics[scale=1]{g36_tree_a_term_3}\\[-15pt]\text{{\small${\color{perm}\{4,6,5,7,8,9\}}$}}\\[-00pt]\end{array}\hspace{-3cm}\vspace{-.2cm}\label{6_particle_nmhv_tree}}
Although the on-shell graphs of each contribution appear quite different, it is easy to see from the permutations that they are all cyclically-related to one another:
\vspace{-.6cm}\eq{\hspace{-2.5cm}\begin{array}{c}\\[-10pt]\includegraphics[scale=1]{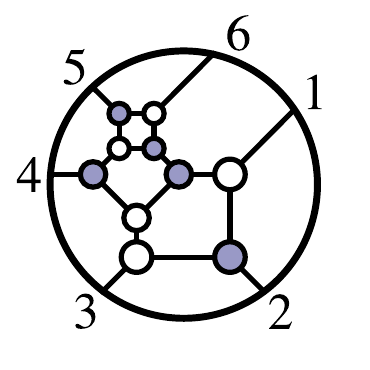}\\[-15pt]\text{{\small${\color{perm}\{3,5,6,7,8,10\}}$}}\\[-00pt]\end{array}\raisebox{-2pt}{$\text{{\LARGE$=$}}$}\begin{array}{c}\\[-10pt]\includegraphics[scale=1]{g36_tree_a_term_2}\\[-15pt]\text{{\small${\color{perm}\{3,5,6,7,8,10\}}$}}\\[-00pt]\end{array}\raisebox{-2pt}{$\text{{\LARGE$=$}}$}\begin{array}{c}\\[-10pt]\includegraphics[scale=1]{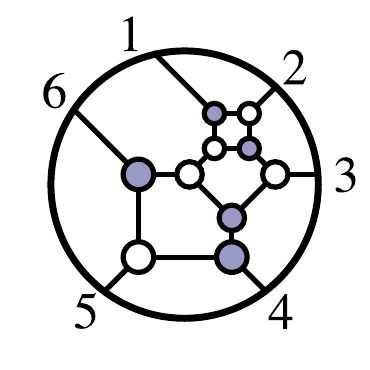}\\[-10pt]\text{{\small${\color{perm}\{3,5,6,7,8,10\}}$}}\\[-00pt]\end{array}\hspace{-3cm}\vspace{-.2cm}}
The on-shell differential form drawn above---labeled by the permutation ${\color{perm}\{3,5,6,7,8,10\}}$ ---was given directly in terms of the kinematical variables $\lambda,\widetilde\lambda$ in equation (\ref{g36_on_shell_function}). Because each term is cyclically-related, if we use `$r$' to denote the operation that `rotates' all particle labels forward by $1$, we can write the entire tree-amplitude as:
\vspace{-.2cm}\eq{\hspace{-4.05cm}\mathcal{A}_6^{(3)}\!\!=\!(1\pl\,r^2\pl\,r^4)\frac{\delta^{3\times4}(C^*\!\!\cdot\widetilde \eta)\,\delta^{2\times2}\big(\lambda\!\cdot\!\widetilde{\lambda}\big)}{\ab{23}\sb{56}(\ab{34}\sb{64}\pl\ab{53}\sb{56})s_{456}(\ab{61}\sb{64}\pl\ab{15}\sb{45})\ab{12}\sb{45}},\hspace{-3cm}\vspace{-.1cm}}
where the matrix $C^*$ was given in (\ref{g36_configuration_cMatrix}).

Although the precise set of on-shell graphs obtained using the BCFW recursion relations can vary considerably depending on which legs of the lower-point amplitudes are used for {\it their} recursion, the number of terms is of course scheme-independent. It is a relatively simple exercise to show that,
\vspace{-.1cm}\eq{\text{\# BCFW terms in the tree-amplitude }\mathcal{A}_n^{(k)}\!:\,\frac{1}{n-3}\binom{n\,\mi\,3}{k\,\mi\,1}\binom{n\,\mi\,3}{k\,\mi\,2}.\vspace{-.2cm}}

\newpage
\subsection{Canonical Coordinates for Loop Integrands}\label{canonical_coordinates_for_loop_integrands}
The all-loop generalization of the BCFW recursion relations was first described in \cite{ArkaniHamed:2010kv} where it was formulated in terms of explicit operations acting directly on the `functions' of momentum-twistor variables obtained after eliminating the auxiliary Grassmannian degrees of freedom. This led to formulae for the `loop integrands' in the form of a `standard' loop-integration measure $d^4\ell$ weighted by some rational function of the loop-momentum $\ell$, or equivalently a function one the space of lines $AB$ in momentum-twistor space with measure $d^4 z_A d^4 z_B/GL(2)$.  When viewed as rational functions in this way, much of the underlying structure is hidden. However, by viewing each loop-momentum's degrees of freedom as arising from canonical coordinates in the auxiliary Grassmannian, the integration measure is automatically generated in a much more illuminating, `canonical' form: as a wedge-product of ``$d\!\log$'' factors. The fact that loop amplitude integrands can be written in such a form---a fact which is essentially obvious using canonical coordinates on the Grassmannian---is far from obvious from any other  method to compute scattering amplitudes.

We will postpone a systematic discussion of the loop amplitude integrands generated by the recursion relations (\ref{bcfw_all_loop_recursion_disc}) until a future work. Here, we merely want to demonstrate its most important physical implications through the context of simple examples. We first describe how one-loop integrands are generated by the recursion, using the case of MHV for illustration. At the end of this subsection, we will briefly describe the features observed for higher-loop amplitudes.

Let us begin with the simplest of all one-loop amplitudes, the $4$-particle MHV amplitude. As there are no $3$-particle one-loop integrands to appear in the `bridge' term of the recursion, the $4$-particle one-loop integrand is entirely generated as the forward-limit of the $6$-particle NMHV tree-amplitude, $\mathcal{A}_6^{(3)}$. Let us denote the two particles identified in the forward-limit by $(A\,B)$, and use these two legs as the pair singled-out in the recursion of the $6$-particle tree. Of the $3$ terms appearing in the tree-amplitude $\mathcal{A}_6^{(3)}$, (\ref{6_particle_nmhv_tree}), only one is non-vanishing in the forward-limit (a fact that we will demonstrate momentarily); the forward-limit of the $\mathcal{A}_4^{(2)}\!\!\otimes\!\mathcal{A}_4^{(2)}$-term is,
\vspace{-.4cm}\eq{\hspace{-3cm}\begin{array}{l}~\\[-10pt]\includegraphics[scale=.9]{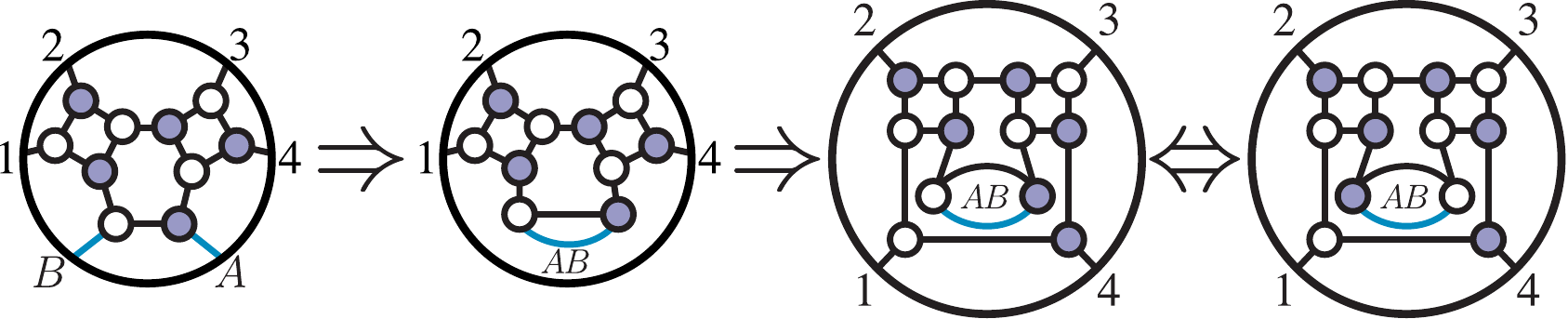}\\[-12pt]\text{{\small$\,\hspace{0.65cm}\mathcal{A}_{4}^{(2)}\!\!\otimes\!\mathcal{A}_4^{(2)}\hspace{0cm}~$}}~\end{array}\vspace{-.2cm}\label{g36_bcfw_term_merging_AB}\nonumber\hspace{-3cm}}
(The last move in this sequence was made only to make subsequent transformations more transparent.) It is easy to see that this diagram has four faces beyond that of the simple box, and thus four extra integration variables. Using reduction, we can of course reduce this diagram to the box, giving us the integrand. We can relate this new form of the integrand to a more familiar form, by identifying the usual loop momentum ``$\ell$'' as,
\vspace{-.2cm}\eq{\ell=\alpha\lambda_1\widetilde\lambda_4+\lambda_{AB}\widetilde{\lambda}_{AB},\vspace{-.2cm}}
where $\lambda_{AB} \widetilde \lambda_{AB}$ is the momentum of the highlighted line in figure above. We can of course determine $\lambda_{AB}, \widetilde \lambda_{AB}$ in terms of the variables associated with the graph, and in this way trade the four `extra' variables for those which parameterize $\ell$.

While this is a straightforward exercise, it is more illuminating to carry out the reduction in a different way. We can use moves to give the on-shell diagram a different representation---as a sequence of BCFW bridges on a core $4$-particle amplitude:
\vspace{-.2cm}\eq{\hspace{-3.cm}\raisebox{-54pt}{\includegraphics[scale=1]{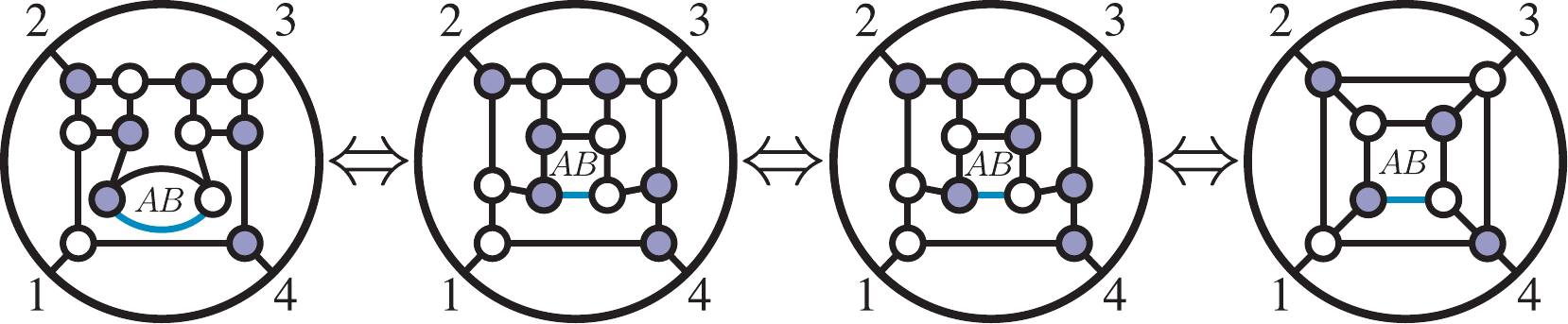}}\label{g24_one_loop_sequence}\hspace{-3cm}\nonumber\vspace{-.2cm}}
(In the last transformation, several mergers were made.) This allows us to think of the object as the usual box, but with `BCFW-shifted kinematical data', given by,
\vspace{0.0cm}\eq{\hspace{-3.cm}\raisebox{-60pt}{\includegraphics[scale=1]{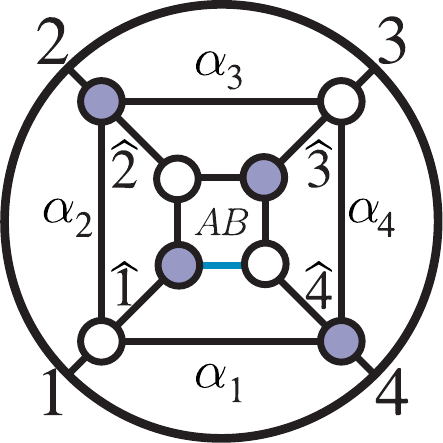}}\quad\begin{array}{|c|l|}\hline\begin{array}{c}\!\text{bridge}\!\end{array}&\multicolumn{1}{|c|}{\text{BCFW shift}}\\[-2pt]\hline(1\,4)&\begin{array}{@{}ll@{$\,$}c@{$\,$}l}\\[-17pt]\lambda_4\!\mapsto\!&\lambda_4&+&\alpha_1\lambda_1\\[-2pt]\widetilde\lambda_1\!\mapsto\!&\widetilde\lambda_1&-&\alpha_1\widetilde\lambda_4\end{array}\\[-2pt]\hline(1\,2)&\begin{array}{@{}ll@{$\,$}c@{$\,$}l}\\[-17pt]\lambda_2\!\mapsto\!&\lambda_2&+&\alpha_2\lambda_1\\[-2pt]\widetilde\lambda_1\!\mapsto\!&\widetilde\lambda_1&-&\alpha_2\widetilde\lambda_2\end{array}\\[-2pt]\hline(3\,2)&\begin{array}{@{}ll@{$\,$}c@{$\,$}l}\\[-17pt]\lambda_2\!\mapsto\!&\lambda_2&+&\alpha_3\lambda_3\\[-2pt]\widetilde\lambda_3\!\mapsto\!&\widetilde\lambda_3&-&\alpha_3\widetilde\lambda_2\end{array}\\[-2pt]\hline(3\,4)&\begin{array}{@{}ll@{$\,$}c@{$\,$}l}\\[-17pt]\lambda_4\!\mapsto\!&\lambda_4&+&\alpha_4\lambda_3\\[-2pt]\widetilde\lambda_3\!\mapsto\!&\widetilde\lambda_3&-&\alpha_4\widetilde\lambda_4\end{array}\\\hline
\end{array}\;\text{{\LARGE$\Rightarrow$}}\;\raisebox{2pt}{$\left\{\rule{0pt}{58pt}\right.$}\begin{array}{c@{$\,$}l}\hat{1}&\left\{\rule{0pt}{15pt}\right.\!\!\!\begin{array}{ll}\lambda_{\hat{1}}=\lambda_1\phantom{+\alpha_1\lambda_4+\alpha_2\lambda_2}\\\widetilde{\lambda}_{\hat{1}}=\widetilde{\lambda}_1-\alpha_1\widetilde\lambda_4-\alpha_2\widetilde\lambda_2\end{array}\!\!\!\left.\rule{0pt}{15pt}\right.\\\hat{2}&\left\{\rule{0pt}{15pt}\right.\!\!\!\begin{array}{ll}\lambda_{\hat{2}}=\lambda_2+\alpha_2\lambda_1+\alpha_3\lambda_3\\\widetilde{\lambda}_{\hat{2}}=\widetilde{\lambda}_2\phantom{-\alpha_2\widetilde{\lambda}_1-\alpha_3\widetilde{\lambda}_3}\end{array}\!\!\!\left.\rule{0pt}{15pt}\right.\\\hat{3}&\left\{\rule{0pt}{15pt}\right.\!\!\!\begin{array}{ll}\lambda_{\hat{3}}=\lambda_3\phantom{+\alpha_3\lambda_2+\alpha_4\lambda_4}\\\widetilde{\lambda}_{\hat{3}}=\widetilde\lambda_3-\alpha_3\widetilde\lambda_2-\alpha_4\widetilde\lambda_4\end{array}\!\!\!\left.\rule{0pt}{15pt}\right.\\\hat{4}&\left\{\rule{0pt}{15pt}\right.\!\!\!\begin{array}{ll}\lambda_{\hat{4}}=\lambda_4+\alpha_1\lambda_1+\alpha_4\lambda_3\\\widetilde\lambda_{\hat{4}}=\widetilde\lambda_4\phantom{-\alpha_1\widetilde\lambda_1-\alpha_4\widetilde\lambda_3}\end{array}\!\!\!\left.\rule{0pt}{15pt}\right.\end{array}\label{four_point_one_loop_with_beta_shifts}\hspace{-3cm}\nonumber\vspace{-.0cm}}

Thus, the integrand is nothing but  $d\!\log(\alpha_{1})\wedge\cdots\wedge d\!\log(\alpha_{4})$ times the shifted four-particle amplitude,
\vspace{-.4cm}\eq{\frac{d\alpha_1}{\alpha_1}\frac{d\alpha_2}{\alpha_2}\frac{d\alpha_3}{\alpha_3}\frac{d\alpha_4}{\alpha_4}\frac{\delta^{2\times4}\big(\,\lambda\!\cdot\!\widetilde\eta\,\big)}{\ab{\hat{1}\,\hat{2}}\ab{\hat{2}\,\hat{3}}\ab{\hat{3}\,\hat{4}}\ab{\hat{4}\,\hat{1}}}.\vspace{-.2cm}}
If we strip-off the (unshifted) Parke-Tyalor prefactor, the integrand for the one-loop ratio function---the loop amplitude divided by the tree-amplitude---is simply \cite{ArkaniHamed:2009si},
\vspace{-.3cm}\eq{\hspace{-3cm}\begin{array}{ll}\displaystyle\frac{\mathcal{A}_{4}^{(2),1}}{\mathcal{A}^{(2),0}_{4}}&=\displaystyle\frac{d\alpha_1}{\alpha_1}\frac{d\alpha_2}{\alpha_2}\frac{d\alpha_3}{\alpha_3}\frac{d\alpha_4}{\alpha_4}\frac{\ab{1\,2}\ab{2\,3}\ab{3\,4}\ab{4\,1}}{\ab{\hat{1}\,\hat{2}}\ab{\hat{2}\,\hat{3}}\ab{\hat{3}\,\hat{4}}\ab{\hat{4}\,\hat{1}}},\\[-16pt]\\&=\displaystyle\frac{d\alpha_1}{\alpha_1}\frac{d\alpha_2}{\alpha_2}\frac{d\alpha_3}{\alpha_3}\frac{d\alpha_4}{\alpha_4}\frac{\ab{12}}{\ab{12}+\alpha_3\ab{13}}\frac{\ab{23}}{\ab{23}+\alpha_2\ab{13}}\frac{\ab{34}}{\ab{34}+\alpha_1\ab{31}}\frac{\ab{41}}{\ab{41}+\alpha_4\ab{31}},\\[10pt]
&=\displaystyle d\!\log\!\left(\!\frac{\alpha_1\ab{34}}{\ab{34}\pl\,\alpha_1\ab{31}}\!\right)\!d\!\log\!\left(\!\frac{\alpha_2\ab{23}}{\ab{23}\pl\,\alpha_2\ab{13}}\!\right)\!d\!\log\!\left(\!\frac{\alpha_3\ab{12}}{\ab{12}\pl\,\alpha_3\ab{13}}\!\right)\!d\!\log\!\left(\!\frac{\alpha_4\ab{41}}{\ab{41}\pl\,\alpha_4\ab{31}}\!\right)\\[0pt]\end{array}\label{g24_one_loop_integrand_bare}\vspace{-.2cm}\hspace{-3cm}\nonumber}
which is manifestly in a `$d\!\log$'-form.

Now, we can determine $\lambda_{AB}$ and $\widetilde \lambda_{AB}$ very simply in terms of the bridge variables. For a general box, the internal momentum on the bridge can be given in terms of the external data according to:
\vspace{-.3cm}\eq{\raisebox{-52pt}{\includegraphics[scale=1]{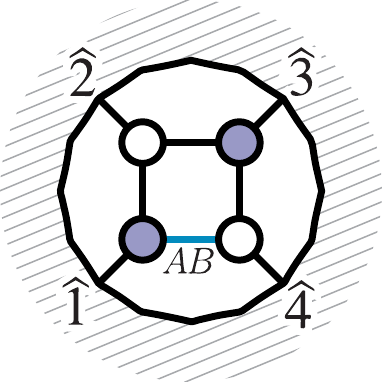}}\vspace{-.3cm}}
which allows us to identify,
\vspace{-.2cm}\eq{\lambda_{AB}\widetilde{\lambda}_{AB}=\frac{\ab{\hat{1}\hat{2}}}{\ab{\hat{4}\hat{2}}}\lambda_{\hat{4}}\widetilde\lambda_{\hat{1}}.\vspace{-.2cm}}
And so in summary, the relation between the BCFW-bridge variables $\alpha_i$ and the usual loop momentum variables is given by,
\vspace{-.1cm}\eq{\ell=\frac{\ab{\hat{1}\hat{2}}}{\ab{\hat{4}\hat{2}}}\lambda_{\hat{4}}\widetilde\lambda_{\hat{1}}+\alpha_1\lambda_{1}\widetilde{\lambda}_4.\vspace{-.2cm}}

Using this change of variables, it is straightforward to re-cast the integrand (\ref{g24_one_loop_integrand_bare}) in the form which we gave earlier in \mbox{section \ref{on_shell_diagrams_section}}:
\vspace{-.2cm}\eq{d\!\log\!\Bigg(\!\frac{\ell^2}{(\ell -\ell^*)^2}\!\Bigg)d\!\log\!\Bigg(\!\frac{(\ell + p_1)^2}{(\ell- \ell^*)^2}\!\Bigg)d\!\log\!\Bigg(\!\frac{(\ell + p_1 + p_2)^2}{(\ell -\ell^*)^2}\!\Bigg)d\!\log\!\Bigg(\!\frac{(\ell-p_4)^2}{(\ell-\ell^*)^2}\!\Bigg)\,,\vspace{-.2cm}}
where $\ell^*=\frac{\ab{12}}{\ab{42}}\lambda_4\widetilde{\lambda}_{1}$.

We can also interpret exactly the same pictures in momentum-twistor space. Recall that a BCFW bridge `$(a\mi1\,a)$'---a white-to-black vertex from $a\mi1$ to $a$---has the effect of shifting the momentum twistor $z_a\mapsto z_a+\alpha z_{a+1}$ where $\alpha$ is the bridge variable. Generally speaking, lines in momentum-twistor space are associated with the faces of the momentum-space on-shell graph; we will not review these ideas here, but let us briefly summarize that the regions of a four-point box are associated with the lines in momentum-twistor space as indicated below:
\vspace{-.2cm}\eq{\raisebox{-52pt}{\includegraphics[scale=1]{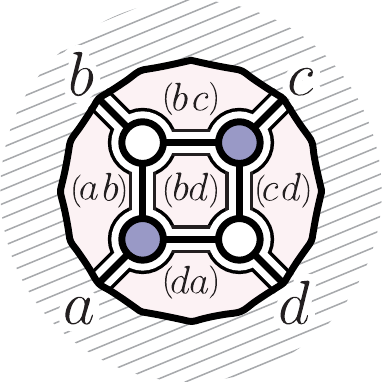}}\quad\mathrm{or}\quad\raisebox{-52pt}{\includegraphics[scale=1]{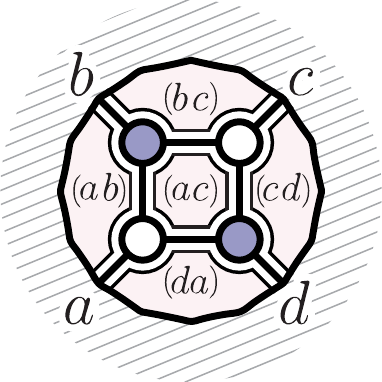}}\vspace{-.2cm}}
Now, this means that if we identify the four unfixed degrees of freedom with the line $(AB)$ in momentum-twistor space, we see that it corresponds to the line $(\hat{2}\,\hat{4})$ in
\vspace{-.2cm}\eq{\raisebox{-52pt}{\includegraphics[scale=1]{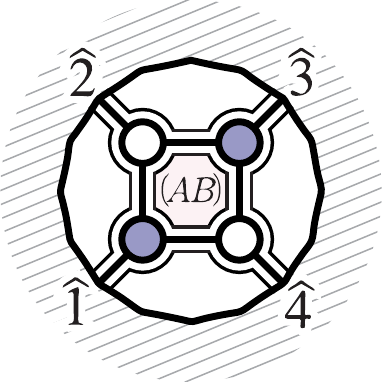}}\vspace{-.2cm}}

Performing the same sequence of shifts as before, but now using momentum-twistor variables, we find:
\vspace{-.0cm}\eq{\hspace{-3.cm}\raisebox{-50pt}{\includegraphics[scale=.825]{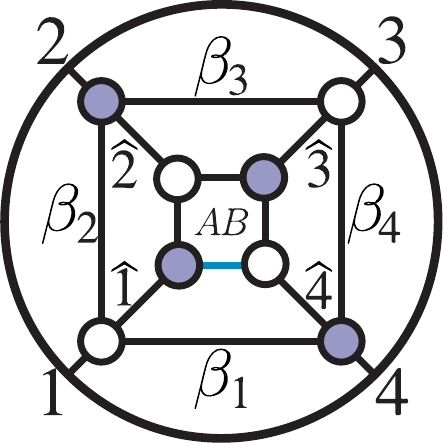}}\qquad\text{{\normalsize$\begin{array}{|c|l|}\hline\begin{array}{c}\!\text{bridge}\!\end{array}&\multicolumn{1}{|c|}{\text{BCFW shift}}\\[-2pt]\hline(1\,4)&z_4\!\mapsto\! z_4\pl\,\beta_1z_3\\[-0pt]\hline(1\,2)&z_2\!\mapsto\! z_2\pl\,\beta_2 z_3\\\hline(3\,2)&z_2\!\mapsto\! z_2\pl\,\beta_3z_1\\\hline(3\,4)&z_4\!\mapsto\! z_4\pl\,\beta_4 z_1\\\hline
\end{array}$}}\;\raisebox{-9pt}{\text{{\LARGE$\Rightarrow$}}}\raisebox{-7pt}{$\left\{\rule{0pt}{34pt}\right.\hspace{-.2cm}$}\raisebox{-6pt}{$\begin{array}{l@{}l}z_{\hat{1}}\!=&z_1\phantom{\pl\beta_1z_2\pl\beta_2 z_4}\\[-2pt] z_{\hat{2}}\!=&z_2\pl\beta_2z_3\pl\beta_3z_1\\[-2pt] z_{\hat{3}}\!=&z_3\phantom{\pl\beta_3 z_4\pl\beta_4z_2}\\[-2pt] z_{\hat{4}}\!=&z_4\pl\beta_1z_3\pl\beta_4z_1\end{array}$}\raisebox{-7pt}{$\left.\hspace{-.2cm}\rule{0pt}{34pt}\right\}$}\nonumber\label{region_momenta_for_one_loop_example}\hspace{-3cm}\nonumber\vspace{-.2cm}}
This uniquely fixes the auxiliary, Grassmannian parameters $\beta_i$ in terms of the momentum-twistor line $(AB)$ according to:
\vspace{-.2cm}\eq{\beta_1=\frac{\ab{AB\,4\,1}}{\ab{AB\,1\,3}},\quad\beta_2=\frac{\ab{AB\,1\,2}}{\ab{AB\,3\,1}},\quad\beta_3=\frac{\ab{AB\,2\,3}}{\ab{AB\,3\,1}},\;\;\mathrm{and}\;\;\beta_4=\frac{\ab{AB\,3\,4}}{\ab{AB\,1\,3}}.\vspace{-.2cm}}
With this identification, we can re-write the integrand in terms of the four auxiliary variables in momentum-twistor space as,
\vspace{-.2cm}\eq{\hspace{-2cm}d\!\log(\beta_1)\cdots d\!\log(\beta_4)=d\!\log\!\Bigg(\!\!\frac{\ab{AB\,41}}{\ab{AB\,13}}\!\!\Bigg)d\!\log\!\Bigg(\!\!\frac{\ab{AB\,12}}{\ab{AB\,31}}\!\!\Bigg)d\!\log\!\Bigg(\!\!\frac{\ab{AB\,23}}{\ab{AB\,31}}\!\!\Bigg)d\!\log\!\Bigg(\!\!\frac{\ab{AB\,34}}{\ab{AB\,13}}\!\!\Bigg).\hspace{-2cm}\vspace{-.2cm}\nonumber}

If we recast this expression as an integration measure on the space of lines $(AB)$ in momentum-twistor space, we find that
\vspace{-.2cm}\eq{\hspace{-2cm}d\!\log\!\Bigg(\!\!\frac{\ab{AB\,41}}{\ab{AB\,13}}\!\!\Bigg)d\!\log\!\Bigg(\!\!\frac{\ab{AB\,12}}{\ab{AB\,31}}\!\!\Bigg)d\!\log\!\Bigg(\!\!\frac{\ab{AB\,23}}{\ab{AB\,31}}\!\!\Bigg)d\!\log\!\Bigg(\!\!\frac{\ab{AB\,34}}{\ab{AB\,13}}\!\!\Bigg)\!=\!\displaystyle\frac{\ab{d^2z_AAB}\ab{d^2z_BAB}\ab{1234}\ab{2341}}{\ab{AB\,12}\ab{AB\,23}\ab{AB\,34}\ab{AB\,41}},\hspace{-2cm}\vspace{-.2cm}\nonumber}
which is precisely the familiar form of the integrand given in reference \cite{ArkaniHamed:2010kv}.

Before moving on to the case of the $n$-particle MHV one-loop integrand, let us go back and understand why only one of the three terms in the $6$-particle NMHV tree amplitude survived the forward-limit, as the reason will prove quite instructive. Let us choose to always represent the $(n\pl2)$-point tree-amplitude appearing in the forward limit using the BCFW recursion which deforms legs $(AB)$. Recall that the tree-amplitude recursion can be broken into three parts as in (\ref{bcfw_tree_recursion_disc}):
\vspace{-.2cm}\eq{\begin{array}{rll}1.&\text{a $k$-preserving inverse-soft factor:}& \mathcal{A}_{n-1}^{(k)}\!\!\otimes\!\mathcal{A}_3^{(1)};\\
2.&\text{a $k$-increasing inverse-soft factor:}&\mathcal{A}_{3}^{(2)}\!\!\otimes\!\mathcal{A}_{n-1}^{(k-1)};\\\mathrm{and}\qquad3.&\text{terms for which $n_L,n_R\geq4$.}\end{array}\vspace{-.2cm}}
Of these, it is not hard to see that if $(AB)$ are the distinguished legs of the bridge, the first two contributions listed above {\it always} vanish. More precisely, any on-shell form for which $A$ {\it or} $B$ is an inverse-soft factor will vanish in the forward-limit. (We should notice that `$A$ or $B$ being an inverse-soft factor' is a {\it sufficient} condition for an on-shell form to vanish in the forward limit, but not a {\it necessary} one.)

Let us now see why any contributions to the lower-loop amplitude where $A$ or $B$ is an inverse-soft factor will vanish. Consider the forward-limit of a term for which $A$ is a $k$-preserving inverse-soft factor (the argument is the same in all other cases):
\vspace{-.2cm}\eq{\raisebox{-47pt}{\includegraphics[scale=1]{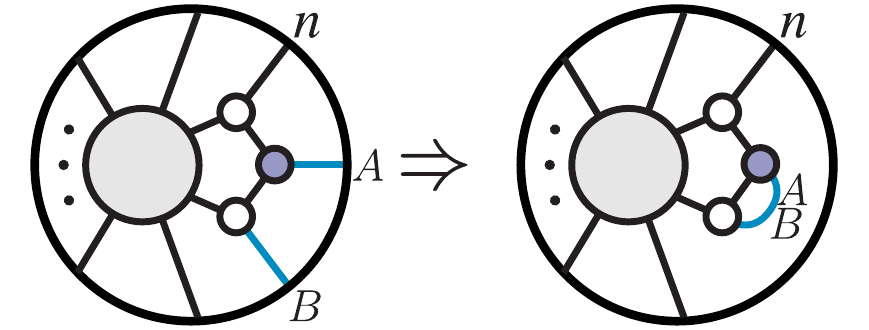}}\vspace{-.2cm}}
Notice that the kinematical constraints associated with the middle---black---vertex requires that $\lambda_A$ be expandable in terms of $\lambda_B$ and $\lambda_n$; but in the forward-limit, we identify $\lambda_A$ with $\lambda_B$, which implies that $\lambda_{AB}\!\propto\!\lambda_n$. As such, the kinematical constraints do not allow for there to be any unfixed degrees of freedom associated with $\lambda_{AB}$ (which should represent loop-integration degrees of freedom).

We are now prepared to determine the $n$-point MHV one-loop integrand in general. The bridge-term always contributes a term $\mathcal{A}_{n-1}^{(2),1}\!\!\otimes\!\mathcal{A}_{3}^{(1)}$, which is simply a {\it $k$-preserving} inverse-soft factor adding $n$ to the $(n\mi1)$-point one-loop amplitude; more interesting are the forward-limit terms. These come from the forward-limit of $\mathcal{A}_{n+2}^{(3),0}$; among the terms that contribute to the higher-point NMHV tree-amplitude, we have seen that only those obtained from bridging $\mathcal{A}_{n_L}^{(2)}\!\!\otimes\!\mathcal{A}_{n_R}^{(2)}$ with $n_L,n_R\geq4$ contribute.

Because $k$-preserving inverse soft factors act trivially in momentum-twistors, and the left- and right-amplitudes appearing in the NMHV tree-amplitudes are trivially chains of inverse-soft factors, it will be useful to define the notion of an ``MHV region'' obtained by any number of successive $k$-preserving inverse-soft factors:
\vspace{-.2cm}\eq{\hspace{-3cm}\raisebox{-40pt}{\includegraphics[scale=.9]{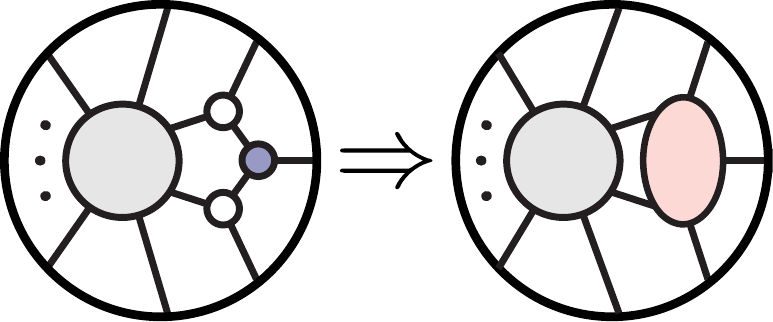}}\quad\text{with}\quad\raisebox{-40pt}{\includegraphics[scale=.9]{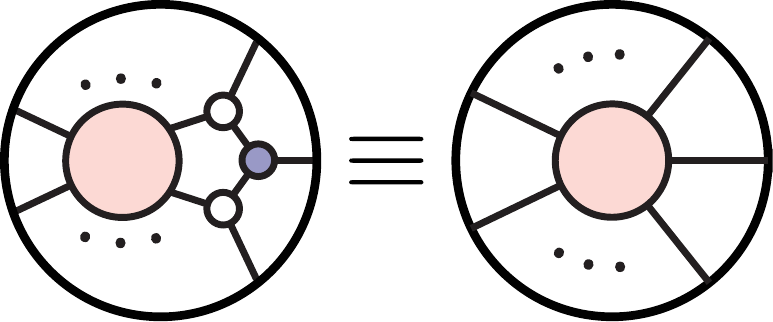}}\hspace{-3cm}\vspace{-.2cm}\nonumber}

Allowing for such MHV regions in our diagrammatic expansion, we see that the one-loop MHV integrand is given by,
\vspace{-.5cm}\eq{\hspace{-3cm}\raisebox{-44pt}{\includegraphics[scale=1]{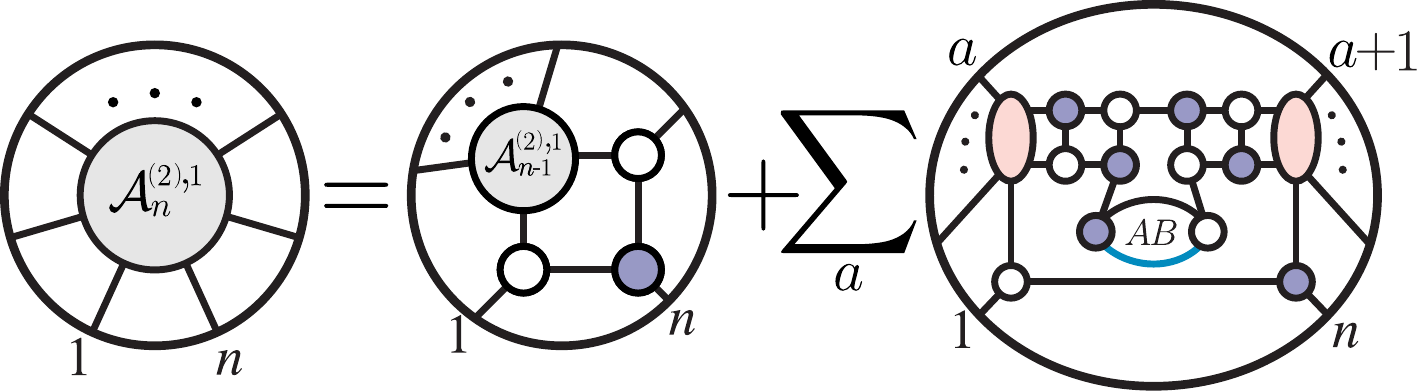}}\hspace{-3cm}\vspace{-.2cm}\nonumber}

We can rearrange the NMHV forward-limit contributions as we did above in order to make manifest the sequence of BCFW-bridges which parameterize the extra degrees of freedom:
\vspace{-.2cm}\eq{\hspace{-3cm}\raisebox{-54pt}{\includegraphics[scale=1]{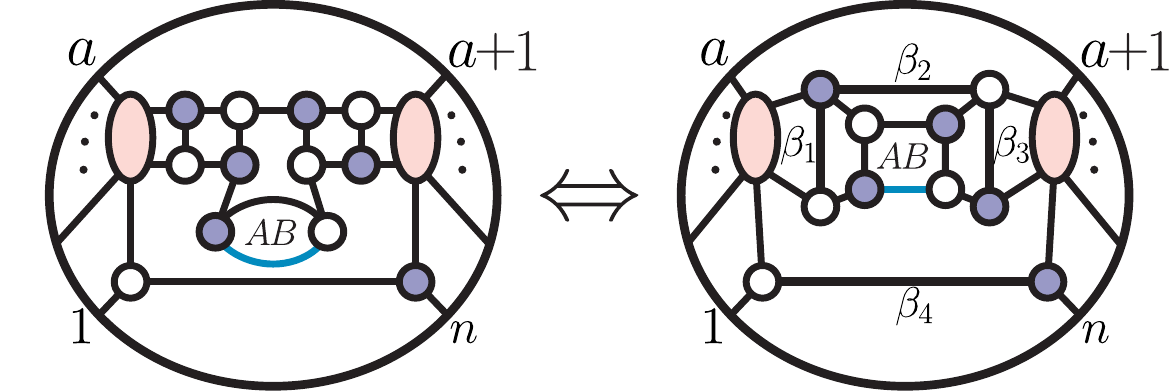}}\hspace{-.0cm}\label{general_nmhv_forward_limit_graph_to_canonical_form}\hspace{-3cm}\nonumber\vspace{-.1cm}}
Proceeding as before, we can identify the shifted momentum-twistors appearing in the box as,
\vspace{-.2cm}\eq{(AB)=(\hat{a}\,\hat{n})\quad\mathrm{with}\quad\left\{\begin{array}{rl}z_{\hat{a}}&=z_a\pl\beta_1z_{a+1}\pl\beta_2z_1\\z_{\hat{n}}&=z_n\pl\beta_3z_1\pl\beta_4z_{n-1}\end{array}\right\},\vspace{-.2cm}}
which allows us to re-cast the BCFW-bridge variables $\beta_i$ in terms of the line $(AB)$:
\vspace{-.2cm}\eq{\beta_1=\frac{\ab{AB\,1\,a}}{\ab{AB\,a\pl1\,1}},\quad\beta_2=\frac{\ab{AB\,a\,a\pl1}}{\ab{AB\,a\pl1\,1}},\quad\beta_3=\frac{\ab{AB\,n\mi1\,n}}{\ab{AB\,1\,n\mi1}},\;\;\mathrm{and}\;\;\beta_4=\frac{\ab{AB\,n\,1}}{\ab{AB\,1\,n\mi1}}.\vspace{-.2cm}\nonumber}

Therefore, we see that the forward-limit terms are given by:
\vspace{-.2cm}\eq{\hspace{-4cm}\begin{array}{rl}\raisebox{-55pt}{\includegraphics[scale=1]{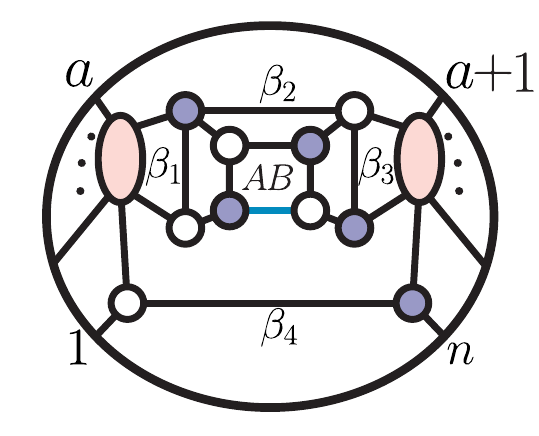}}&\hspace{-.3cm}=d\!\log(\beta_1)d\!\log(\beta_2)d\!\log(\beta_3)d\!\log(\beta_4)\\[-20pt]&\hspace{-2cm}=\displaystyle d\!\log\!\Bigg(\!\!\frac{\ab{AB\,1a}}{\ab{AB\,1a\pl1}}\!\!\Bigg)d\!\log\!\Bigg(\!\!\frac{\ab{AB\,aa\pl1}}{\ab{AB\,1a\pl1}}\!\!\Bigg)d\!\log\!\Bigg(\!\!\frac{\ab{AB\,n\mi1n}}{\ab{AB\,1n\mi1}}\!\!\Bigg)d\!\log\!\Bigg(\!\!\frac{\ab{AB\,n1}}{\ab{AB\,1n\mi1}}\!\!\Bigg).\end{array}\hspace{-.0cm}\hspace{-3cm}\nonumber\vspace{-.2cm}}
Quite amazingly, if we re-cast this integration measure directly in terms of the line $(AB)$, we see that this is equivalent to,
\vspace{-.2cm}\eq{\hspace{-3.85cm}\begin{array}{rl}\raisebox{-55pt}{\includegraphics[scale=1]{general_nmhv_forward_limit_graph_with_bridges}}&\hspace{-.3cm}=\displaystyle\frac{\ab{d^2\!z_A\,AB}\ab{d^2\!z_B\,AB}\ab{AB\,(1\,a\,a\pl1)\newcap(1\,n\mi1\,n)}^2}{\ab{AB\,1\,a}\ab{AB\,a\,a\pl1}\ab{AB\,a\pl1\,1}\ab{AB\,1\,n\mi1}\ab{AB\,n\mi1\,n}\ab{AB\,n\,1}}\\[-20pt]&\hspace{-.3cm}\equiv K[a;n\,\mi\,1].\end{array}\hspace{-.0cm}\hspace{-3cm}\nonumber\vspace{-.2cm}}
We have obtained this result entirely by manipulating pictures of on-shell diagrams; of course the result precisely matches the form obtained by direct computation, using the methods of \cite{ArkaniHamed:2010kv}, where all MHV one-loop integrands were given in the form,
\vspace{-.2cm}\eq{\mathcal{A}_n^{(2),1}=\!\!\sum_{1<a<b<n}\!\!\!\!\!K[a;b].\vspace{-.2cm}}

Before moving on to multi-loop integrands, it is worth mentioning that for one-loop integrands, so long as the forward-limits are taken of tree-amplitudes obtained by BCFW deforming the identified legs $(A\,B)$, it turns out that the obvious $k$-preserving and $k$-increasing inverse-soft factors are the {\it only} terms which vanish in the forward limit; this allows us to conclude that,
\vspace{-.2cm}\eq{\text{\# BCFW terms in the one-loop amplitude }\mathcal{A}_n^{(k),1}\!:\,\binom{n\,\mi\,2}{k}\binom{n\,\mi\,2}{k\,\mi\,2}.\phantom{\frac{1}{n-3}}\!\vspace{-.2cm}}
(It turns out that this counting holds regardless of how the forward-limit terms are recursed---even though it is generally difficult to identify beforehand which terms will vanish if $(AB)$ are not singled-out for the recursion. Beyond one-loop, however, the number of non-vanishing contributions is not invariant, and depends sensitively on how the lower-loop amplitudes are recursed.)

When expressing tree-amplitudes and their forward-limits in terms of canonical coordinates on the auxiliary Grassmannian, it is obvious that all loop integrands can be---and are most naturally---expressed in such a `$d\!\log$'-representation. Although in principle we have all the tools necessary to construct such formulae for all  amplitudes---and although the BCFW recursion relations (\ref{bcfw_all_loop_recursion_disc}) is {\it dramatically} more efficient that any representation obtained using `traditional' methods (such as Feynman diagrams)---even the simplest 2-loop integrands would require more space to write completely than would be warranted for the purpose of illustration.

Let us therefore content ourselves to consider one simple example of a contribution to the $4$-particle $2$-loop integrand which arises as the double forward-limit of a of the contributions to the $8$-particle N$^2$MHV tree-amplitude, that of  $\big(\mathcal{A}_{4}^{(2)}\!\!\otimes\!\mathcal{A}_4^{(2)}\big)\!\otimes\!\mathcal{A}_4^{(2)}$:
\vspace{-.2cm}\eq{\hspace{-2.7cm}\raisebox{-54pt}{\includegraphics[scale=1]{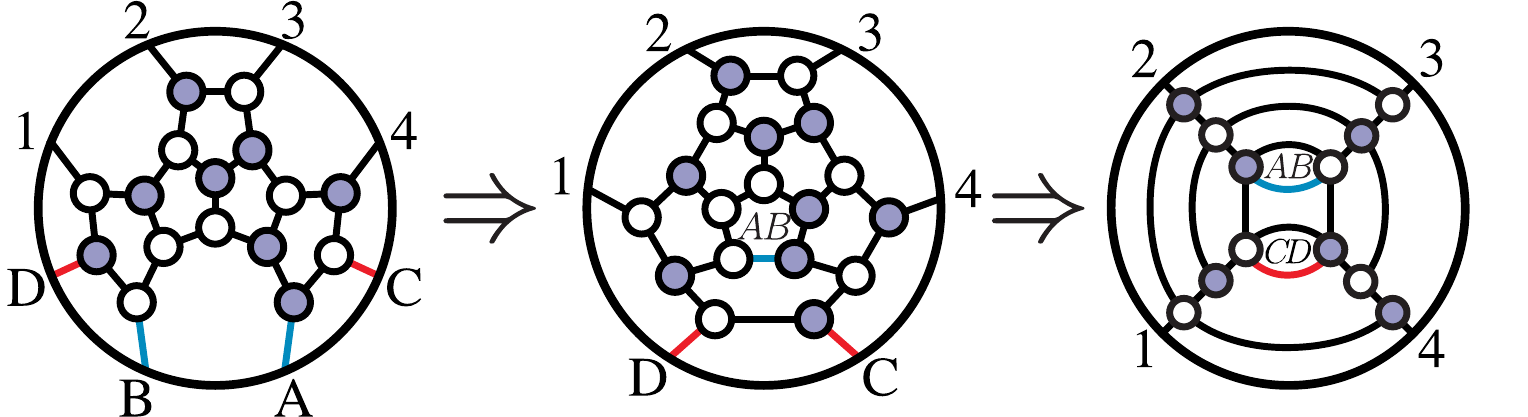}}\label{g48_double_forward_limit_graph_sequence}\hspace{-3cm}\nonumber\vspace{-.2cm}}
In the last step, we have made liberal use of square-moves and merge/un-merge operations to bring it in the form which exposes a sequence of recognizable BCFW-bridges which themselves encode the additional degrees of freedom.

Using the tools described in \cite{ArkaniHamed:2010kv} to compute this contribution directly as a `function' of lines $(AB)$ and $(CD)$ in momentum-twistor space, the following rational integrand is found:
\vspace{-.2cm}\eq{\hspace{-2cm}\text{{\small$\displaystyle\frac{\ab{d^2z_{A}\,AB}\ab{d^2z_B\,AB}\ab{d^2z_C\,CD}\ab{d^2z_D\,CD}\ab{1234}^3\ab{A\!B(C\!D)\!\bigcap(341)1}^2}{\ab{A\!B 14}\ab{A\!B 1(123)\!\bigcap(C\!D)}\ab{A\!B1(234)\!\bigcap(C\!D)}\ab{A\!B(((C\!D(341)\!\bigcap(A\!B))\!\bigcap(12))34)\!\bigcap(C\!D)1}\ab{A\!B C\!D}\ab{C\!D 34}}$}}.\hspace{-2cm}\nonumber\vspace{-.2cm}}
While the expression above is of course obtained in a straight-forward way, it is obviously rather complicated and not particularly illuminating. Moreover, as written in the form given above---as a rational integrand---it is not at all obvious that there exists {\it any} change of variables for which it becomes simply the wedge-product of 8 logarithmic factors. But from our present perspective, the existence of such a change of coordinates is an obvious consequence of the Grassmannian formulation of the initial tree-amplitude; and it will be instructive to see how this remarkable connection is realized.

To be extremely concrete, we want to identify the lines $(AB)$ and $(CD)$ as parameterizing the region-momenta according to,
\vspace{-.2cm}\eq{\raisebox{-52pt}{\includegraphics[scale=1]{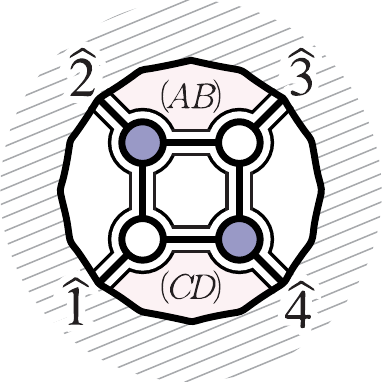}}\quad\mathrm{with}\quad\begin{array}{c}(AB)=(\hat{2}\,\hat{3})\\(CD)=(\hat{4}\,\hat{1})\end{array}\quad\vspace{-.2cm}}
We can find the shifted momentum-twistors $z_{\hat{a}}$ by performing the successive BCFW-shifts obvious from the way the double forward-limit graph is drawn:
\vspace{-.0cm}\eq{\hspace{-3.cm}\raisebox{-60pt}{\includegraphics[scale=1]{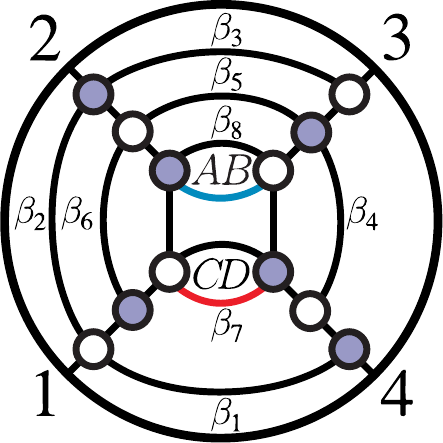}}\,\text{{\small$\begin{array}{|c|l|}\hline\begin{array}{c}\!\text{bridge}\!\end{array}&\multicolumn{1}{|c|}{\text{BCFW shift}}\\[-2pt]\hline(1\,4)&z_4\!\mapsto\! z_4\pl\,\beta_1z_3\\[-0pt]\hline(1\,2)&z_2\!\mapsto\! z_2\pl\,\beta_2 z_3\\\hline(3\,2)&z_2\!\mapsto\! z_2\pl\,\beta_3z_1\\\hline(4\,3)&z_3\!\mapsto\! z_3\pl\,\beta_4 z_2\\\hline(2\,3)&z_3\!\mapsto\! z_3\pl\,\beta_5z_4\\\hline(2\,1)&z_1\!\mapsto\! z_1\pl\,\beta_6z_4\\\hline(1\,4)&z_4\!\mapsto\! z_4\pl\,\beta_7z_3\\\hline(3\,2)&z_2\!\mapsto\! z_2\pl\,\beta_8z_1\\\hline
\end{array}$}}\;\raisebox{-2pt}{{\LARGE$\Rightarrow$}}\left\{\!\!\text{{\small$\begin{array}{l@{}l}z_{\hat{1}}\!=&z_1\pl\beta_5(z_4\pl\beta_1z_3);\\[2pt] z_{\hat{2}}\!=&z_2\pl\beta_2z_3\pl\beta_3z_1\\&\pl\beta_8(z_1\pl\beta_5(z_4\pl\beta_1z_3));\\[2pt] z_{\hat{3}}\!=&z_3\pl\beta_4(z_2\pl\beta_2z_3\pl\beta_3z_1)\\&\pl\beta_6(z_4\pl\beta_1z_3);\\[2pt] z_{\hat{4}}\!=&z_4\pl\beta_1z_3\pl\beta_7\beta_6(z_4\pl\beta_1z_3)\\&\pl\beta_7(z_3\pl\beta_4(z_2\pl\beta_2z_3\pl\beta_3z_1));\end{array}$}}\!\!\right\}\nonumber\label{region_momenta_for_two_loop_example}\hspace{-3cm}\vspace{-.2cm}}
One can readily verify that quite remarkably, with this change of variables, the complicated expression given above becomes simply,
\vspace{-.1cm}\eq{d\!\log(\beta_1)\wedge\cdots \wedge d\!\log(\beta_8).\vspace{-.2cm}}

\subsection{The Transcendentality of Loop Amplitudes}\label{transcendentality_of_loop_amplitudes_subsection}

The integrand obtained from the BCFW recursion relations allows us to draw some important general conclusions about the structure of the final, integrated expressions for the amplitude. Let us start with MHV amplitudes. As we have seen, all the BCFW terms at $L$ loops can be written in the form,
\vspace{-.2cm}\eq{{\cal A}^{(2)}_{n,L} = {\cal A}^{(2)}_{n,0}\times \prod_{i=1}^{4L} d\!\log(\beta_i).\vspace{-.2cm}}
The first and most obvious point to observe is that the integrand has {\it only} logarithmic singularities! There are {\it no} ``{\it sub-leading}'' pieces of the integrand with less than the maximal number of logarithmic singularities. At one-loop, this (together with dual conformal invariance) tells us the famous fact that the loop amplitude only depends on ``box'' integrals, and doesn't involve any triangles, bubbles, or rational pieces \cite{Bern:1992ad,Bern:1993tz}.

As we have stressed a number of times, the fact that the integrand has only logarithmic singularities is not at all obvious from inspection of the actual rational functions involved in sufficiently high loop-amplitude integrands, where there don't seem to be enough ``obvious'' singularities in cutting propagators, and so singularities must emerge as ``composites''. By contrast, the positive Grassmannian story makes this fact completely obvious.  Intuitively, this  guarantees that after integration, the $L$-loop MHV amplitudes can always be expressed as a sum of polylogarithms of transcendentality $2L$. The reason is roughly that discontinuities of the amplitude are related to unitarity cuts that put pairs of particles on-shell; thereby computing partial residues of the integrand. Taking $2L$ discontinuities gives the leading singularity ``1'', which has no further discontinuities. These amplitudes are thus ``pure''---not polluted by lower-transcendentality terms, which would arise from pieces of the integrand without purely logarithmic singularities. This has long been conjectured to be true for MHV amplitudes in connection to the maximal transcendentality principle of \cite{Kotikov:2002ab}. We see that the property needed of the integrand to guarantee this is a trivial consequence of the $d\!\log$ form.

Beyond MHV amplitudes, we know that the integrated amplitudes can involve more complicated functions than polylogarithms. For instance, as pointed out in ref. \cite{CaronHuot:2012ab}, the two-loop, $10$-point N$^3$MHV amplitude includes a contribution from a function whose seven-fold discontinuity is proportional to the following on-shell form:
\vspace{-.7cm}\eq{\begin{array}{c}\text{{\small$\phantom{\{\}}$}}\\[-7pt]\raisebox{-1pt}{\includegraphics[scale=1]{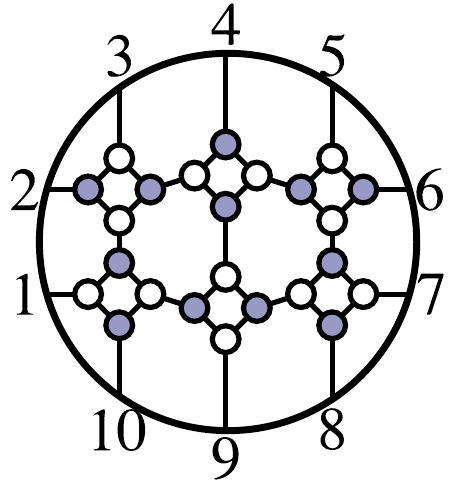}}\\[-7pt]\text{{\footnotesize${\color{perm}\{7,6,10,9,8,12,11,15,14,13\}}$}}\end{array}\vspace{-.2cm}\label{g510_elliptic_integrand}}
This on-shell graph corresponds to a $17$-dimensional cell in $G(5,10)$; the kinematical constraints will fix this to an integral over one degree of freedom (interpreted as the `hepta-cut' of the two-loop integrand). The component amplitude proportional to,
\vspace{-.2cm}\eq{\big(\widetilde\eta_1^{\,1}\widetilde\eta_1^{\,2}\big)\big(\widetilde\eta_2^{\,1}\widetilde\eta_2^{\,2}\big)\big(\widetilde\eta_3^{\,2}\widetilde\eta_3^{\,3}\big)\big(\widetilde\eta_4^{\,2}\widetilde\eta_4^{\,3}\big)\big(\widetilde\eta_5^{\,2}\widetilde\eta_5^{\,3}\big)\big(\widetilde\eta_6^{\,3}\widetilde\eta_6^{\,4}\big)\big(\widetilde\eta_7^{\,3}\widetilde\eta_7^{\,4}\big)\big(\widetilde\eta_8^{\,4}\widetilde\eta_8^{\,1}\big)\big(\widetilde\eta_9^{\,4}\widetilde\eta_9^{\,1}\big)\big(\widetilde\eta_{10}^{\,4}\widetilde\eta_{10}^{\,1}\big),\vspace{-.2cm}}
(a component which vanishes exactly at tree-level and one-loop) vanishes on all the positroid cells in the boundary of (\ref{g510_elliptic_integrand}). Therefore, the only contour integral available must enclose the Jacobian resulting from the kinematical constraints; this Jacobian generically involves the square-root of an irreducible quartic, implying that (at least for this component) the seven-fold discontinuity of the 2-loop integrand is an elliptic integral.

We can understand the difference between MHV and higher-$k$ amplitudes from the Grassmannian. Recall that cells of dimensionality $(2n\,\mi\,4)$ are fully localized by the kinematical constraints. Since for MHV amplitudes, \mbox{$\dim(G(2,n))=(2n\,\mi\,4)$}, {\it all} of the unfixed degrees of freedom associated with `loop-momenta' are associated with faces which can always be removed by reduction (as there no irreducible graphs with more faces than that of the top-cell). Beyond MHV, however, the reduction of on-shell diagrams can result in cells of higher dimensionality than $(2n\,\mi\,4)$. For example, consider the top-cell of $G(3,6)$:
\vspace{-.0cm}\eq{\hspace{-3.cm}\raisebox{-52pt}{\includegraphics[scale=1]{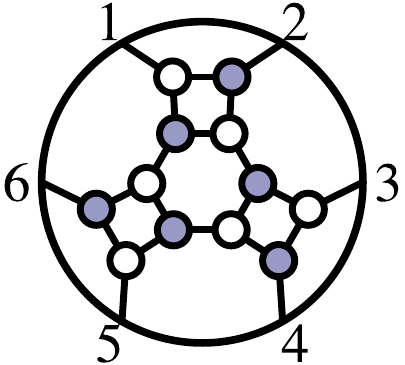}}\label{g36_top_cell}\hspace{-3cm}\vspace{-.2cm}}
Here, we have chosen a representative graph which makes it clear that it can be associated with a triple-cut of the 6-particle amplitude at 1-loop. The \mbox{$9\,\mi\,8 = 1$} degree of freedom of the top-cell which is not fixed by the kinematical constants can always be interpreted as the single integration variable of a triple-cut integral.

Similarly, the top-cell of $G(4,8)$ is 16-dimensional, while the kinematical constraints can be used to isolate only \mbox{$2\!\times\!8\,\mi\,4=12$} degrees of freedom; therefore, the top-cell on $G(4,8)$ can be viewed as an on-shell differential form with {\it four} unfixed auxiliary degrees of freedom---which can in fact be interpreted as the four-degrees of freedom of a `loop-integrand' at one-loop. Indeed, we can represent the top-cell by,
\vspace{-.2cm}\eq{\hspace{-3.cm}\raisebox{-60pt}{\includegraphics[scale=1]{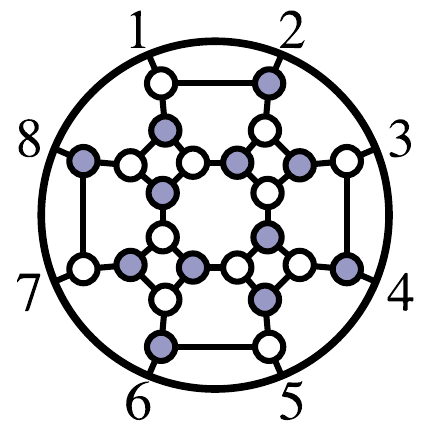}}\label{g48_top_cell}\hspace{-3cm}\vspace{-.2cm}}

Therefore beyond MHV, while the integration measures are purely $d\!\log$'s, some free integration variables are {\it inside} the Grassmannian, and must be localized by the kinematic constraints. This is the reason why more complicated functions can appear after integration. However, it is clear that for fixed $n$ and $k$, the functions can't get arbitrarily more complicated at high loop orders. The reason is that at most $\dim(G(k,n))\mi(2n\,\mi\,4)$ of the integration variables can remain `entangled' in the Grassmannian (meaning that they cannot be pulled-off as overall $d\!\log$ factors in the measure via bubble-reduction); at arbitrarily-high loop order, all but a finite number of these auxiliary degrees of freedom must be associated with the more trivial factors in the measure arising from bubble-reduction.

Actually, it is easy to see that, for NMHV amplitudes, the integrations that are ``stuck'' in the Grassmannian can easily be removed, preserving the $d\!\log$ form, and thus that all NMHV amplitudes are also polylogarithms. Let us illustrate with the top cell of $G(3,6)$; it is convenient to work in momentum-twistor language, where this maps to the top cell of $G(1,6)$. On the support of the (ordinary) $\delta$-functions,
we have a $1$-form which we can represent as,
\vspace{-.2cm}\eq{ [1 \, 2 \, 3 \, 4 \, (5\pl\beta 6)] d\!\log(\beta),\vspace{-.2cm}}
However, we can use the identity among the $5$-brackets, (\ref{nmhv_identity}), to rewrite this as
\vspace{-.2cm}\eq{\begin{array}{c}[1\,2\,3\,4\,5]d\!\log(\beta)+\displaystyle [2\,3\,4\,5\,6]d\!\log\!\left(\beta\,\mi\,\frac{\ab{2345}}{\ab{2346}}\right)+\displaystyle[3\,4\,5\,6\,1]d\!\log\!\left(\beta\,\mi\,\frac{\ab{3451}}{\ab{3461}}\right)\\[12pt]+\displaystyle[4\,5\,6\,1\,2]d\!\log\!\left(\beta\,\mi\,\frac{\ab{4512}}{\ab{4612}}\right)+\displaystyle[5\,6\,1\,2\,3]d\!\log\!\left(\beta\,\mi\,\frac{\ab{5123}}{\ab{6123}}\right)\\[-20pt]\end{array}\vspace{.6cm}}
\noindent In this way, we have removed the integration variable from inside the Grassmannian and decomposed the result into a sum of terms, each of which is in canonical form. The same thing can be done for the top cell of any NMHV amplitude, since the ``internal'' variables always occur linearly. Things can start becoming non-trivial at N$^2$MHV, where square-roots first make an appearance, and as we've seen concretely above, already for 10-particle N$^3$MHV amplitudes, elliptic integrals do make an appearance.

The on-shell, BCFW-representation of loop-integrands delivers them manifestly in a canonical, $d\!\log$-form; but having noted that the integrand can be put in this form, it is natural to wonder if this is a consequence of the BCFW-representation, or a more general result. For instance, in reference \cite{ArkaniHamed:2010gh}, extremely compact, {\it local} forms of many integrands were found; can these also be written in terms of integrands with only logarithmic singularities? The answer yes: the $d\!\log$ form is a {\it general} property of ``pure'' integrands with unit leading singularities.  Let us briefly demonstrate this  fact with two examples: local forms of the MHV $1$- and $2$-loop integrands.

In \cite{ArkaniHamed:2010gh}, the 1-loop MHV integrand was given in the local form,
\vspace{-.4cm}\eq{\hspace{-2.cm}\raisebox{-44pt}{\includegraphics[scale=1]{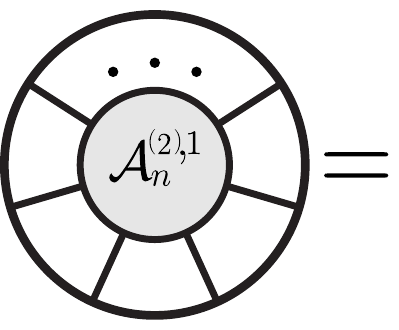}}\text{{\Large$\!\sum_{a<b<a}\!\!\!I_X[a;b],$}}\label{local_mhv_one_loop_expansion}\hspace{-3cm}\vspace{-.3cm}}
where $I_X[a;b]$ denotes the integrand,
\vspace{-.2cm}\eq{I_X[a;b]\equiv\frac{\ab{A\!B d^2 z_A}\ab{A\!B d^2 z_B}\;\ab{A\!B\,(a\,\mi1\,a\,a\pl1)\newcap(b\,\mi1\,b\,b\pl1)}\ab{X\,ab}}{\ab{A\!B\,a\,\mi1\,a}\ab{A\!B\,a\,a\pl1}\ab{A\!B\,b\,\mi1\,b}\ab{A\!B\,b\,b\pl1}\ab{A\!B\,X}},\vspace{-.2cm}}
and where $X$ is an arbitrary reference-line in momentum-twistor space (spanned by any pair of twistors). Remarkably, it turns out that $I_X[a;b]$ {\it can} be expressed in canonical form:
\vspace{-.2cm}\eq{\hspace{-1.5cm}\!d\!\log\!\Bigg(\!\!\frac{\ab{A\!B\,a\mi1a}}{\ab{A\!B\,X}}\!\!\Bigg)\!d\!\log\!\Bigg(\!\!\frac{\ab{A\!B\,aa\pl1}}{\ab{A\!B\,X}}\!\!\Bigg)\!d\!\log\!\Bigg(\!\!\frac{\ab{A\!B\,b\mi1b}}{\ab{A\!B(aX)\newcap(b\mi1bb\pl1)}}\!\!\Bigg)\!d\!\log\!\Bigg(\!\!\frac{\ab{A\!B\,bb\pl1}}{\ab{A\!B(aX)\newcap(b\mi1bb\pl1)}}\!\!\Bigg).\hspace{-2cm}\vspace{-.2cm}\nonumber}
(Note that while this form appears to break the symmetry between $a$ and $b$, the form is of course symmetrical.)

Similarly, it was found in \cite{ArkaniHamed:2010gh} that 2-loop MHV integrand could be written in the following local form:
\vspace{-.2cm}\eq{\hspace{-2.cm}\raisebox{-44pt}{\includegraphics[scale=1]{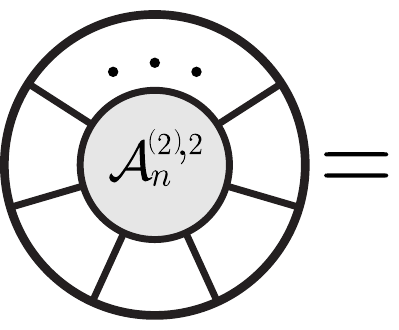}}\raisebox{-0pt}{{\Large$\displaystyle\;\!\!\!\!\!\!\!\sum_{a<b<c<d<a}\!\!\!\!\!\!I[a,b;c,d]$}}\label{local_mhv_two_loop_expansion}\hspace{-3cm}\vspace{-.2cm}}
where $I[a,b;c,d]$ denotes the integrand,
\vspace{-.2cm}\eq{\hspace{-2cm}\frac{\ab{A\!B d^2 z_A}\ab{A\!B d^2 z_B}\ab{C\!D d^2 z_C}\ab{C\!D d^2 z_D}\ab{A\!B\,(a\,\mi1aa\pl1)\newcap(b\,\mi1bb\pl1)}\ab{A\!B\,(c\,\mi1cc\pl1)\newcap(d\,\mi1dd\pl1)}}{\ab{A\!B\,a\,\mi1a}\ab{A\!B\,aa\pl1}\ab{A\!B\,b\,\mi1b}\ab{A\!B\,bb\pl1}\ab{A\!B\,c\,\mi1c}\ab{A\!B\,cc\pl1}\ab{A\!B\,d\,\mi1d}\ab{A\!B\,dd\pl1}}.\hspace{-2cm}\vspace{-.2cm}\nonumber}
But it turns out that this integrand can also be written in canonical form:
\vspace{-.2cm}\eq{\hspace{-2cm}\begin{array}{l}\displaystyle \phantom{\times}d\!\log\!\Bigg(\!\!\frac{\ab{A\!B\,a\,\mi1a}}{\ab{A\!B\,C\!D}}\!\!\Bigg)\!d\!\log\!\Bigg(\!\!\frac{\ab{A\!B\,aa\pl1}}{\ab{A\!B\,C\!D}}\!\!\Bigg)\!d\!\log\!\Bigg(\!\!\frac{\ab{A\!B\,b\,\mi1b}}{\ab{A\!B(aC\!D)\newcap(b\,\mi1bb\pl1)}}\!\!\Bigg)\!d\!\log\!\Bigg(\!\!\frac{\ab{A\!B\,bb\pl1}}{\ab{A\!B(aC\!D)\newcap(b\,\mi1bb\pl1)}}\!\!\Bigg)\\\times \displaystyle
d\!\log\!\Bigg(\!\!\frac{\ab{C\!D\,c\,\mi1c}}{\ab{C\!D\,ab}}\!\!\Bigg)\!d\!\log\!\Bigg(\!\!\frac{\ab{C\!D\,cc\pl1}}{\ab{C\!D\,ab}}\!\!\Bigg)\!d\!\log\!\Bigg(\!\!\frac{\ab{C\!D\,d\,\mi1d}}{\ab{C\!D(abc)\newcap(d\,\mi1dd\pl1)}}\!\!\Bigg)\!d\!\log\!\Bigg(\!\!\frac{\ab{C\!D\,dd\pl1}}{\ab{C\!D(abc)\newcap(d\,\mi1dd\pl1)}}\!\!\Bigg).\end{array}\hspace{-2cm}\vspace{-.2cm}\nonumber}

\newpage
\section{Outlook}\label{outlook_section}

We have explored much of the remarkable physics and mathematics of scattering amplitudes in planar $\mathcal{N}\!=\!4$ SYM, as seen through the lens of on-shell diagrams as the primary objects of study. Let us conclude by making some brief comments on further avenues of research.

One immediate extension of our work is the continued study of theories with $\mathcal{N} < 4$ SUSY, whose most basic features we sketched out in section \ref{less_supersymmetries_section}. For $\mathcal{N} \geq 1$, all-loop BCFW recursion holds just as for $\mathcal{N}\!=\!4$, together with its realization in terms of on-shell diagrams. For $\mathcal{N}\!=\!0$ SUSY, the forward limit of tree amplitudes are singular, and thus don't directly give us the single-cuts of the loop-integrand \cite{CaronHuot:2010zt}. More thought is needed to establish a connection between on-shell diagrams and the full amplitude, though it is likely that fully understanding the on-shell diagrams will continue to play an important role in determining $\mathcal{N}\!=\!0$ amplitudes as well.

The general connection between on-shell diagrams and the Grassmannian has nothing to do with any particular theory, only with the general picture of amalgamating basic three-particle amplitudes, and the connection to the positive Grassmannian in particular holds for any planar theory. Only the form on the Grassmannian changes from theory to theory. As briefly discussed in section \ref{less_supersymmetries_section}, the essential physical novelty of gauge theories with $\mathcal{N} \leq 2$ supersymmetry is the presence of UV-divergences. The most physical, Wilsonian, way to think about UV-divergences makes critical use of off-shell ideas, and so a major challenge is finding the correct way of thinking about such physics in a directly on-shell language. It is fascinating to see that the UV and IR singularities, together with UV/IR decoupling, is reflected directly in on-shell diagrams through simple structures in the Grassmannian. A clear goal would be to understand the physics of the renormalization group along these lines.

Another obvious extension is to push beyond the planar limit, starting already with $\mathcal{N}\!=\!4$ SYM; in this case, there is no longer an obvious notion of ``the loop integrand'',  and thus we must learn how to establish a connection between on-shell diagrams and the full scattering amplitude along the lines of the BCFW construction in the planar limit. It is also very likely that on-shell ideas can be used to determine other observables in gauge theories beyond scattering amplitudes, including all correlation functions. These objects also have discontinuities and cuts, and the on-shell diagrams for leading singularities of form-factors and correlation functions are exactly the same as the (in general non-planar) on-shell diagrams we have been considering. The structure of cuts has already proved to powerful in determining form-factors, \cite{Brandhuber:2011tv}. For scattering amplitudes, we have seen that off-shell notions like virtual loop integration variables can be fully understood in on-shell terms. It is tempting to try and compute completely off-shell objects like correlation functions in the same way.

Moving further afield, as alluded to in \mbox{section \ref{cluster_coordinates_section}}, the basic mathematical structures we have encountered in scattering amplitudes have also recently made an appearance in apparently completely different physical settings, related to conformal blocks for higher Toda theories \cite{FG1,FG4}, wall-crossing \cite{KS,Gaiotto:2011tf}, various versions of the AGT conjecture \cite{Alday:2009aq}, scattering amplitudes at strong coupling \cite{Alday:2010kn}, and soliton solutions of the KP equation \cite{Kodama:2011ht,Kodama:2011iq,Kodama:2012zx}. The identical graphical structure has also appeared in the construction of $\mathcal{N}\!=\!1$ SCFTs associated with quiver gauge theories (see e.g.\ \cite{Heckman:2012jh,Franco:2012wv}). The combinatorial classification of on-shell diagrams and these planar $\mathcal{N}\!=\!1$ SCFTs coincide perfectly. It would be interesting to see if the rest of the structure we have been seeing--especially the connection with the positive Grassmannian--have a natural interpretation as well.

There is a unifying theme running through the physics and mathematics we have been discussing.  We have an object---the positive Grassmannian---which is fundamentally defined by global properties, either as a real space, by demanding all ordered minors are positive, or as a complex space, by specifying linear dependencies between consecutive vectors. However quite remarkably, the best way of building-up these objects (albeit in a highly redundant way) is through the amalgamation of elementary building blocks.

For scattering amplitudes, amalgamation representations have a direct physical interpretation as on-shell diagrams. For $\mathcal{N}\!=\!1$ gauge theories, they correspond to gluing together gauge groups with bi-fundamental content to generate more complicated quiver gauge theories. For scattering amplitudes, it is physically clear why we should be interested in complicated on-shell diagrams, since they are ultimately needed to compute the amplitude to all-loop order. But what is physically important about complicated quiver gauge theories? One possible answer is that precisely these sort of quiver gauge theories, with an infinite number of sites and links, occur in the deconstruction of the still mysterious $(2,0)$ and little string theories in six dimensions,  \cite{ArkaniHamed:2001ie}. It would be fascinating to use the powerful new machinery for studying these quivers to try and learn more about the dynamics of the underlying six-dimensional theories, which would perhaps shed some light on a more direct physical reason for the appearance of the same Grassmannian structure in seemingly vastly different settings.

We have seen that scattering amplitudes in $(1\pl1),$ $(2\pl1)$ and $(3\pl1)$ dimensions are described by various interpretations of permutations and associated structures in the Grassmannian. It is natural to ask whether other variations of these mathematical ideas might have a physical interpretation. There is one natural further specialization of the positive Grassmannian we have not discussed, which in fact goes back to the historical roots of the subject: the study of totally positive matrices. Here, one considers $(n\!\times\!n)$ square matrices $M$ with positive determinant, and studies the space where all its $(m\!\times\!m)$-minors are non-negative. This classical problem was studied by Gantmacher and Krein \cite{GKr} and Schoenberg \cite{Sch} in the 1930's, where the stratification was found to be determined by {\it pairs} of permutations $\sigma_1$ and $\sigma_2$. This theory is a special case of the positive Grassmannian $G(n,2n)$. Consider cells where the first $n$ columns of the $(n\!\times2n)$ $C$ matrix are linearly independent, and also the second $n$ columns are linearly independent. We can then gauge-fix $C$ to the form
\vspace{-.2cm}\eq{C = \left(1_{n \times n} \, |\, M_{n \times n} \right),\vspace{-.2cm}}
where $M$ is a positive matrix. Let us label the first $n$ columns $1, \ldots, n$ and the second $n$ columns by $1^\prime, \ldots, n^\prime$. It is clear that e.g.\ $\sigma(1) = a^\prime$ for some $a^\prime$ in the second set of columns, since $1$ can not be in the span of $\{2, \ldots, n\}$, given that the first $n$ columns are linearly independent. This is true for all the other columns in the first set---i.e.\ $\sigma(a) = b^\prime$. Similarly, $\sigma(a^\prime) = b$. Thus, we see that our permutation naturally breaks into two pieces, mapping $(1, \ldots, n)\!\mapsto\!(1^\prime,\ldots, n^\prime)$ and vice-versa. It would be nice to find a physical interpretation for the subclass of on-shell diagrams associated with these pairs of permutations.

We have also seen a reliable guide to the Grassmannian structure associated with scattering amplitudes is to find a Grassmannian interpretation of the space of external kinematical data. In four dimensions, the $\lambda$- and $\widetilde \lambda$-planes are represented by points in $G(2,n)$. In three dimensions, the $\lambda$-plane is an element of the null orthogonal Grassmannian. What happens in higher dimensions? The description of the external kinematical data in six dimensions is particularly simple \cite{Cheung:2009dc}. The complexified Lorentz group can be taken to be $SO(5,1)\!\sim\!SL(4)$, and a null momentum vector can be represented as an antisymmetric $(4\!\times\!4)$ tensor $p^{IJ}$ of vanishing determinant. The complexified little group is $SL(2)\!\times\!SL(2)$.
As such, we can express the momentum $p_a^{IJ}$ of particle $a$ as,
\vspace{-.2cm}\eq{p_a^{IJ} = \epsilon^{\alpha \beta} \lambda^I_{a\,\alpha} \lambda^J_{a\,\beta} (= \epsilon^{\dot{\alpha} \dot{\beta}} \widetilde \lambda^I_{a\,\dot{\alpha}} \widetilde \lambda^J_{a\,\dot{\beta}}).\vspace{-.2cm}}
Note the similarity to ordinary spinor-helicity variables---except that here, the $\alpha, \dot{\alpha}$ indices aren't Lorentz indices as familiar from four dimensions, but are instead indices of the $SL(2)\times SL(2)$ {\it little group}. We can group all the $\lambda$'s for the particles $a= 1,\ldots,n$ together into a $(4\!\times\!2n)$-matrix,
\vspace{-.2cm}\eq{
\Lambda_A^{I} = \left(\lambda^I_{1\,1} \, \, \lambda^I_{1\,2} \, \, \lambda^I_{2\,1} \, \, \lambda^I_{2\,2} \, \, \cdots\,\, \lambda^I_{n\,1} \, \, \lambda^I_{n\,2} \right).\vspace{-.2cm}}
Momentum conservation is then the statement that,
\vspace{-.2cm}\eq{\Lambda_A^I\Lambda_B^JJ^{AB}=0\quad\mathrm{where}\quad J\equiv\left(\begin{array}{@{}cc@{$\,$}c@{$\!$}cc@{}}0&1&\\[-3pt]\mi1\phantom{\mi}&0&\\[-10pt]&&\ddots\\[-7pt]&&&0&1\\[-3pt]&&&\mi1\phantom{\mi}&0\end{array}\right).\vspace{-.2cm}}
Thus, in close parallel with $(2\pl1)$ dimensions, the external data in $6$ dimensions is associated with a point in the null symplectic Grassmannian, \cite{yinthuang}. It would be interesting to see if this structure has any role to play in six-dimensional physics.

Let us close by returning to a number of concrete open directions of research flowing more directly from the ideas presented here.

In this paper, we have given a complete classification and understanding of all reduced on-shell diagrams, whose invariant content is captured by the permutation associated with the left-right paths. Amongst other things, all terms occurring
in the tree-level BCFW recursion relations are reduced graphs, and indeed, the recursion can be described purely combinatorially as a simple and canonical ``bridging'' of permutations. We have however also seen that non-reduced on-shell diagrams are also physically important, directly giving the loop integrand. Of course, the non-reduced graphs for the loop integrand arise from merging adjacent legs of higher-point reduced graphs, which we understand completely. Nonetheless, it would clearly be interesting and important to try and extend the classification of the on-shell diagrams to non-reduced graphs as well; in other words, we would like to understand all the invariants on non-reduced graphs, that can be related by merges and square moves. Obviously the left-right path permutations are still invariants, but there are clearly further invariants as well. For instance, suppose we have two non-reduced graphs with exactly the same permutation, but where the first graph has a bad double-crossing between the two paths starting at $a$ and $b$, while the second has a bad double-crossing for a different pair of paths starting at $c$ and $d$. Clearly square moves and merges can't connect these two diagrams. It is plausible that the complete set of invariants involves the permutation together with other labels characterizing the pattern of intersections  of the left-right paths. Finding a complete classification will be very important, not least because it would allow us to cast the BCFW construction of all-loop integrand in completely combinatorial terms.

We have seen that the all-loop integrand is naturally presented in a ``$d\!\log$'' form. This form begs to be integrated, indeed most na\"{i}vely of course, these forms integrate to zero! The integrals don't vanish because of  branch cuts in the arguments of the logarithms, on the real contour of integration. This leads to novel ways of performing the loop integrations directly in spacetime, which will be pursued in future work.

Finally, the BCFW construction of scattering amplitudes in the Grassmannian still leaves something to be desired. It is not entirely satisfying to give the scattering amplitude a fundamentally recursive definition. Put another way, we have yet to see locality and unitarity fully emerge from more primitive principles in a completely satisfactory way. We would like to have a direct definition of the amplitude, linked to the Grassmannian, making all the symmetries manifest, and discover their singularities in the form of factorization and forward limits as an emergent property. This is bound to be linked to the ``polytope picture'' studied in \cite{ArkaniHamed:2010gg,Hodges:2009hk}. This line of thought will certainly be taken up again with our vastly improved understanding of the positive Grassmannian in hand. 

\newpage
\section*{Acknowledgements}
N A-H, JB, FC, AG, and JT owe an enormous debt of gratitude to Pierre Deligne, Bob MacPherson, and Mark Goresky for several months of intensive discussions on many of the mathematical topics in this paper---and in particular, for discussions about the stratification of configurations of vectors according to consecutive linear dependencies; and they thank Andrew Hodges and Simon Caron-Huot for many stimulating discussions; and we are grateful to Sebastian Franco, Yu-tin Huang, Tomek Lukowski, Jan Plefka, Matthias Staudacher, Cristian Vergu, and Congkao Wen for comments regarding earlier versions of this manuscript. AG and JB thank the School of Natural Sciences at the Institute of Advanced Study for their generous hospitality. This work was supported in part by US Department of Energy contracts \mbox{DE-FG02-91ER40654} (N A-H) and \mbox{DE-FG02-91ER40654} (JB), the Harvard Society of Fellows and a grant form the Harvard Milton Fund (JB), the NSERC of Canada and MEDT of Ontario (FC), and NSF grants DMS-1100147 (AP), DMS-1059129 (AG), and PHY-0756966 (JT).

~\newpage
\providecommand{\href}[2]{#2}\begingroup\raggedright\endgroup

~\newpage\thispagestyle{empty}
~\newpage\thispagestyle{empty}
~\newpage\thispagestyle{empty}
\thispagestyle{empty}
~\vspace{\fill}\eq{\hspace{-5.0cm}\includegraphics[scale=.85]{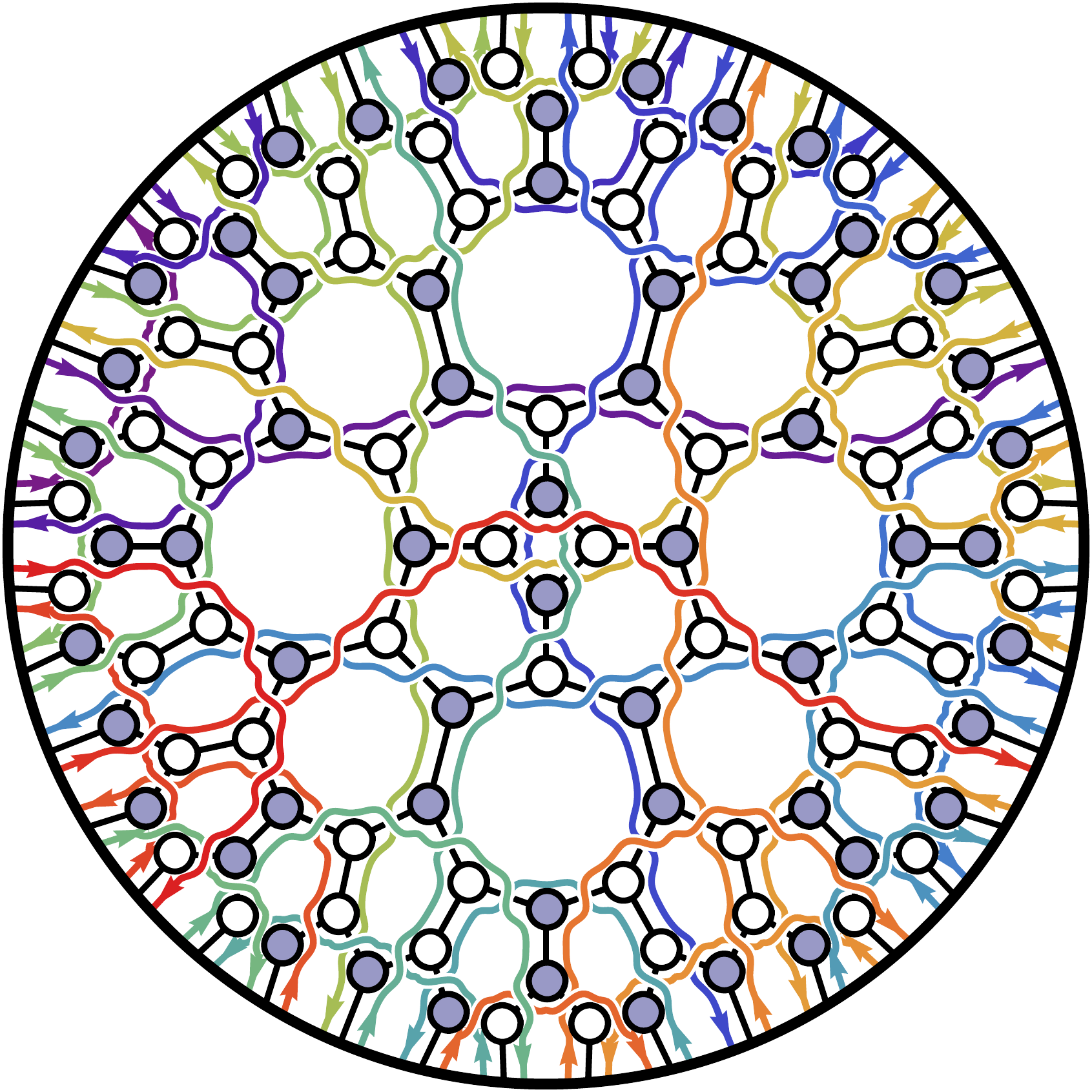}\hspace{-5cm}\nonumber}~\vspace{\fill}

\end{document}